\documentclass[preprint,nofootinbib,floatfix,a4paper,prd,superscriptaddress]{revtex4-1}   	
\usepackage{geometry}                		
\usepackage[colorlinks,citecolor=blue]{hyperref}
\usepackage{amsmath}
\usepackage{mathrsfs}
\usepackage{bbm}
\usepackage{amsfonts}
\usepackage{amssymb}
\usepackage{latexsym}
\usepackage{graphicx}
\usepackage[english]{babel}
\usepackage{multirow}
\usepackage{float}
\usepackage{url}
\usepackage{slashed}
\usepackage{tabularx}
\usepackage{orcidlink}
\usepackage{appendix}
\usepackage{ulem}

\newcommand{\be}{\begin{equation}}
\newcommand{\ee}{\end{equation}}
\newcommand{\ba}{\begin{array}}
\newcommand{\ea}{\end{array}}
\newcommand{\bea}{\begin{eqnarray}}
\newcommand{\eea}{\end{eqnarray}}

\usepackage[utf8]{inputenc}
\usepackage[T1]{fontenc}

\def  \bcen   {\begin{center}}
\def  \ecen   {\end{center}}
\def  \beq    {\begin{equation}}
\def  \eeq    {\end{equation}}

\def\la   {\lambda}

\def\lee { \left( }
\def\rii { \right) }

\def\to {\rightarrow}
\def\lphp {\la^\prime_{H\Phi}}

\newcommand{\TSUa}{\affiliation{\small Tsung-Dao Lee Institute \& School of Physics and Astronomy, Shanghai Jiao Tong University, Shanghai 200240, China }}

\newcommand{\TSUb}{\affiliation{\small Shanghai Key Laboratory for Particle Physics and Cosmology, Key Laboratory for Particle Astrophysics and Cosmology (MOE), Shanghai Jiao Tong University, Shanghai 200240, China }}

\newcommand{\AMH}{\affiliation{\small Amherst Center for Fundamental Interactions, Department of Physics,\\
University of Massachusetts, Amherst,
MA 01003, USA }}

\newcommand{\CAL}{\affiliation{\small Kellogg Radiation Laboratory, California Institute of Technology,\\
Pasadena,
CA 91125, USA}}

\newcommand{\PU}{\affiliation{\small Phenikaa Institute for Advanced Study, Phenikaa University, Yen Nghia, Ha Dong, Hanoi 100000, Vietnam}}

\newcommand{\AS}{\affiliation{\small Institute of Physics, Academia Sinica, Nangang, Taipei 11529, Taiwan}}

\begin{document}

\title{Gravitational Waves and Dark Matter in the Gauged Two-Higgs Doublet Model}

\author{Michael J.~Ramsey-Musolf \orcidlink{0000-0001-8110-2479}} 
\email{mjrm@sjtu.edu.cn, mjrm@physics.umass.edu}  \TSUa \TSUb \AMH \CAL

\author{\\Van Que Tran \orcidlink{0000-0003-4643-4050}}
\email{vqtran@gate.sinica.edu.tw} \TSUa \PU \AS

\author{Tzu-Chiang Yuan \orcidlink{0000-0001-8546-5031} \, }
\email{tcyuan@phys.sinica.edu.tw} \AS


\begin{abstract}
We investigate the possibility of a strong first-order electroweak phase transition during the early universe within the framework of the gauged two-Higgs doublet model (G2HDM) and explore its detectability through stochastic gravitational wave signals. The G2HDM introduces a dark replica of the Standard Model electroweak gauge group, inducing an accidental $Z_2$ symmetry which not only leads to a simple scalar potential at tree-level but also offers a compelling vectorial dark matter candidate.
Using the high temperature expansion in the effective potential that manifests gauge invariance, we find a possible two-step phase transition pattern in the model with a strong first-order transition occurring in the second step at the electroweak scale temperature.
Collider data from the LHC plays a crucial role in constraining the parameter space conducive to this two-step transition. Furthermore, satisfying the nucleation condition necessitates the masses of scalar bosons in the hidden sector to align with the electroweak scale, potentially probed by future collider detectors.
The stochastic gravitational wave energy spectrum associated with the phase transition is computed. The results indicate that forthcoming detectors such as BBO, LISA, DECIGO, TianQin and Taiji could potentially detect the gravitational wave signals generated by the first-order phase transition.
Additionally, we find that the parameter space probed by gravitational waves can also be searched for in future dark matter direct detection experiments, in particular those designed for dark matter masses in the sub-GeV range using the superfluid Helium target detectors.
\end{abstract}

\maketitle

\section{Introduction}
\label{Intro}
Unraveling the thermal history of electroweak symmetry breaking 
is considered a crucial task in both particle physics and cosmology. In the Standard Model (SM), EW symmetry breaking proceeds through a smooth crossover as indicated 
by lattice simulations \cite{Kajantie:1996mn,Rummukainen:1998as,Laine:1998jb,Aoki:1999fi}. However, a first-order electroweak phase transition (FOEWPT) can occur in many BSM theories, such as singlet scalar extensions~\cite{McDonald:1993ey,Profumo:2007wc,Cline:2009sn,Espinosa:2011ax,Cline:2012hg,Alanne:2014bra,Curtin:2014jma,Profumo:2014opa,Huang:2015bta,Kotwal:2016tex,Vaskonen:2016yiu,Ghorbani:2017jls,Cline:2017qpe,Beniwal:2017eik,Kurup:2017dzf,Chiang:2017nmu,Carena:2018cjh,Carena:2018vpt,Grzadkowski:2018nbc,Huang:2018aja,Alves:2018jsw,Ghorbani:2018yfr,Cheng:2018ajh,Alanne:2019bsm,Gould:2019qek,Carena:2019une,Li:2019tfd,Bell:2019mbn}, 
triplet scalar extensions~\cite{Inoue:2015pza,Niemi:2018asa,Chala:2018opy,Zhou:2018zli},
two-Higgs doublet models~\cite{Bochkarev:1990fx, McLerran:1990zh, Bochkarev:1990gb, Turok:1990zg, Cohen:1991iu, Turok:1991uc, Nelson:1991ab, Funakubo:1993jg, Davies:1994id,Cline:1995dg,Funakubo:1995kw, Funakubo:1996iw, Cline:1996mga, Fromme:2006cm, Cline:2011mm, Dorsch:2013wja, Cline:2013bln, Dorsch:2014qja, Fuyuto:2015jha, Chao:2015uoa, Fuyuto:2015ida, Chiang:2016vgf, Haarr:2016qzq, Dorsch:2016nrg, Basler:2016obg, Fuyuto:2017ewj, Dorsch:2017nza, Cherchiglia:2017gko, Basler:2017uxn, Andersen:2017ika, Bernon:2017jgv, Huang:2017rzf, Gorda:2018hvi, Basler:2018cwe, Modak:2018csw, Wang:2018hnw,Wang:2019pet, Kainulainen:2019kyp, Paul:2020wbz, Su:2020pjw}, 
supersymmetric models~\cite{Apreda:2001us,Huber:2015znp,Huber:2007vva,Demidov:2017lzf}, 
left-right symmetric models~\cite{Brdar:2019fur,Li:2020eun}, 
Pati-Salam model~\cite{Huang:2020bbe}, 
Georgi-Machacek model~\cite{Chiang:2014hia,Zhou:2018zli}, 
Zee-Babu model~\cite{Phong:2015vlk, Phong:2021lea},
composite Higgs models~\cite{Espinosa:2011eu,Chala:2016ykx,Chala:2018opy,Bruggisser:2018mus,Bruggisser:2018mrt,Bian:2019kmg,DeCurtis:2019rxl,Xie:2020bkl}, 
neutrino mass models~\cite{DiBari:2021dri,Zhou:2022mlz} 
and hidden sector models involving scalars \cite{Schwaller:2015tja,Baldes:2018emh,Breitbach:2018ddu,Croon:2018erz,Baldes:2017rcu,Croon:2019rqu,Hall:2019rld,Hall:2019ank,Chao:2020adk,Dent:2022bcd},
{\it etc}.
In some cases, the FOEWPT can occur in multiple steps~\cite{Weinberg:1974hy,Land:1992sm,Patel:2012pi,Patel:2013zla,Blinov:2015sna, Niemi:2018asa,Croon:2018new,Morais:2018uou,Morais:2019fnm,Angelescu:2018dkk,Friedrich:2022cak}. 

Apart from being of interest in its own right, the occurrence of FOEWPT fulfills one of the conditions established by Sakharov~\cite{Sakharov:1967dj} for the realization of the electroweak baryogenesis mechanism~\cite{Trodden:1998ym,Cline:2006ts,Morrissey:2012db, White:2016nbo,Garbrecht:2018mrp}
which accounts for the observed cosmological matter-antimatter asymmetry. A FOEWPT also admits a rich array of possible experimentally observable signatures. In particular, the presence of BSM physics related to a FOEWPT can have an impact on the properties of the Higgs boson and predicts the existence of new scalar particles with masses at or below the TeV scale, which may be probed by future collider experiments (see {\it e.g.}~\cite{Ramsey-Musolf:2019lsf, Carena:2022yvx, Wang:2022dkz, Zhang:2023jvh, Wang:2023zys} and references therein).
Apart from undergoing a strong FOEWPT, these new physics models often provide  candidates for dark matter (DM), see {\it e.g.}~\cite{Chowdhury:2011ga, Katz:2015uja, Fabian:2020hny, Bell:2020gug, Chiang:2020yym, Zhang:2023mnu}. 
Furthermore, 
a FOEWPT can result in the formation of bubble nucleation, which can expand and collide, producing stochastic gravitational waves (GWs) (see \cite{Weir:2017wfa} for a brief review). 
This GW signal is potentially within the reach of upcoming space-based laser interferometer GW detectors, such as LISA~\cite{Audley:2017drz, Cornish:2018dyw}, BBO~\cite{Crowder:2005nr}, TianQin~\cite{Luo:2015ght,Hu:2017yoc}, Taiji~\cite{Hu:2017mde,Guo:2018npi} and DECIGO~\cite{Kawamura:2006up,Kawamura:2011zz,Musha:2017usi}.

 \begin{figure}[tb]
    \centering
    \includegraphics[width=1.0\textwidth]{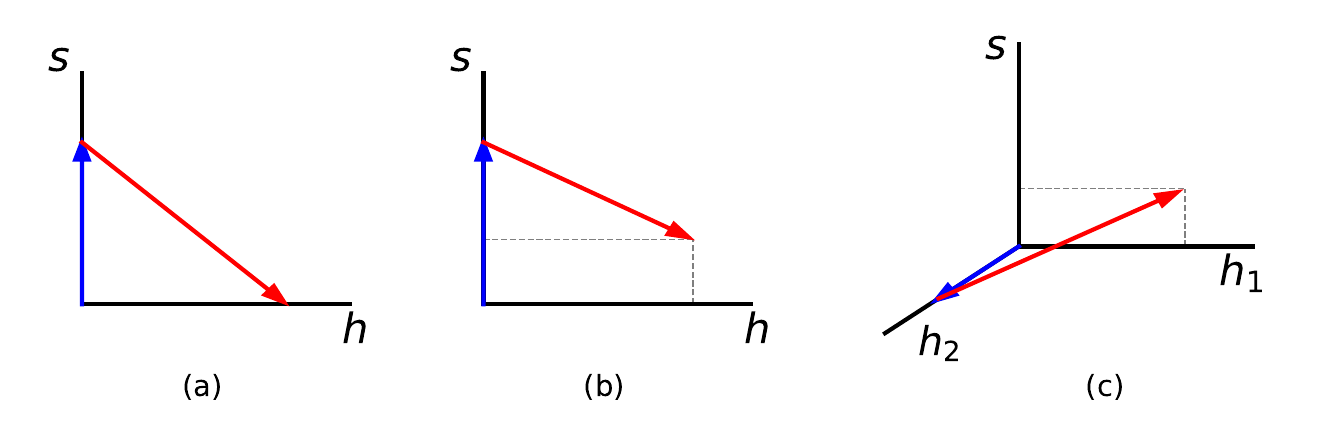}
    \caption{ \label{fig:twostepsketch} Patterns of a two-step transition at finite temperature. The left and center panels are for two background fields scenarios while the right panel is for three background fields scenarios.
    Here, $h \equiv h_1$ indicates the SM Higgs background field while $s$ and $h_2$ represent new scalar boson background fields. The blue and red arrows represent the first and second step transitions, respectively. The end point of the red arrow is the electroweak vacuum value. }
\end{figure}

One of the most straightforward extensions to SM involves the presence of a real singlet scalar field alongside the SM Higgs doublet field, accompanied by the imposition of a discrete $Z_2$ symmetry. 
In such the framework, it can admit a strong FOEWPT in the two-step phase transition $(0,0) \to (0,s) \to (h,0)$ as depicted in the left panel of Fig.~\ref{fig:twostepsketch} --- where $s$ ($h$) represents classical background field of the singlet (SM doublet)~\cite{Ghorbani:2020xqv}
\footnote{Derived from an analysis of the high-temperature expansion of the thermal effective potential.}. 
In the case of spontaneous $Z_2$ breaking~ \cite{Carena:2019une}, the inclusion of the cubic term ($T h^3$) in the effective potential can allow for the occurrence of a FOEWPT through $(0,0) \to (0,s) \to (h,s)$ as depicted in the center panel of Fig.~\ref{fig:twostepsketch}. However, it is worth noting that the presence of this cubic term is subject to the issue of gauge dependence \cite{Patel:2011th}. Nevertheless, to facilitate these transitions, the mass of the singlet scalar must be confined to be light. 
In particular, necessitating the occurrence of a two-step EWPT like the left panel of Fig.~\ref{fig:twostepsketch} requires 
\be
\label{eq:twostepcond}
T_1 > T_2 \sim T_{\rm EW} \; ,
\ee
where $T_1 (T_2)$ is the critical temperature in the first (second) step and $T_{\rm EW}$ is the temperature at electroweak (EW) scale. 
If the $Z_2$ symmetry is preserved at zero temperature (left panel of Fig.~\ref{fig:twostepsketch}),
the scalar singlet acquires its mass solely from the SM VEV. 
The requirement in (\ref{eq:twostepcond}) sets the scalar singlet mass to be below 700 GeV~\cite{Ramsey-Musolf:2019lsf}. 
On the other hand, for the spontaneous $Z_2$ breaking case (center panel of Fig.~\ref{fig:twostepsketch}), the scalar singlet mass is confined to below 50 GeV and the associated GW signals are typically too low to be probed by future GW detectors~\cite{Carena:2019une}.

Furthermore, for the $Z_2$ symmetry preserving model, where the singlet scalar gets zero vacuum expectation value (VEV) at zero temperature, the singlet scalar can remain stable and serve as a DM candidate. 
This straightforward extension of SM incorporating a $Z_2$ symmetric potential allows us to perform gauge-invariant and scale-invariant treatments of EWPT computation.
However, except for small DM mass in the range of $65 \sim 200$ GeV that can be probed by future colliders, this scenario is ruled out by direct detection experiments~\cite{Ghorbani:2020xqv}. 

We thus ask following questions:
\begin{itemize}
\item In what context the upper limit on the new scalar mass due to the requirement of the two-step FOEWPT can be relaxed?
\item Can such the model yield a DM candidate that satisfies the correct DM relic abundance and evades current DM direct detection constraints?
\item Can the GW signals generated from FOEWPT detectable at forthcoming GW detectors and how does it interplay with DM direct detection signals? 
\end{itemize}

In this analysis, we address these questions within the framework of a simplified G2HDM~\cite{Ramos:2021txu, Ramos:2021omo} which is based on the original version proposed in~\cite{Huang:2015wts}. In addition to the SM $SU(2)_L$ doublet $H_1$ and singlet scalars $\Phi_H$, this model incorporates an extra inert doublet $H_2$. While inert at zero temperature, this inert doublet can develop a background field ({\it i.e.}, VEV) at high temperature (high-T), resulting in rich phase transition patterns. Using the high-T expansion approximation in the effective potential that manifests gauge invariance, we find a particularly novel and interesting two-step phase transition, with the first step leading to the inert doublet vacuum and the subsequent step returning to the true EW vacuum with non-zero SM doublet and singlet scalar VEVs as depicted schematically in the right panel of Fig.~\ref{fig:twostepsketch}. The first-order phase transition may take place in the second step.
In the second step the singlet can develop a new VEV $v_\Phi$ along with the SM VEV $v$ from $H_1$, thereby providing additional contributions to the masses of new scalars.
This can lead to a relaxation of the upper bounds on masses of extra scalars in the model while archiving the two-step FOEWPT. 

Moreover, the model naturally accommodates a hidden $h$-parity, if not spontaneously broken at zero temperature, ensuring that the lightest $h$-parity odd particle serves as the DM candidate. The DM candidate in the model can be either a complex scalar~\cite{Chen:2019pnt}, a hidden heavy neutrino, or a non-abelian vector gauge boson --- all of which are electrically neutral and possess odd $h$-parity. In this work, we will focus on the hidden gauge bosons $W^{\prime \, (p,m)}$ as the DM candidate~\cite{Ramos:2021txu,Ramos:2021omo}.

We demonstrate that the stochastic GW signals from the two-step FOEWPT in the simplified G2HDM can be probed 
in future GW experiments. The  predictions of GW signals in this study are derived from the high-T expansion approximation in the one-loop effective potential, which preserves gauge invariance. To attain more accurate results and ensure a realistic perturbative treatment~\cite{Patel:2011th}, it is imperative to incorporate a two-loop finite temperature effective potential which has yet been computed for G2HDM.
Our investigation reveals that the parameter space accessible through GWs can also be explored in upcoming DM direct detection experiments. Specifically, we highlight the relevance of experiments tailored for detecting DM with masses in the sub-GeV range, utilizing superfluid Helium target detectors.

The layout of this paper is as follows. 
In Sec.~\ref{sec:G2HDM}, we will briefly review the simplified version of G2HDM and its phenomenological constraints.
In Sec.~\ref{sec:EWPT}, after a succinct review of the current status of theoretical EWPT computation, we present our analysis in G2HDM 
using the high-T expansion of the finite temperature effective potential in order to maintain gauge invariance. We focus on the two-step transition pattern and identify parameter space where the desired pattern can occur. 
In Sec.~\ref{sec:GW}, we present the stochastic GW prediction signals for the FOEWPT. 
In Sec.~\ref{sec:DMDD}, we discuss the DM direct detection probes. We conclude in Sec.~\ref{sec:cons}. 
Six appendices are compiled for (\ref{app:FDMasses}) field dependent masses, (\ref{app:ThermalMasses}) thermal masses, (\ref{app:ThermalIntegrals}) thermal integrals, (\ref{app:GaugeInvariantMethod}) gauge invariant method beyond leading order computation and scale dependence of the effective potential at finite temperature, (\ref{app:CriticalTemp}) critical temperatures, and (\ref{app:RGE}) renormalization group equations.

\section{The Simplified G2HDM}
\label{sec:G2HDM}
In this section, we will briefly review the simplified version of G2HDM \cite{Ramos:2021txu,Ramos:2021omo}. Unlike the original G2HDM presented in~\cite{Huang:2015wts}, the simplified version has omitted the $SU(2)_H$ triplet scalar $\Delta_H$, thereby simplifying the scalar potential. The scalar sector in the simplified G2HDM consists of the SM Higgs doublet $H_1$ and a second inert Higgs doublet $H_2$ of $SU(2)_L$, as well as a 
hidden doublet $\Phi_H$ of $SU(2)_H$. 
Regarding the two doublets $H_1$ and $H_2$, the model is different from the popular inert Higgs doublet model (IHDM) in that they are combined into a doublet of the hidden $SU(2)_H$.
Their extended electroweak quantum numbers and $h$-parity are summarized in Table~{\ref{tab:quantumnosscalar}} for reference.
\begin{table}[htbp!]
\begin{tabular}{|c|c|c|c|c|c|c|}
\hline
Scalar & $SU(2)_L$ & $SU(2)_H$ & $U(1)_Y$ & $U(1)_X$ & $h$-parity\\
\hline\hline
$H= \begin{pmatrix} H_1 \\ H_2 \end{pmatrix} $ 
& $ \begin{matrix} 2 \\ 2 \end{matrix} $
& 2 & $\frac{1}{2}$ & $\frac{1}{2}$ & $ \begin{matrix} + \\ - \end{matrix} $ \\
$\Phi_H = \begin{pmatrix} G^p_H \\ \Phi_H^0 \end{pmatrix} $ & 1 & 2 & 0 & $\frac{1}{2}$ & $ \begin{matrix} - \\ + \end{matrix} $\\
\hline
\end{tabular}
\caption{Higgs sector in the simplified G2HDM and their quantum number assignments.
Note that $H_{1,2}$ individually transform as SU(2)$_L$ doublets and together as a doublet under SU(2)$_H$, whereas the components of the SU(2)$_H$ doublet transform individually as SU(2)$_L$ singlets. }
\label{tab:quantumnosscalar}
\end{table}

We note that the model also consists of heavy hidden fermions for the cancellations of  gauge and gravitational anomalies. The hidden doublet $\Phi_H$ of $SU(2)_H$ can provide realistic Yukawa couplings and mass spectra for the extra fermions and vector bosons. 
The loop-induced flavor changing neutral current (FCNC) processes within the model have also been investigated~\cite{Tran:2022cwh, Liu:2024nkl}.
Additionally, the G2HDM is capable of addressing the recent high precision measurement of the $W$ boson mass at the CDF II detector~\cite{Tran:2022yrh}. Many other phenomenological aspects of the model have been explored as well in ~\cite{Huang:2015rkj,Arhrib:2018sbz,Huang:2019obt,Huang:2017bto,Chen:2018wjl,Dirgantara:2020lqy}. 
A pure dark gauge-Higgs sector of $SU(2) \times U(1)$ with the dark gauge bosons $W^{\prime \, (p,m)}$ implemented as a self-interacting DM candidate was recently pursued in~\cite{Tran:2023lzv}.

\subsection{Higgs Potential and Spontaneous Symmetry Breaking}
\label{HiggsPotential}

The most general Higgs potential which is invariant under the extended electroweak gauge group $SU(2)_L\times U(1)_Y \times SU(2)_H \times  U(1)_X$  
is given by
\begin{align}\label{eq:V}
V = {}& - \mu^2_H   \left(H^{\alpha i}  H_{\alpha i} \right)
+  \lambda_H \left(H^{\alpha i}  H_{\alpha i} \right)^2  
+ \frac{1}{2} \lambda'_H \epsilon_{\alpha \beta} \epsilon^{\gamma \delta}
\left(H^{ \alpha i}  H_{\gamma  i} \right)  \left(H^{ \beta j}  H_{\delta j} \right)  \nonumber \\
{}&- \mu^2_{\Phi}   \Phi_H^\dag \Phi_H  + \la_\Phi \lee \Phi_H^\dag \Phi_H  \rii^2 
+\lambda_{H\Phi} \lee H^\dag H  \rii  \lee \Phi_H^\dag \Phi_H \rii  
 + \lambda^\prime_{H\Phi} \lee H^\dag \Phi_H  \rii  \lee \Phi_H^\dag H \rii, 
\end{align}
where  ($\alpha$, $\beta$, $\gamma$, $\delta$) and ($i$, $j$) refer to the $SU(2)_H$ and $SU(2)_L$ indices respectively, 
all of which run from 1 to 2, and $H^{\alpha i} = H^*_{\alpha i}$.

To study general spontaneous symmetry breaking (SSB) at finite temperature, we parameterize the fields according to standard practice
\begin{eqnarray}
\label{eq:scalarfields}
H_1 = 
\begin{pmatrix}
h_1^+ \\ H_1^0 = \frac{h_{1c} + h_{\rm SM}}{\sqrt 2} + i \frac{G_1^0}{\sqrt 2}
\end{pmatrix}
\; , \; \; \; 
H_2 = 
\begin{pmatrix}
h_2^+  \\ H_2^0 = \frac{h_{2c} + h_2^0}{\sqrt 2} + i \frac{G_2^0}{\sqrt 2}
\end{pmatrix}
\; , \,
\end{eqnarray}
\begin{eqnarray}
\label{eq:scalarfieldPhiH}
\Phi_H = 
\begin{pmatrix}
G_H^p = \frac{G_H^1 + i G_H^2}{\sqrt 2} \\ \Phi_H^0 = \frac{\phi_{Hc}+ \phi_H}{\sqrt 2} + i \frac{G_H^0}{\sqrt 2}
\end{pmatrix}
 \; ,
\end{eqnarray}
where $h_{ic}$ and $\phi_{Hc}$ are the only nonvanishing temperature dependent classical background field components 
in $H_i$ and $\Phi_{H}$ fields respectively. The temperature dependence in the background fields is implicit.

In terms of the background fields, the effective potential at tree level is
\begin{eqnarray}
\label{Vefftree}
V_{0} (h_{1c},h_{2c},\phi_{Hc}) & = & \frac{1}{4} \Bigl[
-2 \mu_H^2 \left( h_{1c}^2 + h_{2c}^2 \right) + \lambda_H \left( h_{1c}^2 + h_{2c}^2 \right)^2 
-2 \mu_{\Phi_H}^2 \phi_{Hc}^2 + \lambda_{\Phi}  \phi_{Hc}^4 \Bigr. \nonumber \\
 && \Bigl. \;\;\;\; + \lambda_{H\Phi} \left( h_{1c}^2 + h_{2c}^2 \right)  \phi_{Hc}^2 + \lambda_{H\Phi}^\prime h_{2c}^2 \phi_{Hc}^2
\Bigr] \; .
\end{eqnarray}
Note that in the tree level effective potential, 
the $h_{ic} \phi_{Hc}$ term and all the terms involving the coupling $\lambda_H^\prime$ are absent.

\subsection{Mass spectrum and mixing at tree level}
\label{TLMassesMixings}

At zero temperature, the classical background fields are the constant vacuum expectation values (VEVs) {\it i.e.}
$h_{ic}(0) = v_i$ and $\phi_{Hc}(0) = v_\Phi$. Since $H_2$ is odd under $h$-parity, it plays the role as the inert Higgs doublet at zero temperature, 
so $v_2 = 0$.  With this set up, a $h$-parity  protects the stability of DM candidate in the model~\cite{Ramos:2021txu,Ramos:2021omo}. 
The scalar potential at tree level given in Eq.~(\ref{Vefftree}) becomes
\begin{eqnarray}
\label{Vefftree0}
V_{0} (v_1,0,v_{\Phi}) & = & \frac{1}{4} \Bigl[
-2 \mu_H^2  v_{1}^2 + \lambda_H  v_{1}^4 
-2 \mu_{\Phi_H}^2 v_{\Phi}^2 + \lambda_{\Phi} v_{\Phi}^4  + \lambda_{H\Phi} v_{1}^2 v_{\Phi}^2 
\Bigr] \; .
\end{eqnarray}
 
From the $h$-parity assignment given in Table~\ref{tab:quantumnosscalar}, we note that while there is no mixing between $H_1$ and $H_2$, $H_1$ ($H_2$) can mix with the lower (upper) component of $\Phi_H$.
Specifically, the neutral components $h_{\rm SM}$ and $\phi_H$ in $H_1$ and $\Phi_H$, respectively, are both even under $h$-parity and can combine to form two observable Higgs fields, denoted as $h_1$ and $h_2$. This mixing can be given as 
\be
\left(
\begin{matrix}
h_{\rm SM} \\
\phi_H
\end{matrix}
\right)
=
\left( 
\begin{matrix}
 \cos \theta_1  &  \sin \theta_1 \\
- \sin \theta_1  &  \cos \theta_1 
\end{matrix}
\right) \cdot 
\left(
\begin{matrix}
h_1 \\
h_2
\end{matrix}
\right)
\; ,
\ee
where the mixing angle $\theta_1$ is given by
\be
\tan 2 \theta_1 = \frac{\lambda_{H\Phi} v v_\Phi }{ \lambda_\Phi v^2_\Phi - \lambda_H v^2 } \; .
\ee
The masses of $h_1$ and $h_2$ are given by
\bea
\label{h1h2masses}
m_{h_1,h_2}^2 &=& \lambda_H v^2 + \lambda_\Phi v_\Phi^2 \nonumber \\
& \mp &\sqrt{\lambda_H^2 v^4 + \lambda_\Phi^2 v_\Phi^4 + \left( \lambda^2_{H\Phi}  - 2 \lambda_H \lambda_\Phi \right) v^2 v_\Phi^2 }\;.  \hspace{0.6cm}
\eea
The observed Higgs boson (which we denote as $h$) at the LHC is either $h_1$ or $h_2$, depending on their masses (\ref{h1h2masses}) determined by the underlying portal parameters in the scalar potential. Currently, the most precise determination of the Higgs boson mass is $m_h = 125.38 \pm 0.14$ GeV, as reported in Ref.~\cite{CMS:2020xrn}.

Furthermore, the $h$-parity odd fields $H_2^{0}$ and $G^m_H$ can mix to produce two fields: a physical dark Higgs boson denoted as $D$, and an unphysical Goldstone boson $\tilde{G}$ that gets absorbed by the $W^{\prime \, m}$ boson. The mixing is given by 
\be
\label{H20field}
\begin{pmatrix}
G^m_H \\
H_2^{0} 
\end{pmatrix}
=
\begin{pmatrix}
 \cos \theta_2 & \sin \theta_2  \\
- \sin \theta_2  & \cos \theta_2 
\end{pmatrix} 
\cdot
\begin{pmatrix}
 \tilde G \\
 D 
 \end{pmatrix}
 \;,
\ee
where the mixing angle $\theta_2$ satisfies
\be
\label{theta2}
\tan 2 \theta_2 = \frac{2 v v_\Phi}{v^2_\Phi - v^2} \; ,
\ee
and the mass of $D$ is 
\be
\label{mD}
m_D^2 = \frac{1}{2} \lambda^\prime_{H\Phi} \left( v^2 + v_\Phi^2 \right) \; .
\ee
$\tilde G$ $(\tilde G^*)$ is the Goldstone field associated with the $SU(2)_H$ gauge boson $W^{\prime \, m}$ ($W^{\prime \, p}$). Its mass depends on the gauge fixings~\cite{Ramos:2021txu}.
In Feynman-'t Hooft gauge it has the same mass as the $W^{\prime \, m}$ ($W^{\prime \, p}$). In Landau and unitary gauge, its mass is zero and infinite respectively. 

The charged Higgs $H^\pm$, same as $h_2^\pm$ at zero temperature, is also $h$-parity odd and has a mass
\be
\label{mcH}
m^2_{H^\pm} = \frac{1}{2} \left( \lambda^\prime_{H\Phi} v^2_\Phi - \lambda^\prime_H v^2 \right) \; .
\ee
We note that the quartic coupling parameters in the scalar potential that determine the masses of $h$-parity odd states $D$ and $H^\pm$ are different from those of $h$-parity even states $h_1$ and $h_2$.

The $SU(2)_H$ gauge bosons $W^{\prime \, (p,m)}$ have a mass  
\be 
\label{mwprime}
m_{W^\prime}  =  \frac{1}{2} g_H \sqrt{ v^2 + v_\Phi^2 } \; ,
\ee
where $g_H$ is the gauge coupling of $SU(2)_H$. $W^{\prime \, (p,m)}$ are both electrically neutral and odd under $h$-parity. Therefore, they could potentially serve as dark matter candidates, provided that their mass is the lightest among all odd $h$-parity particles.

Finally the SM $Z$ boson mixes with the $W^{\prime}_3$ gauge field associated with the third generator of $SU(2)_H$ and the $U(1)_X$ gauge field $X$ (see Ref.~\cite{Ramos:2021txu,Ramos:2021omo} for the mass matrix). This leads to three physical fields $Z_i$ for $ i=1,2,3$.
We will identify $Z_1$ with the neutral gauge boson resonance $Z$ observed at LEP with a mass of 91.1876 GeV~\cite{Zyla:2020zbs}, while the other two states can be identified as the dark photon ($\gamma^\prime$) and the dark $Z$ boson ($Z^\prime$) with the ordering $m_{\gamma^\prime} < m_{Z^\prime}$.

The fundamental parameters in the scalar potential can be expressed in terms of particle masses by performing an inversion, as demonstrated in Refs.~\cite{Ramos:2021txu,Ramos:2021omo,Tran:2022cwh}: 
\bea
v_\Phi & = & 
\begin{cases}
\begin{matrix}
v  \cot \theta_2 \, , &&  {\rm for} \; \theta_2 > 0 \; , \\
- v \tan \theta_2 \, , &&  {\rm for} \; \theta_2 \leq 0 \; , 
\end{matrix}
\end{cases} 
\\
\lambda_H & = & \frac{1}{2 v^2} \left( m^2_{h_1} \cos^2 \theta_1 + m^2_{h_2} \sin^2 \theta_1 \right) \; , \\
\lambda_\Phi & = & \frac{1}{2 v_\Phi^2} \left( m^2_{h_1} \sin^2 \theta_1 + m^2_{h_2} \cos^2 \theta_1 \right) \; , \\
\lambda_{H\Phi} & = & \frac{1}{2 v v_\Phi} \left( m^2_{h_2} - m^2_{h_1} \right) \sin \left( 2 \theta_1 \right) \; , \\
\lambda^\prime_{H\Phi} & = & \frac{ 2  m_D^2 }{v^2 + v_\Phi^2} \; , \\
\lambda^\prime_{H} & = & \frac{2}{v^2} \left( \frac{m_D^2 v_\Phi^2 }{ v^2 + v_\Phi^2 } - m^2_{H^\pm}\right) \; .
\eea
From (\ref{mwprime}), we also obtain
\beq
\label{gH}
g_H  =   \frac{ 2 m_{W'} } { \sqrt{v^2 + v^2_{\Phi} }}  \;. 
\eeq
Thus one can use $m_{h_2}$, $m_{W^\prime}$, $m_D$, $m_{H^\pm}$, $\theta_1$ and $\theta_2$ as input in our numerical scan. The additional free parameters in the model are the heavy hidden fermion mass $m_{f^{H}}$, the $U(1)_X$ gauge coupling $g_X$ and the Stueckelberg mass $m_X$. 

\subsection{Constraints}
The constraints on the model have been studied in~\cite{Ramos:2021txu,Ramos:2021omo,Tran:2022cwh}, taking into account 
the theoretical constraints on the scalar potential, as well as constraints from Higgs measurements at the LHC, electroweak precision data, dark photon and dark matter searches.
Here we summarize these constraints as follows.

{\underline {\it Theoretical constraints: }}

(a) Vacuum Stability: 
To ensure that the scalar potential has a minimum value, 
we adopt the copositivity conditions suggested by~\cite{Arhrib:2018sbz}, which provide the following set of constraints 
on the scalar potential parameters 
\be
\widetilde \lambda_H (\eta) \geq 0, \;\; \lambda_\Phi \geq 0 \;\; {\rm and} \;\;
\widetilde \lambda_{H\Phi}(\xi) + 2 \sqrt{\widetilde \lambda_H (\eta) \lambda_\Phi}  \geq  0,
\ee
where $\widetilde \lambda_H (\eta) \equiv \lambda_H + \eta \lambda^\prime_H$ and
${\widetilde  \lambda_{H\Phi}}(\xi) \equiv \lambda_{H\Phi} + \xi \lphp $ with $0 \leq \xi \leq 1$ and $-1 \leq \eta \leq 0$.

(b) Perturbative Unitarity: 
The condition for perturbative unitarity, as presented in~\cite{Arhrib:2018sbz, Ramos:2021txu, Ramos:2021omo}, provides the following constraints 
\begin{align}
&\vert \lambda_H \vert, \vert  \lambda_\Phi \vert  \leq  4 \pi \;,
 \vert \lambda_{H\Phi} \vert  \leq  8 \pi \;,
 \vert \lambda^\prime_{H\Phi} \vert \;, \vert \lambda_H^\prime \vert  \leq  8 \sqrt 2 \pi \; , \\
& \vert 2 \lambda_H \pm \lambda^\prime_H \vert \leq 8 \pi \;,
\vert \lambda_{H\Phi} + \lambda_{H\Phi}^\prime \vert \leq  8 \pi \; ,\\
& \left| ( \lambda_H + \lambda_H^\prime/2 +
\lambda_\Phi) \pm \sqrt{2\lambda^{\prime 2}_{H\Phi} +
( \lambda_H + \lambda_H^\prime/2 - \lambda_\Phi)^2} \right|  \leq  8 \pi \;,  \\
& \left| (5 \lambda_H - \lambda_H^\prime/2 + 3 \lambda_\Phi) 
\pm \sqrt{(5 \lambda_H - \lambda_H^\prime/2 - 3\lambda_\Phi)^2 
+ 2 (2 \lambda_{H\Phi} + {\lambda}'_{H\Phi} )^2} \right|  \leq  8 \pi \; .
\end{align}

We note that the perturbativity bound can be crucial in some BSM Higgs scenarios, such as the real triplet scalar extension model \cite{Bell:2020gug}. However, to impose this constraint, two-loop renormalization group equations are required. We defer to future work an evaluation of the two-loop $\beta$-functions and corresponding perturbativity constraints, keeping in mind that the latter may lead to further reductions in the viable parameter space. 

 \begin{figure}[tb]
         \centering
	\includegraphics[width=0.7\textwidth]{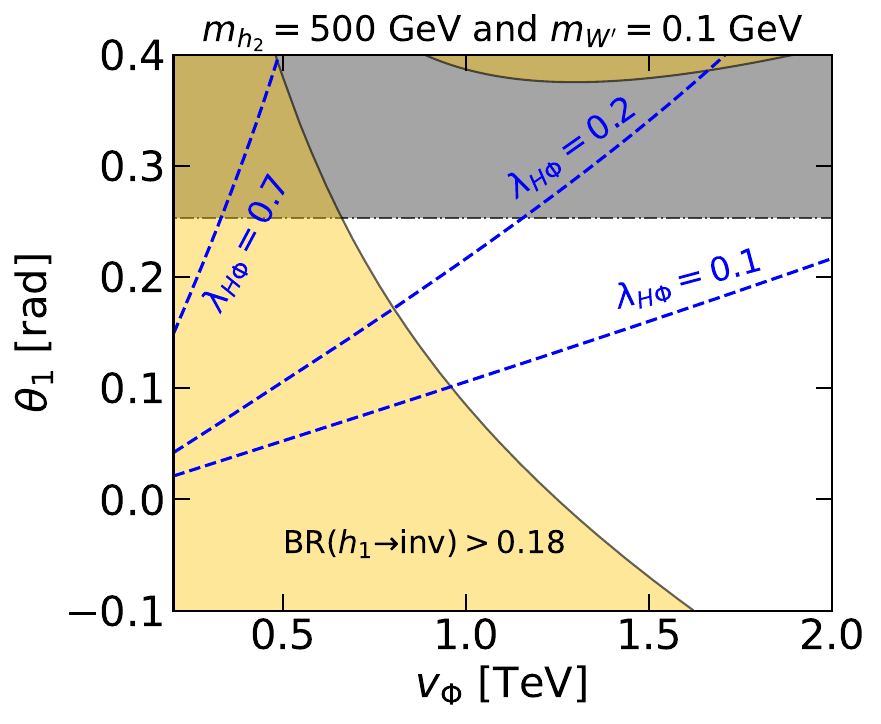}
	\caption{ \label{fig:collider-constraint} The collider constraints projected onto the plane of ($v_\Phi, \theta_1$). The gray shaded region represents the exclusion region from the combined Higgs signal strength data at the LHC~\cite{CMS:2020gsy,ATLAS:2021vrm}, while the yellow shaded region indicates the exclusion from the branching ratio of Higgs invisible decay measurements. The blue dashed lines represent the Higgs portal parameter $\lambda_{H\Phi}$ values. The relevant parameters $m_{h_2} = 500$ GeV and $m_{W'} = 0.1$ GeV are fixed.}
\end{figure}

{\underline{\it Higgs data constraints }}:
The mixing between $h$-parity even scalar bosons, $h_1$ (Higgs boson) and $h_2$, results in modifications on the couplings between the Higgs boson and SM particles as compared with those in the SM.    
Moreover, in addition to contributions from the SM charged particles, the process $h \to \gamma \gamma$ receives contributions from hidden charged fermions $f^{\rm H}$ and charged Higgs $H^\pm$ in the G2HDM. The analytical formulation of this 1-loop process is presented in~\cite{Tran:2022cwh}.
Utilizing the Higgs signal strength data of the decays into $\gamma\gamma, \tau^+ \tau^-, W^+W^-$ and $ZZ$ from CMS measurement \cite{CMS:2020gsy}, we derive a constraint on the mixing angle, finding that $|\sin \theta_1| \lesssim 0.35$. When combining Higgs signal strength data from both CMS~\cite{CMS:2020gsy} and ATLAS~\cite{ATLAS:2021vrm}, a more stringent limit of $|\sin \theta_1| \lesssim 0.25$ is established, which is depicted as a gray shaded area in Fig.~\ref{fig:collider-constraint}.

We consider the gauge boson $W'$ with a sub-GeV mass range as the DM candidate in the model, as studied in~\cite{Ramos:2021txu,Ramos:2021omo}. Therefore, $h_1$ (identified as $h$ here) can decay invisibly into a pair of $W^{\prime (p,m)}$. The invisible branching ratio can be given by 
\be
{\rm BR} ({h_1 \to {\rm inv}}) =\frac{ \Gamma({h_1 \to W^{\prime p} W^{\prime m}})}{\Gamma_{h_1}} \; ,
\ee 
where 
$\Gamma_{h_1} \simeq \cos^2{\theta_1} \Gamma^{\rm SM}_{h} + \Gamma({h_1 \to W^{\prime p} W^{\prime m}})$ is the total decay width of the 125.38 GeV Higgs boson, $\Gamma^{\rm SM}_{h} = 4.1$ MeV is the SM decay width 
and $\Gamma({h_1 \to W^{\prime p} W^{\prime m}})$ is its invisible decay width. The latter is given by
\be
\Gamma ({h_1 \to W^{\prime p} W^{\prime m}}) = \frac{g_H^4 \left( v \cos \theta_1 - v_\Phi \sin \theta_1 \right)^2 }{256 \pi} 
\frac{m_{h_1}^3}{m_{W'}^4} \left( 1- \tau_{W'} +\frac{3}{4}  \tau_{W'}^2\right) 
\sqrt{1- \tau_{W'} } \; ,
\label{eq:invdecaywidth}
\ee
with $\tau_{W'} = 4 m_{W'}^2 / m_{h_1}^2$. 
Recently, CMS set a limit of ${\rm BR}({h_1 \to{ \rm inv}}) < 0.18$ at $95 \%$ C.L., assuming that the Higgs boson production cross-section via vector boson fusion is comparable to the SM prediction~\cite{CMS:2022qva}. The exclusion region from the Higgs invisible decays is shown on the ($v_\Phi, \theta_1$) plane as the yellow shaded region in Fig.~\ref{fig:collider-constraint}.

 Combining the constraints from the Higgs signal strengths and Higgs invisible width yields a lower bound on $v_\Phi$ and an upper bound on the Higgs portal parameter $\lambda_{H\Phi}$, as illustrated in Fig.~\ref{fig:collider-constraint}. Specifically, for fixed values of $m_{h_2} = 500$ GeV and $m_{W' }= 0.1$ GeV, the collider data require $v_\Phi \gtrsim 600$ GeV and $\lambda_{H\Phi} \lesssim 0.5$.
 
As we will see later the Higgs portal parameter $\lambda_{H\Phi}$ can have a significant impact on the tree-level effects of the first-order electroweak phase transition. A relatively large value of $\lambda_{H\Phi}$ is required to have a substantial effect \cite{Blinov:2015sna}. By combining this requirement with the Higgs signal strength data, an upper bound on $v_\Phi$ can be derived. For instance, our results show that if a first-order electroweak phase transition necessitates $\lambda_{H\Phi} > 0.2$, it leads to a constraint of $v_\Phi < 1.2$ TeV, as demonstrated in Fig.~\ref{fig:collider-constraint}.

\underline{{\it Electroweak precision and dark photon/$Z^\prime$ measurements: }}

We consider constraints from electroweak precision measurements at the $Z$ pole~\cite{Zyla:2020zbs}, as well as from  $Z'$~\cite{ATLAS:2019erb} and dark photon physics (see~\cite{Fabbrichesi:2020wbt} for a recent review). 
For the specific range of light dark photon and $Z'$ masses, {\it i.e.}, $m_{A', Z'} < m_Z$, considered in this study, 
these constraints require the new gauge couplings $g_H$ and $g_X$ to be less than $\sim 10^{-3}$~\cite{Ramos:2021txu,Ramos:2021omo}. 
In this analysis, we also follow Ref.~\cite{Tran:2022cwh} to take into account the recent $W$ boson mass measurement at the CDF II~\cite{CDF:2022hxs}. 

\underline{\it DM constraints: }

(a) Relic density: 
The main annihilation process of a (sub-)GeV DM $W'$ is $W'^{p} W'^{m} \to f\bar{f}$ via s-channel, mediated by $Z'$ and $\gamma'$. 
However, due to the smallness of the gauge couplings $g_H$ and $g_X$, the cross sections of such processes are typically suppressed except in the resonant regions where the mediator masses are approximately twice the DM mass, as discussed in~\cite{Ramos:2021txu,Ramos:2021omo}. In fact, the annihilation process with a heavier mediator $Z'$ near the resonance can account for the observed DM relic density, $\Omega_{\rm DM} h^2 = 0.120 \pm 0.001$ from the Planck Collaboration~\cite{Aghanim:2018eyx}. 
Here we utilize {\tt micrOMEGAs} package \cite{Belanger:2018ccd} to calculate the DM relic density. 

(b) Direct detection:
The DM $W'$ can scatter off the target material in underground detectors resulting in a recoil energy that can be detected. 
The scattering cross section between the DM and the nucleon is dominated by the dark photon mediation process, as the dark photon is considered to have the lightest mass among the mediators~\cite{Ramos:2021txu,Ramos:2021omo}. 
We again use {\tt micrOMEGAs} package to calculate the DM-nucleon scattering cross section. 
The results of the cross section are compared against recent upper limits from experiments including
 CRESST III \cite{Angloher:2017sxg}, DarkSide-50 \cite{Agnes:2018ves,DarkSide-50:2022qzh}, XENON1T \cite{XENON:2018voc, Aprile:2019xxb}, {XENONnT \cite{XENON:2023cxc} }, PandaX-4T \cite{PandaX-4T:2021bab} and LZ \cite{LZ:2022lsv}. 

(c) Indirect detection and mono-jet: 
as shown in~\cite{Ramos:2021txu,Ramos:2021omo}, the results of electroweak precision measurements lead to a suppression of the new gauge couplings, resulting in a small DM annihilation cross section that easily satisfies the canonical limits set for various channels by Fermi-LAT data~\cite{Ackermann:2015zua, Fermi-LAT:2016uux}. 
Furthermore, this suppression of the gauge couplings also leads to a minuscule  production cross section for the mono-jet signal at the LHC~\cite{Ramos:2021txu,Ramos:2021omo}. Thus we do not include the constraints from indirect detection and mono-jet in this analysis.

\section{Electroweak Phase Transition}
\label{sec:EWPT}

In the literature, a common approach to calculate the EWPT is through the use of perturbation theory. 
However, this method becomes unreliable at temperatures $T>0$ due to infrared~(IR) bosonic contribution to thermal loops~\cite{Linde:1980ts}, which can cause a breakdown of the theory near the phase transition.
Apart from this inherent shortcoming, the reliability of $T>0$ perturbative computations can be improved using the well-known daisy resummation~\cite{Arnold:1992fb,Parwani:1991gq}, though the convergence remains slow. Recent attempts to improve the thermal resummation have been performed in Refs.~\cite{Curtin:2016urg,Curtin:2022ovx}.
Additionally, the conventional perturbative studies are often challenged by gauge dependence issues~\cite{Patel:2011th,Wainwright:2011qy}. Efforts in Refs.~\cite{Laine:1994zq,Patel:2011th} have been made to remove this gauge dependence which is guaranteed by the Nielsen identities~\cite{Nielsen:1975fs,Fukuda:1975di}. The choice of renormalization scheme and renormalization scale can also introduce additional uncertainties, leading to theoretical predictions that can vary significantly~\cite{Croon:2020cgk, Gould:2021oba, Athron:2022jyi}.

A more systematic approach to include thermal resummation is through the use of the dimensionally reduced 3d effective field theory (EFT)\cite{Kajantie:1995dw, Farakos:1994xh}. Recently, an automated package has been developed for this~\cite{Ekstedt:2022bff}.
The use of the dimensional reduced 3d EFT in the calculation of the EWPT has been studied in models such as the triplet model~\cite{Niemi:2018asa,Friedrich:2022cak}. Efforts have also been made to combine thermal resummation with gauge invariance using the  dimensional reduced 3d EFT~\cite{Kainulainen:2019kyp,Croon:2020cgk,Niemi:2021qvp,Gould:2021oba, Schicho:2022wty}. 
Another combination of thermal resummation and gauge invariance has also been developed recently~\cite{Ekstedt:2020abj}, which uses the 4d perturbation theory with a consistent power counting. Ultimately, one must rely on non-perturbative (lattice) computations to fully include the important bosonic IR contributions.
For recent studies of the EWPT using Monte Carlo simulations, see Refs.~\cite{Gould:2019qek,Niemi:2020hto,Gould:2022ran, Niemi:2024axp}. Importantly, one requires such lattice studies to determine the parameter-space boundary between a {\it bona fide} phase transition and a smooth crossover. In what follows, we rely on perturbation theory to identify the potentially FOWEPT-viable regions of parameter space, and defer a complete non-perturbative study to the future.

\subsection{One Loop Finite Temperature Effective Potential With Daisy Resummation}
\label{1loopEffV}

The finite temperature effective potential $V_{\rm eff}$ at one loop is given by
\begin{equation}
V_{\rm eff} (h_{1c}, h_{2c},\phi_{Hc}, T)  = V_0 (h_{1c}, h_{2c},\phi_{Hc})+ V_{1} (h_{1c}, h_{2c},\phi_{Hc}, T) \;,
\end{equation}
where $V_0$ is given by Eq.~(\ref{Vefftree}).

The one loop correction $V_{1} (h_{1c},h_{2c},\phi_{Hc}, T)$ can be split into several pieces 
\begin{eqnarray}
V_{1} (h_{1c},h_{2c},\phi_{Hc}, T) & = & V_{\rm CW} (h_{1c},h_{2c},\phi_{Hc}) + \Delta V_{1} (h_{1c},h_{2c},\phi_{Hc}, T) \nonumber \\
&& \;\;\;\;\; +  \Delta V_{\rm C.T.} (h_{1c},h_{2c},\phi_{Hc})  \; .
\end{eqnarray}
The first term is the Coleman-Weinberg effective potential~\cite{Coleman:1973jx,Weinberg:1973am} which is temperature independent, 
the second term is finite temperature correction, and the third term 
$\Delta V_{\rm C.T.}$ is the counter term. 
The Coleman-Weinberg effective potential 
in $4-\epsilon$ dimensional regularization can be expressed as
\begin{align}
\label{VCW}
V_{\rm CW} (h_{1c},h_{2c},\phi_{Hc}) & = \frac{1}{64 \pi^2} \sum_i (-1)^{2 s_i} n_i m_i^4 (h_{1c},h_{2c},\phi_{Hc}) \nonumber \\
&\;\;\;\;  \times \left\{ 
\log \left( \frac{m_i^2( h_{1c},h_{2c},\phi_{Hc} )}{\mu^2} \right) - C_i - C_{\rm UV}
\right\} \; ,
\end{align}
where $m_i$ denotes the field dependent mass (See Appendix~\ref{app:FDMasses}) of the particle $i$ with spin $s_i = (0,1/2,1)$ and number of degree of freedom $n_i$;
$C_i = 3/2$ for scalars and fermions and $5/6$ for gauge bosons; 
\beq
\label{CUV}
C_{\rm UV} = \frac{2}{\epsilon} - \gamma_E + \log 4 \pi \; ,
\eeq
with $\gamma_E=0.5772...$ denotes the Euler-Mascheroni constant; 
$\mu$ is a renormalization scale (See Appendix~\ref{app:RGE}).  
The finite temperature correction piece is~\cite{Dolan:1973qd,Weinberg:1974hy}
\begin{align}
\label{VT1loop}
\Delta V_{1} ( h_{1c},h_{2c},\phi_{Hc}, T) & = \frac{T^4}{2 \pi^2} \sum_i (-1)^{2 s_i} n_i {\mathcal J}_i \left( \frac{m_i ( h_{1c},h_{2c},\phi_{Hc} ) }{T} \right) \; .
\end{align}
Here ${\mathcal J}_i(x)$ is the one loop thermal function (with daisy resummation) defined by (\ref{ThermalIntegral_J}) in Appendix~\ref{app:ThermalIntegrals}. 
In the G2HDM, we have to sum over the following 3 sets of particles inside the loop 
\beq
\label{CPscalars}
\left\{ S_1,S_2,S_3,S_4,P_1,P_2,P_3,P_4,G^\pm,H^\pm \right\} \; ,
\eeq
\beq
\label{GaugeBosons}
\left\{ W^\pm_L, W^\pm_T, W^{\prime \, 2}_L,  W^{\prime \,2}_T, 
Z_{1L}, Z_{2L}, Z_{3L}, Z_{4L}, Z_{1T}, Z_{2T}, Z_{3T}, Z_{4T}, \gamma_L , \gamma_T \right\} \; ,
\eeq
\beq
\left\{  l, \nu, q, l^H, \nu^H, q^H \right\} \; .
\eeq
In (\ref{CPscalars}), $S_i (i=1,2,3,4)$ and $P_i (i=1,2,3,4)$  are the CP-even and CP-odd scalars respectively. 
In (\ref{GaugeBosons}), $Z_i (i=1,2,3,4)$ are the mass eigenstates of the massive neutral gauge bosons. For the general mass spectra of the model, see Appendix~\ref{app:FDMasses} for details.
The corresponding degrees of freedom $n_i$ are
\[
\begin{matrix}
n_{S_i} = 1, & n_{P_i} =1, & n_{H^\pm} = 2,  & n_{G^\pm} = 2, & & & \\
n_{W^\pm_L} = 2, & n_{W^\pm_T} = 4, & n_{W^{\prime \, 2}_L} = 1, & n_{W^{\prime \, 2}_T} = 2, & n_{Z_{iL}} = 1, & n_{Z_{iT}} = 2, & n_{\gamma_L} = 1, & n_{\gamma_T} = 2 ,\\
n_l = 4, & n_\nu = 4,  & n_{l^H} = 4, & n_{\nu^H} = 4, & n_q =12, & n_{q^H} = 12. & &
\end{matrix}
\]
 
 We note that due to the thermal mass corrections for the scalars and the longitudinal gauge bosons, the summation in Eq.~(\ref{VT1loop}) has to be treated 
 slightly different from the summation in Eq.~(\ref{VCW}) as given explicitly in (\ref{ThermalIntegral_J}). This is the result of the daisy resummation~\cite{Arnold:1992rz,Carrington:1991hz}.

In order to facilitate the analysis of the phase transition, we limit our consideration to the leading terms of the thermal correction functions ${\mathcal J}_B$ and ${\mathcal J}_F$ for the boson $B$ and fermion $F$ respectively. As a result, the thermal effective potential can be expressed in this high-T expansion as follows:
\be
\label{eq:VHT}
V^{\rm HT}_{\rm eff} (h_{1c}, h_{2c},\phi_{Hc}, T) = V_0 (h_{1c}, h_{2c},\phi_{Hc}) + \frac{1}{2} \Pi_{H_1} (T)  h_{1c}^2 
 + \frac{1}{2} \Pi_{H_2} (T)  h_{2c}^2 + \frac{1}{2} \Pi_{\Phi_H} (T)  \phi_{Hc}^2 \, ,
\ee
where $\Pi_{H_1}(T)$, $\Pi_{H_2}(T)$ and $\Pi_{\Phi_H}(T)$ are the thermal mass corrections for the scalar fields and their formulas are provided by (\ref{PiH1}), (\ref{PiH2}) and (\ref{PiPhiH}) respectively in Appendix~\ref{app:ThermalMasses}.

\subsection{Thermal history}

Given the three possible non-zero VEVs $(h_{1c}, h_{2c}, \phi_{Hc})$, there are eight combinations of possible extrema. 
To ensure the stability of the dark matter candidate, it is necessary that $h_{2c} = 0$ at zero temperature, thereby restoring the $h$-parity in the model \cite{Ramos:2021omo, Ramos:2021txu}.
We first note that, with a high-T expansion of the effective potential given in Eq.~(\ref{eq:VHT}), a single-step transition to the electroweak vacuum, {\it i.e.}, $(0,0,0) \to (h_{1c}, 0, \phi_{Hc})$, cannot be of first-order type. This is due to conflicting requirements that arise from the need for both $h_{1c}$ and $\phi_{Hc}$ to be real and non-zero, as well as the positivity and minimum conditions of the scalar potential at zero temperature \cite{Ghorbani:2020xqv}. 

After thoroughly scanning the parameter space in the model, we find a phase transition pattern that satisfies the electroweak vacuum requirement and leads to a strong first-order phase transition, as illustrated in the right panel of Fig.~\ref{fig:twostepsketch} where $s \equiv \phi_{Hc}$, $h_1 \equiv h_{1c}$ and $h_2 \equiv h_{2c}$. This pattern involves two-steps: $(0,0,0) \to (0,h_{2c}, 0) \to (h_{1c}, 0, \phi_{Hc})$, where the first step is a second-order phase transition, and the second is a first-order phase transition. 
We note that other two-step transition patterns are possible; 
however, these patterns may not result in a strong first-order phase transition. For instance, for the transition patterns of $(0,0,0) \to (0, 0, \phi'_{Hc})  \to (h_{1c}, 0, \phi_{Hc})$ and $(0,0,0) \to (h'_{1c}, 0, 0)  \to (h_{1c}, 0, \phi_{Hc})$, the scalar potential in Eq.~(\ref{eq:VHT}) can be reduced to the one that is the same as in a real singlet scalar extension model with $Z_2$ symmetry, wherein the two-step transition ended up with two non-zero VEVs cannot generate a first-order phase transition~\cite{Ghorbani:2020xqv}.
We would like to mention that multiple-step transitions beyond two-step transitions can occur, and a first-order phase transition can arise among these steps. However, for this analysis, we have restricted ourselves to the two-step EWPT.

 \begin{figure}[tb]
         \centering
	\includegraphics[width=0.495\textwidth]{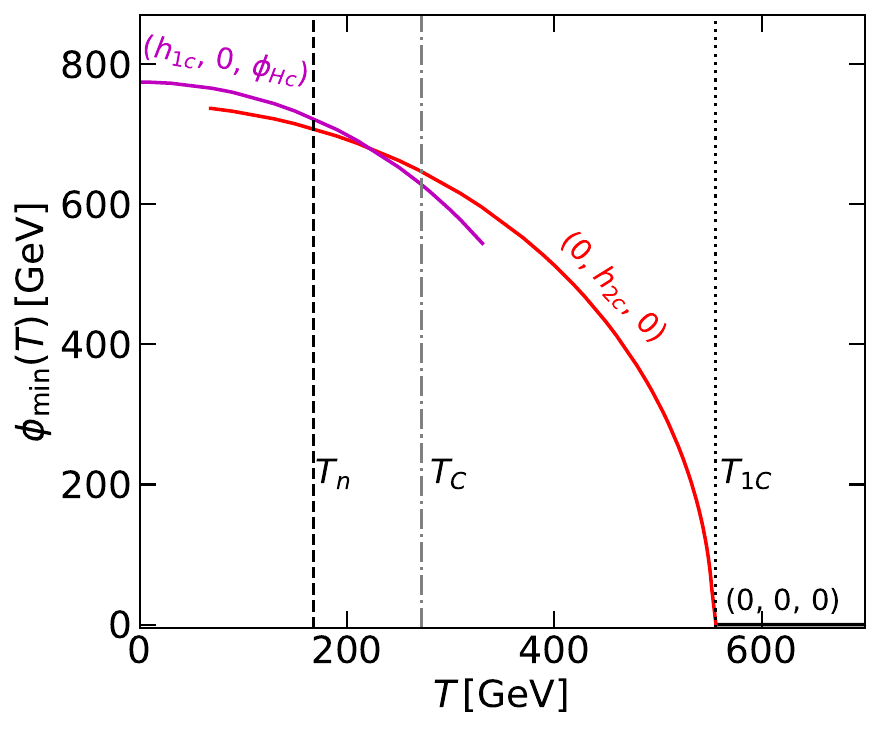}
	\includegraphics[width=0.495\textwidth]{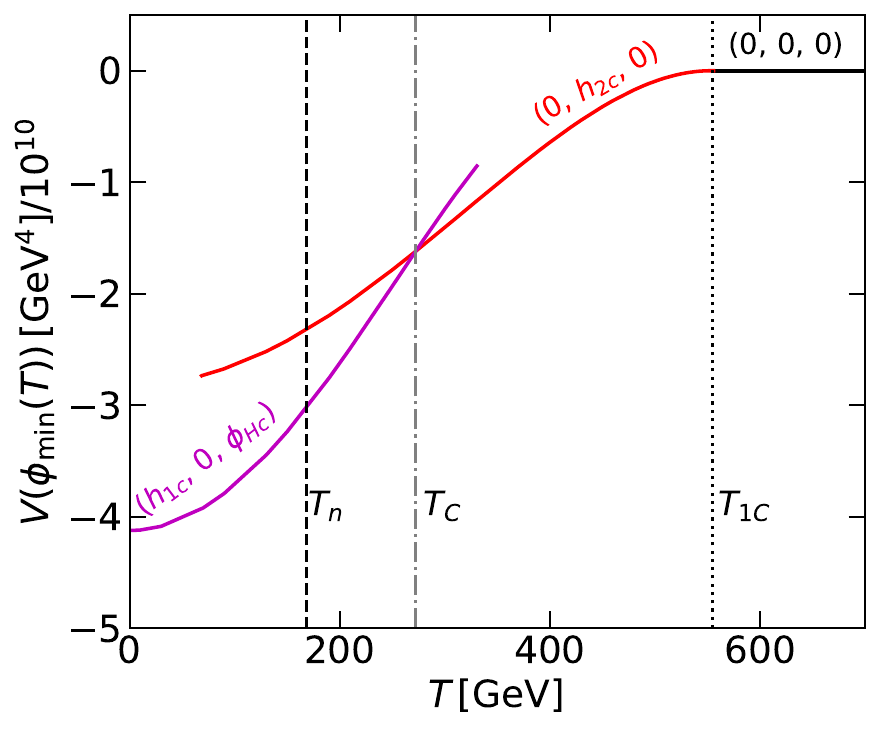}
	\caption{ \label{fig:phases} The evolution of $\phi_{\rm min} (T) = \sqrt{h_{1c}^2 (T) + h_{2c}^2 (T) + \phi_{Hc}^2 (T) }$ (left panel) 
	and the minimal of the effective potential (right panel) as functions of temperature. 
	The solid black, red and purple lines represent the $(0, 0, 0)$, $(0, h_{2c}, 0)$ and $(h_{1c}, 0, \phi_{Hc})$ phases. 
	The dotted, dash-dotted and dashed lines indicate the critical temperatures $T_{1C}$, $T_C$ and nucleation temperature $T_n$ respectively. Here we fix the model parameter space as $m_{h_2} = 745$ GeV,  $m_{H^{\pm}} = 374$ GeV, 
$m_D = 320$ GeV, $m_{W'} = 0.115$ GeV, $\theta_1 = 0.235$ rad, $\theta_2 = 0.32$ rad, $m_{f^H} = 1$ TeV, $g_X = 1.17\times10^{-4}$ and $m_X = 0.25$ GeV. 
}
\end{figure}

Fig.~\ref{fig:phases} illustrates the evolution of electroweak vacuum (left panel) and minimal of the effective potential (right panel) as functions of temperature. Here we introduce the quantity 
\be
\phi_{\rm min} (T) = \sqrt{h_{1c}^2 (T) + h_{2c}^2 (T)  
+ \phi_{Hc}^2 (T) }
\ee
to trace the evolution of the vacuum
and fix the model parameter space as $m_{h_2} = 745$ GeV,  $m_{H^{\pm}} = 374$ GeV, 
$m_D = 320$ GeV, $m_{W'} = 0.115$ GeV, $\theta_1 = 0.235$ rad, $\theta_2 = 0.32$ rad, $m_{f^H} = 1$ TeV, $g_X = 1.17\times10^{-4}$ and $m_X = 0.25$ GeV.  
One can see that $m_X \sim 2 m_{W'}$ so that the DM relic density satisfies the Planck observation through the resonant annihilation~\cite{Ramos:2021omo, Ramos:2021txu, Tran:2022yrh}. 
Hereafter, we label this benchmark point as \textbf{BM}.
 We use {\tt PhaseTracer} package \cite{Athron:2020sbe} to find the phases and calculate the critical temperatures.
At high-T, the stable vacuum is in a symmetric phase $(0, 0, 0)$. 
As the temperature decreases, a continuous transition from the symmetric phase to $(0, h_{2c}, 0)$ phase occurs at 
$T_{1C} = \sqrt{\mu_H^2/\tilde\Pi_{H_2}}$,
where from~(\ref{PiH2})
\be
\label{PitildeH2}
\tilde\Pi_{H_2} = \frac{1}{12} \left( 10 \lambda_H - \lambda_H^\prime + 2 \lambda_{H\Phi} + \lambda^\prime_{H\Phi} \right) 
 + \frac{1}{16} \left( 3 g^2 + g^{\prime\, 2} + 3 g_H^2 + g_X^2 \right)  \; ,
\ee
and $\mu_H^2 = \lambda_H  v^2 + \lambda_{H\Phi}v_\Phi^2/2$ determined by the tadpole condition at tree level. At this benchmark point, $T_{1C} \sim 556$ GeV, as indicated by the dotted line. 
Subsequently, a new minimum $(h_{1c}, 0, \phi_{Hc})$ appears at $T = T_C \simeq 272$ GeV~\footnote{See Appendix~\ref{app:CriticalTemp} for the analytical expression of $T_C$.}, as indicated by the dash-dotted line. These two phases, $(0, h_{2c}, 0)$ and $(h_{1c}, 0, \phi_{Hc})$, are degenerate and separated by a barrier, which is a characteristic feature of a FOEWPT.
As the temperature drops below $T_C$, 
the system undergoes a phase transition through the formation of bubbles, 
with the stable vacuum inside the bubbles being $(h_{1c}, 0, \phi_{Hc})$, 
and the metastable vacuum outside being $(0, h_{2c}, 0)$. 
The black dashed line in Fig.~\ref{fig:phases} represents the nucleation temperature, $T_n \simeq 168$ GeV, 
at which the formation rate of bubbles is about one per Hubble volume. 

The $h$-parity which was broken at the metastable vacuum at high-T has been restored at the stable electroweak vacuum at lower temperature. Thus one has realized symmetry anti-restoration of a discrete symmetry in the model. While the occurrence of symmetry restoration at high-T is quite common in spontaneously broken global and gauge theories, the possible occurrence of symmetry anti-restoration at lower temperature was first pointed out by Weinberg~\cite{Weinberg:1974hy} (See also~\cite{Mohapatra:1979vr,Masiero:1983ux,Salomonson:1984px,Kephart:1988xh,Patel:2013zla}) . It has been used to solve the $U(1)$ magnetic monopole problem in grand unified models~\cite{Langacker:1980kd,Farris:1991rg}.  
Here we have an instance of a discrete symmetry. 

\begin{figure}[tb]
         \centering
         \includegraphics[width=1.0\textwidth]{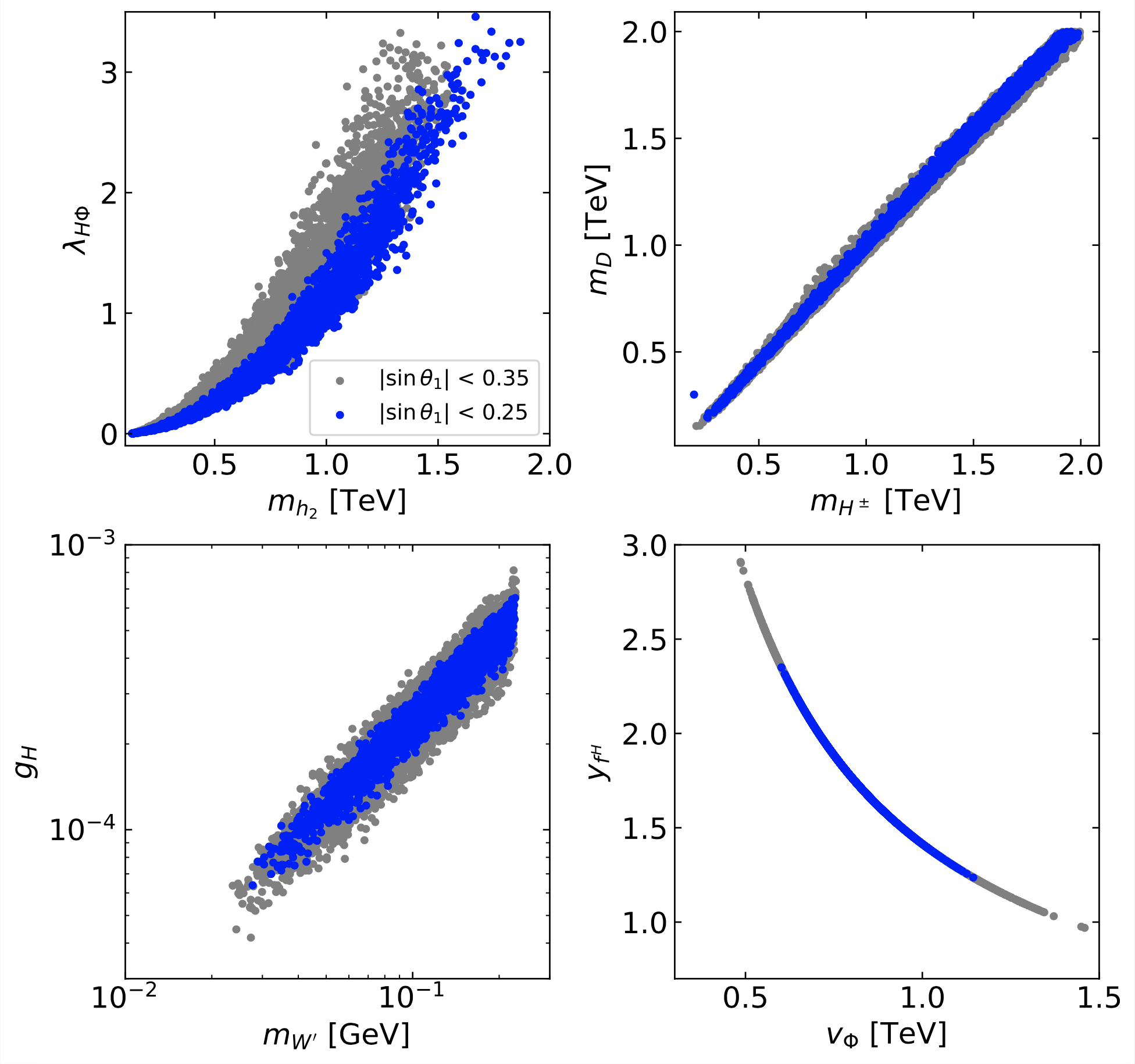}
	\caption{ \label{fig:paraPT-2step-LHCconstraint} Viable model parameter regions for the two-step phase transition with different constraints on the mixing angles $\theta_1$. The gray and blue regions indicate $|\sin \theta_1| < 0.35$ and $|\sin \theta_1| < 0.25$, respectively. 
	}
\end{figure}

\begin{figure}[tb]
         \centering
         \includegraphics[width=1.0\textwidth]{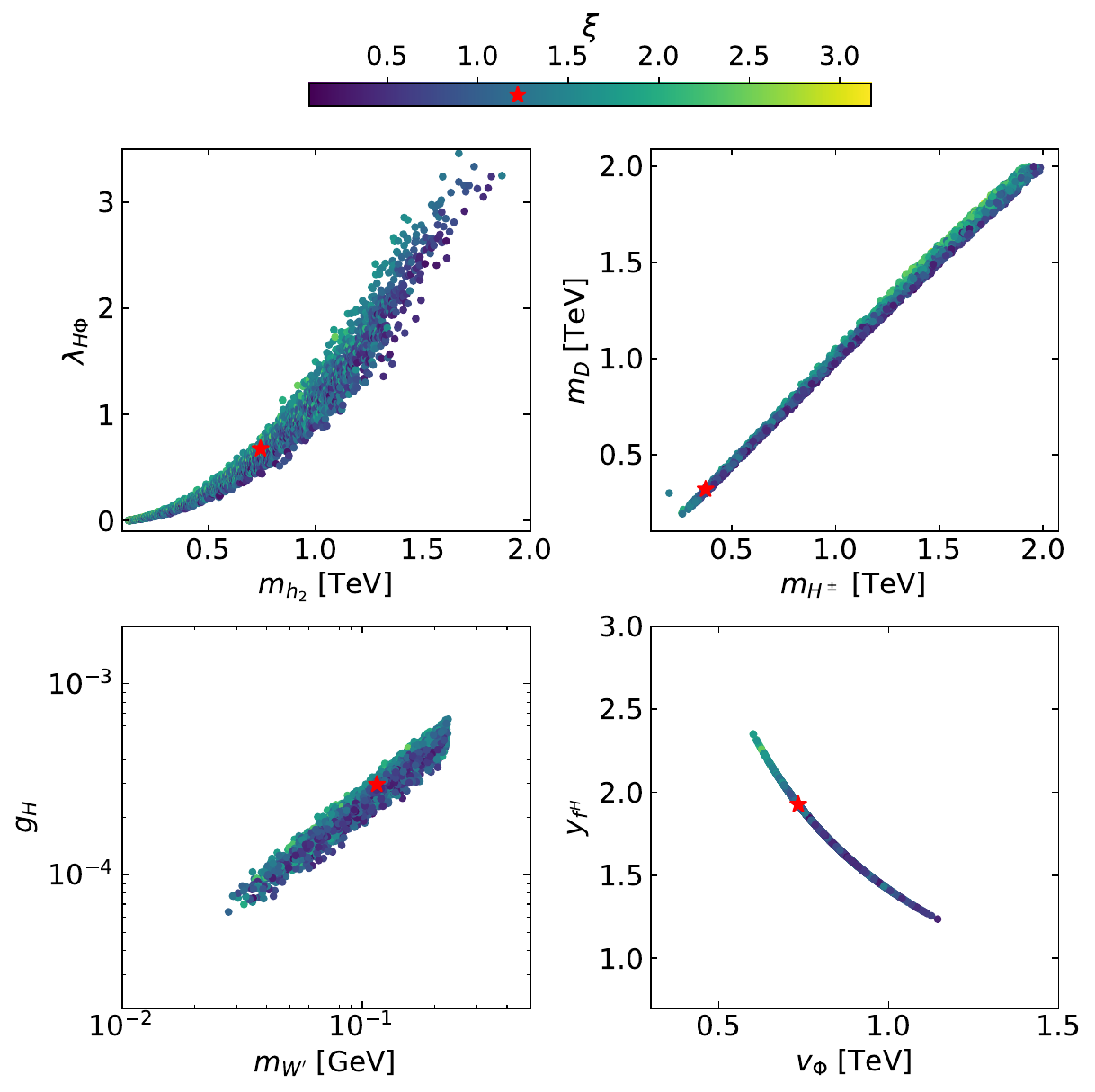}
	\caption{ \label{fig:paraPT-2step-xi} Viable model parameter region for the two-step phase transition, correlated with the phase transition strength. 
	The color legend on the top indicates the value of $\xi = v_C/T_C$ where $v_C = h_{1c}(T_C)$ for the second transition. The red star represents the benchmark point \textbf{BM}.
	}
\end{figure}

\begin{figure}[tb]
         \centering
         \includegraphics[width=1.0\textwidth]{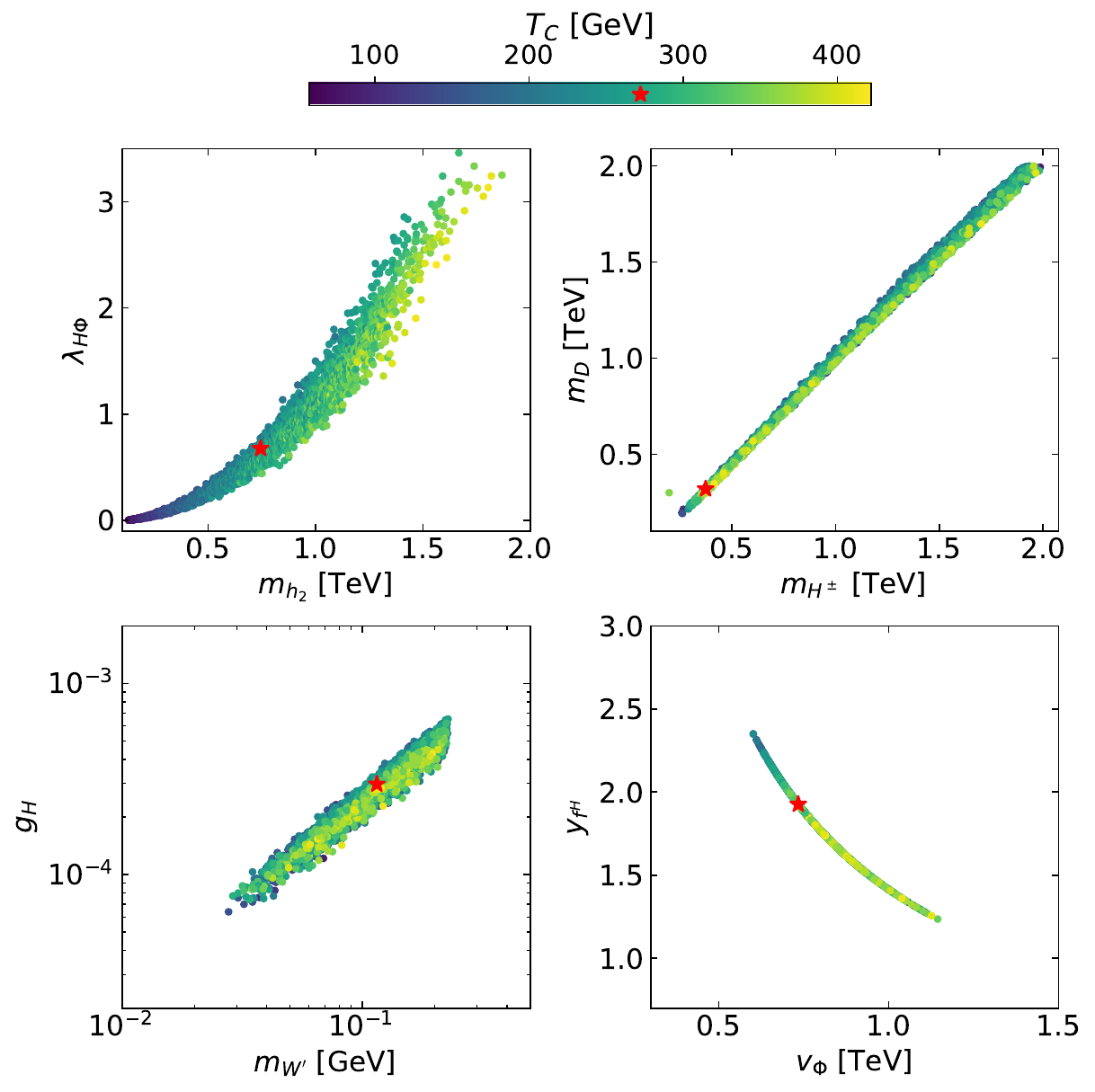}
	\caption{ \label{fig:paraPT-2step-TC} Viable model parameter region for the two-step phase transition, correlated with the critical temperature for the second transition. 
	The color legend on the top indicates the critical temperature $T_C$ for the second transition. The red star represents the benchmark point \textbf{BM}.
	}
\end{figure}

\subsection{Numerical result for two-step phase transition}

We investigate the parameter space of the model that gives rise to the two-step phase transition pattern, as depicted in the right panel of Fig.~\ref{fig:twostepsketch} where $s \equiv \phi_{Hc}$, $h_1 \equiv h_{1c}$ and $h_2 \equiv h_{2c}$.
We first explore the parameter region that satisfies all the theoretical and experimental constraints discussed in the previous section.
Next, we utilize the {\tt PhaseTracer} package~\cite{Athron:2020sbe} to trace the minima of the effective potential as the temperature varies and to calculate the critical temperatures.
To sample the parameter space in the model, we employ MCMC scans using {\tt emcee}~\cite{ForemanMackey:2012ig}.
The scanning ranges of model parameters are set as follows, 
\bea
m_{h_2}/ {\rm GeV} &\in& (130,\, 2000) \,, \\
m_{H^{\pm}} /{\rm GeV}  &\in& (80,\, 2000) \,, \\
m_{D} /{\rm GeV}  &\in& (10,\, 2000) \,, \\
m_{W'}/{\rm GeV}&\in& (0.01,\, 50) \, , \\
\theta_{1, 2}/{\rm rad}  &\in& \left(-\frac{\pi}{2},\, \frac{\pi}{2} \right) \,, \\ 
g_X  &\in& (10^{-6},\, 10^{-2}) \,, \\
M_{X}/{\rm GeV} &\in& (10^{-3},\, 10^{2})  \,,
\eea
and we fix $m_{f^H} = 1$ TeV.

 Given the viable model parameter space, we find that the FOPT is most likely to occur in the second step of a two-step phase transition, as sketched in the right panel of Fig.~\ref{fig:twostepsketch} where $s \equiv \phi_{Hc}$, $h_1 \equiv h_{1c}$ and $h_2 \equiv h_{2c}$.
The parameter space conducive to this two-step phase transition is depicted in Fig.~\ref{fig:paraPT-2step-LHCconstraint} with different choices of the constraint on the mixing angle $\theta_1$. The gray area denotes $|\sin \theta_1| < 0.35$, a constraint derived from Higgs signal strength data measured at the CMS experiment, while the blue area indicates $|\sin \theta_1| < 0.25$, a requirement based on the combined Higgs signal strength data from both the ATLAS and CMS experiments.
We observe that imposing a stringent constraint on $\theta_1$ leads to a significant reduction in the viable parameter space necessary to achieve the two-step phase transition. Specifically, upon comparing the gray and blue regions in Fig.~\ref{fig:paraPT-2step-LHCconstraint}, a tighter constraint on $|\sin \theta_1|$ requires a smaller value for $\lambda_{H\Phi}$ and a narrower region on the $(m_{H^\pm}, m_D)$ and $(m_{W'}, g_H)$ planes. Additionally, it results in stronger upper and lower bounds on $v_\phi$. 

In Fig. \ref{fig:paraPT-2step-xi}, we showcase results for $|\sin \theta_1| < 0.25$. The color within this figure indicates the strength of the EWPT in the second step, denoted here by $\xi = v_C/T_C$~\footnote{Not to be confused with the gauge fixing parameter having the same symbol discussed in Appendix~\ref{app:GaugeInvariantMethod}.}, where $v_C = h_{1c}(T_C)$, which remains gauge invariant under the high-T treatment of the effective scalar potential.
If one takes a glimpse back at the left panel of Fig.~\ref{fig:phases}, $v_C$ is the intersection point between the purple and dash-dotted lines.  
A commonly accepted criterion to avoid the washout of baryon number generated during the phase transition is $\xi \geq 1$~\cite{Patel:2011th}. We find that $\xi \lesssim 3.2$ in our scanning region.  
Similar to Fig. \ref{fig:paraPT-2step-xi}, Fig. \ref{fig:paraPT-2step-TC} show the viable parameter region, but with the color indicating the critical temperature $T_C$. Interestingly, $T_C$ lies within the electroweak scale.

As shown in the top-left panel of Fig. \ref{fig:paraPT-2step-xi}, for a fixed value of $m_{h_2}$, an increase in the quadratic coupling $\lambda_{H\Phi}$ leads to a larger $\xi$ and therefore a stronger phase transition. This occurs due to the presence of a tree-level barrier, the height of which is proportional to $\lambda_{H\Phi}$, in the second step of the phase transition. 
This effect has been extensively studied in SM extensions involving gauge singlets \cite{Profumo:2007wc, Espinosa:2011ax, Curtin:2014jma, Chiang:2017nmu} and real triplets \cite{Patel:2012pi, Blinov:2015sna, Niemi:2020hto}. 
Additionally, an increase in $\lambda_{H\Phi}$ also results in a lower critical temperature, as shown in the top-left panel of Fig. \ref{fig:paraPT-2step-TC}.

In the top-right panel of Fig.~\ref{fig:paraPT-2step-xi} or \ref{fig:paraPT-2step-TC}, the viable parameter space projected on the $(m_{H^\pm}, m_D)$ plane shows a distinct shape, which is due to the recent precision measurement of the $W$ boson mass by the CDF-II collaboration~\cite{Tran:2022yrh}\footnote{If one instead uses the data from the PDG~\cite{ParticleDataGroup:2020ssz}, excluding the CDF-II measurement, the most significant change in the parameter space is the mass splitting between the charged Higgs and the dark Higgs. The PDG data favor a smaller mass splitting~\cite{Tran:2022yrh}.}. Furthermore, this panel indicates that the dark Higgs and charged Higgs must satisfy lower bounds of $m_{H^\pm} \gtrsim 200$ GeV and $m_D \gtrsim 150$ GeV, respectively.

We observe that in this model, even for scalar masses on the order of 1 TeV, the two-step phase transition can occur, which contrasts with the generic upper bound of $~\sim 700$ GeV as discussed in Ref. \cite{Ramsey-Musolf:2019lsf}. This larger range for the new scalar masses arises because the extra scalar boson considered in Ref. \cite{Ramsey-Musolf:2019lsf} obtains its mass solely from the SM doublet VEV $v$, whereas the inert Higgs doublet $H_2$ in G2HDM is also a member of a doublet under a hidden $SU(2)_H$ that is broken by a new VEV $v_\Phi$, thereby generating additional contributions to their masses. For example, two-step transition requires $T_{1C} > T_C$ and thus 
\be
\mu_{H}^2 = \Pi_{H_2}(T_{1C}) > \Pi_{H_2}(T_{C}) \sim \Pi_{H_2}(T_{\rm EW}) \; ,
\ee
where $T_{1C}$ ($T_C$) is the critical temperature for the first (second) step and $T_{\rm EW}$ is the electroweak temperature. Assuming a small mixing between $h$-parity odd scalar fields, one can obtain the dark Higgs boson mass at zero temperature as 
\bea
m_{D} \simeq m_{H_{2}^0} &=& \sqrt{-\mu_{H}^2 + \lambda_H v^2 + \frac{1}{2} \left(\lambda_{H\Phi} + \lambda'_{H\Phi}\right) v_{\Phi}^2} \;  \\
&<& \sqrt{-\Pi_{H_2}(T_{\rm EW}) + \lambda_H v^2 + \frac{1}{2} \left(\lambda_{H\Phi} + \lambda'_{H\Phi}\right) v_{\Phi}^2} \; .
\eea
Taking numerical values of $\lambda_{H} = 0.8$, $\lambda_\Phi = 1.7$, $\lambda_{H\Phi} = 2.1$, $\lambda'_{H\Phi} = 2.9$, $\lambda'_{H} = -3.3$ and $g_{H,X} \ll 1$, we obtain 
\be
m_{D} \simeq m_{H_{2}^0}  \lesssim \sqrt{-1.6 \,T_{\rm EW}^2 + 0.8\, v^2 + 2.5 \,v_{\Phi}^2} \; .
\ee
With $T_{\rm EW} = 200$ GeV and $v_\Phi = 1.3$ TeV, one can get an upper bound of $m_{D} \lesssim 2$ TeV.

The mass of the DM particle, $m_{W'}$, falls in the sub-GeV range of 0.02 GeV to 0.25 GeV, as revealed in the bottom-left panel of Fig.~\ref{fig:paraPT-2step-xi} or \ref{fig:paraPT-2step-TC}. This restriction range of $m_{W'}$ is due to the limits on the new gauge coupling $g_H$, which arises from dark photon searches \cite{Ramos:2021omo, Ramos:2021txu, Tran:2022yrh}, as well as the bounds on $v_\Phi$.
As shown in the bottom-right panel of Fig.~\ref{fig:paraPT-2step-xi} or \ref{fig:paraPT-2step-TC}, $v_\Phi$ falls within the range of $[0.5 \,\rm{TeV}, 1.4\,\rm{TeV}]$. This range represents a combination of collider constraints illustrated in Fig.~\ref{fig:collider-constraint} and the two-step phase transition condition requirement for $\lambda_{H\Phi}$, as shown in the top-left panel of Fig.~\ref{fig:paraPT-2step-xi} or \ref{fig:paraPT-2step-TC}.
With the mass of the heavy hidden fermions fixed at 1 TeV, a relatively large value of the Yukawa coupling $y_{f^H}$ is required in the small region of $v_\Phi$. Nevertheless $y_{f^H}$ is still below $\sqrt{4 \pi}$ allowed by perturbative unitarity arguments.
Additionally, smaller regions of $v_\Phi$ show a trend towards larger values of $\xi$.
Finally, the bottom-right panel of Fig.~\ref{fig:paraPT-2step-TC} shows that $T_C$ increases in regions of larger values of $v_\Phi$.

\section{Stochastic Gravitational Waves}
\label{sec:GW}
During a cosmological FOEWPT, 
the production of GWs is a result of several distinct mechanisms. In particular, primary sources of GW generation are bubble collisions, sound waves, and magneto-hydrodynamic (MHD) turbulence \cite{Caprini:2015zlo, Weir:2017wfa}. 
For bubble collisions, the GW is generated from the stress energy localized at the bubble wall. 
Such GW spectra have been estimated analytically \cite{Jinno:2016vai} and numerically 
under the so-called thin-wall and envelope approximations 
respectively~\cite{Kosowsky:1991ua, Kosowsky:1992rz, Kosowsky:1992vn, Huber:2008hg}. 
On the other hand, the motion of the fluid in the plasma during the phase transition generates sound waves. 
 As the sound waves dissipate, they transfer energy into GWs \cite{Hindmarsh:2013xza, Giblin:2013kea, Giblin:2014qia, Hindmarsh:2015qta}. 
 The GW spectrum from this source typically relies on large scale lattice simulations \cite{Hindmarsh:2013xza, Hindmarsh:2015qta, Hindmarsh:2017gnf, Cutting:2019zws}. 
 However, an analytical modeling can reproduce the spectra from simulations reasonably well based on the sound shell model \cite{Hindmarsh:2016lnk, Hindmarsh:2019phv}.
 Turbulence can also be induced in the plasma by percolation, particularly MHD turbulence, given that the plasma is fully ionized. 
 This also leads to the production of GWs \cite{Caprini:2006jb, Kahniashvili:2008pf, Kahniashvili:2008pe, Kahniashvili:2009mf, Caprini:2009yp}. 
 The significance of each contribution to GW generation is heavily influenced 
 by the characteristics and dynamics of the phase transition. 
 The velocity of the bubble wall is a critical factor in this regard. 
 If the wall velocity is slow, the thermal bath can effectively absorb the energy released during the phase transition, 
 leading to suppression of the GW spectrum. 
 On the other hand, if the bubble wall velocity is relativistic, a considerable amount of energy can be converted into bulk motion.
 In this study we consider a so-called {\it non-runaway bubbles} in which the bubble expansion in the plasma can result in the attainment of a relativistic terminal velocity \cite{Caprini:2015zlo}. 
 In this case, the contributions from the sound waves and the MHD turbulence are dominant. 
 For recent reviews on GW signals and cosmological first order phase transitions, we refer our readers to~\cite{Hindmarsh:2020hop,Athron:2023xlk}.

In order to calculate the GW signals resulting from the FOEWPT, 
the following 4 parameters must be determined: 
the nucleation temperature, $T_n$; 
the fraction of energy released from the phase transition to the total radiation energy density at the nucleation temperature, $\alpha$;
the inverse duration of the phase transition $\beta$ (usually  normalized by $H_\ast$, the Hubble rate at the percolation temperature);  
and the velocity of the bubble wall, $v_w$. The wall velocity can be calculated from micro-dynamics of particle interactions with the Higgs condensate, 
its precise value remains undetermined due to the theoretical uncertainties in the calculations. Here we choose $v_w = 0.95$ in our analysis. 
The sensitivity of GW detectors can significantly depend on values of bubble wall velocity~\cite{Friedrich:2022cak}. Within the viable parameter space in G2HDM, we find that sensitivity of LISA detector is maximized around $v_w \simeq [0.7, 0.8]$. Higher values of $v_w$ slightly reduce sensitivity, while it drops dramatically for smaller $v_w$ ($v_w < 0.6$).

The nucleation temperature is usually determined by solving the following equation
\begin{equation}
\label{eq:cosmoeq}
\int_{T_n}^{\infty} \frac{{\rm d} T}{T} \frac{\Gamma(T)}{H(T)^4} \simeq 1 \; ,
\end{equation}
where $\Gamma(T)$ is the tunneling probability per unit time per unit volume, which is calculated using the following formula
\begin{equation}
\label{eq:gammaT}
\Gamma(T) \simeq T^4\left(\frac{\mathcal S_3}{2 \pi T}\right)^{3 / 2} e^{-\mathcal S_3 / T} \; .
\end{equation}
Here $\mathcal S_3$ represents the three-dimensional Euclidean action corresponding to the critical bubble, 
which is given by
\begin{equation}
\label{eq:action}
\mathcal S_3=\int_0^{\infty} {\rm d} r r^2\left[\frac{1}{2}\left(\frac{{\rm d} \hat \phi(r)}{{\rm d} r}\right)^2+V(\hat \phi, T)\right]\; ,
\end{equation}
where $\hat \phi$ denotes a collection of classical scalar fields~\footnote{In the present context, $\hat \phi=\{ h_{1c}(r,T), h_{2c}(r,T), \phi_{Hc}(r,T) \}$.} that minimizes the action $\mathcal S_3$ and are determined by solving the following equation of motion
\begin{equation}
\label{eq:bounceeq}
\frac{{\rm d}^2 \hat \phi}{{\rm d} r^2}+\frac{2}{r} \frac{{\rm d} \hat \phi}{{\rm d} r}=\frac{{\rm d} V(\hat \phi, T)}{{\rm d} r} \; , 
\end{equation}
with the boundary conditions
\begin{equation}
\lim_{r \rightarrow \infty} \hat \phi(r)=0,\left.\quad \frac{{\rm d} \hat \phi}{{\rm d} r}\right|_{r=0}=0 \; . 
\end{equation}
Here we utilize the {\tt CosmoTransitions} package~\cite{Wainwright:2011kj} to numerically solve the bounce equation in Eq.~(\ref{eq:bounceeq}) and subsequently calculate the action.
A rough estimation of the nucleation temperature, $T_n$, can be obtained by solving the following nucleation condition 
\be
\label{eq:nucleationeq}
\mathcal S_3(T_n)/T_n \simeq 140 \; .
\ee

The inverse duration parameter $\beta$ can be calculated using the following equation
\begin{equation}
\frac{\beta}{H_*} =\left. T_* \frac{{\mathrm d} \left(\mathcal S_3 / T\right)}{{\mathrm d} T}\right|_{T_*} \; ,
\end{equation}
where $T_*$ is the temperature at which GWs are generated and it is expected that $T_* \simeq T_n$, 
and $H_*$ is the Hubble rate at $T_*$. 
The parameter $\alpha$ can be obtained by
\begin{equation}
\alpha=\frac{\rho_{\mathrm{vac}}}{\rho_{\mathrm{rad}}^*}=\left.\frac{1}{\rho_{\mathrm{rad}}^*}\left[T \frac{\partial \Delta V(T)}{\partial T}-\Delta V(T)\right]\right|_{T_*} \; ,
\end{equation}
where
$\rho_{\mathrm{vac}}$ is the vacuum energy density,
$\rho^*_{\mathrm{rad}} = g_* \pi^2 T^4 / 30$ is the total radiation energy density with $g_* = 106.75$ being the number of relativistic degrees of freedom at $T_*$,
and $\Delta V$ is the difference in scalar potential between the lower and higher phases.

With these portal parameters, we utilize the {\tt PTPlot} package \cite{Hindmarsh:2017gnf, Caprini:2019egz} to calculate the GW spectrum. 
To quantify the detectability of the signals, one can define the signal-to-noise ratio (SNR) \cite{Caprini:2019egz}
\begin{equation}
\mathrm{SNR}=\sqrt{\mathcal{T} \int_{f_{\min }}^{f_{\max }} \mathrm{d} f\left[\frac{h^2 \Omega_{\mathrm{GW}}(f)}{h^2 \Omega_{\mathrm{exp}}(f)}\right]^2},
\end{equation}
where $\mathcal{T}$ is the duration of the observation period in years, $h^2 \Omega_{\mathrm{GW}}(f)$ represents the GW energy fraction spectrum from the FOEWPT and $h^2\Omega_{\mathrm{exp}}(f)$ is the sensitivity of the experimental setup. 
It is commonly accepted that a SNR value of 10 is the threshold for detection. 
For the benchmark point \textbf{BM} depicted in Fig.~\ref{fig:phases}, the values of $T_n$, $\alpha$, $\beta/H_n$ and LISA SNR we obtained 
are $168$ GeV, 0.43, 332 and 144 respectively.

\begin{figure}[tb]
         \centering
         \includegraphics[width=1.0\textwidth]{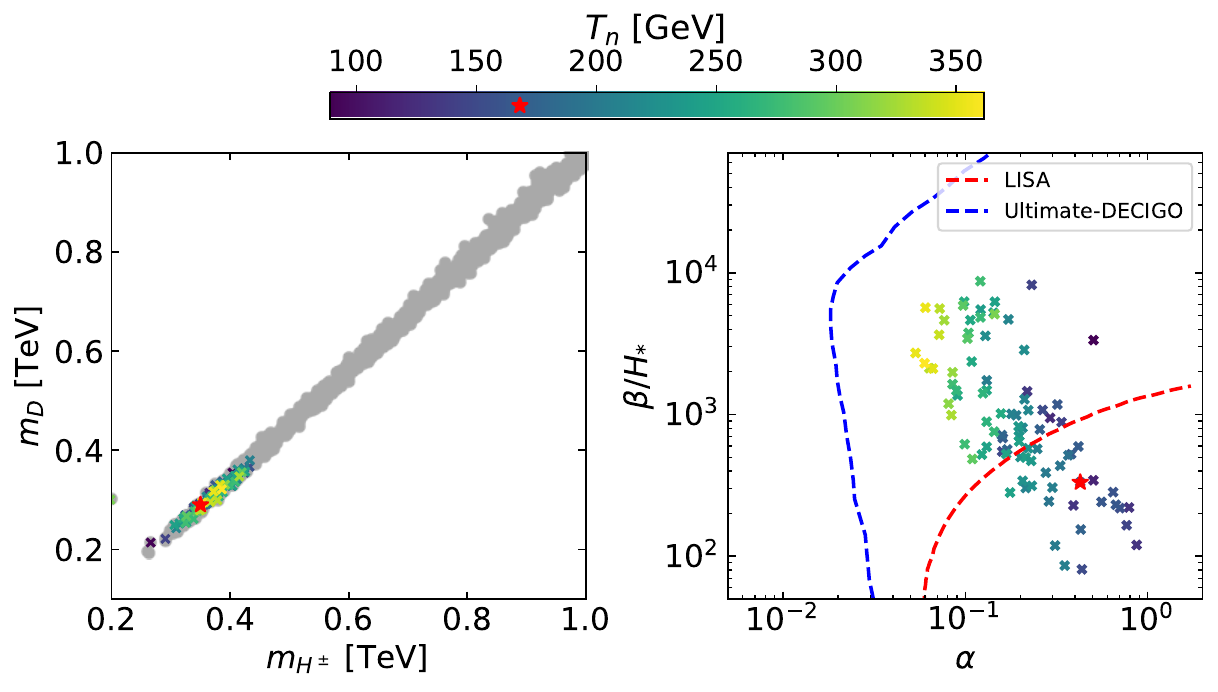}
	\caption{ \label{fig:nucleation-points} Strong FOEWPT points in the two-step phase transition projected on the ($m_{H^\pm}, m_D$) plane (left) and the ($\alpha, \beta/H_*$) plane (right). The light gray points on the left panel indicate the points fulfilling the two-step phase transition, while the colored cross points in both panels indicate the surviving points after applying the nucleation condition. The color legend on the top indicates the nucleation temperature $T_n$. The dashed red and blue lines represent the LISA with a SNR = 10 and Ultimate-DECIGO sensitivity respectively. The red star represents the benchmark point \textbf{BM}. 
	}
\end{figure}

By applying the nucleation condition described in Eq.~(\ref{eq:nucleationeq}) to the two-step phase transition data points, we find that only approximately $3\%$ of the data points meet this criterion. 
While the range of other parameters remains largely unchanged after applying the nucleation condition, 
the upper bound on the masses of the charged Higgs and dark Higgs is significantly altered. 
In particular, as shown in the left panel of Fig.~\ref{fig:nucleation-points}, it is required that $m_{H^\pm} \lesssim 450$ GeV and $m_{D} \lesssim 400$ GeV.
Additionally, as shown in the same panel, 
we can see that the value of the nucleation temperature $T_n$ tends to increase in the regions of larger masses of the charged Higgs and dark Higgs.

The right panel of Fig.~\ref{fig:nucleation-points} shows the surviving points on the ($\alpha, \beta/H_*$) plane after applying the nucleation condition.
As a result, the range of $\beta/H_*$ is approximately ($50,10^4$) and the range of $\alpha$ is approximately ($0.05,1$). All prediction points lie within the sensitivity range of the Ultimate-DECIGO detector (dashed blue line) \cite{Kawamura:2011zz} and 
a part of them lies within the LISA sensitivity region with a SNR > 10 \cite{Caprini:2019egz} 
\footnote{Here, in order to obtain the Ultimate-DECIGO and LISA sensitivity lines, we take $T_n$ = 100 GeV and $v_w = 0.95$ for the purpose of illustration.}.
A larger value of $\alpha$ suggests more energy from the plasma being converted into GWs, 
while a smaller value of $\beta/H_*$ implies a longer duration of the strong first-order phase transition, 
thus leading to an enhancement of the GW energy spectrum.

\begin{figure}[tb]
         \centering
	\includegraphics[width=0.7\textwidth]{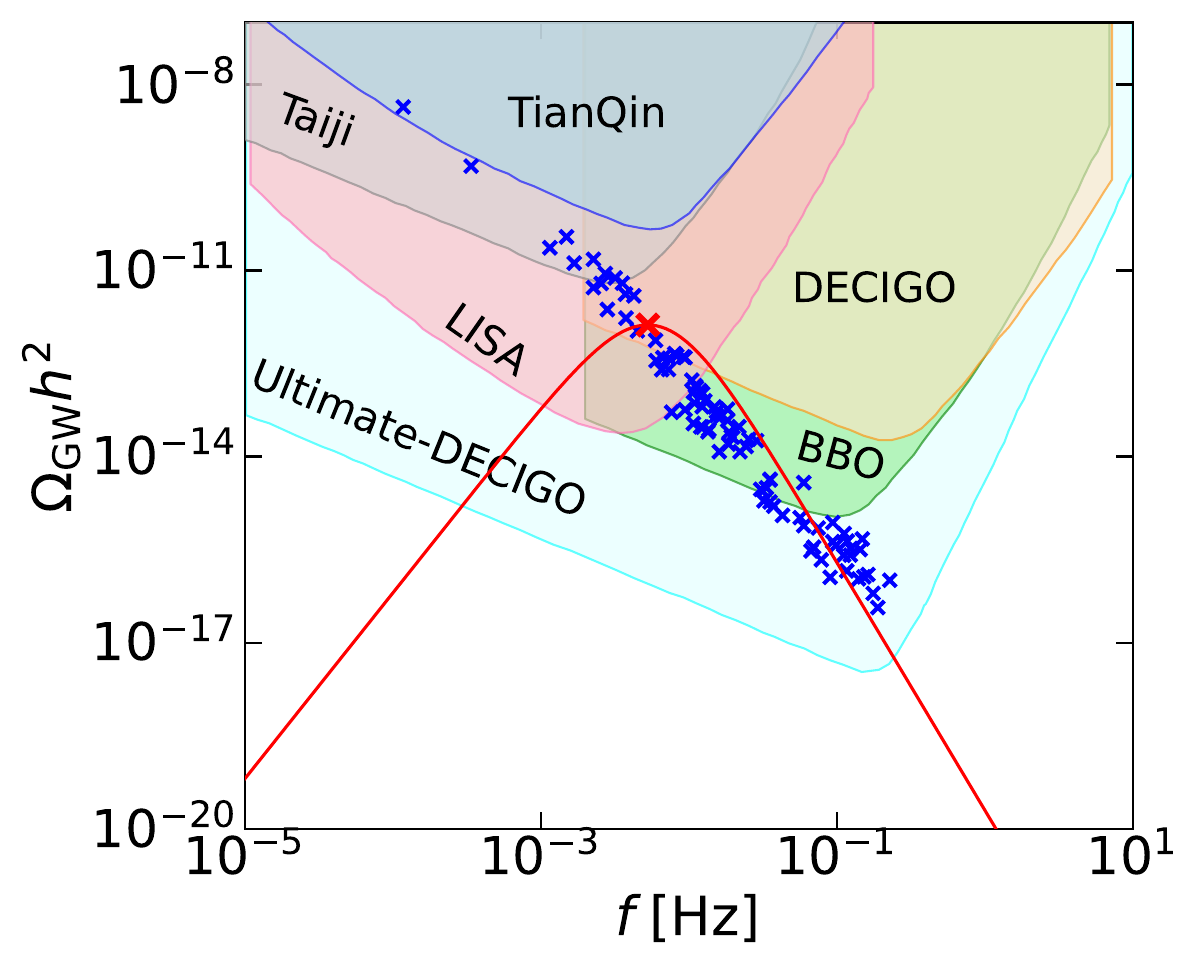}
	\caption{ \label{fig:spectrum}  
 The GW energy spectrum generated from the benchmark point \textbf{BM} is represented by the solid red line with its peak indicated by the red cross. Peaks of the GW energy spectrum   from other surviving points generated from a two-step phase transition are indicated by blue cross points. 
	The experimental sensitivities of various GW detectors, including Taiji, TianQin, DECIGO, Ultimate-DECIGO, BBO, and LISA, are indicated by shaded color regions.  
	}
\end{figure}

The predicted energy spectrum peaks of GWs as a function of frequency are presented in Fig.~\ref{fig:spectrum}. 
As a result, the generated GWs resulting from the FOEWPT in the two-step transition have peak frequencies within the range of $(\sim 10^{-4}, \sim 0.2)$ Hz and peak yields vary in the range $(\sim 10^{-17}, \sim 10^{-9})$. 
These peak yields are expected to be accessible at BBO, LISA, (Ultimate-)DECIGO, TianQin and Taiji detectors in the future.
For demonstration purposes, we show the GW spectrum generated from the benchmark point \textbf{BM}  as the solid red line in Fig.~\ref{fig:spectrum}. 
The benchmark point is selected with a relatively high value of $\alpha$ in order to increase the magnitude of the generated GWs, making it possible to be accessible by BBO, LISA, and (Ultimate-)DECIGO.

\begin{figure}[tb]
         \centering
	\includegraphics[width=0.7\textwidth]{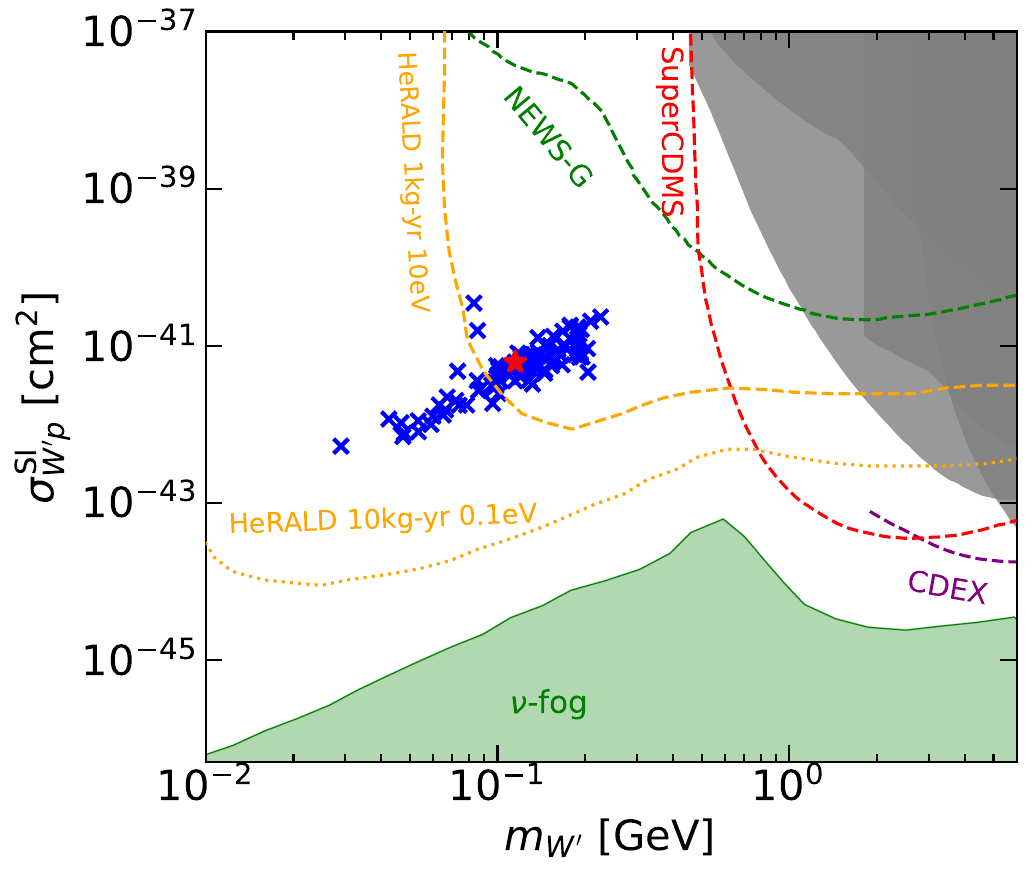}
	\caption{ \label{fig:DMDD} The spin-independent DM-proton scattering cross section as a function of DM mass for the parameter space probed by the GW signals at the Ultimate-DECIGO detector. The gray regions represent the exclusion limits from 
  current DM direct detection experiments. The dashed green, dashed red, dashed purple, dashed orange and dotted orange lines show the expected sensitivities of future DM direct detection experiments at NEWS-G \cite{Durnford:2021mzg}, SuperCDMS \cite{Agnese:2016cpb}, CDEX \cite{Ma:2017nhc}, HeRALD \cite{Hertel:2018aal} with 1 kg-year of exposure and 10 eV of energy threshold, and with 10 kg-year of exposure and 0.1 eV of energy threshold, respectively. The green region represents the neutrino fog background. The red star indicates benchmark point \textbf{BM}. 
	}
\end{figure}

\section{Dark Matter Direct Detection Prospects }
\label{sec:DMDD}
In recent decades, significant advancements have been made in the design and construction of direct-detection experiments aimed at detecting weakly interacting massive particles (WIMPs) that are potential candidates for DM. Notable examples of these experiments include CRESST III \cite{Angloher:2017sxg}, DarkSide-50 \cite{Agnes:2018ves, DarkSide-50:2022qzh}, XENON1T \cite{XENON:2018voc, Aprile:2019xxb}, PandaX-4T \cite{PandaX-4T:2021bab}, and LZ \cite{LZ:2022lsv}. These experiments have led to significant improvements in searches for few keV-scale nuclear recoils, which are indicative of spin-independent scattering of WIMPs with masses greater than $1$ GeV. While the next generation of experiments, such as PandaX30T~\cite{Wang:2023wrr}, XENONnT~\cite{XENON:2020kmp}, DarkSide-20k~\cite{DarkSide-20k:2017zyg}, and DARWIN~\cite{DARWIN:2016hyl}, are expected to explore a large fraction of the theoretically well-motivated parameter space for this mass range and even reach the neutrino fog, there is growing interest in exploring DM with sub-GeV mass ranges \cite{Essig:2022dfa}.

However, the detection of sub-GeV DM remains challenging due to the low nuclear recoil energies involved and the finite detector thresholds. Nevertheless, new technologies capable of achieving lower detector energy thresholds, down to meV$-$eV scales, are being proposed \cite{Essig:2022dfa}. For example, the HeRALD proposal \cite{Hertel:2018aal} involves an array of microchannel plates and photomultiplier tubes that can measure ionization signals resulting from DM interactions with helium nuclei. This approach enables HeRALD to reach energy thresholds as low as a few meV, thus making it suitable for probing DM in the sub-GeV mass region.

As demonstrated before, a two-step transition for achieving a strong FOEWPT in the G2HDM leads to the prediction that the DM mass falls within the sub-GeV range. This prediction renders the DM candidate a suitable candidate for exploration in upcoming DM direct detection experiments.
Fig. \ref{fig:DMDD} shows the spin-independent DM-proton scattering cross-section as a function of DM mass for the parameter space being probed by gravitational wave detectors. While NEWS-G \cite{Durnford:2021mzg}, SuperCDMS \cite{Agnese:2016cpb}, and CDEX \cite{Ma:2017nhc} are designed to probe DM with masses around and above $1$ GeV, HeRALD with 1 kg-year of exposure and 10 eV of detector energy threshold can probe part of our parameter space that predicts gravitational wave signals. HeRALD can fully cover our parameter space with 10 kg-year of exposure and a detector energy threshold of 0.1 eV.
Therefore, it highlights an interesting interplay between future DM direct detection and GW signal searches.

\section{Conclusions}
\label{sec:cons}

We investigate the electroweak phase transition and the possibility of detecting GW signals resulting from the phase transition in the simplified G2HDM. The simplified G2HDM is a well-motivated dark matter model with a hidden non-abelian gauge boson $W^{\prime \, (p,m)}$ serving as a DM candidate. The stability of the DM candidate is protected by an  accidental $h$-parity in the model. We take into account both theoretical and experimental constraints, with the latter including Higgs experimental data, electroweak precision measurements, dark photon and DM direct search experiments.

We derive the effective potential in the simplified G2HDM, including the one-loop finite temperature correction with daisy resummation. The effective potential consists of three background fields ($h_{1c}$, $h_{2c}$, $\phi_{Hc}$) that give rise to rich patterns of phase transition. We require $h_{2c}=0$ at zero temperature to anti-restore the $h$-parity and analyze the possible first-order electroweak phase transition in the early universe. 

We identify several rather distinctive features of the simplified G2HDM as follows: 
\begin{itemize}
    \item A two-step phase transition pattern is possible, where the symmetric phase transfers to the inert doublet $h_{2c}$ direction first and subsequently to the electroweak vacuum phase with a non-zero hidden doublet VEV $v_\Phi$ in addition to the SM doublet VEV $v$ as the temperature decreases. Using the high-T approximation in the effective potential that manifests gauge invariance, we found that the former transition is a second-order phase transition, while the latter can be first-order due to a tree-level barrier. 
    The first step breaks $h$-parity and the second step restores it.
    \item Unlike the conventional $Z_2$ symmetry model, where the two-step FOEWPT imposes strict upper bounds on new scalar masses, these bounds can be relaxed in G2HDM. In G2HDM, the new scalars receive contributions to their masses from a hidden sector VEV, allowing for somewhat larger BSM scalar masses to be compatible with a FOEWPT.
    However, due to the perturbative limit on the couplings and the bubble nucleation condition for GWs, scalar masses can only reach $\mathcal O$(1 TeV). In particular, the extra $h$-parity even scalar boson mass $m_{h_2}$ can reach 1.8 TeV, the $h$-parity odd scalars $D$ and $H^{\pm}$ can only reach up to 500 GeV. 
    Our findings also suggest that the Higgs portal coupling $\lambda_{H\Phi}$ needs to be sizable and the VEV $v_\Phi$ of the $SU(2)_L$ singlet/$SU(2)_H$ doublet scalar $\Phi_H$ needs to be within the range of 0.5 TeV to 1.4 TeV.
    \item The predicted GW energy spectrum from two-step FOEWPT in this model can be probed in the next-generation GW detectors. In particular, we find that the peak frequencies lie within the range of $(\sim 10^{-4}, \sim 0.2)$ Hz, and the peak yields vary from $\sim 10^{-17}$ to $\sim 10^{-9}$. These frequencies and yields can be covered by BBO, LISA, (Ultimate-)DECIGO, TianQin, and Taiji in the future.
    \item Interestingly, these GW probed regions can also be searched for in the future dark matter direct detection experiments, such as the superfluid-He target detectors. While it is possible for the DM candidate to be heavy ~\cite{Tran:2022yrh}, we have concentrated on the light mass region due to its potential accessibility by future direct detection experiments.
    \item Albeit the above distinctive features are obtained here in a specific model of G2HDM, it is conceivable that they might be  present in other class of models, for instance in certain limit of the IHDM, augmented further by a scalar singlet. After all, the scalar potential of G2HDM can be regarded as a gauged version of the one from IHDM, supplemented by a hidden $SU(2)_H$ doublet $\Phi_H$ which is nevertheless a SM singlet.
\end{itemize}

Finally, we note that the results in this work are presented at leading order, {\it i.e.}, using the 
high-T approximation in the effective potential. Higher-order corrections that include both a systematical thermal resummation and a properly gauge-invariant treatment are needed to be taken into account to obtain more accurate predictions. This can be done by using the state-of-the-art dimensional reduction method reducing the simplified G2HDM to a 3d effective field theory. To fully resolve the IR problem and properly identify the boundary between FOEWPT-viable and crossover regions of the parameter space, lattice studies are required. We will defer this exploration to the future. 

\newpage

\section*{Acknowledgments}

We would like to thank Eibun Senaha for useful discussions.
This work was supported in part by the National Natural Science Foundation of China, grant Nos. 19Z103010239, 12350410369 (VQT) and 11975150, 12375094 (MJRM) and the NSTC of Taiwan, grant Nos. 111-2112-M-001-035, 113-2112-M-001-001 (TCY) and 112-2811-M-001-089 (VQT). 
VQT would like to thank the Medium and High Energy Physics Group at the Institute of Physics, Academia Sinica, Taiwan for their hospitality during the course of this work.

\appendix


\section{Field Dependent Masses}
\label{app:FDMasses}

We will work in the Landau gauge where all the Goldstone bosons are massless and the ghost contributions can be ignored.
The thermal mass corrections for the scalar fields ($\Pi_{H_1} (T)$, $\Pi_{H_2} (T)$ and $\Pi_{\Phi_H} (T)$) and the longitudinal components of the gauge 
fields ($\Pi_W (T),\Pi_B (T),\Pi_{W^\prime} (T)$ and $\Pi_X (T)$) are also included in this work for the daisy resummation. 
These thermal mass corrections are given in Appendix~\ref{app:ThermalMasses}.


\subsection{Gauge Bosons}
\label{gaugebosonmasses}

In what follows, we will suppress the temperature dependence of the constant background fields
$h_{ic}(T)$ and $\phi_{Hc}(T)$ and denote them simply as $h_{ic}$ and $\phi_{Hc}$ unless stated otherwise.
For convenience, we define the following combinations of the background fields
\bea
h_\pm^2 & \equiv & h_{1c}^2  \pm h_{2c}^2 \; , \\
h_{\pm \Phi}^2 & \equiv & h_\pm^2 \pm \phi_{Hc}^2 \; .
\eea

There is only one electric charged vector boson, the SM $W^\pm$, with field dependent mass given by
\bea
m_{W^\pm_T}^2 & = & \frac{1}{4} g^2  h_+^2  \; , \\
\overline m_{W^\pm_L}^2 & = & m_{W^\pm_T}^2  + \Pi_W (T) \; .
\eea

In general, $W^{\prime\, 2}$ doesn't mix with other gauge fields and its field dependent mass is given by
\bea
m_{W^{\prime\,2}_T}^2 & = & \frac{1}{4} g_H^2 h_{+\Phi}^2  \; , \\
\overline m_{W^{\prime\,2}_L}^2 & = & m_{W^{\prime\,2}_T}^2 + \Pi_{W^\prime}  (T) \; .
\eea
 All the other 5 gauge bosons associated with the remaining 5 generators in G2HDM are electrically neutral and in general mixed together.
In the basis of $\{B, W^3, X, W^{\prime \, 3}, W^{\prime \, 1} \}$,
one finds the transverse mass matrix
\begin{eqnarray}
\label{MassVT}
M_{N_T}^2 
&=& 
\begin{pmatrix}
\frac{1}{4} g^{\prime\, 2} h_+^2 & - \frac{1}{4} g^\prime g h_+^2 &  \frac{1}{4} g^\prime g_X h_+^2 
& + \frac{1}{4} g^\prime g_H h_-^2 & \frac{1}{2} g^\prime g_H h_{1c} h_{2c}   \\
- \frac{1}{4} g g^{\prime} h_+^2 &  \frac{1}{4} g^2 h_+^2 & - \frac{1}{4} g g_X h_+^2 
& - \frac{1}{4} g g_H h_-^2 & - \frac{1}{2} g g_H h_{1c} h_{2c}  \\
 \frac{1}{4} g_X g^{\prime} h_+^2 &  - \frac{1}{4} g_X g h_+^2 &  \frac{1}{4} g_X^2 h_{+\Phi}^2  + m_X^2 
&  \frac{1}{4} g_X g_H h_{-\Phi}^2  & \frac{1}{2} g_X g_H h_{1c} h_{2c}  \\
 \frac{1}{4} g_H g^{\prime} h_-^2 &  - \frac{1}{4} g_H g h_-^2 & \frac{1}{4} g_H g_X h_{-\Phi}^2 
&  \frac{1}{4} g_H^2 h_{+\Phi}^2  & 0  \\
 \frac{1}{2} g_H g^{\prime} h_{1c} h_{2c} &  - \frac{1}{2} g_H g h_{1c} h_{2c} & \frac{1}{2} g_H g_X h_{1c} h_{2c} 
&  0 & \frac{1}{4} g_H^2 h_{+\Phi}^2  
\end{pmatrix} \; .\nonumber \\
\end{eqnarray}
It is clear that for $h_{1c} h_{2c} \neq 0$, $W^{\prime \, 1}$ can mix with $B$, $W^3$ and $X$. 
Thus in general $W^{\prime \,1}$ and $W^{\prime\, 2}$ have different masses and
do not combine into two complex conjugate fields (denoted by $W^{\prime \, (p,m)} = (W^{\prime\,1} \mp i W^{\prime\, 2} )/\sqrt 2$ in previous studies)
unless one of the $h_{1c}$ and $h_{2c}$ or both vanishes! 
Note that we have included a Stueckelberg mass $m_X$ for the $U(1)_X$ while for the 
$U(1)_Y$ we set its Stueckelberg mass $M_Y$ to be zero due to phenomenological reasons
discussed in earlier works. We also do not include kinetic mixing among the two $U(1)$s.

Despite its unappealing look, the determinant of Eq.~(\ref{MassVT}) vanishes for arbitrary values of $h_{1c}$, $h_{2c}$ and $\phi_{Hc}$,
all of which can be different from their  
vacuum values of $v_1$, $v_2$ and $v_\Phi$ respectively at zero temperature.
So it has at least one zero eigenvalue which can be identified as the photon~\footnote{Including a Stueckelberg mass $M_Y$ for the hypercharge $U(1)_Y$
gauge field $B$ would break this statement.}.
The exact results of the eigenvalues of  $\mathcal M_{N_T}^2$ depend on the detailed symmetry breaking patterns.
We will denote the eigenstates of Eq.~(\ref{MassVT}) as the transverse $\gamma_T$ and $Z_{iT}$ with eigenvalues $m^2_{\gamma_T} = 0$ and $m^2_{Z_{iT}}$ for $i=1,2,3$, and 4 respectively.

For the longitudinal components of the neutral gauge bosons, we have to include their thermal masses as well. The mass matrix now becomes
\beq
\label{MassVL}
\overline M_{N_L}^2 = M_{N_T}^2 + 
\begin{pmatrix}
\Pi_B  (T) & 0 & 0 & 0 & 0 \\
0 & \Pi_W  (T) & 0 & 0 & 0 \\
0 & 0 & \Pi_X  (T) & 0 & 0 \\
0 & 0 & 0 & \Pi_{W^\prime}  (T) & 0  \\
0 & 0 & 0 & 0 & \Pi_{W^\prime}  (T)
\end{pmatrix} \; .
\eeq 
We will denote the eigenstates of $\overline M_{N_L}^2$ in (\ref{MassVL}) as the longitudinal $\gamma_L$ and $Z_{iL}$ with eigenvalues $\overline m^2_{\gamma_L}$ and $\overline m^2_{Z_{iL}}$ for  
$i=1,2,3,$ and 4 respectively.

Setting all the thermal masses to be zeros, $h_{1c} \to v$, $h_{2c} \to 0$ and $\phi_{Hc} \to v_\Phi$ in the above formulae, one can reproduce the tree level mass spectrum and mixings for the neutral gauge bosons presented in 
Sec.~{\ref{TLMassesMixings}} for the zero temperature case. For instance, $W^{\prime}_1$ and $W^{\prime}_2$ will have the same mass and can combine into two complex fields $W^{\prime \, (p,m)}=(W^{\prime}_1 \mp i W^{\prime}_2)/\sqrt 2$ which will be considered as sub-GeV DM candidate in this work.

\subsection{Scalar Bosons}
\label{scalarmasses}

In G2HDM, the charged scalar $h_1^+$ and $h_2^+$ in~(\ref{eq:scalarfields}) mix  according to the following mass matrix
\beq
M_C^2 = 
\begin{pmatrix}
-\mu_H^2 + \lambda_H h_+^2 - \frac{1}{2} \left( \lambda_H^\prime  h_{2c}^2 - \lambda_{H\Phi} \phi_{Hc}^2 \right) & \frac{1}{2} \lambda_H^\prime h_{1c} h_{2c} \\
 \frac{1}{2} \lambda_H^\prime h_{1c} h_{2c} & -\mu_H^2 + \lambda_H h_+^2 - \frac{1}{2} \left( \lambda_H^\prime  h_{1c}^2 - \widetilde\lambda_{H\Phi} \phi_{Hc}^2 \right)
\end{pmatrix} 
\eeq
Here we have defined $\widetilde\lambda_{H\Phi} \equiv \lambda_{H\Phi} + \lambda_{H\Phi} ^\prime$.
Note that $M_C^2$ depends on the values of $\mu_H^2$, its eigenvalues also depend on the detail of the
symmetry breaking patterns. 
Including the thermal mass corrections, we have
\beq
\overline M_C^2 = M_C^2  + 
\begin{pmatrix}
\Pi_{H_1} (T) & 0 \\
0 & \Pi_{H_2}  (T)
\end{pmatrix} \; .
\eeq
We will denote the two eigenstates as $G^\pm$ and $H^\pm$ and their corresponding eigenvalues of $M_C^2$ ($\overline M_C^2$) are 
$m^2_{G^\pm}$ and $m^2_{H^\pm}$ ($\overline m^2_{G^\pm}$ and $\overline m^2_{H^\pm}$) respectively.
Explicitly, we have
\bea
\label{MChargedHiggs}
\overline m^2_{G^\pm,H^\pm} & = & \frac{1}{4} \biggl\{
- 2 \left( 2 \mu_H^2 -  \Pi_{H_+}  \right) + \left( 4 \lambda_H - \lambda_H^\prime \right) h_+^2 + \left( \lambda_{H\Phi} + \widetilde\lambda_{H\Phi} \right) \phi_{Hc}^2
\biggr. \nonumber \\
&  \mp & \biggl. \biggl[ \lambda_H^{\prime \, 2} h_+^4 - 2 \lambda_H^\prime  h_-^2 \left( \lambda_{H\Phi}^\prime \phi_{Hc}^2  - 2 \Pi_{H_-}  \right) 
+  \left( \lambda_{H\Phi}^\prime \phi_{Hc}^2  - 2 \Pi_{H_-}  \right)^2
\biggr]^{1/2} \biggr\} \; , \nonumber \\
\eea
with $\Pi_{H_\pm} = ( \Pi_{H_1} \pm \Pi_{H_2}) $ and $m^2_{G^\pm,H^\pm}$ is given by (\ref{MChargedHiggs}) with $\Pi_{H_{\pm}}$ set to $0$.

In the general spontaneous symmetry breaking in G2HDM, all the other scalar fields in the three doublets mix together according to their CP properties. 
This is expected since the parameters in the scalar potential of G2HDM are all real and the scalar potential is CP conserving. 
The real and imaginary fields have opposite CP properties and they should not mix!

For the CP-even scalars in the basis of $\{ h_{\rm SM}, h_{2}^0, \phi_H , G_H^1 \}$,
the mass matrix is given by 
\begin{eqnarray}
\label{MSsq}
M_{S}^2 = 
\begin{pmatrix}
 (\sharp)_{11} & 2 \lambda_H h_{1c} h_{2c} & \lambda_{H\Phi} h_{1c} \phi_{Hc}  & \frac{1}{2} \lambda_{H\Phi}^\prime h_{2c} \phi_{Hc}  \\
2 \lambda_H h_{1c} h_{2c}  &  (\sharp)_{22} & \widetilde \lambda_{H\Phi} h_{2c} \phi_{Hc}  &  \frac{1}{2} \lambda_{H\Phi}^\prime h_{1c} \phi_{Hc}   \\
\lambda_{H\Phi} h_{1c} \phi_{Hc}  & \widetilde\lambda_{H\Phi} h_{2c} \phi_{Hc}  & (\sharp)_{33} & \frac{1}{2} \lambda_{H\Phi}^\prime h_{1c} h_{2c}  \\
\frac{1}{2} \lambda_{H\Phi}^\prime h_{2c} \phi_{Hc} & \frac{1}{2} \lambda_{H\Phi}^\prime h_{1c} \phi_{Hc} & \frac{1}{2} \lambda_{H\Phi}^\prime h_{1c} h_{2c} & (\sharp)_{44} \\
\end{pmatrix} \; ,
\end{eqnarray}
with the diagonal elements defined as follows
\begin{eqnarray}
\label{MSesqDiag}
(\sharp)_{11} & = & -\mu_H^2 + \lambda_H (3 h_{1c}^2 + h_{2c}^2 ) + \frac{1}{2}\lambda_{H\Phi} \phi_{Hc}^2 \;, \\
(\sharp)_{22} & = & -\mu_H^2 + \lambda_H ( h_{1c}^2 + 3 h_{2c}^2 ) + \frac{1}{2}\widetilde \lambda_{H\Phi} \phi_{Hc}^2 \;,\\
(\sharp)_{33} & = &  - \mu_\Phi^2 + 3 \lambda_\Phi \phi_{Hc}^2  + \frac{1}{2} \left(  \lambda_{H\Phi} h_+^2 
+ \lambda_{H\Phi}^\prime h_{2c}^2 \right) \;,\\
(\sharp)_{44} & = & -\mu_\Phi^2 + \lambda_\Phi \phi_{Hc}^2 + \frac{1}{2} \left( \lambda_{H\Phi} h_+^2 + \lambda_{H\Phi}^\prime h_{1c}^2 \right) \;.
\end{eqnarray}
Including the thermal masses, we have
\beq
\label{MSsqThermal}
\overline M_{S}^2 =  M_{S}^2 + 
\begin{pmatrix}
 \Pi_{H_1} (T) & 0 & 0 & 0 \\
 0 & \Pi_{H_2} (T)  & 0 & 0 \\
 0 & 0 & \Pi_{\Phi_H} (T)  & 0 \\
 0 & 0 & 0 & \Pi_{\Phi_H} (T)  
\end{pmatrix}
\; .
\eeq
We denote the eigenvalues of $M_S^2$ and $\overline M_S^2$ as $m^2_{S_i}$ and $\overline m^2_{S_i}$ respectively for the CP-even states $S_i$ with $i=1,2,3,$ and $4$.

For the CP-odd scalars in the basis $\{ G_1^0, G_2^0, G_H^0, G_H^2  \}$, 
the mass matrix is
\begin{eqnarray}
\label{MSosq}
M_{P}^2 = 
\begin{pmatrix}
(\natural)_{11} & 0 & 0 & \frac{1}{2} \lambda_{H\Phi}^\prime h_{2c} \phi_{Hc}  \\
0 & (\natural)_{22} & 0 &  -\frac{1}{2} \lambda_{H\Phi}^\prime h_{1c} \phi_{Hc} \\
0 & 0 & (\natural)_{22} & \frac{1}{2} \lambda_{H\Phi}^\prime h_{1c} h_{2c} \\
\frac{1}{2} \lambda_{H\Phi}^\prime h_{2c} \phi_{Hc} & -\frac{1}{2} \lambda_{H\Phi}^\prime h_{1c} \phi_{Hc}  & \frac{1}{2} \lambda_{H\Phi}^\prime h_{1c} h_{2c} & (\natural)_{44} 
\end{pmatrix} \; ,
\end{eqnarray}
with the following diagonal elements
\begin{eqnarray}
\label{MSosqDiag}
(\natural)_{11} & = & -\mu_H^2 + \lambda_H h_+^2 + \frac{1}{2} \lambda_{H\Phi} \phi_{Hc}^2\;,\\
(\natural)_{22} & = & -\mu_H^2 + \lambda_H h_+^2 + \frac{1}{2} \widetilde\lambda_{H\Phi} \phi_{Hc}^2 \;,\\
(\natural)_{33} & = & -\mu_\Phi^2 + \lambda_\Phi \phi_{Hc}^2 + \frac{1}{2} \left( \lambda_{H\Phi} h_+^2  + \lambda_{H\Phi}^\prime h_{2c}^2 \right) \;,\\
(\natural)_{44} & = &  -\mu_\Phi^2 +  \lambda_\Phi \phi_{Hc}^2 + \frac{1}{2} \left( \lambda_{H\Phi} h_+^2 + \lambda_{H\Phi}^\prime h_{1c}^2 \right) \; .
\end{eqnarray}
Similarly, including the thermal masses, we have
\beq
\label{MPsqThermal}
\overline M_{P}^2 =  M_{P}^2 + 
\begin{pmatrix}
 \Pi_{H_1} (T) & 0 & 0 & 0 \\
 0 & \Pi_{H_2} (T)  & 0 & 0 \\
 0 & 0 & \Pi_{\Phi_H} (T)  & 0 \\
 0 & 0 & 0 & \Pi_{\Phi_H} (T)  
\end{pmatrix}
\; .
\eeq
We denote the eigenvalues of $M_P^2$ and $\overline M_P^2$ as $m^2_{P_i}$ and $\overline m^2_{P_i}$ respectively for the CP-odd states $P_i$ with $i=1,2,3,$ and $4$.

Since the matrix elements of $M_S^2 (\overline M_S^2)$ and  $M_P^2 (\overline M_P^2) $ 
depend on the values of $\mu_H^2$ and $\mu_\Phi^2$, the scalar mass spectrum depends on the detail of the
symmetry breaking patterns just like the gauge bosons. 
Similarly one can reproduce the zero temperature mass spectra and mixings for the scalars presented in Sec.~\ref{TLMassesMixings} by ignoring the thermal masses and setting 
$h_{1c}\to v$, $h_{2c} \to 0$ and $\phi_{Hc} \to v_\Phi$ in the above formulae. 
For example, $G_H^1$ and $G_H^2$ will have the same mass and they can be combined into two complex
Goldstone bosons $G^{(p,m)}_H=(G_H^1 \pm i G_H^2 )/\sqrt 2$ absorbed by 
the longitudinal components of $W^{\prime \, (p,m)}$, and $h_{\rm SM}$ and $\phi_H$ can be combined into $h_1$ and $h_2$, and so on.

\subsection{Fermions}
\label{FermionMasses}

The gauge invariant Yukawa couplings in G2HDM lead to the following fermionic mass terms in the Lagrangian~\cite{Huang:2015wts}
\begin{eqnarray}
\label{fermionmassmatrix}
\mathcal L_{\rm mass}^f & = & -\frac{1}{\sqrt 2} \left( \bar u_L, \bar u_L^H \right) 
\begin{pmatrix}
v_1 \mathcal Y_u & v_2 \mathcal Y_u \\
0 & v_\Phi \mathcal Y_u^\prime 
\end{pmatrix} 
\begin{pmatrix}
u_R \\
u^H_R
\end{pmatrix} + {\rm H.c.} \nonumber \\
 && -\frac{1}{\sqrt 2} \left( \bar d_L, \bar d_L^H \right) 
\begin{pmatrix}
v_1 \mathcal Y_d & - v_2 \mathcal Y_d \\
0 & v_\Phi \mathcal Y_d^\prime 
\end{pmatrix} 
\begin{pmatrix}
d_R \\
d^H_R
\end{pmatrix} + {\rm H.c.} \nonumber \\
 && -\frac{1}{\sqrt 2} \left( \bar \nu_L, \bar \nu_L^H \right) 
\begin{pmatrix}
v_1 \mathcal Y_\nu &  v_2 \mathcal Y_\nu \\
0 & v_\Phi \mathcal Y_\nu^\prime 
\end{pmatrix} 
\begin{pmatrix}
\nu_R \\
\nu^H_R
\end{pmatrix} + {\rm H.c.} \nonumber \\
 && -\frac{1}{\sqrt 2} \left( \bar e_L, \bar e_L^H \right) 
\begin{pmatrix}
v_1 \mathcal Y_e & - v_2 \mathcal Y_e \\
0 & v_\Phi \mathcal Y_e^\prime 
\end{pmatrix} 
\begin{pmatrix}
e_R \\
e^H_R
\end{pmatrix} + {\rm H.c.}
\end{eqnarray} 
Since the lower left blocks in the mass matrices vanish, the SM fermions $f_R$ will not mix with the hidden fermions $f^H_L$ in G2HDM even if $v_2 \neq 0$. The Yukawa matrices $\mathcal Y_f$ and $\mathcal Y^\prime_f$ can be blocked diagonalized independently. 
Since fermion loops do not suffer from the infrared problems associated with the zero Matsubara frequencies in the bosonic case, daisy resummation is not necessary. The leading temperature effects for the fermion masses are usually captured via the following prescriptions
\bea
m^2_f (v_1)  =  \frac{1}{2} y_f^2 v_1^2 & \rightarrow & m^2_f (h_{1c})  =  \frac{1}{2} y_f^2 h_{1c}(T)^2 = m^2_f (v_1) \frac{h_{1c}(T)^2}{v_1^2}\; , \\
m^2_{f^H} (v_\Phi)  =  \frac{1}{2} y_{f^H}^2 v_\Phi^2 & \rightarrow & m^2_{f^H} (\phi_{Hc})  =  \frac{1}{2} y_{f^H}^{ 2} \phi_{Hc}(T)^2 = m^2_{f^H} (v_\Phi) \frac{\phi_{Hc}(T)^2}{v_\Phi^2}\; , \;\;\;
\eea
where $y_f^2$ and $y^2_{f^H}$ are the eigenvalues of the  
Yukawa matrices 
$\mathcal{Y}_f \mathcal{Y}_f^\dagger$ (or $\mathcal{Y}_f^\dagger \mathcal{Y}_f$) and $ \mathcal{Y}^\prime_f \mathcal{Y}_f^{\prime \dagger} $  
(or $ \mathcal{Y}^{\prime \dagger}_f \mathcal{Y}_f^{\prime} $) respectively.
The temperature dependence for the fermion masses are coming through the classical fields $h_{1c}(T)$ and $\phi_{Hc}(T)$ only.
For simplicity, we will consider just the heavy top quark $t$ and its hidden heavy partner $t^H$ from the third generation in the numerical work since they represent the most dominant contribution from the fermion sector.

\section{Thermal Masses}
\label{app:ThermalMasses}

In this appendix, we list the thermal mass (Debye screening mass) for all the scalar and vector bosons in G2HDM. 
Including these masses are essential for the resummation of the daisy diagrams in the finite temperature effective potential.


For the two doublets $H_1$ and $H_2$, using the field components defined in Eq.~(\ref{eq:scalarfields}),
we found
\begin{eqnarray}
\label{PiH1}
\Pi_{H_1} (T) & \equiv & \Pi_{h_1}  (T)  =  \Pi_{G_1^0}  (T) = \Pi_{h_1^\pm}  (T) = \tilde \Pi_{H_1} T^2 \nonumber \\
  & = &  \frac{1}{12} \left( 10 \lambda_H - \lambda_H^\prime + 2 \lambda_{H\Phi} + \lambda^\prime_{H\Phi} \right) T^2  \\
&&  + \frac{1}{16} \left( 3 g^2 + g^{\prime\, 2} + 3 g_H^2 + g_X^2 \right) T^2 
+  \frac{1}{4} y_t^2 T^2 \; , \nonumber \\
\label{PiH2}
\Pi_{H_2} (T) & \equiv &  \Pi_{h_2}  (T) = \Pi_{G_2^0}  (T) = \Pi_{h_2^\pm}  (T)  = \tilde \Pi_{H_2} T^2 \nonumber \\
  & = &  \frac{1}{12} \left( 10 \lambda_H - \lambda_H^\prime + 2 \lambda_{H\Phi} + \lambda^\prime_{H\Phi} \right) T^2  \nonumber \\
  && + \frac{1}{16} \left( 3 g^2 + g^{\prime\, 2} + 3 g_H^2 + g_X^2 \right) T^2  \; , 
\end{eqnarray}
and for the hidden doublet $\Phi_H$, using Eq.~(\ref{eq:scalarfieldPhiH}), we got
\begin{eqnarray}
\label{PiPhiH}
\Pi_{\Phi_H} (T) & \equiv & \Pi_{\phi_H}  (T)  =  \Pi_{G_H^0}  (T) = \Pi_{G_H^{(p,m)}}   (T)  = \tilde \Pi_{\Phi_H} T^2 \nonumber \\
  & = & 
 \frac{1}{6} \left( 3 \lambda_{\Phi} + 2 \lambda_{H\Phi} + \lambda^\prime_{H\Phi} \right) T^2 
 + \frac{1}{16} \left( 3 g_H^2 + g_X^2 \right) T^2 
+  \frac{1}{4} y^{\prime \, 2}_t T^2 \; .
\end{eqnarray}
For the Yukawa couplings, we will keep only the contributions from the SM top quark ($y_t$) and its heavy partner ($y_t'$) in G2HDM. 

For the SM gauge bosons, we have
\begin{eqnarray}
\Pi_W  (T) & \equiv & \Pi_{W^i}   (T)=  2 g^2 T^2 \; , \; \;\; \; (i =1,2,3) \\
\Pi_B  (T) & = & \frac{14}{3} g^{\prime \, 2} T^2 \; .
\end{eqnarray}
For the hidden gauge bosons, we have
\begin{eqnarray}
\Pi_{W^\prime}  (T) & \equiv & \Pi_{W^{\prime \, i}}   (T) =  \frac{19}{6 } g_H^2 T^2 \; , \; \;\; \; (i =1,2,3) \\
\Pi_X  (T) & = & \frac{17}{2} g^2_X T^2 \; .
\end{eqnarray}
In the thermal masses of the gauge bosons, we have summed over all three generations of quarks and leptons in G2HDM.

\section{Thermal Integrals}
\label{app:ThermalIntegrals}

The definition of $J_i (x)$ depends on whether $i$ is a boson $(B)$ or a fermion $(F)$, {\it i.e.}
\begin{equation}
J_{B/F} (x) \equiv \int_0^\infty {\rm d} y \, y^2 \log \left[ 1 \mp \exp \left( - \sqrt{x^2 + y^2 } \right) \right] \; .
\end{equation}
Note that in our notation, $x \equiv m_{B/F} / T$. In the literature, some authors' $x$ is our $x^2$.

For high-T where $x \ll 1$, one has the following series expansions~\cite{Quiros:1999jp}
\begin{align}
\label{JB}
J_B( x ) & = - \frac{\pi^4}{45} + \frac{\pi^2}{12} x^2 - \frac{\pi}{6} \left( x^2 \right) ^{3/2} 
- \frac{1}{32} x^4 \log \left( \frac{x^2}{c_B} \right) \nonumber \\
& - \frac{1}{ 8 \pi^{1/2} } x^4 \sum_{l=1}^{\infty} (-1)^l \frac{\zeta (2 l + 1) }{(l+1)!} 
\Gamma \left (l+\frac{1}{2} \right) \left( \frac{x^2}{4 \pi^2} \right)^{l} \; , \\
\label{JF}
J_F( x ) & = \frac{7}{8} \cdot \frac{\pi^4}{45} - \frac{\pi^2}{24} x^2 
- \frac{1}{32} x^4 \log \left( \frac{x^2}{c_F} \right) \nonumber \\
& -  \frac{1}{4 \pi^{1/2}} x^4 \sum_{l=1}^{\infty}   (-1)^l  \left(1 - \frac{1}{2^{2 l +1}} \right) \frac{\zeta (2 l + 1) }{(l+1)!} 
\Gamma\left( l+\frac{1}{2} \right) \left( \frac{x^2}{ \pi^2} \right)^{l} \; ,
\end{align}
 where $c_B = 16 \pi^2 \exp \left( \frac{3}{2} - 2 \gamma_E \right)$, $c_F = c_B/16$, and 
 $\Gamma(z)$ and $\zeta(z)$ are the Euler $\Gamma$-function and Riemann $\zeta$-function respectively.
 Numerically, we have
 \begin{align}
 \gamma_E & = 0.5772156649015329\cdots \; ,\nonumber \\
 c_B & = 223.0993446886696\cdots \; ,\\ 
 c_F & = 13.94370904304185\cdots \nonumber \; . 
 \end{align}
 Note that there is no $\left( x^2 \right)^{3/2}$ term in $J_F$! For high-T, keeping the first lines in (\ref{JB}) and (\ref{JF}) 
 are usually sufficient!

On the other hand, we also have the following series representations~\cite{Anderson:1991zb}
\begin{align}
J_{B} (x) & = - \sum_{n=1}^\infty \frac{1}{n^2} x^2 K_2 (n x) \; , \\
J_{F} (x) & = - \sum_{n=1}^\infty \frac{1}{n^2}  \left( -1 \right)^n  x^2 K_2 (n x) \; , 
\end{align}
where $K_2(z)$ is the modified Bessel function of second kind which falls off exponentially for large positive values of $z$.
For high mass or low temperature where $x \gg 1$, these series representations converge rapidly. Usually, taking the first five 
terms in the series is sufficient~\cite{Bernon:2017jgv,Fabian:2020hny}.
 
The thermal integral ${\mathcal J}_i (x_i)$ in (\ref{VT1loop}) is defined by~\cite{Arnold:1992rz,Carrington:1991hz} 
\begin{equation}
\label{ThermalIntegral_J}
{\mathcal J}_i \left( \frac{m_i}{T} \right) = \left\{ 
\begin{array}{ll}
J_B ( \frac{m_i}{T} ) - \frac{\pi}{6} \left( \frac{ \overline m_i^{3} }{T^3} - \frac{ m_i^{3} }{T^3} \right) \, , 
 &  i \in \left\{ S_{1},S_{2},S_{3},S_{4},P_{1},P_{2},P_{3},P_{4},G^\pm,H^\pm \right.\\
 & \;\;\;\;\;\;\;\; W^\pm_L, W^{\prime\,2}_L, Z_{1L}, Z_{2L}, Z_{3L}, Z_{4L}, \gamma_L \} \; ; \\
J_B ( \frac{m_i}{T} )\, , & i \in \{ W^\pm_T, W^{\prime\,2}_T, Z_{1T}, Z_{2T}, Z_{3T}, Z_{4T}, \gamma_T \} \; ; \\
J_F ( \frac{m_i}{T} ) \, , & i \in \{ l, \nu, l^H, \nu^H, q, q^H \}.
\end{array}
\right. 
\end{equation}
Note that for the scalars and longitudinal components of the gauge bosons we have included the resummation effects from the
daisy (ring) diagrams from higher loops in the thermal corrections.  
$\overline m_i$ and $m_i$ are the field dependent masses (mass eigenvalues in case of mixings) 
with and without the thermal mass correction respectively for particle $i$.

It has been shown recently~\cite{Curtin:2016urg} that at high-T the daisy resummation can be quite poor 
for large values of the quartic coupling in BSM and does not shown how decoupling occurs when the internal mass of the BSM particle 
is large compared with the temperature. More accurate calculation would have to include the superdaisy diagrams which will lead one 
to solve the gap equations~\cite{Curtin:2016urg}. 
We will not get into this business here. Also ignored here is the two-loop corrections to the finite temperature effective potential 
which have been shown quite relevant to determine the strength of the phase transition in MSSM~\cite{Espinosa:1996qw}.

\newpage

\section{Gauge-Invariant Method Beyond Leading Order and Scale Dependence}
\label{app:GaugeInvariantMethod}

The determination of the phase transition quantities such as $v_C$ and $T_C$ in a gauge theory can lead to the gauge fixing parameter dependence if the effective potential is calculated beyond the leading high-T approximation.  However one can treat the gauge dependence issue using the $\hbar$-expansion method proposed in Ref.\cite{Patel:2011th}. Such method is based on the Nielsen-Fukuda-Kugo (NFK) identity~\cite{Nielsen:1975fs,Fukuda:1975di}, 
which states that the energies at the stationary points of the effective potential are independent of the gauge fixing parameter $\xi$.
The NFK identity can be described by
\begin{align}
\label{eq:NFKidentity}
\frac{\partial V_{\text{eff}}(\varphi)}{\partial \xi}  = -C(\varphi,\xi)\frac{\partial V_{\text{eff}}(\varphi)}{\partial \varphi} \; ,
\end{align}
where the functional $C(\varphi, \xi)$ can be found explicitly in Ref.~\cite{Nielsen:1975fs}.

By expanding both the effective potential and the functional $C(\varphi, \xi)$ in powers of $\hbar$,
\begin{align}
V_{\text{eff}}(\varphi) &= V_0(\varphi)+\hbar V_1(\varphi)+\hbar^2 V_2(\varphi)+\cdots, \\
C(\varphi,\xi) &= c_0+\hbar c_1(\varphi)+\hbar^2 c_2(\varphi)+\cdots, 
\end{align}
and, for example, considering the first order with respect to $\hbar$, one can obtain from (\ref{eq:NFKidentity})
\begin{align}
\frac{\partial V_1}{\partial \xi} &= -c_1\frac{\partial V_0}{\partial\varphi} \; ,
\end{align}
which implies that the $\xi$-dependence of $V_1$ can be eliminated at the extremal points of the tree-level potential. 
Therefore, the critical temperature to $\mathcal{O}(\hbar)$ can be determined by evaluating 
the degeneracy condition of the effective potential at the stationary points of the tree-level potential.
For the transition from the $(0, h_{2c}, 0)$ phase to $(h_{1c}, 0, \phi_{Hc})$ phase in G2HDM as depicted in the right panel of Fig.~\ref{fig:twostepsketch} where $s \equiv \phi_{Hc}$, $h_1 \equiv h_{1c}$ and $h_2 \equiv h_{2c}$, 
the critical temperature $T_C$ can be determined by solving the following equation
\begin{align}
V_0 (0, v_2, 0 )+V_1 (0, v_2, 0, T_C)
= V_0(v_1, 0, v_\Phi)+V_1(v_1, 0, v_\Phi, T_C) \; ,
\label{eq:Tc}
\end{align}
where $(0, v_2, 0)$ and $(v_1, 0, v_\Phi)$ are the extremal points at the tree-level potential. 
On the other hand, the value of $v_C$ is determined by utilizing the high-T expansion effective potential. 

Another issue that arises in the calculation beyond the high-T expansion effective potential is the dependence on the renormalization scale $\mu$ in the Coleman-Weinberg effective potential $V_{\rm CW}$.
This dependence can have a significant impact on the properties of phase transitions and the associated predicted GW signals \cite{Croon:2020cgk, Gould:2021oba}. To mitigate this dependence, renormalization group (RG) improvement can be applied to the effective potential \cite{Kastening:1991gv, Bando:1992np, Ford:1992mv, Funakubo:2023cyv, Funakubo:2023eic}. This can be done by running the parameters in the tree-level effective potential using the $\beta$ functions given in Appendix~\ref{app:RGE}. At leading order, the RG improved potential satisfies
\begin{equation}
\label{eq:RG-improve}
\mu \frac{\rm d}{\rm d \mu} \left[ V_0 (\mu) + V_{\rm CW}(\mu) \right] = 0 \; ,
\end{equation}
which ensures independence from the choice of $\mu$. 

\vfill

\section{The Critical Temperatures}
\label{app:CriticalTemp}

In this section, we show the analytical expressions for the critical temperatures in the two-step phase transition. 
The critical temperature for the first step [$(0,0,0) \to (0,h_{2c},0)$] can be easily obtained by solving the degenerate condition 
$V_{\rm eff}^{\rm HT} (0,0,0, T_{1C}) = V_{\rm eff}^{\rm HT} (0,h_{2c},0, T_{1C})$. 
This gives the result of 
\be
T_{1C} = \sqrt{\mu_H^2/\tilde\Pi_{H_2}} \; ,
\ee
where $\mu_H^2 = \lambda_H v^2 + \lambda_{H\Phi}v_\Phi^2 /2 $ 
fixed by the tree level tadpole condition
and $\tilde\Pi_{H_2}$ is given by (\ref{PitildeH2}).

For the second step $[ (0,h_{2c},0) \to (h_{1c},0,\phi_{Hc})]$, by solving the following degenerate condition
\be
V_{\rm eff}^{\rm HT} (0,h_{2c},0, T_{C}) = V_{\rm eff}^{\rm HT} (h_{1c},0,\phi_{Hc}, T_{C}) \; ,
\ee
one obtains 
\be
T_C = \sqrt{\frac{\left(\lambda_{H\Phi}^2 - 4\lambda_H \lambda_\Phi\right) v_{\Phi}^4} {2 \left(\lambda_{H\Phi} \tilde\Pi_{H_2} - 2 \lambda_{H} \tilde\Pi_{\Phi_H} \right) v_{\Phi}^2 - \lambda_H y_t^2 v^2 - \kappa } } \; ,
\ee
where 
\bea
\kappa &=& y_t \lambda_{H}^{1/2} \left\{ \left[4 \lambda_{H\Phi} \tilde\Pi_{\Phi_H} - \lambda_\Phi (8 \tilde\Pi_{H_2} +  y_t^2) \right]v_\Phi^4 \right. \nonumber \\
& & \;\;\;\;\; \left. + \, 4\left(2 \lambda_H \tilde\Pi_{\Phi_H} -  \lambda_{H\Phi} \tilde\Pi_{H_2} \right) v^2 v_{\Phi}^2 + \lambda_H y_t^2 v^4 \right\}^{1/2}  \; , \\ 
\tilde\Pi_{\Phi_H} &=& \frac{1}{6} \left( 3 \lambda_{\Phi} + 2 \lambda_{H\Phi} + \lambda^\prime_{H\Phi} \right)
 + \frac{1}{16} \left( 3 g_H^2 + g_X^2 \right) 
+  \frac{1}{4} y^{\prime \, 2}_t \;.
\eea
We note that with the viable parameter space in the model, the critical temperature $T_C$ stays in the electroweak scale as shown in Fig.~\ref{fig:paraPT-2step-TC}.

\vfill

\section{Renormalization Group Equations}
\label{app:RGE}

Here we present the one loop renormalization group equations in G2HDM. 
The $\beta$-function, computed with the help of \texttt{SARAH}~\cite{Staub:2013tta}, is defined as 
$\beta_X \equiv \mu {\mathrm d} X/{\mathrm d} {\mu}$, 
where $X$ denotes a generic coupling/mass and $\mu$ is the renormalization scale.

Gauge couplings:
\bea
\beta_{g'} &=& \frac{1}{16 \pi^2 } \left( \frac{53}{3} g^{\prime 3} \right) \; ,  \\
\beta_{g} &=& \frac{1}{16 \pi^2 } \left( -3 g^3 \right) \; ,  \\
\beta_{g_s} &=& \frac{1}{16 \pi^2 } \left( -3 g_s^3 \right) \; ,  \\
\beta_{g_X} &=& \frac{1}{16 \pi^2 } \left( \frac{17}{2} g_X^3 \right) \; ,  \\
\beta_{g_H} &=& \frac{1}{16 \pi^2 } \left( \frac{7}{6} g_H^3 \right) \; .
\eea

Scalar masses and quartic couplings:
\bea
\beta_{\mu_H^2} &=& \frac{1}{16 \pi^2 } \biggl[
  4\mu_{H}^2 \left(5\lambda_{H} - \frac{1}{2}\lambda_{H}^{\prime}\right)  
  - \frac{3}{2}\mu_{H}^2 \left(3g^2 + g^{\prime 2} + 3g_{H}^2 + g_{X}^2 - 4y_t^2\right) \biggr. \nonumber \\
&&   + \biggl. \; 2 \mu_{\Phi}^2 \left(2\lambda_{H\Phi} +  \lambda_{H\Phi}^{\prime} \right) \biggr] \; ,\\
 \beta_{\mu_\Phi^2} &=&  \frac{1}{16 \pi^2 } \biggl[ 12 \mu_{\Phi}^2 \lambda_{\Phi} - \frac{3}{2} \mu_{\Phi}^2 \left(3 g_H^2 + g_X^2 - 4 y_{t^H}^2 \right) + 4 \mu_{H}^2 \left(2 \lambda_{H\Phi} + \lambda^{\prime}_{H\Phi}\right) \biggr] \; , \\
 \nonumber \\
\beta_{\lambda_H} &=& \frac{1}{16 \pi^2 } \biggl\{
32 \lambda_H^2 + \lambda_H^{'2}  + 2 \lambda_{H\Phi}^2  +  \lambda_{H\Phi}^{'2} - 4 \lambda_H \lambda'_H +  2 \lambda_{H\Phi}  \lambda'_{H\Phi} \biggr.  \nonumber \\
 && 
 - \; 3 \, \lambda_H \left(  3 g^2 +g^{\prime2} + 3 g_H^2  + g_X^2 \right)  \nonumber \\
&&  
 + \; \frac{3}{8} \biggl[ g^4 + 3 g_H^4 +  2 g_H^2 \left(g^{\prime2} + g_X^2\right) 
 + \left(g^{\prime2} + g_X^2\right)^2 + 2 g^2 \left(g^{\prime 2} + g_H^2 + g_X^2\right) \biggr]
  \nonumber \\
&& 
 + \biggl. \; 6 y_t^2 \left( 2 \lambda_H  -  y_t^2 \right) \biggr\} \; , \\
%
 %
 \beta_{\lambda'_H} &=& \frac{1}{16 \pi^2 } \biggl[
  2\left( 2 \lambda_H^{\prime2} - \lambda_{H\Phi}^{\prime2} + 12 \lambda_H \lambda'_H \right)
 - 3 \lambda'_H \left( 3 g^2  + g^{\prime2}  + 3 g_H^2  + g_X^2 \right) \biggr. \nonumber \\
 &&
 + \biggl. \; 3 g_H^2 \left( 2 g^2 - g^{\prime2} - g_X^2 \right)  -3 g^2 \left( g^{\prime 2} + g_X^2 \right) 
 +12 \left( 2 \lambda_H + 3 \lambda'_H \right) y_t^2 \biggr] \; ,
\eea

\bea
 \beta_{\lambda_\Phi} &=& \frac{1}{16 \pi^2 } \biggl[
  4 \left(  6 \lambda_{\Phi}^2 + \lambda_{H\Phi}^2 + \lambda_{H\Phi} \lambda_{H\Phi}^{\prime} + \frac{1}{2} \lambda_{H\Phi}^{\prime 2} \right) - 3 \lambda_{\Phi} \left(3 g_H^2 + g_X^2 \right) \biggr. \nonumber \\
&& \biggl. 
+\;\frac{3}{8} \left(3 g_H^4 + g_X^4 + 2 g_H^2 g_X^2 \right) + 6y_{t^H}^2 \left(2 \lambda_{\Phi}  - y_{t^H}^2 \right) \biggr] \; , \\ %
 %
  %
  \beta_{\lambda_{H \Phi}} &=&  \frac{1}{16 \pi^2 } \biggl[ 2\left( 10\lambda_H\lambda_{H\Phi} + 2\lambda_{H\Phi}^2 + 4\lambda_H \lambda^{\prime}_{H\Phi} + \lambda^{\prime 2}_{H\Phi} \right. \biggr. \nonumber \\
  && \;\;\; \; \;\; \;\;\; \; \; \; \left.
    - \; \lambda_H \lambda^{\prime}_{H} - \lambda^{\prime}_{H\Phi}\lambda^{\prime}_{H} + 6\lambda_{H\Phi}\lambda_{\Phi} + 2\lambda^{\prime}_{H\Phi} \lambda_{\Phi} \right)  \nonumber \\ 
 && 
 - \; \frac{3}{2} \left( g^{\prime2} + 3 g^2 + 6 g_H^2 + 2 g_X^2\right) \lambda_{H\Phi}  
 + \frac{3}{4} \left(3 g_H^4 + g_X^4 - 2 g_H^2 g_X^2  \right) 
 \nonumber \\ 
&&  
 + \biggl. \; 6 \lambda_{H\Phi} \left( y_t^2 + y_{t^H}^2 \right) \biggr] \; , \\ 
  %
  %
 \beta_{\lambda_{H \Phi}^\prime} &=&  \frac{1}{16 \pi^2 } \biggl[
2  \lambda'_{H\Phi}  \left(  2  \lambda'_{H\Phi} + 2 \lambda_H + \lambda'_H + 2 \lambda_\Phi  + 4  \lambda_{H\Phi}   \right) \nonumber \\
&& 
- \; \frac{3}{2} \left(3 g^2 + 6 g_H^2 + g^{\prime2} + 2 g_X^2\right) \lambda'_{H\Phi}   \nonumber \\
&& + \biggl. \; 3 g_H^2 g_X^2 + 6 \lambda'_{H\Phi} \left( y_t^2 +  y_{t^H} \right) \biggr] \; .
\eea

Yukawa couplings:
\bea
\beta_{y_t} &=& \frac{1}{16 \pi^2 } \biggl[ 18 y_t^3  - y_t \left( - 2 y_{t^H}^2 + \frac{17}{2} g_s^2 + \frac{9}{4} g^2 + \frac{17}{12}  g^{\prime2} + \frac{9}{4} g_H^2 + \frac{3}{4} g_X^2 \right) \biggr] \; , \;\; \;\; \\
\beta_{y_{t^H}} &=& \frac{1}{16 \pi^2 } \biggl[ 16 y_{t^H}^3  - y_{t^H} \left( - 4 y_t^2 + \frac{17}{2} g_s^2 + \frac{8}{3}  g^{\prime2} + \frac{9}{4} g_H^2 + \frac{3}{4} g_X^2 \right) \biggr] \; .
\eea

\vfill 

\newpage


\allowdisplaybreaks
\bibliographystyle{apsrev4-1}
\bibliography{refs}

\begin{thebibliography}{255}%
\makeatletter
\providecommand \@ifxundefined [1]{%
 \@ifx{#1\undefined}
}%
\providecommand \@ifnum [1]{%
 \ifnum #1\expandafter \@firstoftwo
 \else \expandafter \@secondoftwo
 \fi
}%
\providecommand \@ifx [1]{%
 \ifx #1\expandafter \@firstoftwo
 \else \expandafter \@secondoftwo
 \fi
}%
\providecommand \natexlab [1]{#1}%
\providecommand \enquote  [1]{``#1''}%
\providecommand \bibnamefont  [1]{#1}%
\providecommand \bibfnamefont [1]{#1}%
\providecommand \citenamefont [1]{#1}%
\providecommand \href@noop [0]{\@secondoftwo}%
\providecommand \href [0]{\begingroup \@sanitize@url \@href}%
\providecommand \@href[1]{\@@startlink{#1}\@@href}%
\providecommand \@@href[1]{\endgroup#1\@@endlink}%
\providecommand \@sanitize@url [0]{\catcode `\\12\catcode `\$12\catcode
  `\&12\catcode `\#12\catcode `\^12\catcode `\_12\catcode `\%12\relax}%
\providecommand \@@startlink[1]{}%
\providecommand \@@endlink[0]{}%
\providecommand \url  [0]{\begingroup\@sanitize@url \@url }%
\providecommand \@url [1]{\endgroup\@href {#1}{\urlprefix }}%
\providecommand \urlprefix  [0]{URL }%
\providecommand \Eprint [0]{\href }%
\providecommand \doibase [0]{http://dx.doi.org/}%
\providecommand \selectlanguage [0]{\@gobble}%
\providecommand \bibinfo  [0]{\@secondoftwo}%
\providecommand \bibfield  [0]{\@secondoftwo}%
\providecommand \translation [1]{[#1]}%
\providecommand \BibitemOpen [0]{}%
\providecommand \bibitemStop [0]{}%
\providecommand \bibitemNoStop [0]{.\EOS\space}%
\providecommand \EOS [0]{\spacefactor3000\relax}%
\providecommand \BibitemShut  [1]{\csname bibitem#1\endcsname}%
\let\auto@bib@innerbib\@empty
\bibitem [{\citenamefont {Kajantie}\ \emph
  {et~al.}(1996{\natexlab{a}})\citenamefont {Kajantie}, \citenamefont {Laine},
  \citenamefont {Rummukainen},\ and\ \citenamefont
  {Shaposhnikov}}]{Kajantie:1996mn}%
  \BibitemOpen
  \bibfield  {author} {\bibinfo {author} {\bibfnamefont {K.}~\bibnamefont
  {Kajantie}}, \bibinfo {author} {\bibfnamefont {M.}~\bibnamefont {Laine}},
  \bibinfo {author} {\bibfnamefont {K.}~\bibnamefont {Rummukainen}}, \ and\
  \bibinfo {author} {\bibfnamefont {M.~E.}\ \bibnamefont {Shaposhnikov}},\
  }\href {\doibase 10.1103/PhysRevLett.77.2887} {\bibfield  {journal} {\bibinfo
   {journal} {Phys. Rev. Lett.}\ }\textbf {\bibinfo {volume} {77}},\ \bibinfo
  {pages} {2887} (\bibinfo {year} {1996}{\natexlab{a}})},\ \Eprint
  {http://arxiv.org/abs/hep-ph/9605288} {arXiv:hep-ph/9605288} \BibitemShut
  {NoStop}%
\bibitem [{\citenamefont {Rummukainen}\ \emph {et~al.}(1998)\citenamefont
  {Rummukainen}, \citenamefont {Tsypin}, \citenamefont {Kajantie},
  \citenamefont {Laine},\ and\ \citenamefont
  {Shaposhnikov}}]{Rummukainen:1998as}%
  \BibitemOpen
  \bibfield  {author} {\bibinfo {author} {\bibfnamefont {K.}~\bibnamefont
  {Rummukainen}}, \bibinfo {author} {\bibfnamefont {M.}~\bibnamefont {Tsypin}},
  \bibinfo {author} {\bibfnamefont {K.}~\bibnamefont {Kajantie}}, \bibinfo
  {author} {\bibfnamefont {M.}~\bibnamefont {Laine}}, \ and\ \bibinfo {author}
  {\bibfnamefont {M.~E.}\ \bibnamefont {Shaposhnikov}},\ }\href {\doibase
  10.1016/S0550-3213(98)00494-5} {\bibfield  {journal} {\bibinfo  {journal}
  {Nucl. Phys. B}\ }\textbf {\bibinfo {volume} {532}},\ \bibinfo {pages} {283}
  (\bibinfo {year} {1998})},\ \Eprint {http://arxiv.org/abs/hep-lat/9805013}
  {arXiv:hep-lat/9805013} \BibitemShut {NoStop}%
\bibitem [{\citenamefont {Laine}\ and\ \citenamefont
  {Rummukainen}(1999)}]{Laine:1998jb}%
  \BibitemOpen
  \bibfield  {author} {\bibinfo {author} {\bibfnamefont {M.}~\bibnamefont
  {Laine}}\ and\ \bibinfo {author} {\bibfnamefont {K.}~\bibnamefont
  {Rummukainen}},\ }\href {\doibase 10.1016/S0920-5632(99)85017-8} {\bibfield
  {journal} {\bibinfo  {journal} {Nucl. Phys. B Proc. Suppl.}\ }\textbf
  {\bibinfo {volume} {73}},\ \bibinfo {pages} {180} (\bibinfo {year} {1999})},\
  \Eprint {http://arxiv.org/abs/hep-lat/9809045} {arXiv:hep-lat/9809045}
  \BibitemShut {NoStop}%
\bibitem [{\citenamefont {Aoki}\ \emph {et~al.}(1999)\citenamefont {Aoki},
  \citenamefont {Csikor}, \citenamefont {Fodor},\ and\ \citenamefont
  {Ukawa}}]{Aoki:1999fi}%
  \BibitemOpen
  \bibfield  {author} {\bibinfo {author} {\bibfnamefont {Y.}~\bibnamefont
  {Aoki}}, \bibinfo {author} {\bibfnamefont {F.}~\bibnamefont {Csikor}},
  \bibinfo {author} {\bibfnamefont {Z.}~\bibnamefont {Fodor}}, \ and\ \bibinfo
  {author} {\bibfnamefont {A.}~\bibnamefont {Ukawa}},\ }\href {\doibase
  10.1103/PhysRevD.60.013001} {\bibfield  {journal} {\bibinfo  {journal} {Phys.
  Rev. D}\ }\textbf {\bibinfo {volume} {60}},\ \bibinfo {pages} {013001}
  (\bibinfo {year} {1999})},\ \Eprint {http://arxiv.org/abs/hep-lat/9901021}
  {arXiv:hep-lat/9901021} \BibitemShut {NoStop}%
\bibitem [{\citenamefont {McDonald}(1994)}]{McDonald:1993ey}%
  \BibitemOpen
  \bibfield  {author} {\bibinfo {author} {\bibfnamefont {J.}~\bibnamefont
  {McDonald}},\ }\href {\doibase 10.1016/0370-2693(94)91229-7} {\bibfield
  {journal} {\bibinfo  {journal} {Phys. Lett. B}\ }\textbf {\bibinfo {volume}
  {323}},\ \bibinfo {pages} {339} (\bibinfo {year} {1994})}\BibitemShut
  {NoStop}%
\bibitem [{\citenamefont {Profumo}\ \emph {et~al.}(2007)\citenamefont
  {Profumo}, \citenamefont {Ramsey-Musolf},\ and\ \citenamefont
  {Shaughnessy}}]{Profumo:2007wc}%
  \BibitemOpen
  \bibfield  {author} {\bibinfo {author} {\bibfnamefont {S.}~\bibnamefont
  {Profumo}}, \bibinfo {author} {\bibfnamefont {M.~J.}\ \bibnamefont
  {Ramsey-Musolf}}, \ and\ \bibinfo {author} {\bibfnamefont {G.}~\bibnamefont
  {Shaughnessy}},\ }\href {\doibase 10.1088/1126-6708/2007/08/010} {\bibfield
  {journal} {\bibinfo  {journal} {JHEP}\ }\textbf {\bibinfo {volume} {08}},\
  \bibinfo {pages} {010} (\bibinfo {year} {2007})},\ \Eprint
  {http://arxiv.org/abs/0705.2425} {arXiv:0705.2425 [hep-ph]} \BibitemShut
  {NoStop}%
\bibitem [{\citenamefont {Cline}\ \emph {et~al.}(2009)\citenamefont {Cline},
  \citenamefont {Laporte}, \citenamefont {Yamashita},\ and\ \citenamefont
  {Kraml}}]{Cline:2009sn}%
  \BibitemOpen
  \bibfield  {author} {\bibinfo {author} {\bibfnamefont {J.~M.}\ \bibnamefont
  {Cline}}, \bibinfo {author} {\bibfnamefont {G.}~\bibnamefont {Laporte}},
  \bibinfo {author} {\bibfnamefont {H.}~\bibnamefont {Yamashita}}, \ and\
  \bibinfo {author} {\bibfnamefont {S.}~\bibnamefont {Kraml}},\ }\href
  {\doibase 10.1088/1126-6708/2009/07/040} {\bibfield  {journal} {\bibinfo
  {journal} {JHEP}\ }\textbf {\bibinfo {volume} {07}},\ \bibinfo {pages} {040}
  (\bibinfo {year} {2009})},\ \Eprint {http://arxiv.org/abs/0905.2559}
  {arXiv:0905.2559 [hep-ph]} \BibitemShut {NoStop}%
\bibitem [{\citenamefont {Espinosa}\ \emph
  {et~al.}(2012{\natexlab{a}})\citenamefont {Espinosa}, \citenamefont
  {Konstandin},\ and\ \citenamefont {Riva}}]{Espinosa:2011ax}%
  \BibitemOpen
  \bibfield  {author} {\bibinfo {author} {\bibfnamefont {J.~R.}\ \bibnamefont
  {Espinosa}}, \bibinfo {author} {\bibfnamefont {T.}~\bibnamefont
  {Konstandin}}, \ and\ \bibinfo {author} {\bibfnamefont {F.}~\bibnamefont
  {Riva}},\ }\href {\doibase 10.1016/j.nuclphysb.2011.09.010} {\bibfield
  {journal} {\bibinfo  {journal} {Nucl. Phys. B}\ }\textbf {\bibinfo {volume}
  {854}},\ \bibinfo {pages} {592} (\bibinfo {year} {2012}{\natexlab{a}})},\
  \Eprint {http://arxiv.org/abs/1107.5441} {arXiv:1107.5441 [hep-ph]}
  \BibitemShut {NoStop}%
\bibitem [{\citenamefont {Cline}\ and\ \citenamefont
  {Kainulainen}(2013{\natexlab{a}})}]{Cline:2012hg}%
  \BibitemOpen
  \bibfield  {author} {\bibinfo {author} {\bibfnamefont {J.~M.}\ \bibnamefont
  {Cline}}\ and\ \bibinfo {author} {\bibfnamefont {K.}~\bibnamefont
  {Kainulainen}},\ }\href {\doibase 10.1088/1475-7516/2013/01/012} {\bibfield
  {journal} {\bibinfo  {journal} {JCAP}\ }\textbf {\bibinfo {volume} {01}},\
  \bibinfo {pages} {012} (\bibinfo {year} {2013}{\natexlab{a}})},\ \Eprint
  {http://arxiv.org/abs/1210.4196} {arXiv:1210.4196 [hep-ph]} \BibitemShut
  {NoStop}%
\bibitem [{\citenamefont {Alanne}\ \emph {et~al.}(2014)\citenamefont {Alanne},
  \citenamefont {Tuominen},\ and\ \citenamefont {Vaskonen}}]{Alanne:2014bra}%
  \BibitemOpen
  \bibfield  {author} {\bibinfo {author} {\bibfnamefont {T.}~\bibnamefont
  {Alanne}}, \bibinfo {author} {\bibfnamefont {K.}~\bibnamefont {Tuominen}}, \
  and\ \bibinfo {author} {\bibfnamefont {V.}~\bibnamefont {Vaskonen}},\ }\href
  {\doibase 10.1016/j.nuclphysb.2014.11.001} {\bibfield  {journal} {\bibinfo
  {journal} {Nucl. Phys. B}\ }\textbf {\bibinfo {volume} {889}},\ \bibinfo
  {pages} {692} (\bibinfo {year} {2014})},\ \Eprint
  {http://arxiv.org/abs/1407.0688} {arXiv:1407.0688 [hep-ph]} \BibitemShut
  {NoStop}%
\bibitem [{\citenamefont {Curtin}\ \emph {et~al.}(2014)\citenamefont {Curtin},
  \citenamefont {Meade},\ and\ \citenamefont {Yu}}]{Curtin:2014jma}%
  \BibitemOpen
  \bibfield  {author} {\bibinfo {author} {\bibfnamefont {D.}~\bibnamefont
  {Curtin}}, \bibinfo {author} {\bibfnamefont {P.}~\bibnamefont {Meade}}, \
  and\ \bibinfo {author} {\bibfnamefont {C.-T.}\ \bibnamefont {Yu}},\ }\href
  {\doibase 10.1007/JHEP11(2014)127} {\bibfield  {journal} {\bibinfo  {journal}
  {JHEP}\ }\textbf {\bibinfo {volume} {11}},\ \bibinfo {pages} {127} (\bibinfo
  {year} {2014})},\ \Eprint {http://arxiv.org/abs/1409.0005} {arXiv:1409.0005
  [hep-ph]} \BibitemShut {NoStop}%
\bibitem [{\citenamefont {Profumo}\ \emph {et~al.}(2015)\citenamefont
  {Profumo}, \citenamefont {Ramsey-Musolf}, \citenamefont {Wainwright},\ and\
  \citenamefont {Winslow}}]{Profumo:2014opa}%
  \BibitemOpen
  \bibfield  {author} {\bibinfo {author} {\bibfnamefont {S.}~\bibnamefont
  {Profumo}}, \bibinfo {author} {\bibfnamefont {M.~J.}\ \bibnamefont
  {Ramsey-Musolf}}, \bibinfo {author} {\bibfnamefont {C.~L.}\ \bibnamefont
  {Wainwright}}, \ and\ \bibinfo {author} {\bibfnamefont {P.}~\bibnamefont
  {Winslow}},\ }\href {\doibase 10.1103/PhysRevD.91.035018} {\bibfield
  {journal} {\bibinfo  {journal} {Phys. Rev. D}\ }\textbf {\bibinfo {volume}
  {91}},\ \bibinfo {pages} {035018} (\bibinfo {year} {2015})},\ \Eprint
  {http://arxiv.org/abs/1407.5342} {arXiv:1407.5342 [hep-ph]} \BibitemShut
  {NoStop}%
\bibitem [{\citenamefont {Huang}\ and\ \citenamefont
  {Li}(2015)}]{Huang:2015bta}%
  \BibitemOpen
  \bibfield  {author} {\bibinfo {author} {\bibfnamefont {F.~P.}\ \bibnamefont
  {Huang}}\ and\ \bibinfo {author} {\bibfnamefont {C.~S.}\ \bibnamefont {Li}},\
  }\href {\doibase 10.1103/PhysRevD.92.075014} {\bibfield  {journal} {\bibinfo
  {journal} {Phys. Rev. D}\ }\textbf {\bibinfo {volume} {92}},\ \bibinfo
  {pages} {075014} (\bibinfo {year} {2015})},\ \Eprint
  {http://arxiv.org/abs/1507.08168} {arXiv:1507.08168 [hep-ph]} \BibitemShut
  {NoStop}%
\bibitem [{\citenamefont {Kotwal}\ \emph {et~al.}(2016)\citenamefont {Kotwal},
  \citenamefont {Ramsey-Musolf}, \citenamefont {No},\ and\ \citenamefont
  {Winslow}}]{Kotwal:2016tex}%
  \BibitemOpen
  \bibfield  {author} {\bibinfo {author} {\bibfnamefont {A.~V.}\ \bibnamefont
  {Kotwal}}, \bibinfo {author} {\bibfnamefont {M.~J.}\ \bibnamefont
  {Ramsey-Musolf}}, \bibinfo {author} {\bibfnamefont {J.~M.}\ \bibnamefont
  {No}}, \ and\ \bibinfo {author} {\bibfnamefont {P.}~\bibnamefont {Winslow}},\
  }\href {\doibase 10.1103/PhysRevD.94.035022} {\bibfield  {journal} {\bibinfo
  {journal} {Phys. Rev. D}\ }\textbf {\bibinfo {volume} {94}},\ \bibinfo
  {pages} {035022} (\bibinfo {year} {2016})},\ \Eprint
  {http://arxiv.org/abs/1605.06123} {arXiv:1605.06123 [hep-ph]} \BibitemShut
  {NoStop}%
\bibitem [{\citenamefont {Vaskonen}(2017)}]{Vaskonen:2016yiu}%
  \BibitemOpen
  \bibfield  {author} {\bibinfo {author} {\bibfnamefont {V.}~\bibnamefont
  {Vaskonen}},\ }\href {\doibase 10.1103/PhysRevD.95.123515} {\bibfield
  {journal} {\bibinfo  {journal} {Phys. Rev. D}\ }\textbf {\bibinfo {volume}
  {95}},\ \bibinfo {pages} {123515} (\bibinfo {year} {2017})},\ \Eprint
  {http://arxiv.org/abs/1611.02073} {arXiv:1611.02073 [hep-ph]} \BibitemShut
  {NoStop}%
\bibitem [{\citenamefont {Ghorbani}(2017)}]{Ghorbani:2017jls}%
  \BibitemOpen
  \bibfield  {author} {\bibinfo {author} {\bibfnamefont {P.~H.}\ \bibnamefont
  {Ghorbani}},\ }\href {\doibase 10.1007/JHEP08(2017)058} {\bibfield  {journal}
  {\bibinfo  {journal} {JHEP}\ }\textbf {\bibinfo {volume} {08}},\ \bibinfo
  {pages} {058} (\bibinfo {year} {2017})},\ \Eprint
  {http://arxiv.org/abs/1703.06506} {arXiv:1703.06506 [hep-ph]} \BibitemShut
  {NoStop}%
\bibitem [{\citenamefont {Cline}\ \emph {et~al.}(2017)\citenamefont {Cline},
  \citenamefont {Kainulainen},\ and\ \citenamefont
  {Tucker-Smith}}]{Cline:2017qpe}%
  \BibitemOpen
  \bibfield  {author} {\bibinfo {author} {\bibfnamefont {J.~M.}\ \bibnamefont
  {Cline}}, \bibinfo {author} {\bibfnamefont {K.}~\bibnamefont {Kainulainen}},
  \ and\ \bibinfo {author} {\bibfnamefont {D.}~\bibnamefont {Tucker-Smith}},\
  }\href {\doibase 10.1103/PhysRevD.95.115006} {\bibfield  {journal} {\bibinfo
  {journal} {Phys. Rev. D}\ }\textbf {\bibinfo {volume} {95}},\ \bibinfo
  {pages} {115006} (\bibinfo {year} {2017})},\ \Eprint
  {http://arxiv.org/abs/1702.08909} {arXiv:1702.08909 [hep-ph]} \BibitemShut
  {NoStop}%
\bibitem [{\citenamefont {Beniwal}\ \emph {et~al.}(2017)\citenamefont
  {Beniwal}, \citenamefont {Lewicki}, \citenamefont {Wells}, \citenamefont
  {White},\ and\ \citenamefont {Williams}}]{Beniwal:2017eik}%
  \BibitemOpen
  \bibfield  {author} {\bibinfo {author} {\bibfnamefont {A.}~\bibnamefont
  {Beniwal}}, \bibinfo {author} {\bibfnamefont {M.}~\bibnamefont {Lewicki}},
  \bibinfo {author} {\bibfnamefont {J.~D.}\ \bibnamefont {Wells}}, \bibinfo
  {author} {\bibfnamefont {M.}~\bibnamefont {White}}, \ and\ \bibinfo {author}
  {\bibfnamefont {A.~G.}\ \bibnamefont {Williams}},\ }\href {\doibase
  10.1007/JHEP08(2017)108} {\bibfield  {journal} {\bibinfo  {journal} {JHEP}\
  }\textbf {\bibinfo {volume} {08}},\ \bibinfo {pages} {108} (\bibinfo {year}
  {2017})},\ \Eprint {http://arxiv.org/abs/1702.06124} {arXiv:1702.06124
  [hep-ph]} \BibitemShut {NoStop}%
\bibitem [{\citenamefont {Kurup}\ and\ \citenamefont
  {Perelstein}(2017)}]{Kurup:2017dzf}%
  \BibitemOpen
  \bibfield  {author} {\bibinfo {author} {\bibfnamefont {G.}~\bibnamefont
  {Kurup}}\ and\ \bibinfo {author} {\bibfnamefont {M.}~\bibnamefont
  {Perelstein}},\ }\href {\doibase 10.1103/PhysRevD.96.015036} {\bibfield
  {journal} {\bibinfo  {journal} {Phys. Rev. D}\ }\textbf {\bibinfo {volume}
  {96}},\ \bibinfo {pages} {015036} (\bibinfo {year} {2017})},\ \Eprint
  {http://arxiv.org/abs/1704.03381} {arXiv:1704.03381 [hep-ph]} \BibitemShut
  {NoStop}%
\bibitem [{\citenamefont {Chiang}\ \emph {et~al.}(2018)\citenamefont {Chiang},
  \citenamefont {Ramsey-Musolf},\ and\ \citenamefont
  {Senaha}}]{Chiang:2017nmu}%
  \BibitemOpen
  \bibfield  {author} {\bibinfo {author} {\bibfnamefont {C.-W.}\ \bibnamefont
  {Chiang}}, \bibinfo {author} {\bibfnamefont {M.~J.}\ \bibnamefont
  {Ramsey-Musolf}}, \ and\ \bibinfo {author} {\bibfnamefont {E.}~\bibnamefont
  {Senaha}},\ }\href {\doibase 10.1103/PhysRevD.97.015005} {\bibfield
  {journal} {\bibinfo  {journal} {Phys. Rev. D}\ }\textbf {\bibinfo {volume}
  {97}},\ \bibinfo {pages} {015005} (\bibinfo {year} {2018})},\ \Eprint
  {http://arxiv.org/abs/1707.09960} {arXiv:1707.09960 [hep-ph]} \BibitemShut
  {NoStop}%
\bibitem [{\citenamefont {Carena}\ \emph {et~al.}(2019)\citenamefont {Carena},
  \citenamefont {Quir\'os},\ and\ \citenamefont {Zhang}}]{Carena:2018cjh}%
  \BibitemOpen
  \bibfield  {author} {\bibinfo {author} {\bibfnamefont {M.}~\bibnamefont
  {Carena}}, \bibinfo {author} {\bibfnamefont {M.}~\bibnamefont {Quir\'os}}, \
  and\ \bibinfo {author} {\bibfnamefont {Y.}~\bibnamefont {Zhang}},\ }\href
  {\doibase 10.1103/PhysRevLett.122.201802} {\bibfield  {journal} {\bibinfo
  {journal} {Phys. Rev. Lett.}\ }\textbf {\bibinfo {volume} {122}},\ \bibinfo
  {pages} {201802} (\bibinfo {year} {2019})},\ \Eprint
  {http://arxiv.org/abs/1811.09719} {arXiv:1811.09719 [hep-ph]} \BibitemShut
  {NoStop}%
\bibitem [{\citenamefont {Carena}\ \emph {et~al.}(2018)\citenamefont {Carena},
  \citenamefont {Liu},\ and\ \citenamefont {Riembau}}]{Carena:2018vpt}%
  \BibitemOpen
  \bibfield  {author} {\bibinfo {author} {\bibfnamefont {M.}~\bibnamefont
  {Carena}}, \bibinfo {author} {\bibfnamefont {Z.}~\bibnamefont {Liu}}, \ and\
  \bibinfo {author} {\bibfnamefont {M.}~\bibnamefont {Riembau}},\ }\href
  {\doibase 10.1103/PhysRevD.97.095032} {\bibfield  {journal} {\bibinfo
  {journal} {Phys. Rev. D}\ }\textbf {\bibinfo {volume} {97}},\ \bibinfo
  {pages} {095032} (\bibinfo {year} {2018})},\ \Eprint
  {http://arxiv.org/abs/1801.00794} {arXiv:1801.00794 [hep-ph]} \BibitemShut
  {NoStop}%
\bibitem [{\citenamefont {Grzadkowski}\ and\ \citenamefont
  {Huang}(2018)}]{Grzadkowski:2018nbc}%
  \BibitemOpen
  \bibfield  {author} {\bibinfo {author} {\bibfnamefont {B.}~\bibnamefont
  {Grzadkowski}}\ and\ \bibinfo {author} {\bibfnamefont {D.}~\bibnamefont
  {Huang}},\ }\href {\doibase 10.1007/JHEP08(2018)135} {\bibfield  {journal}
  {\bibinfo  {journal} {JHEP}\ }\textbf {\bibinfo {volume} {08}},\ \bibinfo
  {pages} {135} (\bibinfo {year} {2018})},\ \Eprint
  {http://arxiv.org/abs/1807.06987} {arXiv:1807.06987 [hep-ph]} \BibitemShut
  {NoStop}%
\bibitem [{\citenamefont {Huang}\ \emph
  {et~al.}(2018{\natexlab{a}})\citenamefont {Huang}, \citenamefont {Qian},\
  and\ \citenamefont {Zhang}}]{Huang:2018aja}%
  \BibitemOpen
  \bibfield  {author} {\bibinfo {author} {\bibfnamefont {F.~P.}\ \bibnamefont
  {Huang}}, \bibinfo {author} {\bibfnamefont {Z.}~\bibnamefont {Qian}}, \ and\
  \bibinfo {author} {\bibfnamefont {M.}~\bibnamefont {Zhang}},\ }\href
  {\doibase 10.1103/PhysRevD.98.015014} {\bibfield  {journal} {\bibinfo
  {journal} {Phys. Rev. D}\ }\textbf {\bibinfo {volume} {98}},\ \bibinfo
  {pages} {015014} (\bibinfo {year} {2018}{\natexlab{a}})},\ \Eprint
  {http://arxiv.org/abs/1804.06813} {arXiv:1804.06813 [hep-ph]} \BibitemShut
  {NoStop}%
\bibitem [{\citenamefont {Alves}\ \emph {et~al.}(2019)\citenamefont {Alves},
  \citenamefont {Ghosh}, \citenamefont {Guo}, \citenamefont {Sinha},\ and\
  \citenamefont {Vagie}}]{Alves:2018jsw}%
  \BibitemOpen
  \bibfield  {author} {\bibinfo {author} {\bibfnamefont {A.}~\bibnamefont
  {Alves}}, \bibinfo {author} {\bibfnamefont {T.}~\bibnamefont {Ghosh}},
  \bibinfo {author} {\bibfnamefont {H.-K.}\ \bibnamefont {Guo}}, \bibinfo
  {author} {\bibfnamefont {K.}~\bibnamefont {Sinha}}, \ and\ \bibinfo {author}
  {\bibfnamefont {D.}~\bibnamefont {Vagie}},\ }\href {\doibase
  10.1007/JHEP04(2019)052} {\bibfield  {journal} {\bibinfo  {journal} {JHEP}\
  }\textbf {\bibinfo {volume} {04}},\ \bibinfo {pages} {052} (\bibinfo {year}
  {2019})},\ \Eprint {http://arxiv.org/abs/1812.09333} {arXiv:1812.09333
  [hep-ph]} \BibitemShut {NoStop}%
\bibitem [{\citenamefont {Ghorbani}\ and\ \citenamefont
  {Ghorbani}(2020)}]{Ghorbani:2018yfr}%
  \BibitemOpen
  \bibfield  {author} {\bibinfo {author} {\bibfnamefont {K.}~\bibnamefont
  {Ghorbani}}\ and\ \bibinfo {author} {\bibfnamefont {P.~H.}\ \bibnamefont
  {Ghorbani}},\ }\href {\doibase 10.1088/1361-6471/ab4823} {\bibfield
  {journal} {\bibinfo  {journal} {J. Phys. G}\ }\textbf {\bibinfo {volume}
  {47}},\ \bibinfo {pages} {015201} (\bibinfo {year} {2020})},\ \Eprint
  {http://arxiv.org/abs/1804.05798} {arXiv:1804.05798 [hep-ph]} \BibitemShut
  {NoStop}%
\bibitem [{\citenamefont {Cheng}\ and\ \citenamefont
  {Bian}(2018)}]{Cheng:2018ajh}%
  \BibitemOpen
  \bibfield  {author} {\bibinfo {author} {\bibfnamefont {W.}~\bibnamefont
  {Cheng}}\ and\ \bibinfo {author} {\bibfnamefont {L.}~\bibnamefont {Bian}},\
  }\href {\doibase 10.1103/PhysRevD.98.023524} {\bibfield  {journal} {\bibinfo
  {journal} {Phys. Rev. D}\ }\textbf {\bibinfo {volume} {98}},\ \bibinfo
  {pages} {023524} (\bibinfo {year} {2018})},\ \Eprint
  {http://arxiv.org/abs/1801.00662} {arXiv:1801.00662 [hep-ph]} \BibitemShut
  {NoStop}%
\bibitem [{\citenamefont {Alanne}\ \emph {et~al.}(2020)\citenamefont {Alanne},
  \citenamefont {Hugle}, \citenamefont {Platscher},\ and\ \citenamefont
  {Schmitz}}]{Alanne:2019bsm}%
  \BibitemOpen
  \bibfield  {author} {\bibinfo {author} {\bibfnamefont {T.}~\bibnamefont
  {Alanne}}, \bibinfo {author} {\bibfnamefont {T.}~\bibnamefont {Hugle}},
  \bibinfo {author} {\bibfnamefont {M.}~\bibnamefont {Platscher}}, \ and\
  \bibinfo {author} {\bibfnamefont {K.}~\bibnamefont {Schmitz}},\ }\href
  {\doibase 10.1007/JHEP03(2020)004} {\bibfield  {journal} {\bibinfo  {journal}
  {JHEP}\ }\textbf {\bibinfo {volume} {03}},\ \bibinfo {pages} {004} (\bibinfo
  {year} {2020})},\ \Eprint {http://arxiv.org/abs/1909.11356} {arXiv:1909.11356
  [hep-ph]} \BibitemShut {NoStop}%
\bibitem [{\citenamefont {Gould}\ \emph {et~al.}(2019)\citenamefont {Gould},
  \citenamefont {Kozaczuk}, \citenamefont {Niemi}, \citenamefont
  {Ramsey-Musolf}, \citenamefont {Tenkanen},\ and\ \citenamefont
  {Weir}}]{Gould:2019qek}%
  \BibitemOpen
  \bibfield  {author} {\bibinfo {author} {\bibfnamefont {O.}~\bibnamefont
  {Gould}}, \bibinfo {author} {\bibfnamefont {J.}~\bibnamefont {Kozaczuk}},
  \bibinfo {author} {\bibfnamefont {L.}~\bibnamefont {Niemi}}, \bibinfo
  {author} {\bibfnamefont {M.~J.}\ \bibnamefont {Ramsey-Musolf}}, \bibinfo
  {author} {\bibfnamefont {T.~V.~I.}\ \bibnamefont {Tenkanen}}, \ and\ \bibinfo
  {author} {\bibfnamefont {D.~J.}\ \bibnamefont {Weir}},\ }\href {\doibase
  10.1103/PhysRevD.100.115024} {\bibfield  {journal} {\bibinfo  {journal}
  {Phys. Rev. D}\ }\textbf {\bibinfo {volume} {100}},\ \bibinfo {pages}
  {115024} (\bibinfo {year} {2019})},\ \Eprint
  {http://arxiv.org/abs/1903.11604} {arXiv:1903.11604 [hep-ph]} \BibitemShut
  {NoStop}%
\bibitem [{\citenamefont {Carena}\ \emph {et~al.}(2020)\citenamefont {Carena},
  \citenamefont {Liu},\ and\ \citenamefont {Wang}}]{Carena:2019une}%
  \BibitemOpen
  \bibfield  {author} {\bibinfo {author} {\bibfnamefont {M.}~\bibnamefont
  {Carena}}, \bibinfo {author} {\bibfnamefont {Z.}~\bibnamefont {Liu}}, \ and\
  \bibinfo {author} {\bibfnamefont {Y.}~\bibnamefont {Wang}},\ }\href {\doibase
  10.1007/JHEP08(2020)107} {\bibfield  {journal} {\bibinfo  {journal} {JHEP}\
  }\textbf {\bibinfo {volume} {08}},\ \bibinfo {pages} {107} (\bibinfo {year}
  {2020})},\ \Eprint {http://arxiv.org/abs/1911.10206} {arXiv:1911.10206
  [hep-ph]} \BibitemShut {NoStop}%
\bibitem [{\citenamefont {Li}\ \emph {et~al.}(2019)\citenamefont {Li},
  \citenamefont {Ramsey-Musolf},\ and\ \citenamefont {Willocq}}]{Li:2019tfd}%
  \BibitemOpen
  \bibfield  {author} {\bibinfo {author} {\bibfnamefont {H.-L.}\ \bibnamefont
  {Li}}, \bibinfo {author} {\bibfnamefont {M.}~\bibnamefont {Ramsey-Musolf}}, \
  and\ \bibinfo {author} {\bibfnamefont {S.}~\bibnamefont {Willocq}},\ }\href
  {\doibase 10.1103/PhysRevD.100.075035} {\bibfield  {journal} {\bibinfo
  {journal} {Phys. Rev. D}\ }\textbf {\bibinfo {volume} {100}},\ \bibinfo
  {pages} {075035} (\bibinfo {year} {2019})},\ \Eprint
  {http://arxiv.org/abs/1906.05289} {arXiv:1906.05289 [hep-ph]} \BibitemShut
  {NoStop}%
\bibitem [{\citenamefont {Bell}\ \emph {et~al.}(2019)\citenamefont {Bell},
  \citenamefont {Dolan}, \citenamefont {Friedrich}, \citenamefont
  {Ramsey-Musolf},\ and\ \citenamefont {Volkas}}]{Bell:2019mbn}%
  \BibitemOpen
  \bibfield  {author} {\bibinfo {author} {\bibfnamefont {N.~F.}\ \bibnamefont
  {Bell}}, \bibinfo {author} {\bibfnamefont {M.~J.}\ \bibnamefont {Dolan}},
  \bibinfo {author} {\bibfnamefont {L.~S.}\ \bibnamefont {Friedrich}}, \bibinfo
  {author} {\bibfnamefont {M.~J.}\ \bibnamefont {Ramsey-Musolf}}, \ and\
  \bibinfo {author} {\bibfnamefont {R.~R.}\ \bibnamefont {Volkas}},\ }\href
  {\doibase 10.1007/JHEP09(2019)012} {\bibfield  {journal} {\bibinfo  {journal}
  {JHEP}\ }\textbf {\bibinfo {volume} {09}},\ \bibinfo {pages} {012} (\bibinfo
  {year} {2019})},\ \Eprint {http://arxiv.org/abs/1903.11255} {arXiv:1903.11255
  [hep-ph]} \BibitemShut {NoStop}%
\bibitem [{\citenamefont {Inoue}\ \emph {et~al.}(2016)\citenamefont {Inoue},
  \citenamefont {Ovanesyan},\ and\ \citenamefont
  {Ramsey-Musolf}}]{Inoue:2015pza}%
  \BibitemOpen
  \bibfield  {author} {\bibinfo {author} {\bibfnamefont {S.}~\bibnamefont
  {Inoue}}, \bibinfo {author} {\bibfnamefont {G.}~\bibnamefont {Ovanesyan}}, \
  and\ \bibinfo {author} {\bibfnamefont {M.~J.}\ \bibnamefont
  {Ramsey-Musolf}},\ }\href {\doibase 10.1103/PhysRevD.93.015013} {\bibfield
  {journal} {\bibinfo  {journal} {Phys. Rev. D}\ }\textbf {\bibinfo {volume}
  {93}},\ \bibinfo {pages} {015013} (\bibinfo {year} {2016})},\ \Eprint
  {http://arxiv.org/abs/1508.05404} {arXiv:1508.05404 [hep-ph]} \BibitemShut
  {NoStop}%
\bibitem [{\citenamefont {Niemi}\ \emph {et~al.}(2019)\citenamefont {Niemi},
  \citenamefont {Patel}, \citenamefont {Ramsey-Musolf}, \citenamefont
  {Tenkanen},\ and\ \citenamefont {Weir}}]{Niemi:2018asa}%
  \BibitemOpen
  \bibfield  {author} {\bibinfo {author} {\bibfnamefont {L.}~\bibnamefont
  {Niemi}}, \bibinfo {author} {\bibfnamefont {H.~H.}\ \bibnamefont {Patel}},
  \bibinfo {author} {\bibfnamefont {M.~J.}\ \bibnamefont {Ramsey-Musolf}},
  \bibinfo {author} {\bibfnamefont {T.~V.~I.}\ \bibnamefont {Tenkanen}}, \ and\
  \bibinfo {author} {\bibfnamefont {D.~J.}\ \bibnamefont {Weir}},\ }\href
  {\doibase 10.1103/PhysRevD.100.035002} {\bibfield  {journal} {\bibinfo
  {journal} {Phys. Rev. D}\ }\textbf {\bibinfo {volume} {100}},\ \bibinfo
  {pages} {035002} (\bibinfo {year} {2019})},\ \Eprint
  {http://arxiv.org/abs/1802.10500} {arXiv:1802.10500 [hep-ph]} \BibitemShut
  {NoStop}%
\bibitem [{\citenamefont {Chala}\ \emph {et~al.}(2019)\citenamefont {Chala},
  \citenamefont {Ramos},\ and\ \citenamefont {Spannowsky}}]{Chala:2018opy}%
  \BibitemOpen
  \bibfield  {author} {\bibinfo {author} {\bibfnamefont {M.}~\bibnamefont
  {Chala}}, \bibinfo {author} {\bibfnamefont {M.}~\bibnamefont {Ramos}}, \ and\
  \bibinfo {author} {\bibfnamefont {M.}~\bibnamefont {Spannowsky}},\ }\href
  {\doibase 10.1140/epjc/s10052-019-6655-1} {\bibfield  {journal} {\bibinfo
  {journal} {Eur. Phys. J. C}\ }\textbf {\bibinfo {volume} {79}},\ \bibinfo
  {pages} {156} (\bibinfo {year} {2019})},\ \Eprint
  {http://arxiv.org/abs/1812.01901} {arXiv:1812.01901 [hep-ph]} \BibitemShut
  {NoStop}%
\bibitem [{\citenamefont {Zhou}\ \emph {et~al.}(2019)\citenamefont {Zhou},
  \citenamefont {Cheng}, \citenamefont {Deng}, \citenamefont {Bian},\ and\
  \citenamefont {Wu}}]{Zhou:2018zli}%
  \BibitemOpen
  \bibfield  {author} {\bibinfo {author} {\bibfnamefont {R.}~\bibnamefont
  {Zhou}}, \bibinfo {author} {\bibfnamefont {W.}~\bibnamefont {Cheng}},
  \bibinfo {author} {\bibfnamefont {X.}~\bibnamefont {Deng}}, \bibinfo {author}
  {\bibfnamefont {L.}~\bibnamefont {Bian}}, \ and\ \bibinfo {author}
  {\bibfnamefont {Y.}~\bibnamefont {Wu}},\ }\href {\doibase
  10.1007/JHEP01(2019)216} {\bibfield  {journal} {\bibinfo  {journal} {JHEP}\
  }\textbf {\bibinfo {volume} {01}},\ \bibinfo {pages} {216} (\bibinfo {year}
  {2019})},\ \Eprint {http://arxiv.org/abs/1812.06217} {arXiv:1812.06217
  [hep-ph]} \BibitemShut {NoStop}%
\bibitem [{\citenamefont {Bochkarev}\ \emph {et~al.}(1990)\citenamefont
  {Bochkarev}, \citenamefont {Kuzmin},\ and\ \citenamefont
  {Shaposhnikov}}]{Bochkarev:1990fx}%
  \BibitemOpen
  \bibfield  {author} {\bibinfo {author} {\bibfnamefont {A.~I.}\ \bibnamefont
  {Bochkarev}}, \bibinfo {author} {\bibfnamefont {S.~V.}\ \bibnamefont
  {Kuzmin}}, \ and\ \bibinfo {author} {\bibfnamefont {M.~E.}\ \bibnamefont
  {Shaposhnikov}},\ }\href {\doibase 10.1016/0370-2693(90)90069-I} {\bibfield
  {journal} {\bibinfo  {journal} {Phys. Lett. B}\ }\textbf {\bibinfo {volume}
  {244}},\ \bibinfo {pages} {275} (\bibinfo {year} {1990})}\BibitemShut
  {NoStop}%
\bibitem [{\citenamefont {McLerran}\ \emph {et~al.}(1991)\citenamefont
  {McLerran}, \citenamefont {Shaposhnikov}, \citenamefont {Turok},\ and\
  \citenamefont {Voloshin}}]{McLerran:1990zh}%
  \BibitemOpen
  \bibfield  {author} {\bibinfo {author} {\bibfnamefont {L.~D.}\ \bibnamefont
  {McLerran}}, \bibinfo {author} {\bibfnamefont {M.~E.}\ \bibnamefont
  {Shaposhnikov}}, \bibinfo {author} {\bibfnamefont {N.}~\bibnamefont {Turok}},
  \ and\ \bibinfo {author} {\bibfnamefont {M.~B.}\ \bibnamefont {Voloshin}},\
  }\href {\doibase 10.1016/0370-2693(91)91794-V} {\bibfield  {journal}
  {\bibinfo  {journal} {Phys. Lett. B}\ }\textbf {\bibinfo {volume} {256}},\
  \bibinfo {pages} {451} (\bibinfo {year} {1991})}\BibitemShut {NoStop}%
\bibitem [{\citenamefont {Bochkarev}\ \emph {et~al.}(1991)\citenamefont
  {Bochkarev}, \citenamefont {Kuzmin},\ and\ \citenamefont
  {Shaposhnikov}}]{Bochkarev:1990gb}%
  \BibitemOpen
  \bibfield  {author} {\bibinfo {author} {\bibfnamefont {A.~I.}\ \bibnamefont
  {Bochkarev}}, \bibinfo {author} {\bibfnamefont {S.~V.}\ \bibnamefont
  {Kuzmin}}, \ and\ \bibinfo {author} {\bibfnamefont {M.~E.}\ \bibnamefont
  {Shaposhnikov}},\ }\href {\doibase 10.1103/PhysRevD.43.369} {\bibfield
  {journal} {\bibinfo  {journal} {Phys. Rev. D}\ }\textbf {\bibinfo {volume}
  {43}},\ \bibinfo {pages} {369} (\bibinfo {year} {1991})}\BibitemShut
  {NoStop}%
\bibitem [{\citenamefont {Turok}\ and\ \citenamefont
  {Zadrozny}(1991)}]{Turok:1990zg}%
  \BibitemOpen
  \bibfield  {author} {\bibinfo {author} {\bibfnamefont {N.}~\bibnamefont
  {Turok}}\ and\ \bibinfo {author} {\bibfnamefont {J.}~\bibnamefont
  {Zadrozny}},\ }\href {\doibase 10.1016/0550-3213(91)90356-3} {\bibfield
  {journal} {\bibinfo  {journal} {Nucl. Phys. B}\ }\textbf {\bibinfo {volume}
  {358}},\ \bibinfo {pages} {471} (\bibinfo {year} {1991})}\BibitemShut
  {NoStop}%
\bibitem [{\citenamefont {Cohen}\ \emph {et~al.}(1991)\citenamefont {Cohen},
  \citenamefont {Kaplan},\ and\ \citenamefont {Nelson}}]{Cohen:1991iu}%
  \BibitemOpen
  \bibfield  {author} {\bibinfo {author} {\bibfnamefont {A.~G.}\ \bibnamefont
  {Cohen}}, \bibinfo {author} {\bibfnamefont {D.~B.}\ \bibnamefont {Kaplan}}, \
  and\ \bibinfo {author} {\bibfnamefont {A.~E.}\ \bibnamefont {Nelson}},\
  }\href {\doibase 10.1016/0370-2693(91)91711-4} {\bibfield  {journal}
  {\bibinfo  {journal} {Phys. Lett. B}\ }\textbf {\bibinfo {volume} {263}},\
  \bibinfo {pages} {86} (\bibinfo {year} {1991})}\BibitemShut {NoStop}%
\bibitem [{\citenamefont {Turok}\ and\ \citenamefont
  {Zadrozny}(1992)}]{Turok:1991uc}%
  \BibitemOpen
  \bibfield  {author} {\bibinfo {author} {\bibfnamefont {N.}~\bibnamefont
  {Turok}}\ and\ \bibinfo {author} {\bibfnamefont {J.}~\bibnamefont
  {Zadrozny}},\ }\href {\doibase 10.1016/0550-3213(92)90284-I} {\bibfield
  {journal} {\bibinfo  {journal} {Nucl. Phys. B}\ }\textbf {\bibinfo {volume}
  {369}},\ \bibinfo {pages} {729} (\bibinfo {year} {1992})}\BibitemShut
  {NoStop}%
\bibitem [{\citenamefont {Nelson}\ \emph {et~al.}(1992)\citenamefont {Nelson},
  \citenamefont {Kaplan},\ and\ \citenamefont {Cohen}}]{Nelson:1991ab}%
  \BibitemOpen
  \bibfield  {author} {\bibinfo {author} {\bibfnamefont {A.~E.}\ \bibnamefont
  {Nelson}}, \bibinfo {author} {\bibfnamefont {D.~B.}\ \bibnamefont {Kaplan}},
  \ and\ \bibinfo {author} {\bibfnamefont {A.~G.}\ \bibnamefont {Cohen}},\
  }\href {\doibase 10.1016/0550-3213(92)90440-M} {\bibfield  {journal}
  {\bibinfo  {journal} {Nucl. Phys. B}\ }\textbf {\bibinfo {volume} {373}},\
  \bibinfo {pages} {453} (\bibinfo {year} {1992})}\BibitemShut {NoStop}%
\bibitem [{\citenamefont {Funakubo}\ \emph {et~al.}(1994)\citenamefont
  {Funakubo}, \citenamefont {Kakuto},\ and\ \citenamefont
  {Takenaga}}]{Funakubo:1993jg}%
  \BibitemOpen
  \bibfield  {author} {\bibinfo {author} {\bibfnamefont {K.}~\bibnamefont
  {Funakubo}}, \bibinfo {author} {\bibfnamefont {A.}~\bibnamefont {Kakuto}}, \
  and\ \bibinfo {author} {\bibfnamefont {K.}~\bibnamefont {Takenaga}},\ }\href
  {\doibase 10.1143/PTP.91.341} {\bibfield  {journal} {\bibinfo  {journal}
  {Prog. Theor. Phys.}\ }\textbf {\bibinfo {volume} {91}},\ \bibinfo {pages}
  {341} (\bibinfo {year} {1994})},\ \Eprint
  {http://arxiv.org/abs/hep-ph/9310267} {arXiv:hep-ph/9310267} \BibitemShut
  {NoStop}%
\bibitem [{\citenamefont {Davies}\ \emph {et~al.}(1994)\citenamefont {Davies},
  \citenamefont {froggatt}, \citenamefont {Jenkins},\ and\ \citenamefont
  {Moorhouse}}]{Davies:1994id}%
  \BibitemOpen
  \bibfield  {author} {\bibinfo {author} {\bibfnamefont {A.~T.}\ \bibnamefont
  {Davies}}, \bibinfo {author} {\bibfnamefont {C.~D.}\ \bibnamefont
  {froggatt}}, \bibinfo {author} {\bibfnamefont {G.}~\bibnamefont {Jenkins}}, \
  and\ \bibinfo {author} {\bibfnamefont {R.~G.}\ \bibnamefont {Moorhouse}},\
  }\href {\doibase 10.1016/0370-2693(94)90559-2} {\bibfield  {journal}
  {\bibinfo  {journal} {Phys. Lett. B}\ }\textbf {\bibinfo {volume} {336}},\
  \bibinfo {pages} {464} (\bibinfo {year} {1994})}\BibitemShut {NoStop}%
\bibitem [{\citenamefont {Cline}\ \emph {et~al.}(1996)\citenamefont {Cline},
  \citenamefont {Kainulainen},\ and\ \citenamefont {Vischer}}]{Cline:1995dg}%
  \BibitemOpen
  \bibfield  {author} {\bibinfo {author} {\bibfnamefont {J.~M.}\ \bibnamefont
  {Cline}}, \bibinfo {author} {\bibfnamefont {K.}~\bibnamefont {Kainulainen}},
  \ and\ \bibinfo {author} {\bibfnamefont {A.~P.}\ \bibnamefont {Vischer}},\
  }\href {\doibase 10.1103/PhysRevD.54.2451} {\bibfield  {journal} {\bibinfo
  {journal} {Phys. Rev. D}\ }\textbf {\bibinfo {volume} {54}},\ \bibinfo
  {pages} {2451} (\bibinfo {year} {1996})},\ \Eprint
  {http://arxiv.org/abs/hep-ph/9506284} {arXiv:hep-ph/9506284} \BibitemShut
  {NoStop}%
\bibitem [{\citenamefont {Funakubo}\ \emph {et~al.}(1995)\citenamefont
  {Funakubo}, \citenamefont {Kakuto}, \citenamefont {Otsuki}, \citenamefont
  {Takenaga},\ and\ \citenamefont {Toyoda}}]{Funakubo:1995kw}%
  \BibitemOpen
  \bibfield  {author} {\bibinfo {author} {\bibfnamefont {K.}~\bibnamefont
  {Funakubo}}, \bibinfo {author} {\bibfnamefont {A.}~\bibnamefont {Kakuto}},
  \bibinfo {author} {\bibfnamefont {S.}~\bibnamefont {Otsuki}}, \bibinfo
  {author} {\bibfnamefont {K.}~\bibnamefont {Takenaga}}, \ and\ \bibinfo
  {author} {\bibfnamefont {F.}~\bibnamefont {Toyoda}},\ }\href {\doibase
  10.1143/PTP.94.845} {\bibfield  {journal} {\bibinfo  {journal} {Prog. Theor.
  Phys.}\ }\textbf {\bibinfo {volume} {94}},\ \bibinfo {pages} {845} (\bibinfo
  {year} {1995})},\ \Eprint {http://arxiv.org/abs/hep-ph/9507452}
  {arXiv:hep-ph/9507452} \BibitemShut {NoStop}%
\bibitem [{\citenamefont {Funakubo}\ \emph {et~al.}(1996)\citenamefont
  {Funakubo}, \citenamefont {Kakuto}, \citenamefont {Otsuki},\ and\
  \citenamefont {Toyoda}}]{Funakubo:1996iw}%
  \BibitemOpen
  \bibfield  {author} {\bibinfo {author} {\bibfnamefont {K.}~\bibnamefont
  {Funakubo}}, \bibinfo {author} {\bibfnamefont {A.}~\bibnamefont {Kakuto}},
  \bibinfo {author} {\bibfnamefont {S.}~\bibnamefont {Otsuki}}, \ and\ \bibinfo
  {author} {\bibfnamefont {F.}~\bibnamefont {Toyoda}},\ }\href {\doibase
  10.1143/PTP.96.771} {\bibfield  {journal} {\bibinfo  {journal} {Prog. Theor.
  Phys.}\ }\textbf {\bibinfo {volume} {96}},\ \bibinfo {pages} {771} (\bibinfo
  {year} {1996})},\ \Eprint {http://arxiv.org/abs/hep-ph/9606282}
  {arXiv:hep-ph/9606282} \BibitemShut {NoStop}%
\bibitem [{\citenamefont {Cline}\ and\ \citenamefont
  {Lemieux}(1997)}]{Cline:1996mga}%
  \BibitemOpen
  \bibfield  {author} {\bibinfo {author} {\bibfnamefont {J.~M.}\ \bibnamefont
  {Cline}}\ and\ \bibinfo {author} {\bibfnamefont {P.-A.}\ \bibnamefont
  {Lemieux}},\ }\href {\doibase 10.1103/PhysRevD.55.3873} {\bibfield  {journal}
  {\bibinfo  {journal} {Phys. Rev. D}\ }\textbf {\bibinfo {volume} {55}},\
  \bibinfo {pages} {3873} (\bibinfo {year} {1997})},\ \Eprint
  {http://arxiv.org/abs/hep-ph/9609240} {arXiv:hep-ph/9609240} \BibitemShut
  {NoStop}%
\bibitem [{\citenamefont {Fromme}\ \emph {et~al.}(2006)\citenamefont {Fromme},
  \citenamefont {Huber},\ and\ \citenamefont {Seniuch}}]{Fromme:2006cm}%
  \BibitemOpen
  \bibfield  {author} {\bibinfo {author} {\bibfnamefont {L.}~\bibnamefont
  {Fromme}}, \bibinfo {author} {\bibfnamefont {S.~J.}\ \bibnamefont {Huber}}, \
  and\ \bibinfo {author} {\bibfnamefont {M.}~\bibnamefont {Seniuch}},\ }\href
  {\doibase 10.1088/1126-6708/2006/11/038} {\bibfield  {journal} {\bibinfo
  {journal} {JHEP}\ }\textbf {\bibinfo {volume} {11}},\ \bibinfo {pages} {038}
  (\bibinfo {year} {2006})},\ \Eprint {http://arxiv.org/abs/hep-ph/0605242}
  {arXiv:hep-ph/0605242} \BibitemShut {NoStop}%
\bibitem [{\citenamefont {Cline}\ \emph {et~al.}(2011)\citenamefont {Cline},
  \citenamefont {Kainulainen},\ and\ \citenamefont {Trott}}]{Cline:2011mm}%
  \BibitemOpen
  \bibfield  {author} {\bibinfo {author} {\bibfnamefont {J.~M.}\ \bibnamefont
  {Cline}}, \bibinfo {author} {\bibfnamefont {K.}~\bibnamefont {Kainulainen}},
  \ and\ \bibinfo {author} {\bibfnamefont {M.}~\bibnamefont {Trott}},\ }\href
  {\doibase 10.1007/JHEP11(2011)089} {\bibfield  {journal} {\bibinfo  {journal}
  {JHEP}\ }\textbf {\bibinfo {volume} {11}},\ \bibinfo {pages} {089} (\bibinfo
  {year} {2011})},\ \Eprint {http://arxiv.org/abs/1107.3559} {arXiv:1107.3559
  [hep-ph]} \BibitemShut {NoStop}%
\bibitem [{\citenamefont {Dorsch}\ \emph {et~al.}(2013)\citenamefont {Dorsch},
  \citenamefont {Huber},\ and\ \citenamefont {No}}]{Dorsch:2013wja}%
  \BibitemOpen
  \bibfield  {author} {\bibinfo {author} {\bibfnamefont {G.~C.}\ \bibnamefont
  {Dorsch}}, \bibinfo {author} {\bibfnamefont {S.~J.}\ \bibnamefont {Huber}}, \
  and\ \bibinfo {author} {\bibfnamefont {J.~M.}\ \bibnamefont {No}},\ }\href
  {\doibase 10.1007/JHEP10(2013)029} {\bibfield  {journal} {\bibinfo  {journal}
  {JHEP}\ }\textbf {\bibinfo {volume} {10}},\ \bibinfo {pages} {029} (\bibinfo
  {year} {2013})},\ \Eprint {http://arxiv.org/abs/1305.6610} {arXiv:1305.6610
  [hep-ph]} \BibitemShut {NoStop}%
\bibitem [{\citenamefont {Cline}\ and\ \citenamefont
  {Kainulainen}(2013{\natexlab{b}})}]{Cline:2013bln}%
  \BibitemOpen
  \bibfield  {author} {\bibinfo {author} {\bibfnamefont {J.~M.}\ \bibnamefont
  {Cline}}\ and\ \bibinfo {author} {\bibfnamefont {K.}~\bibnamefont
  {Kainulainen}},\ }\href {\doibase 10.1103/PhysRevD.87.071701} {\bibfield
  {journal} {\bibinfo  {journal} {Phys. Rev. D}\ }\textbf {\bibinfo {volume}
  {87}},\ \bibinfo {pages} {071701} (\bibinfo {year} {2013}{\natexlab{b}})},\
  \Eprint {http://arxiv.org/abs/1302.2614} {arXiv:1302.2614 [hep-ph]}
  \BibitemShut {NoStop}%
\bibitem [{\citenamefont {Dorsch}\ \emph {et~al.}(2014)\citenamefont {Dorsch},
  \citenamefont {Huber}, \citenamefont {Mimasu},\ and\ \citenamefont
  {No}}]{Dorsch:2014qja}%
  \BibitemOpen
  \bibfield  {author} {\bibinfo {author} {\bibfnamefont {G.~C.}\ \bibnamefont
  {Dorsch}}, \bibinfo {author} {\bibfnamefont {S.~J.}\ \bibnamefont {Huber}},
  \bibinfo {author} {\bibfnamefont {K.}~\bibnamefont {Mimasu}}, \ and\ \bibinfo
  {author} {\bibfnamefont {J.~M.}\ \bibnamefont {No}},\ }\href {\doibase
  10.1103/PhysRevLett.113.211802} {\bibfield  {journal} {\bibinfo  {journal}
  {Phys. Rev. Lett.}\ }\textbf {\bibinfo {volume} {113}},\ \bibinfo {pages}
  {211802} (\bibinfo {year} {2014})},\ \Eprint {http://arxiv.org/abs/1405.5537}
  {arXiv:1405.5537 [hep-ph]} \BibitemShut {NoStop}%
\bibitem [{\citenamefont {Fuyuto}\ and\ \citenamefont
  {Senaha}(2015)}]{Fuyuto:2015jha}%
  \BibitemOpen
  \bibfield  {author} {\bibinfo {author} {\bibfnamefont {K.}~\bibnamefont
  {Fuyuto}}\ and\ \bibinfo {author} {\bibfnamefont {E.}~\bibnamefont
  {Senaha}},\ }\href {\doibase 10.1016/j.physletb.2015.05.061} {\bibfield
  {journal} {\bibinfo  {journal} {Phys. Lett. B}\ }\textbf {\bibinfo {volume}
  {747}},\ \bibinfo {pages} {152} (\bibinfo {year} {2015})},\ \Eprint
  {http://arxiv.org/abs/1504.04291} {arXiv:1504.04291 [hep-ph]} \BibitemShut
  {NoStop}%
\bibitem [{\citenamefont {Chao}\ and\ \citenamefont
  {Ramsey-Musolf}(2015)}]{Chao:2015uoa}%
  \BibitemOpen
  \bibfield  {author} {\bibinfo {author} {\bibfnamefont {W.}~\bibnamefont
  {Chao}}\ and\ \bibinfo {author} {\bibfnamefont {M.~J.}\ \bibnamefont
  {Ramsey-Musolf}},\ }\href@noop {} {\  (\bibinfo {year} {2015})},\ \Eprint
  {http://arxiv.org/abs/1503.00028} {arXiv:1503.00028 [hep-ph]} \BibitemShut
  {NoStop}%
\bibitem [{\citenamefont {Fuyuto}\ \emph {et~al.}(2016)\citenamefont {Fuyuto},
  \citenamefont {Hisano},\ and\ \citenamefont {Senaha}}]{Fuyuto:2015ida}%
  \BibitemOpen
  \bibfield  {author} {\bibinfo {author} {\bibfnamefont {K.}~\bibnamefont
  {Fuyuto}}, \bibinfo {author} {\bibfnamefont {J.}~\bibnamefont {Hisano}}, \
  and\ \bibinfo {author} {\bibfnamefont {E.}~\bibnamefont {Senaha}},\ }\href
  {\doibase 10.1016/j.physletb.2016.02.053} {\bibfield  {journal} {\bibinfo
  {journal} {Phys. Lett. B}\ }\textbf {\bibinfo {volume} {755}},\ \bibinfo
  {pages} {491} (\bibinfo {year} {2016})},\ \Eprint
  {http://arxiv.org/abs/1510.04485} {arXiv:1510.04485 [hep-ph]} \BibitemShut
  {NoStop}%
\bibitem [{\citenamefont {Chiang}\ \emph {et~al.}(2016)\citenamefont {Chiang},
  \citenamefont {Fuyuto},\ and\ \citenamefont {Senaha}}]{Chiang:2016vgf}%
  \BibitemOpen
  \bibfield  {author} {\bibinfo {author} {\bibfnamefont {C.-W.}\ \bibnamefont
  {Chiang}}, \bibinfo {author} {\bibfnamefont {K.}~\bibnamefont {Fuyuto}}, \
  and\ \bibinfo {author} {\bibfnamefont {E.}~\bibnamefont {Senaha}},\ }\href
  {\doibase 10.1016/j.physletb.2016.09.052} {\bibfield  {journal} {\bibinfo
  {journal} {Phys. Lett. B}\ }\textbf {\bibinfo {volume} {762}},\ \bibinfo
  {pages} {315} (\bibinfo {year} {2016})},\ \Eprint
  {http://arxiv.org/abs/1607.07316} {arXiv:1607.07316 [hep-ph]} \BibitemShut
  {NoStop}%
\bibitem [{\citenamefont {Haarr}\ \emph {et~al.}(2016)\citenamefont {Haarr},
  \citenamefont {Kvellestad},\ and\ \citenamefont {Petersen}}]{Haarr:2016qzq}%
  \BibitemOpen
  \bibfield  {author} {\bibinfo {author} {\bibfnamefont {A.}~\bibnamefont
  {Haarr}}, \bibinfo {author} {\bibfnamefont {A.}~\bibnamefont {Kvellestad}}, \
  and\ \bibinfo {author} {\bibfnamefont {T.~C.}\ \bibnamefont {Petersen}},\
  }\href@noop {} {\  (\bibinfo {year} {2016})},\ \Eprint
  {http://arxiv.org/abs/1611.05757} {arXiv:1611.05757 [hep-ph]} \BibitemShut
  {NoStop}%
\bibitem [{\citenamefont {Dorsch}\ \emph
  {et~al.}(2017{\natexlab{a}})\citenamefont {Dorsch}, \citenamefont {Huber},
  \citenamefont {Konstandin},\ and\ \citenamefont {No}}]{Dorsch:2016nrg}%
  \BibitemOpen
  \bibfield  {author} {\bibinfo {author} {\bibfnamefont {G.~C.}\ \bibnamefont
  {Dorsch}}, \bibinfo {author} {\bibfnamefont {S.~J.}\ \bibnamefont {Huber}},
  \bibinfo {author} {\bibfnamefont {T.}~\bibnamefont {Konstandin}}, \ and\
  \bibinfo {author} {\bibfnamefont {J.~M.}\ \bibnamefont {No}},\ }\href
  {\doibase 10.1088/1475-7516/2017/05/052} {\bibfield  {journal} {\bibinfo
  {journal} {JCAP}\ }\textbf {\bibinfo {volume} {05}},\ \bibinfo {pages} {052}
  (\bibinfo {year} {2017}{\natexlab{a}})},\ \Eprint
  {http://arxiv.org/abs/1611.05874} {arXiv:1611.05874 [hep-ph]} \BibitemShut
  {NoStop}%
\bibitem [{\citenamefont {Basler}\ \emph {et~al.}(2017)\citenamefont {Basler},
  \citenamefont {Krause}, \citenamefont {Muhlleitner}, \citenamefont
  {Wittbrodt},\ and\ \citenamefont {Wlotzka}}]{Basler:2016obg}%
  \BibitemOpen
  \bibfield  {author} {\bibinfo {author} {\bibfnamefont {P.}~\bibnamefont
  {Basler}}, \bibinfo {author} {\bibfnamefont {M.}~\bibnamefont {Krause}},
  \bibinfo {author} {\bibfnamefont {M.}~\bibnamefont {Muhlleitner}}, \bibinfo
  {author} {\bibfnamefont {J.}~\bibnamefont {Wittbrodt}}, \ and\ \bibinfo
  {author} {\bibfnamefont {A.}~\bibnamefont {Wlotzka}},\ }\href {\doibase
  10.1007/JHEP02(2017)121} {\bibfield  {journal} {\bibinfo  {journal} {JHEP}\
  }\textbf {\bibinfo {volume} {02}},\ \bibinfo {pages} {121} (\bibinfo {year}
  {2017})},\ \Eprint {http://arxiv.org/abs/1612.04086} {arXiv:1612.04086
  [hep-ph]} \BibitemShut {NoStop}%
\bibitem [{\citenamefont {Fuyuto}\ \emph {et~al.}(2018)\citenamefont {Fuyuto},
  \citenamefont {Hou},\ and\ \citenamefont {Senaha}}]{Fuyuto:2017ewj}%
  \BibitemOpen
  \bibfield  {author} {\bibinfo {author} {\bibfnamefont {K.}~\bibnamefont
  {Fuyuto}}, \bibinfo {author} {\bibfnamefont {W.-S.}\ \bibnamefont {Hou}}, \
  and\ \bibinfo {author} {\bibfnamefont {E.}~\bibnamefont {Senaha}},\ }\href
  {\doibase 10.1016/j.physletb.2017.11.073} {\bibfield  {journal} {\bibinfo
  {journal} {Phys. Lett. B}\ }\textbf {\bibinfo {volume} {776}},\ \bibinfo
  {pages} {402} (\bibinfo {year} {2018})},\ \Eprint
  {http://arxiv.org/abs/1705.05034} {arXiv:1705.05034 [hep-ph]} \BibitemShut
  {NoStop}%
\bibitem [{\citenamefont {Dorsch}\ \emph
  {et~al.}(2017{\natexlab{b}})\citenamefont {Dorsch}, \citenamefont {Huber},
  \citenamefont {Mimasu},\ and\ \citenamefont {No}}]{Dorsch:2017nza}%
  \BibitemOpen
  \bibfield  {author} {\bibinfo {author} {\bibfnamefont {G.~C.}\ \bibnamefont
  {Dorsch}}, \bibinfo {author} {\bibfnamefont {S.~J.}\ \bibnamefont {Huber}},
  \bibinfo {author} {\bibfnamefont {K.}~\bibnamefont {Mimasu}}, \ and\ \bibinfo
  {author} {\bibfnamefont {J.~M.}\ \bibnamefont {No}},\ }\href {\doibase
  10.1007/JHEP12(2017)086} {\bibfield  {journal} {\bibinfo  {journal} {JHEP}\
  }\textbf {\bibinfo {volume} {12}},\ \bibinfo {pages} {086} (\bibinfo {year}
  {2017}{\natexlab{b}})},\ \Eprint {http://arxiv.org/abs/1705.09186}
  {arXiv:1705.09186 [hep-ph]} \BibitemShut {NoStop}%
\bibitem [{\citenamefont {Cherchiglia}\ and\ \citenamefont
  {Nishi}(2017)}]{Cherchiglia:2017gko}%
  \BibitemOpen
  \bibfield  {author} {\bibinfo {author} {\bibfnamefont {A.~L.}\ \bibnamefont
  {Cherchiglia}}\ and\ \bibinfo {author} {\bibfnamefont {C.~C.}\ \bibnamefont
  {Nishi}},\ }\href {\doibase 10.1007/JHEP11(2017)106} {\bibfield  {journal}
  {\bibinfo  {journal} {JHEP}\ }\textbf {\bibinfo {volume} {11}},\ \bibinfo
  {pages} {106} (\bibinfo {year} {2017})},\ \Eprint
  {http://arxiv.org/abs/1707.04595} {arXiv:1707.04595 [hep-ph]} \BibitemShut
  {NoStop}%
\bibitem [{\citenamefont {Basler}\ \emph {et~al.}(2018)\citenamefont {Basler},
  \citenamefont {M\"uhlleitner},\ and\ \citenamefont
  {Wittbrodt}}]{Basler:2017uxn}%
  \BibitemOpen
  \bibfield  {author} {\bibinfo {author} {\bibfnamefont {P.}~\bibnamefont
  {Basler}}, \bibinfo {author} {\bibfnamefont {M.}~\bibnamefont
  {M\"uhlleitner}}, \ and\ \bibinfo {author} {\bibfnamefont {J.}~\bibnamefont
  {Wittbrodt}},\ }\href {\doibase 10.1007/JHEP03(2018)061} {\bibfield
  {journal} {\bibinfo  {journal} {JHEP}\ }\textbf {\bibinfo {volume} {03}},\
  \bibinfo {pages} {061} (\bibinfo {year} {2018})},\ \Eprint
  {http://arxiv.org/abs/1711.04097} {arXiv:1711.04097 [hep-ph]} \BibitemShut
  {NoStop}%
\bibitem [{\citenamefont {Andersen}\ \emph {et~al.}(2018)\citenamefont
  {Andersen}, \citenamefont {Gorda}, \citenamefont {Helset}, \citenamefont
  {Niemi}, \citenamefont {Tenkanen}, \citenamefont {Tranberg}, \citenamefont
  {Vuorinen},\ and\ \citenamefont {Weir}}]{Andersen:2017ika}%
  \BibitemOpen
  \bibfield  {author} {\bibinfo {author} {\bibfnamefont {J.~O.}\ \bibnamefont
  {Andersen}}, \bibinfo {author} {\bibfnamefont {T.}~\bibnamefont {Gorda}},
  \bibinfo {author} {\bibfnamefont {A.}~\bibnamefont {Helset}}, \bibinfo
  {author} {\bibfnamefont {L.}~\bibnamefont {Niemi}}, \bibinfo {author}
  {\bibfnamefont {T.~V.~I.}\ \bibnamefont {Tenkanen}}, \bibinfo {author}
  {\bibfnamefont {A.}~\bibnamefont {Tranberg}}, \bibinfo {author}
  {\bibfnamefont {A.}~\bibnamefont {Vuorinen}}, \ and\ \bibinfo {author}
  {\bibfnamefont {D.~J.}\ \bibnamefont {Weir}},\ }\href {\doibase
  10.1103/PhysRevLett.121.191802} {\bibfield  {journal} {\bibinfo  {journal}
  {Phys. Rev. Lett.}\ }\textbf {\bibinfo {volume} {121}},\ \bibinfo {pages}
  {191802} (\bibinfo {year} {2018})},\ \Eprint
  {http://arxiv.org/abs/1711.09849} {arXiv:1711.09849 [hep-ph]} \BibitemShut
  {NoStop}%
\bibitem [{\citenamefont {Bernon}\ \emph {et~al.}(2018)\citenamefont {Bernon},
  \citenamefont {Bian},\ and\ \citenamefont {Jiang}}]{Bernon:2017jgv}%
  \BibitemOpen
  \bibfield  {author} {\bibinfo {author} {\bibfnamefont {J.}~\bibnamefont
  {Bernon}}, \bibinfo {author} {\bibfnamefont {L.}~\bibnamefont {Bian}}, \ and\
  \bibinfo {author} {\bibfnamefont {Y.}~\bibnamefont {Jiang}},\ }\href
  {\doibase 10.1007/JHEP05(2018)151} {\bibfield  {journal} {\bibinfo  {journal}
  {JHEP}\ }\textbf {\bibinfo {volume} {05}},\ \bibinfo {pages} {151} (\bibinfo
  {year} {2018})},\ \Eprint {http://arxiv.org/abs/1712.08430} {arXiv:1712.08430
  [hep-ph]} \BibitemShut {NoStop}%
\bibitem [{\citenamefont {Huang}\ and\ \citenamefont
  {Yu}(2018)}]{Huang:2017rzf}%
  \BibitemOpen
  \bibfield  {author} {\bibinfo {author} {\bibfnamefont {F.~P.}\ \bibnamefont
  {Huang}}\ and\ \bibinfo {author} {\bibfnamefont {J.-H.}\ \bibnamefont {Yu}},\
  }\href {\doibase 10.1103/PhysRevD.98.095022} {\bibfield  {journal} {\bibinfo
  {journal} {Phys. Rev. D}\ }\textbf {\bibinfo {volume} {98}},\ \bibinfo
  {pages} {095022} (\bibinfo {year} {2018})},\ \Eprint
  {http://arxiv.org/abs/1704.04201} {arXiv:1704.04201 [hep-ph]} \BibitemShut
  {NoStop}%
\bibitem [{\citenamefont {Gorda}\ \emph {et~al.}(2019)\citenamefont {Gorda},
  \citenamefont {Helset}, \citenamefont {Niemi}, \citenamefont {Tenkanen},\
  and\ \citenamefont {Weir}}]{Gorda:2018hvi}%
  \BibitemOpen
  \bibfield  {author} {\bibinfo {author} {\bibfnamefont {T.}~\bibnamefont
  {Gorda}}, \bibinfo {author} {\bibfnamefont {A.}~\bibnamefont {Helset}},
  \bibinfo {author} {\bibfnamefont {L.}~\bibnamefont {Niemi}}, \bibinfo
  {author} {\bibfnamefont {T.~V.~I.}\ \bibnamefont {Tenkanen}}, \ and\ \bibinfo
  {author} {\bibfnamefont {D.~J.}\ \bibnamefont {Weir}},\ }\href {\doibase
  10.1007/JHEP02(2019)081} {\bibfield  {journal} {\bibinfo  {journal} {JHEP}\
  }\textbf {\bibinfo {volume} {02}},\ \bibinfo {pages} {081} (\bibinfo {year}
  {2019})},\ \Eprint {http://arxiv.org/abs/1802.05056} {arXiv:1802.05056
  [hep-ph]} \BibitemShut {NoStop}%
\bibitem [{\citenamefont {Basler}\ and\ \citenamefont
  {M\"uhlleitner}(2019)}]{Basler:2018cwe}%
  \BibitemOpen
  \bibfield  {author} {\bibinfo {author} {\bibfnamefont {P.}~\bibnamefont
  {Basler}}\ and\ \bibinfo {author} {\bibfnamefont {M.}~\bibnamefont
  {M\"uhlleitner}},\ }\href {\doibase 10.1016/j.cpc.2018.11.006} {\bibfield
  {journal} {\bibinfo  {journal} {Comput. Phys. Commun.}\ }\textbf {\bibinfo
  {volume} {237}},\ \bibinfo {pages} {62} (\bibinfo {year} {2019})},\ \Eprint
  {http://arxiv.org/abs/1803.02846} {arXiv:1803.02846 [hep-ph]} \BibitemShut
  {NoStop}%
\bibitem [{\citenamefont {Modak}\ and\ \citenamefont
  {Senaha}(2019)}]{Modak:2018csw}%
  \BibitemOpen
  \bibfield  {author} {\bibinfo {author} {\bibfnamefont {T.}~\bibnamefont
  {Modak}}\ and\ \bibinfo {author} {\bibfnamefont {E.}~\bibnamefont {Senaha}},\
  }\href {\doibase 10.1103/PhysRevD.99.115022} {\bibfield  {journal} {\bibinfo
  {journal} {Phys. Rev. D}\ }\textbf {\bibinfo {volume} {99}},\ \bibinfo
  {pages} {115022} (\bibinfo {year} {2019})},\ \Eprint
  {http://arxiv.org/abs/1811.08088} {arXiv:1811.08088 [hep-ph]} \BibitemShut
  {NoStop}%
\bibitem [{\citenamefont {Wang}\ \emph {et~al.}(2019)\citenamefont {Wang},
  \citenamefont {Yang}, \citenamefont {Zhang},\ and\ \citenamefont
  {Zhang}}]{Wang:2018hnw}%
  \BibitemOpen
  \bibfield  {author} {\bibinfo {author} {\bibfnamefont {L.}~\bibnamefont
  {Wang}}, \bibinfo {author} {\bibfnamefont {J.~M.}\ \bibnamefont {Yang}},
  \bibinfo {author} {\bibfnamefont {M.}~\bibnamefont {Zhang}}, \ and\ \bibinfo
  {author} {\bibfnamefont {Y.}~\bibnamefont {Zhang}},\ }\href {\doibase
  10.1016/j.physletb.2018.11.045} {\bibfield  {journal} {\bibinfo  {journal}
  {Phys. Lett. B}\ }\textbf {\bibinfo {volume} {788}},\ \bibinfo {pages} {519}
  (\bibinfo {year} {2019})},\ \Eprint {http://arxiv.org/abs/1809.05857}
  {arXiv:1809.05857 [hep-ph]} \BibitemShut {NoStop}%
\bibitem [{\citenamefont {Wang}\ \emph {et~al.}(2020)\citenamefont {Wang},
  \citenamefont {Huang},\ and\ \citenamefont {Zhang}}]{Wang:2019pet}%
  \BibitemOpen
  \bibfield  {author} {\bibinfo {author} {\bibfnamefont {X.}~\bibnamefont
  {Wang}}, \bibinfo {author} {\bibfnamefont {F.~P.}\ \bibnamefont {Huang}}, \
  and\ \bibinfo {author} {\bibfnamefont {X.}~\bibnamefont {Zhang}},\ }\href
  {\doibase 10.1103/PhysRevD.101.015015} {\bibfield  {journal} {\bibinfo
  {journal} {Phys. Rev. D}\ }\textbf {\bibinfo {volume} {101}},\ \bibinfo
  {pages} {015015} (\bibinfo {year} {2020})},\ \Eprint
  {http://arxiv.org/abs/1909.02978} {arXiv:1909.02978 [hep-ph]} \BibitemShut
  {NoStop}%
\bibitem [{\citenamefont {Kainulainen}\ \emph {et~al.}(2019)\citenamefont
  {Kainulainen}, \citenamefont {Keus}, \citenamefont {Niemi}, \citenamefont
  {Rummukainen}, \citenamefont {Tenkanen},\ and\ \citenamefont
  {Vaskonen}}]{Kainulainen:2019kyp}%
  \BibitemOpen
  \bibfield  {author} {\bibinfo {author} {\bibfnamefont {K.}~\bibnamefont
  {Kainulainen}}, \bibinfo {author} {\bibfnamefont {V.}~\bibnamefont {Keus}},
  \bibinfo {author} {\bibfnamefont {L.}~\bibnamefont {Niemi}}, \bibinfo
  {author} {\bibfnamefont {K.}~\bibnamefont {Rummukainen}}, \bibinfo {author}
  {\bibfnamefont {T.~V.~I.}\ \bibnamefont {Tenkanen}}, \ and\ \bibinfo {author}
  {\bibfnamefont {V.}~\bibnamefont {Vaskonen}},\ }\href {\doibase
  10.1007/JHEP06(2019)075} {\bibfield  {journal} {\bibinfo  {journal} {JHEP}\
  }\textbf {\bibinfo {volume} {06}},\ \bibinfo {pages} {075} (\bibinfo {year}
  {2019})},\ \Eprint {http://arxiv.org/abs/1904.01329} {arXiv:1904.01329
  [hep-ph]} \BibitemShut {NoStop}%
\bibitem [{\citenamefont {Paul}\ \emph {et~al.}(2021)\citenamefont {Paul},
  \citenamefont {Mukhopadhyay},\ and\ \citenamefont {Majumdar}}]{Paul:2020wbz}%
  \BibitemOpen
  \bibfield  {author} {\bibinfo {author} {\bibfnamefont {A.}~\bibnamefont
  {Paul}}, \bibinfo {author} {\bibfnamefont {U.}~\bibnamefont {Mukhopadhyay}},
  \ and\ \bibinfo {author} {\bibfnamefont {D.}~\bibnamefont {Majumdar}},\
  }\href {\doibase 10.1007/JHEP05(2021)223} {\bibfield  {journal} {\bibinfo
  {journal} {JHEP}\ }\textbf {\bibinfo {volume} {05}},\ \bibinfo {pages} {223}
  (\bibinfo {year} {2021})},\ \Eprint {http://arxiv.org/abs/2010.03439}
  {arXiv:2010.03439 [hep-ph]} \BibitemShut {NoStop}%
\bibitem [{\citenamefont {Su}\ \emph {et~al.}(2021)\citenamefont {Su},
  \citenamefont {Williams},\ and\ \citenamefont {Zhang}}]{Su:2020pjw}%
  \BibitemOpen
  \bibfield  {author} {\bibinfo {author} {\bibfnamefont {W.}~\bibnamefont
  {Su}}, \bibinfo {author} {\bibfnamefont {A.~G.}\ \bibnamefont {Williams}}, \
  and\ \bibinfo {author} {\bibfnamefont {M.}~\bibnamefont {Zhang}},\ }\href
  {\doibase 10.1007/JHEP04(2021)219} {\bibfield  {journal} {\bibinfo  {journal}
  {JHEP}\ }\textbf {\bibinfo {volume} {04}},\ \bibinfo {pages} {219} (\bibinfo
  {year} {2021})},\ \Eprint {http://arxiv.org/abs/2011.04540} {arXiv:2011.04540
  [hep-ph]} \BibitemShut {NoStop}%
\bibitem [{\citenamefont {Apreda}\ \emph {et~al.}(2002)\citenamefont {Apreda},
  \citenamefont {Maggiore}, \citenamefont {Nicolis},\ and\ \citenamefont
  {Riotto}}]{Apreda:2001us}%
  \BibitemOpen
  \bibfield  {author} {\bibinfo {author} {\bibfnamefont {R.}~\bibnamefont
  {Apreda}}, \bibinfo {author} {\bibfnamefont {M.}~\bibnamefont {Maggiore}},
  \bibinfo {author} {\bibfnamefont {A.}~\bibnamefont {Nicolis}}, \ and\
  \bibinfo {author} {\bibfnamefont {A.}~\bibnamefont {Riotto}},\ }\href
  {\doibase 10.1016/S0550-3213(02)00264-X} {\bibfield  {journal} {\bibinfo
  {journal} {Nucl. Phys. B}\ }\textbf {\bibinfo {volume} {631}},\ \bibinfo
  {pages} {342} (\bibinfo {year} {2002})},\ \Eprint
  {http://arxiv.org/abs/gr-qc/0107033} {arXiv:gr-qc/0107033} \BibitemShut
  {NoStop}%
\bibitem [{\citenamefont {Huber}\ \emph {et~al.}(2016)\citenamefont {Huber},
  \citenamefont {Konstandin}, \citenamefont {Nardini},\ and\ \citenamefont
  {Rues}}]{Huber:2015znp}%
  \BibitemOpen
  \bibfield  {author} {\bibinfo {author} {\bibfnamefont {S.~J.}\ \bibnamefont
  {Huber}}, \bibinfo {author} {\bibfnamefont {T.}~\bibnamefont {Konstandin}},
  \bibinfo {author} {\bibfnamefont {G.}~\bibnamefont {Nardini}}, \ and\
  \bibinfo {author} {\bibfnamefont {I.}~\bibnamefont {Rues}},\ }\href {\doibase
  10.1088/1475-7516/2016/03/036} {\bibfield  {journal} {\bibinfo  {journal}
  {JCAP}\ }\textbf {\bibinfo {volume} {03}},\ \bibinfo {pages} {036} (\bibinfo
  {year} {2016})},\ \Eprint {http://arxiv.org/abs/1512.06357} {arXiv:1512.06357
  [hep-ph]} \BibitemShut {NoStop}%
\bibitem [{\citenamefont {Huber}\ and\ \citenamefont
  {Konstandin}(2008{\natexlab{a}})}]{Huber:2007vva}%
  \BibitemOpen
  \bibfield  {author} {\bibinfo {author} {\bibfnamefont {S.~J.}\ \bibnamefont
  {Huber}}\ and\ \bibinfo {author} {\bibfnamefont {T.}~\bibnamefont
  {Konstandin}},\ }\href {\doibase 10.1088/1475-7516/2008/05/017} {\bibfield
  {journal} {\bibinfo  {journal} {JCAP}\ }\textbf {\bibinfo {volume} {05}},\
  \bibinfo {pages} {017} (\bibinfo {year} {2008}{\natexlab{a}})},\ \Eprint
  {http://arxiv.org/abs/0709.2091} {arXiv:0709.2091 [hep-ph]} \BibitemShut
  {NoStop}%
\bibitem [{\citenamefont {Demidov}\ \emph {et~al.}(2018)\citenamefont
  {Demidov}, \citenamefont {Gorbunov},\ and\ \citenamefont
  {Kirpichnikov}}]{Demidov:2017lzf}%
  \BibitemOpen
  \bibfield  {author} {\bibinfo {author} {\bibfnamefont {S.~V.}\ \bibnamefont
  {Demidov}}, \bibinfo {author} {\bibfnamefont {D.~S.}\ \bibnamefont
  {Gorbunov}}, \ and\ \bibinfo {author} {\bibfnamefont {D.~V.}\ \bibnamefont
  {Kirpichnikov}},\ }\href {\doibase 10.1016/j.physletb.2018.02.007} {\bibfield
   {journal} {\bibinfo  {journal} {Phys. Lett. B}\ }\textbf {\bibinfo {volume}
  {779}},\ \bibinfo {pages} {191} (\bibinfo {year} {2018})},\ \Eprint
  {http://arxiv.org/abs/1712.00087} {arXiv:1712.00087 [hep-ph]} \BibitemShut
  {NoStop}%
\bibitem [{\citenamefont {Brdar}\ \emph {et~al.}(2019)\citenamefont {Brdar},
  \citenamefont {Graf}, \citenamefont {Helmboldt},\ and\ \citenamefont
  {Xu}}]{Brdar:2019fur}%
  \BibitemOpen
  \bibfield  {author} {\bibinfo {author} {\bibfnamefont {V.}~\bibnamefont
  {Brdar}}, \bibinfo {author} {\bibfnamefont {L.}~\bibnamefont {Graf}},
  \bibinfo {author} {\bibfnamefont {A.~J.}\ \bibnamefont {Helmboldt}}, \ and\
  \bibinfo {author} {\bibfnamefont {X.-J.}\ \bibnamefont {Xu}},\ }\href
  {\doibase 10.1088/1475-7516/2019/12/027} {\bibfield  {journal} {\bibinfo
  {journal} {JCAP}\ }\textbf {\bibinfo {volume} {12}},\ \bibinfo {pages} {027}
  (\bibinfo {year} {2019})},\ \Eprint {http://arxiv.org/abs/1909.02018}
  {arXiv:1909.02018 [hep-ph]} \BibitemShut {NoStop}%
\bibitem [{\citenamefont {Li}\ \emph {et~al.}(2021)\citenamefont {Li},
  \citenamefont {Yan}, \citenamefont {Zhang},\ and\ \citenamefont
  {Zhao}}]{Li:2020eun}%
  \BibitemOpen
  \bibfield  {author} {\bibinfo {author} {\bibfnamefont {M.}~\bibnamefont
  {Li}}, \bibinfo {author} {\bibfnamefont {Q.-S.}\ \bibnamefont {Yan}},
  \bibinfo {author} {\bibfnamefont {Y.}~\bibnamefont {Zhang}}, \ and\ \bibinfo
  {author} {\bibfnamefont {Z.}~\bibnamefont {Zhao}},\ }\href {\doibase
  10.1007/JHEP03(2021)267} {\bibfield  {journal} {\bibinfo  {journal} {JHEP}\
  }\textbf {\bibinfo {volume} {03}},\ \bibinfo {pages} {267} (\bibinfo {year}
  {2021})},\ \Eprint {http://arxiv.org/abs/2012.13686} {arXiv:2012.13686
  [hep-ph]} \BibitemShut {NoStop}%
\bibitem [{\citenamefont {Huang}\ \emph {et~al.}(2020)\citenamefont {Huang},
  \citenamefont {Sannino},\ and\ \citenamefont {Wang}}]{Huang:2020bbe}%
  \BibitemOpen
  \bibfield  {author} {\bibinfo {author} {\bibfnamefont {W.-C.}\ \bibnamefont
  {Huang}}, \bibinfo {author} {\bibfnamefont {F.}~\bibnamefont {Sannino}}, \
  and\ \bibinfo {author} {\bibfnamefont {Z.-W.}\ \bibnamefont {Wang}},\ }\href
  {\doibase 10.1103/PhysRevD.102.095025} {\bibfield  {journal} {\bibinfo
  {journal} {Phys. Rev. D}\ }\textbf {\bibinfo {volume} {102}},\ \bibinfo
  {pages} {095025} (\bibinfo {year} {2020})},\ \Eprint
  {http://arxiv.org/abs/2004.02332} {arXiv:2004.02332 [hep-ph]} \BibitemShut
  {NoStop}%
\bibitem [{\citenamefont {Chiang}\ and\ \citenamefont
  {Yamada}(2014)}]{Chiang:2014hia}%
  \BibitemOpen
  \bibfield  {author} {\bibinfo {author} {\bibfnamefont {C.-W.}\ \bibnamefont
  {Chiang}}\ and\ \bibinfo {author} {\bibfnamefont {T.}~\bibnamefont
  {Yamada}},\ }\href {\doibase 10.1016/j.physletb.2014.06.048} {\bibfield
  {journal} {\bibinfo  {journal} {Phys. Lett. B}\ }\textbf {\bibinfo {volume}
  {735}},\ \bibinfo {pages} {295} (\bibinfo {year} {2014})},\ \Eprint
  {http://arxiv.org/abs/1404.5182} {arXiv:1404.5182 [hep-ph]} \BibitemShut
  {NoStop}%
\bibitem [{\citenamefont {Phong}\ \emph {et~al.}(2018)\citenamefont {Phong},
  \citenamefont {Thao},\ and\ \citenamefont {Long}}]{Phong:2015vlk}%
  \BibitemOpen
  \bibfield  {author} {\bibinfo {author} {\bibfnamefont {V.~Q.}\ \bibnamefont
  {Phong}}, \bibinfo {author} {\bibfnamefont {N.~C.}\ \bibnamefont {Thao}}, \
  and\ \bibinfo {author} {\bibfnamefont {H.~N.}\ \bibnamefont {Long}},\ }\href
  {\doibase 10.1103/PhysRevD.97.115008} {\bibfield  {journal} {\bibinfo
  {journal} {Phys. Rev. D}\ }\textbf {\bibinfo {volume} {97}},\ \bibinfo
  {pages} {115008} (\bibinfo {year} {2018})},\ \Eprint
  {http://arxiv.org/abs/1511.00579} {arXiv:1511.00579 [hep-ph]} \BibitemShut
  {NoStop}%
\bibitem [{\citenamefont {Phong}\ \emph {et~al.}(2022)\citenamefont {Phong},
  \citenamefont {Thao},\ and\ \citenamefont {Long}}]{Phong:2021lea}%
  \BibitemOpen
  \bibfield  {author} {\bibinfo {author} {\bibfnamefont {V.~Q.}\ \bibnamefont
  {Phong}}, \bibinfo {author} {\bibfnamefont {N.~C.}\ \bibnamefont {Thao}}, \
  and\ \bibinfo {author} {\bibfnamefont {H.~N.}\ \bibnamefont {Long}},\ }\href
  {\doibase 10.1140/epjc/s10052-022-10961-2} {\bibfield  {journal} {\bibinfo
  {journal} {Eur. Phys. J. C}\ }\textbf {\bibinfo {volume} {82}},\ \bibinfo
  {pages} {1005} (\bibinfo {year} {2022})},\ \Eprint
  {http://arxiv.org/abs/2107.13823} {arXiv:2107.13823 [hep-ph]} \BibitemShut
  {NoStop}%
\bibitem [{\citenamefont {Espinosa}\ \emph
  {et~al.}(2012{\natexlab{b}})\citenamefont {Espinosa}, \citenamefont
  {Gripaios}, \citenamefont {Konstandin},\ and\ \citenamefont
  {Riva}}]{Espinosa:2011eu}%
  \BibitemOpen
  \bibfield  {author} {\bibinfo {author} {\bibfnamefont {J.~R.}\ \bibnamefont
  {Espinosa}}, \bibinfo {author} {\bibfnamefont {B.}~\bibnamefont {Gripaios}},
  \bibinfo {author} {\bibfnamefont {T.}~\bibnamefont {Konstandin}}, \ and\
  \bibinfo {author} {\bibfnamefont {F.}~\bibnamefont {Riva}},\ }\href {\doibase
  10.1088/1475-7516/2012/01/012} {\bibfield  {journal} {\bibinfo  {journal}
  {JCAP}\ }\textbf {\bibinfo {volume} {01}},\ \bibinfo {pages} {012} (\bibinfo
  {year} {2012}{\natexlab{b}})},\ \Eprint {http://arxiv.org/abs/1110.2876}
  {arXiv:1110.2876 [hep-ph]} \BibitemShut {NoStop}%
\bibitem [{\citenamefont {Chala}\ \emph {et~al.}(2016)\citenamefont {Chala},
  \citenamefont {Nardini},\ and\ \citenamefont {Sobolev}}]{Chala:2016ykx}%
  \BibitemOpen
  \bibfield  {author} {\bibinfo {author} {\bibfnamefont {M.}~\bibnamefont
  {Chala}}, \bibinfo {author} {\bibfnamefont {G.}~\bibnamefont {Nardini}}, \
  and\ \bibinfo {author} {\bibfnamefont {I.}~\bibnamefont {Sobolev}},\ }\href
  {\doibase 10.1103/PhysRevD.94.055006} {\bibfield  {journal} {\bibinfo
  {journal} {Phys. Rev. D}\ }\textbf {\bibinfo {volume} {94}},\ \bibinfo
  {pages} {055006} (\bibinfo {year} {2016})},\ \Eprint
  {http://arxiv.org/abs/1605.08663} {arXiv:1605.08663 [hep-ph]} \BibitemShut
  {NoStop}%
\bibitem [{\citenamefont {Bruggisser}\ \emph
  {et~al.}(2018{\natexlab{a}})\citenamefont {Bruggisser}, \citenamefont
  {Von~Harling}, \citenamefont {Matsedonskyi},\ and\ \citenamefont
  {Servant}}]{Bruggisser:2018mus}%
  \BibitemOpen
  \bibfield  {author} {\bibinfo {author} {\bibfnamefont {S.}~\bibnamefont
  {Bruggisser}}, \bibinfo {author} {\bibfnamefont {B.}~\bibnamefont
  {Von~Harling}}, \bibinfo {author} {\bibfnamefont {O.}~\bibnamefont
  {Matsedonskyi}}, \ and\ \bibinfo {author} {\bibfnamefont {G.}~\bibnamefont
  {Servant}},\ }\href {\doibase 10.1103/PhysRevLett.121.131801} {\bibfield
  {journal} {\bibinfo  {journal} {Phys. Rev. Lett.}\ }\textbf {\bibinfo
  {volume} {121}},\ \bibinfo {pages} {131801} (\bibinfo {year}
  {2018}{\natexlab{a}})},\ \Eprint {http://arxiv.org/abs/1803.08546}
  {arXiv:1803.08546 [hep-ph]} \BibitemShut {NoStop}%
\bibitem [{\citenamefont {Bruggisser}\ \emph
  {et~al.}(2018{\natexlab{b}})\citenamefont {Bruggisser}, \citenamefont
  {Von~Harling}, \citenamefont {Matsedonskyi},\ and\ \citenamefont
  {Servant}}]{Bruggisser:2018mrt}%
  \BibitemOpen
  \bibfield  {author} {\bibinfo {author} {\bibfnamefont {S.}~\bibnamefont
  {Bruggisser}}, \bibinfo {author} {\bibfnamefont {B.}~\bibnamefont
  {Von~Harling}}, \bibinfo {author} {\bibfnamefont {O.}~\bibnamefont
  {Matsedonskyi}}, \ and\ \bibinfo {author} {\bibfnamefont {G.}~\bibnamefont
  {Servant}},\ }\href {\doibase 10.1007/JHEP12(2018)099} {\bibfield  {journal}
  {\bibinfo  {journal} {JHEP}\ }\textbf {\bibinfo {volume} {12}},\ \bibinfo
  {pages} {099} (\bibinfo {year} {2018}{\natexlab{b}})},\ \Eprint
  {http://arxiv.org/abs/1804.07314} {arXiv:1804.07314 [hep-ph]} \BibitemShut
  {NoStop}%
\bibitem [{\citenamefont {Bian}\ \emph {et~al.}(2019)\citenamefont {Bian},
  \citenamefont {Wu},\ and\ \citenamefont {Xie}}]{Bian:2019kmg}%
  \BibitemOpen
  \bibfield  {author} {\bibinfo {author} {\bibfnamefont {L.}~\bibnamefont
  {Bian}}, \bibinfo {author} {\bibfnamefont {Y.}~\bibnamefont {Wu}}, \ and\
  \bibinfo {author} {\bibfnamefont {K.-P.}\ \bibnamefont {Xie}},\ }\href
  {\doibase 10.1007/JHEP12(2019)028} {\bibfield  {journal} {\bibinfo  {journal}
  {JHEP}\ }\textbf {\bibinfo {volume} {12}},\ \bibinfo {pages} {028} (\bibinfo
  {year} {2019})},\ \Eprint {http://arxiv.org/abs/1909.02014} {arXiv:1909.02014
  [hep-ph]} \BibitemShut {NoStop}%
\bibitem [{\citenamefont {De~Curtis}\ \emph {et~al.}(2019)\citenamefont
  {De~Curtis}, \citenamefont {Delle~Rose},\ and\ \citenamefont
  {Panico}}]{DeCurtis:2019rxl}%
  \BibitemOpen
  \bibfield  {author} {\bibinfo {author} {\bibfnamefont {S.}~\bibnamefont
  {De~Curtis}}, \bibinfo {author} {\bibfnamefont {L.}~\bibnamefont
  {Delle~Rose}}, \ and\ \bibinfo {author} {\bibfnamefont {G.}~\bibnamefont
  {Panico}},\ }\href {\doibase 10.1007/JHEP12(2019)149} {\bibfield  {journal}
  {\bibinfo  {journal} {JHEP}\ }\textbf {\bibinfo {volume} {12}},\ \bibinfo
  {pages} {149} (\bibinfo {year} {2019})},\ \Eprint
  {http://arxiv.org/abs/1909.07894} {arXiv:1909.07894 [hep-ph]} \BibitemShut
  {NoStop}%
\bibitem [{\citenamefont {Xie}\ \emph {et~al.}(2020)\citenamefont {Xie},
  \citenamefont {Bian},\ and\ \citenamefont {Wu}}]{Xie:2020bkl}%
  \BibitemOpen
  \bibfield  {author} {\bibinfo {author} {\bibfnamefont {K.-P.}\ \bibnamefont
  {Xie}}, \bibinfo {author} {\bibfnamefont {L.}~\bibnamefont {Bian}}, \ and\
  \bibinfo {author} {\bibfnamefont {Y.}~\bibnamefont {Wu}},\ }\href {\doibase
  10.1007/JHEP12(2020)047} {\bibfield  {journal} {\bibinfo  {journal} {JHEP}\
  }\textbf {\bibinfo {volume} {12}},\ \bibinfo {pages} {047} (\bibinfo {year}
  {2020})},\ \Eprint {http://arxiv.org/abs/2005.13552} {arXiv:2005.13552
  [hep-ph]} \BibitemShut {NoStop}%
\bibitem [{\citenamefont {Di~Bari}\ \emph {et~al.}(2021)\citenamefont
  {Di~Bari}, \citenamefont {Marfatia},\ and\ \citenamefont
  {Zhou}}]{DiBari:2021dri}%
  \BibitemOpen
  \bibfield  {author} {\bibinfo {author} {\bibfnamefont {P.}~\bibnamefont
  {Di~Bari}}, \bibinfo {author} {\bibfnamefont {D.}~\bibnamefont {Marfatia}}, \
  and\ \bibinfo {author} {\bibfnamefont {Y.-L.}\ \bibnamefont {Zhou}},\ }\href
  {\doibase 10.1007/JHEP10(2021)193} {\bibfield  {journal} {\bibinfo  {journal}
  {JHEP}\ }\textbf {\bibinfo {volume} {10}},\ \bibinfo {pages} {193} (\bibinfo
  {year} {2021})},\ \Eprint {http://arxiv.org/abs/2106.00025} {arXiv:2106.00025
  [hep-ph]} \BibitemShut {NoStop}%
\bibitem [{\citenamefont {Zhou}\ \emph {et~al.}(2022)\citenamefont {Zhou},
  \citenamefont {Bian},\ and\ \citenamefont {Du}}]{Zhou:2022mlz}%
  \BibitemOpen
  \bibfield  {author} {\bibinfo {author} {\bibfnamefont {R.}~\bibnamefont
  {Zhou}}, \bibinfo {author} {\bibfnamefont {L.}~\bibnamefont {Bian}}, \ and\
  \bibinfo {author} {\bibfnamefont {Y.}~\bibnamefont {Du}},\ }\href {\doibase
  10.1007/JHEP08(2022)205} {\bibfield  {journal} {\bibinfo  {journal} {JHEP}\
  }\textbf {\bibinfo {volume} {08}},\ \bibinfo {pages} {205} (\bibinfo {year}
  {2022})},\ \Eprint {http://arxiv.org/abs/2203.01561} {arXiv:2203.01561
  [hep-ph]} \BibitemShut {NoStop}%
\bibitem [{\citenamefont {Schwaller}(2015)}]{Schwaller:2015tja}%
  \BibitemOpen
  \bibfield  {author} {\bibinfo {author} {\bibfnamefont {P.}~\bibnamefont
  {Schwaller}},\ }\href {\doibase 10.1103/PhysRevLett.115.181101} {\bibfield
  {journal} {\bibinfo  {journal} {Phys. Rev. Lett.}\ }\textbf {\bibinfo
  {volume} {115}},\ \bibinfo {pages} {181101} (\bibinfo {year} {2015})},\
  \Eprint {http://arxiv.org/abs/1504.07263} {arXiv:1504.07263 [hep-ph]}
  \BibitemShut {NoStop}%
\bibitem [{\citenamefont {Baldes}\ and\ \citenamefont
  {Garcia-Cely}(2019)}]{Baldes:2018emh}%
  \BibitemOpen
  \bibfield  {author} {\bibinfo {author} {\bibfnamefont {I.}~\bibnamefont
  {Baldes}}\ and\ \bibinfo {author} {\bibfnamefont {C.}~\bibnamefont
  {Garcia-Cely}},\ }\href {\doibase 10.1007/JHEP05(2019)190} {\bibfield
  {journal} {\bibinfo  {journal} {JHEP}\ }\textbf {\bibinfo {volume} {05}},\
  \bibinfo {pages} {190} (\bibinfo {year} {2019})},\ \Eprint
  {http://arxiv.org/abs/1809.01198} {arXiv:1809.01198 [hep-ph]} \BibitemShut
  {NoStop}%
\bibitem [{\citenamefont {Breitbach}\ \emph {et~al.}(2019)\citenamefont
  {Breitbach}, \citenamefont {Kopp}, \citenamefont {Madge}, \citenamefont
  {Opferkuch},\ and\ \citenamefont {Schwaller}}]{Breitbach:2018ddu}%
  \BibitemOpen
  \bibfield  {author} {\bibinfo {author} {\bibfnamefont {M.}~\bibnamefont
  {Breitbach}}, \bibinfo {author} {\bibfnamefont {J.}~\bibnamefont {Kopp}},
  \bibinfo {author} {\bibfnamefont {E.}~\bibnamefont {Madge}}, \bibinfo
  {author} {\bibfnamefont {T.}~\bibnamefont {Opferkuch}}, \ and\ \bibinfo
  {author} {\bibfnamefont {P.}~\bibnamefont {Schwaller}},\ }\href {\doibase
  10.1088/1475-7516/2019/07/007} {\bibfield  {journal} {\bibinfo  {journal}
  {JCAP}\ }\textbf {\bibinfo {volume} {07}},\ \bibinfo {pages} {007} (\bibinfo
  {year} {2019})},\ \Eprint {http://arxiv.org/abs/1811.11175} {arXiv:1811.11175
  [hep-ph]} \BibitemShut {NoStop}%
\bibitem [{\citenamefont {Croon}\ \emph {et~al.}(2018)\citenamefont {Croon},
  \citenamefont {Sanz},\ and\ \citenamefont {White}}]{Croon:2018erz}%
  \BibitemOpen
  \bibfield  {author} {\bibinfo {author} {\bibfnamefont {D.}~\bibnamefont
  {Croon}}, \bibinfo {author} {\bibfnamefont {V.}~\bibnamefont {Sanz}}, \ and\
  \bibinfo {author} {\bibfnamefont {G.}~\bibnamefont {White}},\ }\href
  {\doibase 10.1007/JHEP08(2018)203} {\bibfield  {journal} {\bibinfo  {journal}
  {JHEP}\ }\textbf {\bibinfo {volume} {08}},\ \bibinfo {pages} {203} (\bibinfo
  {year} {2018})},\ \Eprint {http://arxiv.org/abs/1806.02332} {arXiv:1806.02332
  [hep-ph]} \BibitemShut {NoStop}%
\bibitem [{\citenamefont {Baldes}(2017)}]{Baldes:2017rcu}%
  \BibitemOpen
  \bibfield  {author} {\bibinfo {author} {\bibfnamefont {I.}~\bibnamefont
  {Baldes}},\ }\href {\doibase 10.1088/1475-7516/2017/05/028} {\bibfield
  {journal} {\bibinfo  {journal} {JCAP}\ }\textbf {\bibinfo {volume} {05}},\
  \bibinfo {pages} {028} (\bibinfo {year} {2017})},\ \Eprint
  {http://arxiv.org/abs/1702.02117} {arXiv:1702.02117 [hep-ph]} \BibitemShut
  {NoStop}%
\bibitem [{\citenamefont {Croon}\ \emph {et~al.}(2020)\citenamefont {Croon},
  \citenamefont {Kusenko}, \citenamefont {Mazumdar},\ and\ \citenamefont
  {White}}]{Croon:2019rqu}%
  \BibitemOpen
  \bibfield  {author} {\bibinfo {author} {\bibfnamefont {D.}~\bibnamefont
  {Croon}}, \bibinfo {author} {\bibfnamefont {A.}~\bibnamefont {Kusenko}},
  \bibinfo {author} {\bibfnamefont {A.}~\bibnamefont {Mazumdar}}, \ and\
  \bibinfo {author} {\bibfnamefont {G.}~\bibnamefont {White}},\ }\href
  {\doibase 10.1103/PhysRevD.101.085010} {\bibfield  {journal} {\bibinfo
  {journal} {Phys. Rev. D}\ }\textbf {\bibinfo {volume} {101}},\ \bibinfo
  {pages} {085010} (\bibinfo {year} {2020})},\ \Eprint
  {http://arxiv.org/abs/1910.09562} {arXiv:1910.09562 [hep-ph]} \BibitemShut
  {NoStop}%
\bibitem [{\citenamefont {Hall}\ \emph {et~al.}(2023)\citenamefont {Hall},
  \citenamefont {Konstandin}, \citenamefont {McGehee},\ and\ \citenamefont
  {Murayama}}]{Hall:2019rld}%
  \BibitemOpen
  \bibfield  {author} {\bibinfo {author} {\bibfnamefont {E.}~\bibnamefont
  {Hall}}, \bibinfo {author} {\bibfnamefont {T.}~\bibnamefont {Konstandin}},
  \bibinfo {author} {\bibfnamefont {R.}~\bibnamefont {McGehee}}, \ and\
  \bibinfo {author} {\bibfnamefont {H.}~\bibnamefont {Murayama}},\ }\href
  {\doibase 10.1103/PhysRevD.107.055011} {\bibfield  {journal} {\bibinfo
  {journal} {Phys. Rev. D}\ }\textbf {\bibinfo {volume} {107}},\ \bibinfo
  {pages} {055011} (\bibinfo {year} {2023})},\ \Eprint
  {http://arxiv.org/abs/1911.12342} {arXiv:1911.12342 [hep-ph]} \BibitemShut
  {NoStop}%
\bibitem [{\citenamefont {Hall}\ \emph {et~al.}(2020)\citenamefont {Hall},
  \citenamefont {Konstandin}, \citenamefont {McGehee}, \citenamefont
  {Murayama},\ and\ \citenamefont {Servant}}]{Hall:2019ank}%
  \BibitemOpen
  \bibfield  {author} {\bibinfo {author} {\bibfnamefont {E.}~\bibnamefont
  {Hall}}, \bibinfo {author} {\bibfnamefont {T.}~\bibnamefont {Konstandin}},
  \bibinfo {author} {\bibfnamefont {R.}~\bibnamefont {McGehee}}, \bibinfo
  {author} {\bibfnamefont {H.}~\bibnamefont {Murayama}}, \ and\ \bibinfo
  {author} {\bibfnamefont {G.}~\bibnamefont {Servant}},\ }\href {\doibase
  10.1007/JHEP04(2020)042} {\bibfield  {journal} {\bibinfo  {journal} {JHEP}\
  }\textbf {\bibinfo {volume} {04}},\ \bibinfo {pages} {042} (\bibinfo {year}
  {2020})},\ \Eprint {http://arxiv.org/abs/1910.08068} {arXiv:1910.08068
  [hep-ph]} \BibitemShut {NoStop}%
\bibitem [{\citenamefont {Chao}\ \emph {et~al.}(2021)\citenamefont {Chao},
  \citenamefont {Li},\ and\ \citenamefont {Wang}}]{Chao:2020adk}%
  \BibitemOpen
  \bibfield  {author} {\bibinfo {author} {\bibfnamefont {W.}~\bibnamefont
  {Chao}}, \bibinfo {author} {\bibfnamefont {X.-F.}\ \bibnamefont {Li}}, \ and\
  \bibinfo {author} {\bibfnamefont {L.}~\bibnamefont {Wang}},\ }\href {\doibase
  10.1088/1475-7516/2021/06/038} {\bibfield  {journal} {\bibinfo  {journal}
  {JCAP}\ }\textbf {\bibinfo {volume} {06}},\ \bibinfo {pages} {038} (\bibinfo
  {year} {2021})},\ \Eprint {http://arxiv.org/abs/2012.15113} {arXiv:2012.15113
  [hep-ph]} \BibitemShut {NoStop}%
\bibitem [{\citenamefont {Dent}\ \emph {et~al.}(2022)\citenamefont {Dent},
  \citenamefont {Dutta}, \citenamefont {Ghosh}, \citenamefont {Kumar},\ and\
  \citenamefont {Runburg}}]{Dent:2022bcd}%
  \BibitemOpen
  \bibfield  {author} {\bibinfo {author} {\bibfnamefont {J.~B.}\ \bibnamefont
  {Dent}}, \bibinfo {author} {\bibfnamefont {B.}~\bibnamefont {Dutta}},
  \bibinfo {author} {\bibfnamefont {S.}~\bibnamefont {Ghosh}}, \bibinfo
  {author} {\bibfnamefont {J.}~\bibnamefont {Kumar}}, \ and\ \bibinfo {author}
  {\bibfnamefont {J.}~\bibnamefont {Runburg}},\ }\href {\doibase
  10.1007/JHEP08(2022)300} {\bibfield  {journal} {\bibinfo  {journal} {JHEP}\
  }\textbf {\bibinfo {volume} {08}},\ \bibinfo {pages} {300} (\bibinfo {year}
  {2022})},\ \Eprint {http://arxiv.org/abs/2203.11736} {arXiv:2203.11736
  [hep-ph]} \BibitemShut {NoStop}%
\bibitem [{\citenamefont {Weinberg}(1974)}]{Weinberg:1974hy}%
  \BibitemOpen
  \bibfield  {author} {\bibinfo {author} {\bibfnamefont {S.}~\bibnamefont
  {Weinberg}},\ }\href {\doibase 10.1103/PhysRevD.9.3357} {\bibfield  {journal}
  {\bibinfo  {journal} {Phys. Rev. D}\ }\textbf {\bibinfo {volume} {9}},\
  \bibinfo {pages} {3357} (\bibinfo {year} {1974})}\BibitemShut {NoStop}%
\bibitem [{\citenamefont {Land}\ and\ \citenamefont
  {Carlson}(1992)}]{Land:1992sm}%
  \BibitemOpen
  \bibfield  {author} {\bibinfo {author} {\bibfnamefont {D.}~\bibnamefont
  {Land}}\ and\ \bibinfo {author} {\bibfnamefont {E.~D.}\ \bibnamefont
  {Carlson}},\ }\href {\doibase 10.1016/0370-2693(92)90616-C} {\bibfield
  {journal} {\bibinfo  {journal} {Phys. Lett. B}\ }\textbf {\bibinfo {volume}
  {292}},\ \bibinfo {pages} {107} (\bibinfo {year} {1992})},\ \Eprint
  {http://arxiv.org/abs/hep-ph/9208227} {arXiv:hep-ph/9208227} \BibitemShut
  {NoStop}%
\bibitem [{\citenamefont {Patel}\ and\ \citenamefont
  {Ramsey-Musolf}(2013)}]{Patel:2012pi}%
  \BibitemOpen
  \bibfield  {author} {\bibinfo {author} {\bibfnamefont {H.~H.}\ \bibnamefont
  {Patel}}\ and\ \bibinfo {author} {\bibfnamefont {M.~J.}\ \bibnamefont
  {Ramsey-Musolf}},\ }\href {\doibase 10.1103/PhysRevD.88.035013} {\bibfield
  {journal} {\bibinfo  {journal} {Phys. Rev. D}\ }\textbf {\bibinfo {volume}
  {88}},\ \bibinfo {pages} {035013} (\bibinfo {year} {2013})},\ \Eprint
  {http://arxiv.org/abs/1212.5652} {arXiv:1212.5652 [hep-ph]} \BibitemShut
  {NoStop}%
\bibitem [{\citenamefont {Patel}\ \emph {et~al.}(2013)\citenamefont {Patel},
  \citenamefont {Ramsey-Musolf},\ and\ \citenamefont {Wise}}]{Patel:2013zla}%
  \BibitemOpen
  \bibfield  {author} {\bibinfo {author} {\bibfnamefont {H.~H.}\ \bibnamefont
  {Patel}}, \bibinfo {author} {\bibfnamefont {M.~J.}\ \bibnamefont
  {Ramsey-Musolf}}, \ and\ \bibinfo {author} {\bibfnamefont {M.~B.}\
  \bibnamefont {Wise}},\ }\href {\doibase 10.1103/PhysRevD.88.015003}
  {\bibfield  {journal} {\bibinfo  {journal} {Phys. Rev. D}\ }\textbf {\bibinfo
  {volume} {88}},\ \bibinfo {pages} {015003} (\bibinfo {year} {2013})},\
  \Eprint {http://arxiv.org/abs/1303.1140} {arXiv:1303.1140 [hep-ph]}
  \BibitemShut {NoStop}%
\bibitem [{\citenamefont {Blinov}\ \emph {et~al.}(2015)\citenamefont {Blinov},
  \citenamefont {Kozaczuk}, \citenamefont {Morrissey},\ and\ \citenamefont
  {Tamarit}}]{Blinov:2015sna}%
  \BibitemOpen
  \bibfield  {author} {\bibinfo {author} {\bibfnamefont {N.}~\bibnamefont
  {Blinov}}, \bibinfo {author} {\bibfnamefont {J.}~\bibnamefont {Kozaczuk}},
  \bibinfo {author} {\bibfnamefont {D.~E.}\ \bibnamefont {Morrissey}}, \ and\
  \bibinfo {author} {\bibfnamefont {C.}~\bibnamefont {Tamarit}},\ }\href
  {\doibase 10.1103/PhysRevD.92.035012} {\bibfield  {journal} {\bibinfo
  {journal} {Phys. Rev. D}\ }\textbf {\bibinfo {volume} {92}},\ \bibinfo
  {pages} {035012} (\bibinfo {year} {2015})},\ \Eprint
  {http://arxiv.org/abs/1504.05195} {arXiv:1504.05195 [hep-ph]} \BibitemShut
  {NoStop}%
\bibitem [{\citenamefont {Croon}\ and\ \citenamefont
  {White}(2018)}]{Croon:2018new}%
  \BibitemOpen
  \bibfield  {author} {\bibinfo {author} {\bibfnamefont {D.}~\bibnamefont
  {Croon}}\ and\ \bibinfo {author} {\bibfnamefont {G.}~\bibnamefont {White}},\
  }\href {\doibase 10.1007/JHEP05(2018)210} {\bibfield  {journal} {\bibinfo
  {journal} {JHEP}\ }\textbf {\bibinfo {volume} {05}},\ \bibinfo {pages} {210}
  (\bibinfo {year} {2018})},\ \Eprint {http://arxiv.org/abs/1803.05438}
  {arXiv:1803.05438 [hep-ph]} \BibitemShut {NoStop}%
\bibitem [{\citenamefont {Morais}\ \emph {et~al.}(2020)\citenamefont {Morais},
  \citenamefont {Pasechnik},\ and\ \citenamefont {Vieu}}]{Morais:2018uou}%
  \BibitemOpen
  \bibfield  {author} {\bibinfo {author} {\bibfnamefont {A.~P.}\ \bibnamefont
  {Morais}}, \bibinfo {author} {\bibfnamefont {R.}~\bibnamefont {Pasechnik}}, \
  and\ \bibinfo {author} {\bibfnamefont {T.}~\bibnamefont {Vieu}},\ }\href
  {\doibase 10.22323/1.364.0054} {\bibfield  {journal} {\bibinfo  {journal}
  {PoS}\ }\textbf {\bibinfo {volume} {EPS-HEP2019}},\ \bibinfo {pages} {054}
  (\bibinfo {year} {2020})},\ \Eprint {http://arxiv.org/abs/1802.10109}
  {arXiv:1802.10109 [hep-ph]} \BibitemShut {NoStop}%
\bibitem [{\citenamefont {Morais}\ and\ \citenamefont
  {Pasechnik}(2020)}]{Morais:2019fnm}%
  \BibitemOpen
  \bibfield  {author} {\bibinfo {author} {\bibfnamefont {A.~P.}\ \bibnamefont
  {Morais}}\ and\ \bibinfo {author} {\bibfnamefont {R.}~\bibnamefont
  {Pasechnik}},\ }\href {\doibase 10.1088/1475-7516/2020/04/036} {\bibfield
  {journal} {\bibinfo  {journal} {JCAP}\ }\textbf {\bibinfo {volume} {04}},\
  \bibinfo {pages} {036} (\bibinfo {year} {2020})},\ \Eprint
  {http://arxiv.org/abs/1910.00717} {arXiv:1910.00717 [hep-ph]} \BibitemShut
  {NoStop}%
\bibitem [{\citenamefont {Angelescu}\ and\ \citenamefont
  {Huang}(2019)}]{Angelescu:2018dkk}%
  \BibitemOpen
  \bibfield  {author} {\bibinfo {author} {\bibfnamefont {A.}~\bibnamefont
  {Angelescu}}\ and\ \bibinfo {author} {\bibfnamefont {P.}~\bibnamefont
  {Huang}},\ }\href {\doibase 10.1103/PhysRevD.99.055023} {\bibfield  {journal}
  {\bibinfo  {journal} {Phys. Rev. D}\ }\textbf {\bibinfo {volume} {99}},\
  \bibinfo {pages} {055023} (\bibinfo {year} {2019})},\ \Eprint
  {http://arxiv.org/abs/1812.08293} {arXiv:1812.08293 [hep-ph]} \BibitemShut
  {NoStop}%
\bibitem [{\citenamefont {Friedrich}\ \emph {et~al.}(2022)\citenamefont
  {Friedrich}, \citenamefont {Ramsey-Musolf}, \citenamefont {Tenkanen},\ and\
  \citenamefont {Tran}}]{Friedrich:2022cak}%
  \BibitemOpen
  \bibfield  {author} {\bibinfo {author} {\bibfnamefont {L.~S.}\ \bibnamefont
  {Friedrich}}, \bibinfo {author} {\bibfnamefont {M.~J.}\ \bibnamefont
  {Ramsey-Musolf}}, \bibinfo {author} {\bibfnamefont {T.~V.~I.}\ \bibnamefont
  {Tenkanen}}, \ and\ \bibinfo {author} {\bibfnamefont {V.~Q.}\ \bibnamefont
  {Tran}},\ }\href@noop {} {\  (\bibinfo {year} {2022})},\ \Eprint
  {http://arxiv.org/abs/2203.05889} {arXiv:2203.05889 [hep-ph]} \BibitemShut
  {NoStop}%
\bibitem [{\citenamefont {Sakharov}(1967)}]{Sakharov:1967dj}%
  \BibitemOpen
  \bibfield  {author} {\bibinfo {author} {\bibfnamefont {A.~D.}\ \bibnamefont
  {Sakharov}},\ }\href {\doibase 10.1070/PU1991v034n05ABEH002497} {\bibfield
  {journal} {\bibinfo  {journal} {Pisma Zh. Eksp. Teor. Fiz.}\ }\textbf
  {\bibinfo {volume} {5}},\ \bibinfo {pages} {32} (\bibinfo {year}
  {1967})}\BibitemShut {NoStop}%
\bibitem [{\citenamefont {Trodden}(1999)}]{Trodden:1998ym}%
  \BibitemOpen
  \bibfield  {author} {\bibinfo {author} {\bibfnamefont {M.}~\bibnamefont
  {Trodden}},\ }\href {\doibase 10.1103/RevModPhys.71.1463} {\bibfield
  {journal} {\bibinfo  {journal} {Rev. Mod. Phys.}\ }\textbf {\bibinfo {volume}
  {71}},\ \bibinfo {pages} {1463} (\bibinfo {year} {1999})},\ \Eprint
  {http://arxiv.org/abs/hep-ph/9803479} {arXiv:hep-ph/9803479} \BibitemShut
  {NoStop}%
\bibitem [{\citenamefont {Cline}(2006)}]{Cline:2006ts}%
  \BibitemOpen
  \bibfield  {author} {\bibinfo {author} {\bibfnamefont {J.~M.}\ \bibnamefont
  {Cline}},\ }in\ \href@noop {} {\emph {\bibinfo {booktitle} {{Les Houches
  Summer School - Session 86: Particle Physics and Cosmology: The Fabric of
  Spacetime}}}}\ (\bibinfo {year} {2006})\ \Eprint
  {http://arxiv.org/abs/hep-ph/0609145} {arXiv:hep-ph/0609145} \BibitemShut
  {NoStop}%
\bibitem [{\citenamefont {Morrissey}\ and\ \citenamefont
  {Ramsey-Musolf}(2012)}]{Morrissey:2012db}%
  \BibitemOpen
  \bibfield  {author} {\bibinfo {author} {\bibfnamefont {D.~E.}\ \bibnamefont
  {Morrissey}}\ and\ \bibinfo {author} {\bibfnamefont {M.~J.}\ \bibnamefont
  {Ramsey-Musolf}},\ }\href {\doibase 10.1088/1367-2630/14/12/125003}
  {\bibfield  {journal} {\bibinfo  {journal} {New J. Phys.}\ }\textbf {\bibinfo
  {volume} {14}},\ \bibinfo {pages} {125003} (\bibinfo {year} {2012})},\
  \Eprint {http://arxiv.org/abs/1206.2942} {arXiv:1206.2942 [hep-ph]}
  \BibitemShut {NoStop}%
\bibitem [{\citenamefont {White}(2016)}]{White:2016nbo}%
  \BibitemOpen
  \bibfield  {author} {\bibinfo {author} {\bibfnamefont {G.~A.}\ \bibnamefont
  {White}},\ }\href {\doibase 10.1088/978-1-6817-4457-5} {\  (\bibinfo {year}
  {2016}),\ 10.1088/978-1-6817-4457-5}\BibitemShut {NoStop}%
\bibitem [{\citenamefont {Garbrecht}(2020)}]{Garbrecht:2018mrp}%
  \BibitemOpen
  \bibfield  {author} {\bibinfo {author} {\bibfnamefont {B.}~\bibnamefont
  {Garbrecht}},\ }\href {\doibase 10.1016/j.ppnp.2019.103727} {\bibfield
  {journal} {\bibinfo  {journal} {Prog. Part. Nucl. Phys.}\ }\textbf {\bibinfo
  {volume} {110}},\ \bibinfo {pages} {103727} (\bibinfo {year} {2020})},\
  \Eprint {http://arxiv.org/abs/1812.02651} {arXiv:1812.02651 [hep-ph]}
  \BibitemShut {NoStop}%
\bibitem [{\citenamefont {Ramsey-Musolf}(2020)}]{Ramsey-Musolf:2019lsf}%
  \BibitemOpen
  \bibfield  {author} {\bibinfo {author} {\bibfnamefont {M.~J.}\ \bibnamefont
  {Ramsey-Musolf}},\ }\href {\doibase 10.1007/JHEP09(2020)179} {\bibfield
  {journal} {\bibinfo  {journal} {JHEP}\ }\textbf {\bibinfo {volume} {09}},\
  \bibinfo {pages} {179} (\bibinfo {year} {2020})},\ \Eprint
  {http://arxiv.org/abs/1912.07189} {arXiv:1912.07189 [hep-ph]} \BibitemShut
  {NoStop}%
\bibitem [{\citenamefont {Carena}\ \emph {et~al.}(2023)\citenamefont {Carena},
  \citenamefont {Kozaczuk}, \citenamefont {Liu}, \citenamefont {Ou},
  \citenamefont {Ramsey-Musolf}, \citenamefont {Shelton}, \citenamefont
  {Wang},\ and\ \citenamefont {Xie}}]{Carena:2022yvx}%
  \BibitemOpen
  \bibfield  {author} {\bibinfo {author} {\bibfnamefont {M.}~\bibnamefont
  {Carena}}, \bibinfo {author} {\bibfnamefont {J.}~\bibnamefont {Kozaczuk}},
  \bibinfo {author} {\bibfnamefont {Z.}~\bibnamefont {Liu}}, \bibinfo {author}
  {\bibfnamefont {T.}~\bibnamefont {Ou}}, \bibinfo {author} {\bibfnamefont
  {M.~J.}\ \bibnamefont {Ramsey-Musolf}}, \bibinfo {author} {\bibfnamefont
  {J.}~\bibnamefont {Shelton}}, \bibinfo {author} {\bibfnamefont
  {Y.}~\bibnamefont {Wang}}, \ and\ \bibinfo {author} {\bibfnamefont {K.-P.}\
  \bibnamefont {Xie}},\ }\href {\doibase 10.31526/lhep.2023.432} {\bibfield
  {journal} {\bibinfo  {journal} {LHEP}\ }\textbf {\bibinfo {volume} {2023}},\
  \bibinfo {pages} {432} (\bibinfo {year} {2023})},\ \Eprint
  {http://arxiv.org/abs/2203.08206} {arXiv:2203.08206 [hep-ph]} \BibitemShut
  {NoStop}%
\bibitem [{\citenamefont {Wang}\ \emph {et~al.}(2022)\citenamefont {Wang},
  \citenamefont {Zhu}, \citenamefont {Khoda}, \citenamefont {Hsu},
  \citenamefont {Konstantinidis}, \citenamefont {Li}, \citenamefont {Li},
  \citenamefont {Ramsey-Musolf}, \citenamefont {Wu},\ and\ \citenamefont
  {Zhang}}]{Wang:2022dkz}%
  \BibitemOpen
  \bibfield  {author} {\bibinfo {author} {\bibfnamefont {Z.}~\bibnamefont
  {Wang}}, \bibinfo {author} {\bibfnamefont {X.}~\bibnamefont {Zhu}}, \bibinfo
  {author} {\bibfnamefont {E.~E.}\ \bibnamefont {Khoda}}, \bibinfo {author}
  {\bibfnamefont {S.-C.}\ \bibnamefont {Hsu}}, \bibinfo {author} {\bibfnamefont
  {N.}~\bibnamefont {Konstantinidis}}, \bibinfo {author} {\bibfnamefont
  {K.}~\bibnamefont {Li}}, \bibinfo {author} {\bibfnamefont {S.}~\bibnamefont
  {Li}}, \bibinfo {author} {\bibfnamefont {M.~J.}\ \bibnamefont
  {Ramsey-Musolf}}, \bibinfo {author} {\bibfnamefont {Y.}~\bibnamefont {Wu}}, \
  and\ \bibinfo {author} {\bibfnamefont {Y.~E.}\ \bibnamefont {Zhang}},\ }in\
  \href@noop {} {\emph {\bibinfo {booktitle} {{Snowmass 2021}}}}\ (\bibinfo
  {year} {2022})\ \Eprint {http://arxiv.org/abs/2203.10184} {arXiv:2203.10184
  [hep-ex]} \BibitemShut {NoStop}%
\bibitem [{\citenamefont {Zhang}\ \emph {et~al.}(2023)\citenamefont {Zhang},
  \citenamefont {Li}, \citenamefont {Liu}, \citenamefont {Ramsey-Musolf},
  \citenamefont {Zeng},\ and\ \citenamefont {Arunasalam}}]{Zhang:2023jvh}%
  \BibitemOpen
  \bibfield  {author} {\bibinfo {author} {\bibfnamefont {W.}~\bibnamefont
  {Zhang}}, \bibinfo {author} {\bibfnamefont {H.-L.}\ \bibnamefont {Li}},
  \bibinfo {author} {\bibfnamefont {K.}~\bibnamefont {Liu}}, \bibinfo {author}
  {\bibfnamefont {M.~J.}\ \bibnamefont {Ramsey-Musolf}}, \bibinfo {author}
  {\bibfnamefont {Y.}~\bibnamefont {Zeng}}, \ and\ \bibinfo {author}
  {\bibfnamefont {S.}~\bibnamefont {Arunasalam}},\ }\href {\doibase
  10.1007/JHEP12(2023)018} {\bibfield  {journal} {\bibinfo  {journal} {JHEP}\
  }\textbf {\bibinfo {volume} {12}},\ \bibinfo {pages} {018} (\bibinfo {year}
  {2023})},\ \Eprint {http://arxiv.org/abs/2303.03612} {arXiv:2303.03612
  [hep-ph]} \BibitemShut {NoStop}%
\bibitem [{\citenamefont {Wang}\ \emph
  {et~al.}(2023{\natexlab{a}})\citenamefont {Wang}, \citenamefont {Zhu},
  \citenamefont {Khoda}, \citenamefont {Hsu}, \citenamefont {Konstantinidis},
  \citenamefont {Li}, \citenamefont {Li}, \citenamefont {Ramsey-Musolf},
  \citenamefont {Wu},\ and\ \citenamefont {Zhang}}]{Wang:2023zys}%
  \BibitemOpen
  \bibfield  {author} {\bibinfo {author} {\bibfnamefont {Z.}~\bibnamefont
  {Wang}}, \bibinfo {author} {\bibfnamefont {X.}~\bibnamefont {Zhu}}, \bibinfo
  {author} {\bibfnamefont {E.~E.}\ \bibnamefont {Khoda}}, \bibinfo {author}
  {\bibfnamefont {S.-C.}\ \bibnamefont {Hsu}}, \bibinfo {author} {\bibfnamefont
  {N.}~\bibnamefont {Konstantinidis}}, \bibinfo {author} {\bibfnamefont
  {K.}~\bibnamefont {Li}}, \bibinfo {author} {\bibfnamefont {S.}~\bibnamefont
  {Li}}, \bibinfo {author} {\bibfnamefont {M.~J.}\ \bibnamefont
  {Ramsey-Musolf}}, \bibinfo {author} {\bibfnamefont {Y.}~\bibnamefont {Wu}}, \
  and\ \bibinfo {author} {\bibfnamefont {Y.~E.}\ \bibnamefont {Zhang}},\ }\href
  {\doibase 10.31526/lhep.2023.436} {\bibfield  {journal} {\bibinfo  {journal}
  {LHEP}\ }\textbf {\bibinfo {volume} {2023}},\ \bibinfo {pages} {436}
  (\bibinfo {year} {2023}{\natexlab{a}})}\BibitemShut {NoStop}%
\bibitem [{\citenamefont {Chowdhury}\ \emph {et~al.}(2012)\citenamefont
  {Chowdhury}, \citenamefont {Nemevsek}, \citenamefont {Senjanovic},\ and\
  \citenamefont {Zhang}}]{Chowdhury:2011ga}%
  \BibitemOpen
  \bibfield  {author} {\bibinfo {author} {\bibfnamefont {T.~A.}\ \bibnamefont
  {Chowdhury}}, \bibinfo {author} {\bibfnamefont {M.}~\bibnamefont {Nemevsek}},
  \bibinfo {author} {\bibfnamefont {G.}~\bibnamefont {Senjanovic}}, \ and\
  \bibinfo {author} {\bibfnamefont {Y.}~\bibnamefont {Zhang}},\ }\href
  {\doibase 10.1088/1475-7516/2012/02/029} {\bibfield  {journal} {\bibinfo
  {journal} {JCAP}\ }\textbf {\bibinfo {volume} {02}},\ \bibinfo {pages} {029}
  (\bibinfo {year} {2012})},\ \Eprint {http://arxiv.org/abs/1110.5334}
  {arXiv:1110.5334 [hep-ph]} \BibitemShut {NoStop}%
\bibitem [{\citenamefont {Katz}\ \emph {et~al.}(2015)\citenamefont {Katz},
  \citenamefont {Perelstein}, \citenamefont {Ramsey-Musolf},\ and\
  \citenamefont {Winslow}}]{Katz:2015uja}%
  \BibitemOpen
  \bibfield  {author} {\bibinfo {author} {\bibfnamefont {A.}~\bibnamefont
  {Katz}}, \bibinfo {author} {\bibfnamefont {M.}~\bibnamefont {Perelstein}},
  \bibinfo {author} {\bibfnamefont {M.~J.}\ \bibnamefont {Ramsey-Musolf}}, \
  and\ \bibinfo {author} {\bibfnamefont {P.}~\bibnamefont {Winslow}},\ }\href
  {\doibase 10.1103/PhysRevD.92.095019} {\bibfield  {journal} {\bibinfo
  {journal} {Phys. Rev. D}\ }\textbf {\bibinfo {volume} {92}},\ \bibinfo
  {pages} {095019} (\bibinfo {year} {2015})},\ \Eprint
  {http://arxiv.org/abs/1509.02934} {arXiv:1509.02934 [hep-ph]} \BibitemShut
  {NoStop}%
\bibitem [{\citenamefont {Fabian}\ \emph {et~al.}(2021)\citenamefont {Fabian},
  \citenamefont {Goertz},\ and\ \citenamefont {Jiang}}]{Fabian:2020hny}%
  \BibitemOpen
  \bibfield  {author} {\bibinfo {author} {\bibfnamefont {S.}~\bibnamefont
  {Fabian}}, \bibinfo {author} {\bibfnamefont {F.}~\bibnamefont {Goertz}}, \
  and\ \bibinfo {author} {\bibfnamefont {Y.}~\bibnamefont {Jiang}},\ }\href
  {\doibase 10.1088/1475-7516/2021/09/011} {\bibfield  {journal} {\bibinfo
  {journal} {JCAP}\ }\textbf {\bibinfo {volume} {09}},\ \bibinfo {pages} {011}
  (\bibinfo {year} {2021})},\ \Eprint {http://arxiv.org/abs/2012.12847}
  {arXiv:2012.12847 [hep-ph]} \BibitemShut {NoStop}%
\bibitem [{\citenamefont {Bell}\ \emph {et~al.}(2020)\citenamefont {Bell},
  \citenamefont {Dolan}, \citenamefont {Friedrich}, \citenamefont
  {Ramsey-Musolf},\ and\ \citenamefont {Volkas}}]{Bell:2020gug}%
  \BibitemOpen
  \bibfield  {author} {\bibinfo {author} {\bibfnamefont {N.~F.}\ \bibnamefont
  {Bell}}, \bibinfo {author} {\bibfnamefont {M.~J.}\ \bibnamefont {Dolan}},
  \bibinfo {author} {\bibfnamefont {L.~S.}\ \bibnamefont {Friedrich}}, \bibinfo
  {author} {\bibfnamefont {M.~J.}\ \bibnamefont {Ramsey-Musolf}}, \ and\
  \bibinfo {author} {\bibfnamefont {R.~R.}\ \bibnamefont {Volkas}},\ }\href
  {\doibase 10.1007/JHEP05(2020)050} {\bibfield  {journal} {\bibinfo  {journal}
  {JHEP}\ }\textbf {\bibinfo {volume} {05}},\ \bibinfo {pages} {050} (\bibinfo
  {year} {2020})},\ \Eprint {http://arxiv.org/abs/2001.05335} {arXiv:2001.05335
  [hep-ph]} \BibitemShut {NoStop}%
\bibitem [{\citenamefont {Chiang}\ \emph {et~al.}(2021)\citenamefont {Chiang},
  \citenamefont {Huang},\ and\ \citenamefont {Lu}}]{Chiang:2020yym}%
  \BibitemOpen
  \bibfield  {author} {\bibinfo {author} {\bibfnamefont {C.-W.}\ \bibnamefont
  {Chiang}}, \bibinfo {author} {\bibfnamefont {D.}~\bibnamefont {Huang}}, \
  and\ \bibinfo {author} {\bibfnamefont {B.-Q.}\ \bibnamefont {Lu}},\ }\href
  {\doibase 10.1088/1475-7516/2021/01/035} {\bibfield  {journal} {\bibinfo
  {journal} {JCAP}\ }\textbf {\bibinfo {volume} {01}},\ \bibinfo {pages} {035}
  (\bibinfo {year} {2021})},\ \Eprint {http://arxiv.org/abs/2009.08635}
  {arXiv:2009.08635 [hep-ph]} \BibitemShut {NoStop}%
\bibitem [{\citenamefont {Zhang}\ \emph {et~al.}(2024)\citenamefont {Zhang},
  \citenamefont {Cai}, \citenamefont {Ramsey-Musolf},\ and\ \citenamefont
  {Zhang}}]{Zhang:2023mnu}%
  \BibitemOpen
  \bibfield  {author} {\bibinfo {author} {\bibfnamefont {W.}~\bibnamefont
  {Zhang}}, \bibinfo {author} {\bibfnamefont {Y.}~\bibnamefont {Cai}}, \bibinfo
  {author} {\bibfnamefont {M.~J.}\ \bibnamefont {Ramsey-Musolf}}, \ and\
  \bibinfo {author} {\bibfnamefont {L.}~\bibnamefont {Zhang}},\ }\href
  {\doibase 10.1007/JHEP01(2024)051} {\bibfield  {journal} {\bibinfo  {journal}
  {JHEP}\ }\textbf {\bibinfo {volume} {01}},\ \bibinfo {pages} {051} (\bibinfo
  {year} {2024})},\ \Eprint {http://arxiv.org/abs/2307.01615} {arXiv:2307.01615
  [hep-ph]} \BibitemShut {NoStop}%
\bibitem [{\citenamefont {Weir}(2018)}]{Weir:2017wfa}%
  \BibitemOpen
  \bibfield  {author} {\bibinfo {author} {\bibfnamefont {D.~J.}\ \bibnamefont
  {Weir}},\ }\href {\doibase 10.1098/rsta.2017.0126} {\bibfield  {journal}
  {\bibinfo  {journal} {Phil. Trans. Roy. Soc. Lond. A}\ }\textbf {\bibinfo
  {volume} {376}},\ \bibinfo {pages} {20170126} (\bibinfo {year} {2018})},\
  \Eprint {http://arxiv.org/abs/1705.01783} {arXiv:1705.01783 [hep-ph]}
  \BibitemShut {NoStop}%
\bibitem [{\citenamefont {Amaro-Seoane}\ \emph {et~al.}(2017)\citenamefont
  {Amaro-Seoane} \emph {et~al.}}]{Audley:2017drz}%
  \BibitemOpen
  \bibfield  {author} {\bibinfo {author} {\bibfnamefont {P.}~\bibnamefont
  {Amaro-Seoane}} \emph {et~al.} (\bibinfo {collaboration} {LISA}),\
  }\href@noop {} {\  (\bibinfo {year} {2017})},\ \Eprint
  {http://arxiv.org/abs/1702.00786} {arXiv:1702.00786 [astro-ph.IM]}
  \BibitemShut {NoStop}%
\bibitem [{\citenamefont {Robson}\ \emph {et~al.}(2019)\citenamefont {Robson},
  \citenamefont {Cornish},\ and\ \citenamefont {Liu}}]{Cornish:2018dyw}%
  \BibitemOpen
  \bibfield  {author} {\bibinfo {author} {\bibfnamefont {T.}~\bibnamefont
  {Robson}}, \bibinfo {author} {\bibfnamefont {N.~J.}\ \bibnamefont {Cornish}},
  \ and\ \bibinfo {author} {\bibfnamefont {C.}~\bibnamefont {Liu}},\ }\href
  {\doibase 10.1088/1361-6382/ab1101} {\bibfield  {journal} {\bibinfo
  {journal} {Class. Quant. Grav.}\ }\textbf {\bibinfo {volume} {36}},\ \bibinfo
  {pages} {105011} (\bibinfo {year} {2019})},\ \Eprint
  {http://arxiv.org/abs/1803.01944} {arXiv:1803.01944 [astro-ph.HE]}
  \BibitemShut {NoStop}%
\bibitem [{\citenamefont {Crowder}\ and\ \citenamefont
  {Cornish}(2005)}]{Crowder:2005nr}%
  \BibitemOpen
  \bibfield  {author} {\bibinfo {author} {\bibfnamefont {J.}~\bibnamefont
  {Crowder}}\ and\ \bibinfo {author} {\bibfnamefont {N.~J.}\ \bibnamefont
  {Cornish}},\ }\href {\doibase 10.1103/PhysRevD.72.083005} {\bibfield
  {journal} {\bibinfo  {journal} {Phys. Rev. D}\ }\textbf {\bibinfo {volume}
  {72}},\ \bibinfo {pages} {083005} (\bibinfo {year} {2005})},\ \Eprint
  {http://arxiv.org/abs/gr-qc/0506015} {arXiv:gr-qc/0506015} \BibitemShut
  {NoStop}%
\bibitem [{\citenamefont {Luo}\ \emph {et~al.}(2016)\citenamefont {Luo} \emph
  {et~al.}}]{Luo:2015ght}%
  \BibitemOpen
  \bibfield  {author} {\bibinfo {author} {\bibfnamefont {J.}~\bibnamefont
  {Luo}} \emph {et~al.} (\bibinfo {collaboration} {TianQin}),\ }\href {\doibase
  10.1088/0264-9381/33/3/035010} {\bibfield  {journal} {\bibinfo  {journal}
  {Class. Quant. Grav.}\ }\textbf {\bibinfo {volume} {33}},\ \bibinfo {pages}
  {035010} (\bibinfo {year} {2016})},\ \Eprint
  {http://arxiv.org/abs/1512.02076} {arXiv:1512.02076 [astro-ph.IM]}
  \BibitemShut {NoStop}%
\bibitem [{\citenamefont {Hu}\ \emph {et~al.}(2017)\citenamefont {Hu},
  \citenamefont {Mei},\ and\ \citenamefont {Luo}}]{Hu:2017yoc}%
  \BibitemOpen
  \bibfield  {author} {\bibinfo {author} {\bibfnamefont {Y.-M.}\ \bibnamefont
  {Hu}}, \bibinfo {author} {\bibfnamefont {J.}~\bibnamefont {Mei}}, \ and\
  \bibinfo {author} {\bibfnamefont {J.}~\bibnamefont {Luo}},\ }\href {\doibase
  10.1093/nsr/nwx115} {\bibfield  {journal} {\bibinfo  {journal} {Natl. Sci.
  Rev.}\ }\textbf {\bibinfo {volume} {4}},\ \bibinfo {pages} {683} (\bibinfo
  {year} {2017})}\BibitemShut {NoStop}%
\bibitem [{\citenamefont {Hu}\ and\ \citenamefont {Wu}(2017)}]{Hu:2017mde}%
  \BibitemOpen
  \bibfield  {author} {\bibinfo {author} {\bibfnamefont {W.-R.}\ \bibnamefont
  {Hu}}\ and\ \bibinfo {author} {\bibfnamefont {Y.-L.}\ \bibnamefont {Wu}},\
  }\href {\doibase 10.1093/nsr/nwx116} {\bibfield  {journal} {\bibinfo
  {journal} {Natl. Sci. Rev.}\ }\textbf {\bibinfo {volume} {4}},\ \bibinfo
  {pages} {685} (\bibinfo {year} {2017})}\BibitemShut {NoStop}%
\bibitem [{\citenamefont {Ruan}\ \emph {et~al.}(2020)\citenamefont {Ruan},
  \citenamefont {Guo}, \citenamefont {Cai},\ and\ \citenamefont
  {Zhang}}]{Guo:2018npi}%
  \BibitemOpen
  \bibfield  {author} {\bibinfo {author} {\bibfnamefont {W.-H.}\ \bibnamefont
  {Ruan}}, \bibinfo {author} {\bibfnamefont {Z.-K.}\ \bibnamefont {Guo}},
  \bibinfo {author} {\bibfnamefont {R.-G.}\ \bibnamefont {Cai}}, \ and\
  \bibinfo {author} {\bibfnamefont {Y.-Z.}\ \bibnamefont {Zhang}},\ }\href
  {\doibase 10.1142/S0217751X2050075X} {\bibfield  {journal} {\bibinfo
  {journal} {Int. J. Mod. Phys. A}\ }\textbf {\bibinfo {volume} {35}},\
  \bibinfo {pages} {2050075} (\bibinfo {year} {2020})},\ \Eprint
  {http://arxiv.org/abs/1807.09495} {arXiv:1807.09495 [gr-qc]} \BibitemShut
  {NoStop}%
\bibitem [{\citenamefont {Kawamura}\ \emph {et~al.}(2006)\citenamefont
  {Kawamura} \emph {et~al.}}]{Kawamura:2006up}%
  \BibitemOpen
  \bibfield  {author} {\bibinfo {author} {\bibfnamefont {S.}~\bibnamefont
  {Kawamura}} \emph {et~al.},\ }\href {\doibase 10.1088/0264-9381/23/8/S17}
  {\bibfield  {journal} {\bibinfo  {journal} {Class. Quant. Grav.}\ }\textbf
  {\bibinfo {volume} {23}},\ \bibinfo {pages} {S125} (\bibinfo {year}
  {2006})}\BibitemShut {NoStop}%
\bibitem [{\citenamefont {Kawamura}\ \emph {et~al.}(2011)\citenamefont
  {Kawamura} \emph {et~al.}}]{Kawamura:2011zz}%
  \BibitemOpen
  \bibfield  {author} {\bibinfo {author} {\bibfnamefont {S.}~\bibnamefont
  {Kawamura}} \emph {et~al.},\ }\href {\doibase 10.1088/0264-9381/28/9/094011}
  {\bibfield  {journal} {\bibinfo  {journal} {Class. Quant. Grav.}\ }\textbf
  {\bibinfo {volume} {28}},\ \bibinfo {pages} {094011} (\bibinfo {year}
  {2011})}\BibitemShut {NoStop}%
\bibitem [{\citenamefont {Musha}(2017)}]{Musha:2017usi}%
  \BibitemOpen
  \bibfield  {author} {\bibinfo {author} {\bibfnamefont {M.}~\bibnamefont
  {Musha}} (\bibinfo {collaboration} {DECIGO Working group}),\ }\href {\doibase
  10.1117/12.2296050} {\bibfield  {journal} {\bibinfo  {journal} {Proc. SPIE
  Int. Soc. Opt. Eng.}\ }\textbf {\bibinfo {volume} {10562}},\ \bibinfo {pages}
  {105623T} (\bibinfo {year} {2017})}\BibitemShut {NoStop}%
\bibitem [{\citenamefont {Ghorbani}(2021)}]{Ghorbani:2020xqv}%
  \BibitemOpen
  \bibfield  {author} {\bibinfo {author} {\bibfnamefont {P.}~\bibnamefont
  {Ghorbani}},\ }\href {\doibase 10.1016/j.dark.2021.100861} {\bibfield
  {journal} {\bibinfo  {journal} {Phys. Dark Univ.}\ }\textbf {\bibinfo
  {volume} {33}},\ \bibinfo {pages} {100861} (\bibinfo {year} {2021})},\
  \Eprint {http://arxiv.org/abs/2010.15708} {arXiv:2010.15708 [hep-ph]}
  \BibitemShut {NoStop}%
\bibitem [{\citenamefont {Patel}\ and\ \citenamefont
  {Ramsey-Musolf}(2011)}]{Patel:2011th}%
  \BibitemOpen
  \bibfield  {author} {\bibinfo {author} {\bibfnamefont {H.~H.}\ \bibnamefont
  {Patel}}\ and\ \bibinfo {author} {\bibfnamefont {M.~J.}\ \bibnamefont
  {Ramsey-Musolf}},\ }\href {\doibase 10.1007/JHEP07(2011)029} {\bibfield
  {journal} {\bibinfo  {journal} {JHEP}\ }\textbf {\bibinfo {volume} {07}},\
  \bibinfo {pages} {029} (\bibinfo {year} {2011})},\ \Eprint
  {http://arxiv.org/abs/1101.4665} {arXiv:1101.4665 [hep-ph]} \BibitemShut
  {NoStop}%
\bibitem [{\citenamefont {Ramos}\ \emph
  {et~al.}(2021{\natexlab{a}})\citenamefont {Ramos}, \citenamefont {Tran},\
  and\ \citenamefont {Yuan}}]{Ramos:2021txu}%
  \BibitemOpen
  \bibfield  {author} {\bibinfo {author} {\bibfnamefont {R.}~\bibnamefont
  {Ramos}}, \bibinfo {author} {\bibfnamefont {V.~Q.}\ \bibnamefont {Tran}}, \
  and\ \bibinfo {author} {\bibfnamefont {T.-C.}\ \bibnamefont {Yuan}},\ }\href
  {\doibase 10.1007/JHEP11(2021)112} {\bibfield  {journal} {\bibinfo  {journal}
  {JHEP}\ }\textbf {\bibinfo {volume} {11}},\ \bibinfo {pages} {112} (\bibinfo
  {year} {2021}{\natexlab{a}})},\ \Eprint {http://arxiv.org/abs/2109.03185}
  {arXiv:2109.03185 [hep-ph]} \BibitemShut {NoStop}%
\bibitem [{\citenamefont {Ramos}\ \emph
  {et~al.}(2021{\natexlab{b}})\citenamefont {Ramos}, \citenamefont {Tran},\
  and\ \citenamefont {Yuan}}]{Ramos:2021omo}%
  \BibitemOpen
  \bibfield  {author} {\bibinfo {author} {\bibfnamefont {R.}~\bibnamefont
  {Ramos}}, \bibinfo {author} {\bibfnamefont {V.~Q.}\ \bibnamefont {Tran}}, \
  and\ \bibinfo {author} {\bibfnamefont {T.-C.}\ \bibnamefont {Yuan}},\ }\href
  {\doibase 10.1103/PhysRevD.103.075021} {\bibfield  {journal} {\bibinfo
  {journal} {Phys. Rev. D}\ }\textbf {\bibinfo {volume} {103}},\ \bibinfo
  {pages} {075021} (\bibinfo {year} {2021}{\natexlab{b}})},\ \Eprint
  {http://arxiv.org/abs/2101.07115} {arXiv:2101.07115 [hep-ph]} \BibitemShut
  {NoStop}%
\bibitem [{\citenamefont {Huang}\ \emph
  {et~al.}(2016{\natexlab{a}})\citenamefont {Huang}, \citenamefont {Tsai},\
  and\ \citenamefont {Yuan}}]{Huang:2015wts}%
  \BibitemOpen
  \bibfield  {author} {\bibinfo {author} {\bibfnamefont {W.-C.}\ \bibnamefont
  {Huang}}, \bibinfo {author} {\bibfnamefont {Y.-L.~S.}\ \bibnamefont {Tsai}},
  \ and\ \bibinfo {author} {\bibfnamefont {T.-C.}\ \bibnamefont {Yuan}},\
  }\href {\doibase 10.1007/JHEP04(2016)019} {\bibfield  {journal} {\bibinfo
  {journal} {JHEP}\ }\textbf {\bibinfo {volume} {04}},\ \bibinfo {pages} {019}
  (\bibinfo {year} {2016}{\natexlab{a}})},\ \Eprint
  {http://arxiv.org/abs/1512.00229} {arXiv:1512.00229 [hep-ph]} \BibitemShut
  {NoStop}%
\bibitem [{\citenamefont {Chen}\ \emph {et~al.}(2020)\citenamefont {Chen},
  \citenamefont {Lin}, \citenamefont {Nugroho}, \citenamefont {Ramos},
  \citenamefont {Tsai},\ and\ \citenamefont {Yuan}}]{Chen:2019pnt}%
  \BibitemOpen
  \bibfield  {author} {\bibinfo {author} {\bibfnamefont {C.-R.}\ \bibnamefont
  {Chen}}, \bibinfo {author} {\bibfnamefont {Y.-X.}\ \bibnamefont {Lin}},
  \bibinfo {author} {\bibfnamefont {C.~S.}\ \bibnamefont {Nugroho}}, \bibinfo
  {author} {\bibfnamefont {R.}~\bibnamefont {Ramos}}, \bibinfo {author}
  {\bibfnamefont {Y.-L.~S.}\ \bibnamefont {Tsai}}, \ and\ \bibinfo {author}
  {\bibfnamefont {T.-C.}\ \bibnamefont {Yuan}},\ }\href {\doibase
  10.1103/PhysRevD.101.035037} {\bibfield  {journal} {\bibinfo  {journal}
  {Phys. Rev. D}\ }\textbf {\bibinfo {volume} {101}},\ \bibinfo {pages}
  {035037} (\bibinfo {year} {2020})},\ \Eprint
  {http://arxiv.org/abs/1910.13138} {arXiv:1910.13138 [hep-ph]} \BibitemShut
  {NoStop}%
\bibitem [{\citenamefont {Tran}\ and\ \citenamefont
  {Yuan}(2023)}]{Tran:2022cwh}%
  \BibitemOpen
  \bibfield  {author} {\bibinfo {author} {\bibfnamefont {V.~Q.}\ \bibnamefont
  {Tran}}\ and\ \bibinfo {author} {\bibfnamefont {T.-C.}\ \bibnamefont
  {Yuan}},\ }\href {\doibase 10.1007/JHEP02(2023)117} {\bibfield  {journal}
  {\bibinfo  {journal} {JHEP}\ }\textbf {\bibinfo {volume} {02}},\ \bibinfo
  {pages} {117} (\bibinfo {year} {2023})},\ \Eprint
  {http://arxiv.org/abs/2212.02333} {arXiv:2212.02333 [hep-ph]} \BibitemShut
  {NoStop}%
\bibitem [{\citenamefont {Liu}\ \emph {et~al.}(2024)\citenamefont {Liu},
  \citenamefont {Tran}, \citenamefont {Wen}, \citenamefont {Xu},\ and\
  \citenamefont {Yuan}}]{Liu:2024nkl}%
  \BibitemOpen
  \bibfield  {author} {\bibinfo {author} {\bibfnamefont {C.~H.}\ \bibnamefont
  {Liu}}, \bibinfo {author} {\bibfnamefont {V.~Q.}\ \bibnamefont {Tran}},
  \bibinfo {author} {\bibfnamefont {Q.}~\bibnamefont {Wen}}, \bibinfo {author}
  {\bibfnamefont {F.}~\bibnamefont {Xu}}, \ and\ \bibinfo {author}
  {\bibfnamefont {T.-C.}\ \bibnamefont {Yuan}},\ }\href@noop {} {\  (\bibinfo
  {year} {2024})},\ \Eprint {http://arxiv.org/abs/2404.06397} {arXiv:2404.06397
  [hep-ph]} \BibitemShut {NoStop}%
\bibitem [{\citenamefont {Tran}\ \emph {et~al.}(2023)\citenamefont {Tran},
  \citenamefont {Nguyen},\ and\ \citenamefont {Yuan}}]{Tran:2022yrh}%
  \BibitemOpen
  \bibfield  {author} {\bibinfo {author} {\bibfnamefont {V.~Q.}\ \bibnamefont
  {Tran}}, \bibinfo {author} {\bibfnamefont {T.~T.~Q.}\ \bibnamefont {Nguyen}},
  \ and\ \bibinfo {author} {\bibfnamefont {T.-C.}\ \bibnamefont {Yuan}},\
  }\href {\doibase 10.1140/epjc/s10052-023-11495-x} {\bibfield  {journal}
  {\bibinfo  {journal} {Eur. Phys. J. C}\ }\textbf {\bibinfo {volume} {83}},\
  \bibinfo {pages} {346} (\bibinfo {year} {2023})},\ \Eprint
  {http://arxiv.org/abs/2208.10971} {arXiv:2208.10971 [hep-ph]} \BibitemShut
  {NoStop}%
\bibitem [{\citenamefont {Huang}\ \emph
  {et~al.}(2016{\natexlab{b}})\citenamefont {Huang}, \citenamefont {Tsai},\
  and\ \citenamefont {Yuan}}]{Huang:2015rkj}%
  \BibitemOpen
  \bibfield  {author} {\bibinfo {author} {\bibfnamefont {W.-C.}\ \bibnamefont
  {Huang}}, \bibinfo {author} {\bibfnamefont {Y.-L.~S.}\ \bibnamefont {Tsai}},
  \ and\ \bibinfo {author} {\bibfnamefont {T.-C.}\ \bibnamefont {Yuan}},\
  }\href {\doibase 10.1016/j.nuclphysb.2016.05.002} {\bibfield  {journal}
  {\bibinfo  {journal} {Nucl. Phys. B}\ }\textbf {\bibinfo {volume} {909}},\
  \bibinfo {pages} {122} (\bibinfo {year} {2016}{\natexlab{b}})},\ \Eprint
  {http://arxiv.org/abs/1512.07268} {arXiv:1512.07268 [hep-ph]} \BibitemShut
  {NoStop}%
\bibitem [{\citenamefont {Arhrib}\ \emph {et~al.}(2018)\citenamefont {Arhrib},
  \citenamefont {Huang}, \citenamefont {Ramos}, \citenamefont {Tsai},\ and\
  \citenamefont {Yuan}}]{Arhrib:2018sbz}%
  \BibitemOpen
  \bibfield  {author} {\bibinfo {author} {\bibfnamefont {A.}~\bibnamefont
  {Arhrib}}, \bibinfo {author} {\bibfnamefont {W.-C.}\ \bibnamefont {Huang}},
  \bibinfo {author} {\bibfnamefont {R.}~\bibnamefont {Ramos}}, \bibinfo
  {author} {\bibfnamefont {Y.-L.~S.}\ \bibnamefont {Tsai}}, \ and\ \bibinfo
  {author} {\bibfnamefont {T.-C.}\ \bibnamefont {Yuan}},\ }\href {\doibase
  10.1103/PhysRevD.98.095006} {\bibfield  {journal} {\bibinfo  {journal} {Phys.
  Rev. D}\ }\textbf {\bibinfo {volume} {98}},\ \bibinfo {pages} {095006}
  (\bibinfo {year} {2018})},\ \Eprint {http://arxiv.org/abs/1806.05632}
  {arXiv:1806.05632 [hep-ph]} \BibitemShut {NoStop}%
\bibitem [{\citenamefont {Huang}\ \emph {et~al.}(2019)\citenamefont {Huang},
  \citenamefont {Ramos}, \citenamefont {Tran}, \citenamefont {Tsai},\ and\
  \citenamefont {Yuan}}]{Huang:2019obt}%
  \BibitemOpen
  \bibfield  {author} {\bibinfo {author} {\bibfnamefont {C.-T.}\ \bibnamefont
  {Huang}}, \bibinfo {author} {\bibfnamefont {R.}~\bibnamefont {Ramos}},
  \bibinfo {author} {\bibfnamefont {V.~Q.}\ \bibnamefont {Tran}}, \bibinfo
  {author} {\bibfnamefont {Y.-L.~S.}\ \bibnamefont {Tsai}}, \ and\ \bibinfo
  {author} {\bibfnamefont {T.-C.}\ \bibnamefont {Yuan}},\ }\href {\doibase
  10.1007/JHEP09(2019)048} {\bibfield  {journal} {\bibinfo  {journal} {JHEP}\
  }\textbf {\bibinfo {volume} {09}},\ \bibinfo {pages} {048} (\bibinfo {year}
  {2019})},\ \Eprint {http://arxiv.org/abs/1905.02396} {arXiv:1905.02396
  [hep-ph]} \BibitemShut {NoStop}%
\bibitem [{\citenamefont {Huang}\ \emph
  {et~al.}(2018{\natexlab{b}})\citenamefont {Huang}, \citenamefont {Ishida},
  \citenamefont {Lu}, \citenamefont {Tsai},\ and\ \citenamefont
  {Yuan}}]{Huang:2017bto}%
  \BibitemOpen
  \bibfield  {author} {\bibinfo {author} {\bibfnamefont {W.-C.}\ \bibnamefont
  {Huang}}, \bibinfo {author} {\bibfnamefont {H.}~\bibnamefont {Ishida}},
  \bibinfo {author} {\bibfnamefont {C.-T.}\ \bibnamefont {Lu}}, \bibinfo
  {author} {\bibfnamefont {Y.-L.~S.}\ \bibnamefont {Tsai}}, \ and\ \bibinfo
  {author} {\bibfnamefont {T.-C.}\ \bibnamefont {Yuan}},\ }\href {\doibase
  10.1140/epjc/s10052-018-6067-7} {\bibfield  {journal} {\bibinfo  {journal}
  {Eur. Phys. J. C}\ }\textbf {\bibinfo {volume} {78}},\ \bibinfo {pages} {613}
  (\bibinfo {year} {2018}{\natexlab{b}})},\ \Eprint
  {http://arxiv.org/abs/1708.02355} {arXiv:1708.02355 [hep-ph]} \BibitemShut
  {NoStop}%
\bibitem [{\citenamefont {Chen}\ \emph {et~al.}(2019)\citenamefont {Chen},
  \citenamefont {Lin}, \citenamefont {Tran},\ and\ \citenamefont
  {Yuan}}]{Chen:2018wjl}%
  \BibitemOpen
  \bibfield  {author} {\bibinfo {author} {\bibfnamefont {C.-R.}\ \bibnamefont
  {Chen}}, \bibinfo {author} {\bibfnamefont {Y.-X.}\ \bibnamefont {Lin}},
  \bibinfo {author} {\bibfnamefont {V.~Q.}\ \bibnamefont {Tran}}, \ and\
  \bibinfo {author} {\bibfnamefont {T.-C.}\ \bibnamefont {Yuan}},\ }\href
  {\doibase 10.1103/PhysRevD.99.075027} {\bibfield  {journal} {\bibinfo
  {journal} {Phys. Rev. D}\ }\textbf {\bibinfo {volume} {99}},\ \bibinfo
  {pages} {075027} (\bibinfo {year} {2019})},\ \Eprint
  {http://arxiv.org/abs/1810.04837} {arXiv:1810.04837 [hep-ph]} \BibitemShut
  {NoStop}%
\bibitem [{\citenamefont {Dirgantara}\ and\ \citenamefont
  {Nugroho}(2022)}]{Dirgantara:2020lqy}%
  \BibitemOpen
  \bibfield  {author} {\bibinfo {author} {\bibfnamefont {B.}~\bibnamefont
  {Dirgantara}}\ and\ \bibinfo {author} {\bibfnamefont {C.~S.}\ \bibnamefont
  {Nugroho}},\ }\href {\doibase 10.1140/epjc/s10052-022-10051-3} {\bibfield
  {journal} {\bibinfo  {journal} {Eur. Phys. J. C}\ }\textbf {\bibinfo {volume}
  {82}},\ \bibinfo {pages} {142} (\bibinfo {year} {2022})},\ \Eprint
  {http://arxiv.org/abs/2012.13170} {arXiv:2012.13170 [hep-ph]} \BibitemShut
  {NoStop}%
\bibitem [{\citenamefont {Tran}\ \emph {et~al.}(2024)\citenamefont {Tran},
  \citenamefont {Nguyen},\ and\ \citenamefont {Yuan}}]{Tran:2023lzv}%
  \BibitemOpen
  \bibfield  {author} {\bibinfo {author} {\bibfnamefont {V.~Q.}\ \bibnamefont
  {Tran}}, \bibinfo {author} {\bibfnamefont {T.~T.~Q.}\ \bibnamefont {Nguyen}},
  \ and\ \bibinfo {author} {\bibfnamefont {T.-C.}\ \bibnamefont {Yuan}},\
  }\href {\doibase 10.1088/1475-7516/2024/05/015} {\bibfield  {journal}
  {\bibinfo  {journal} {JCAP}\ }\textbf {\bibinfo {volume} {05}},\ \bibinfo
  {pages} {015} (\bibinfo {year} {2024})},\ \Eprint
  {http://arxiv.org/abs/2312.10785} {arXiv:2312.10785 [hep-ph]} \BibitemShut
  {NoStop}%
\bibitem [{\citenamefont {Sirunyan}\ \emph {et~al.}(2020)\citenamefont
  {Sirunyan} \emph {et~al.}}]{CMS:2020xrn}%
  \BibitemOpen
  \bibfield  {author} {\bibinfo {author} {\bibfnamefont {A.~M.}\ \bibnamefont
  {Sirunyan}} \emph {et~al.} (\bibinfo {collaboration} {CMS}),\ }\href
  {\doibase 10.1016/j.physletb.2020.135425} {\bibfield  {journal} {\bibinfo
  {journal} {Phys. Lett. B}\ }\textbf {\bibinfo {volume} {805}},\ \bibinfo
  {pages} {135425} (\bibinfo {year} {2020})},\ \Eprint
  {http://arxiv.org/abs/2002.06398} {arXiv:2002.06398 [hep-ex]} \BibitemShut
  {NoStop}%
\bibitem [{\citenamefont {Zyla}\ \emph
  {et~al.}(2020{\natexlab{a}})\citenamefont {Zyla} \emph
  {et~al.}}]{Zyla:2020zbs}%
  \BibitemOpen
  \bibfield  {author} {\bibinfo {author} {\bibfnamefont {P.~A.}\ \bibnamefont
  {Zyla}} \emph {et~al.} (\bibinfo {collaboration} {Particle Data Group}),\
  }\href {\doibase 10.1093/ptep/ptaa104} {\bibfield  {journal} {\bibinfo
  {journal} {PTEP}\ }\textbf {\bibinfo {volume} {2020}},\ \bibinfo {pages}
  {083C01} (\bibinfo {year} {2020}{\natexlab{a}})}\BibitemShut {NoStop}%
\bibitem [{\citenamefont {CMS-Collaboration}(2020)}]{CMS:2020gsy}%
  \BibitemOpen
  \bibfield  {author} {\bibinfo {author} {\bibnamefont {CMS-Collaboration}}
  (\bibinfo {collaboration} {CMS}),\ }\href@noop {} {\  (\bibinfo {year}
  {2020})},\ \Eprint {http://arxiv.org/abs/CMS-PAS-HIG-19-005}
  {CMS-PAS-HIG-19-005} \BibitemShut {NoStop}%
\bibitem [{ATL(2021)}]{ATLAS:2021vrm}%
  \BibitemOpen
  \href@noop {} {\  (\bibinfo {year} {2021})}\BibitemShut {NoStop}%
\bibitem [{\citenamefont {Tumasyan}\ \emph {et~al.}(2022)\citenamefont
  {Tumasyan} \emph {et~al.}}]{CMS:2022qva}%
  \BibitemOpen
  \bibfield  {author} {\bibinfo {author} {\bibfnamefont {A.}~\bibnamefont
  {Tumasyan}} \emph {et~al.} (\bibinfo {collaboration} {CMS}),\ }\href
  {\doibase 10.1103/PhysRevD.105.092007} {\bibfield  {journal} {\bibinfo
  {journal} {Phys. Rev. D}\ }\textbf {\bibinfo {volume} {105}},\ \bibinfo
  {pages} {092007} (\bibinfo {year} {2022})},\ \Eprint
  {http://arxiv.org/abs/2201.11585} {arXiv:2201.11585 [hep-ex]} \BibitemShut
  {NoStop}%
\bibitem [{\citenamefont {Aad}\ \emph {et~al.}(2019)\citenamefont {Aad} \emph
  {et~al.}}]{ATLAS:2019erb}%
  \BibitemOpen
  \bibfield  {author} {\bibinfo {author} {\bibfnamefont {G.}~\bibnamefont
  {Aad}} \emph {et~al.} (\bibinfo {collaboration} {ATLAS}),\ }\href {\doibase
  10.1016/j.physletb.2019.07.016} {\bibfield  {journal} {\bibinfo  {journal}
  {Phys. Lett. B}\ }\textbf {\bibinfo {volume} {796}},\ \bibinfo {pages} {68}
  (\bibinfo {year} {2019})},\ \Eprint {http://arxiv.org/abs/1903.06248}
  {arXiv:1903.06248 [hep-ex]} \BibitemShut {NoStop}%
\bibitem [{\citenamefont {Fabbrichesi}\ \emph {et~al.}(2020)\citenamefont
  {Fabbrichesi}, \citenamefont {Gabrielli},\ and\ \citenamefont
  {Lanfranchi}}]{Fabbrichesi:2020wbt}%
  \BibitemOpen
  \bibfield  {author} {\bibinfo {author} {\bibfnamefont {M.}~\bibnamefont
  {Fabbrichesi}}, \bibinfo {author} {\bibfnamefont {E.}~\bibnamefont
  {Gabrielli}}, \ and\ \bibinfo {author} {\bibfnamefont {G.}~\bibnamefont
  {Lanfranchi}},\ }\href {\doibase 10.1007/978-3-030-62519-1} {\  (\bibinfo
  {year} {2020}),\ 10.1007/978-3-030-62519-1},\ \Eprint
  {http://arxiv.org/abs/2005.01515} {arXiv:2005.01515 [hep-ph]} \BibitemShut
  {NoStop}%
\bibitem [{\citenamefont {Aaltonen}\ \emph {et~al.}(2022)\citenamefont
  {Aaltonen} \emph {et~al.}}]{CDF:2022hxs}%
  \BibitemOpen
  \bibfield  {author} {\bibinfo {author} {\bibfnamefont {T.}~\bibnamefont
  {Aaltonen}} \emph {et~al.} (\bibinfo {collaboration} {CDF}),\ }\href
  {\doibase 10.1126/science.abk1781} {\bibfield  {journal} {\bibinfo  {journal}
  {Science}\ }\textbf {\bibinfo {volume} {376}},\ \bibinfo {pages} {170}
  (\bibinfo {year} {2022})}\BibitemShut {NoStop}%
\bibitem [{\citenamefont {Aghanim}\ \emph {et~al.}(2020)\citenamefont {Aghanim}
  \emph {et~al.}}]{Aghanim:2018eyx}%
  \BibitemOpen
  \bibfield  {author} {\bibinfo {author} {\bibfnamefont {N.}~\bibnamefont
  {Aghanim}} \emph {et~al.} (\bibinfo {collaboration} {Planck}),\ }\href
  {\doibase 10.1051/0004-6361/201833910} {\bibfield  {journal} {\bibinfo
  {journal} {Astron. Astrophys.}\ }\textbf {\bibinfo {volume} {641}},\ \bibinfo
  {pages} {A6} (\bibinfo {year} {2020})},\ \bibinfo {note} {[Erratum:
  Astron.Astrophys. 652, C4 (2021)]},\ \Eprint
  {http://arxiv.org/abs/1807.06209} {arXiv:1807.06209 [astro-ph.CO]}
  \BibitemShut {NoStop}%
\bibitem [{\citenamefont {B\'elanger}\ \emph {et~al.}(2018)\citenamefont
  {B\'elanger}, \citenamefont {Boudjema}, \citenamefont {Goudelis},
  \citenamefont {Pukhov},\ and\ \citenamefont {Zaldivar}}]{Belanger:2018ccd}%
  \BibitemOpen
  \bibfield  {author} {\bibinfo {author} {\bibfnamefont {G.}~\bibnamefont
  {B\'elanger}}, \bibinfo {author} {\bibfnamefont {F.}~\bibnamefont
  {Boudjema}}, \bibinfo {author} {\bibfnamefont {A.}~\bibnamefont {Goudelis}},
  \bibinfo {author} {\bibfnamefont {A.}~\bibnamefont {Pukhov}}, \ and\ \bibinfo
  {author} {\bibfnamefont {B.}~\bibnamefont {Zaldivar}},\ }\href {\doibase
  10.1016/j.cpc.2018.04.027} {\bibfield  {journal} {\bibinfo  {journal}
  {Comput. Phys. Commun.}\ }\textbf {\bibinfo {volume} {231}},\ \bibinfo
  {pages} {173} (\bibinfo {year} {2018})},\ \Eprint
  {http://arxiv.org/abs/1801.03509} {arXiv:1801.03509 [hep-ph]} \BibitemShut
  {NoStop}%
\bibitem [{\citenamefont {Angloher}\ \emph {et~al.}(2017)\citenamefont
  {Angloher} \emph {et~al.}}]{Angloher:2017sxg}%
  \BibitemOpen
  \bibfield  {author} {\bibinfo {author} {\bibfnamefont {G.}~\bibnamefont
  {Angloher}} \emph {et~al.} (\bibinfo {collaboration} {CRESST}),\ }\href
  {\doibase 10.1140/epjc/s10052-017-5223-9} {\bibfield  {journal} {\bibinfo
  {journal} {Eur. Phys. J. C}\ }\textbf {\bibinfo {volume} {77}},\ \bibinfo
  {pages} {637} (\bibinfo {year} {2017})},\ \Eprint
  {http://arxiv.org/abs/1707.06749} {arXiv:1707.06749 [astro-ph.CO]}
  \BibitemShut {NoStop}%
\bibitem [{\citenamefont {Agnes}\ \emph {et~al.}(2018)\citenamefont {Agnes}
  \emph {et~al.}}]{Agnes:2018ves}%
  \BibitemOpen
  \bibfield  {author} {\bibinfo {author} {\bibfnamefont {P.}~\bibnamefont
  {Agnes}} \emph {et~al.} (\bibinfo {collaboration} {DarkSide}),\ }\href
  {\doibase 10.1103/PhysRevLett.121.081307} {\bibfield  {journal} {\bibinfo
  {journal} {Phys. Rev. Lett.}\ }\textbf {\bibinfo {volume} {121}},\ \bibinfo
  {pages} {081307} (\bibinfo {year} {2018})},\ \Eprint
  {http://arxiv.org/abs/1802.06994} {arXiv:1802.06994 [astro-ph.HE]}
  \BibitemShut {NoStop}%
\bibitem [{\citenamefont {Agnes}\ \emph {et~al.}(2023)\citenamefont {Agnes}
  \emph {et~al.}}]{DarkSide-50:2022qzh}%
  \BibitemOpen
  \bibfield  {author} {\bibinfo {author} {\bibfnamefont {P.}~\bibnamefont
  {Agnes}} \emph {et~al.} (\bibinfo {collaboration} {DarkSide-50}),\ }\href
  {\doibase 10.1103/PhysRevD.107.063001} {\bibfield  {journal} {\bibinfo
  {journal} {Phys. Rev. D}\ }\textbf {\bibinfo {volume} {107}},\ \bibinfo
  {pages} {063001} (\bibinfo {year} {2023})},\ \Eprint
  {http://arxiv.org/abs/2207.11966} {arXiv:2207.11966 [hep-ex]} \BibitemShut
  {NoStop}%
\bibitem [{\citenamefont {Aprile}\ \emph {et~al.}(2018)\citenamefont {Aprile}
  \emph {et~al.}}]{XENON:2018voc}%
  \BibitemOpen
  \bibfield  {author} {\bibinfo {author} {\bibfnamefont {E.}~\bibnamefont
  {Aprile}} \emph {et~al.} (\bibinfo {collaboration} {XENON}),\ }\href
  {\doibase 10.1103/PhysRevLett.121.111302} {\bibfield  {journal} {\bibinfo
  {journal} {Phys. Rev. Lett.}\ }\textbf {\bibinfo {volume} {121}},\ \bibinfo
  {pages} {111302} (\bibinfo {year} {2018})},\ \Eprint
  {http://arxiv.org/abs/1805.12562} {arXiv:1805.12562 [astro-ph.CO]}
  \BibitemShut {NoStop}%
\bibitem [{\citenamefont {Aprile}\ \emph {et~al.}(2019)\citenamefont {Aprile}
  \emph {et~al.}}]{Aprile:2019xxb}%
  \BibitemOpen
  \bibfield  {author} {\bibinfo {author} {\bibfnamefont {E.}~\bibnamefont
  {Aprile}} \emph {et~al.} (\bibinfo {collaboration} {XENON}),\ }\href
  {\doibase 10.1103/PhysRevLett.123.251801} {\bibfield  {journal} {\bibinfo
  {journal} {Phys. Rev. Lett.}\ }\textbf {\bibinfo {volume} {123}},\ \bibinfo
  {pages} {251801} (\bibinfo {year} {2019})},\ \Eprint
  {http://arxiv.org/abs/1907.11485} {arXiv:1907.11485 [hep-ex]} \BibitemShut
  {NoStop}%
\bibitem [{\citenamefont {Aprile}\ \emph {et~al.}(2023)\citenamefont {Aprile}
  \emph {et~al.}}]{XENON:2023cxc}%
  \BibitemOpen
  \bibfield  {author} {\bibinfo {author} {\bibfnamefont {E.}~\bibnamefont
  {Aprile}} \emph {et~al.} (\bibinfo {collaboration} {XENON}),\ }\href
  {\doibase 10.1103/PhysRevLett.131.041003} {\bibfield  {journal} {\bibinfo
  {journal} {Phys. Rev. Lett.}\ }\textbf {\bibinfo {volume} {131}},\ \bibinfo
  {pages} {041003} (\bibinfo {year} {2023})},\ \Eprint
  {http://arxiv.org/abs/2303.14729} {arXiv:2303.14729 [hep-ex]} \BibitemShut
  {NoStop}%
\bibitem [{\citenamefont {Meng}\ \emph {et~al.}(2021)\citenamefont {Meng} \emph
  {et~al.}}]{PandaX-4T:2021bab}%
  \BibitemOpen
  \bibfield  {author} {\bibinfo {author} {\bibfnamefont {Y.}~\bibnamefont
  {Meng}} \emph {et~al.} (\bibinfo {collaboration} {PandaX-4T}),\ }\href
  {\doibase 10.1103/PhysRevLett.127.261802} {\bibfield  {journal} {\bibinfo
  {journal} {Phys. Rev. Lett.}\ }\textbf {\bibinfo {volume} {127}},\ \bibinfo
  {pages} {261802} (\bibinfo {year} {2021})},\ \Eprint
  {http://arxiv.org/abs/2107.13438} {arXiv:2107.13438 [hep-ex]} \BibitemShut
  {NoStop}%
\bibitem [{\citenamefont {Aalbers}\ \emph {et~al.}(2023)\citenamefont {Aalbers}
  \emph {et~al.}}]{LZ:2022lsv}%
  \BibitemOpen
  \bibfield  {author} {\bibinfo {author} {\bibfnamefont {J.}~\bibnamefont
  {Aalbers}} \emph {et~al.} (\bibinfo {collaboration} {LZ}),\ }\href {\doibase
  10.1103/PhysRevLett.131.041002} {\bibfield  {journal} {\bibinfo  {journal}
  {Phys. Rev. Lett.}\ }\textbf {\bibinfo {volume} {131}},\ \bibinfo {pages}
  {041002} (\bibinfo {year} {2023})},\ \Eprint
  {http://arxiv.org/abs/2207.03764} {arXiv:2207.03764 [hep-ex]} \BibitemShut
  {NoStop}%
\bibitem [{\citenamefont {Ackermann}\ \emph {et~al.}(2015)\citenamefont
  {Ackermann} \emph {et~al.}}]{Ackermann:2015zua}%
  \BibitemOpen
  \bibfield  {author} {\bibinfo {author} {\bibfnamefont {M.}~\bibnamefont
  {Ackermann}} \emph {et~al.} (\bibinfo {collaboration} {Fermi-LAT}),\ }\href
  {\doibase 10.1103/PhysRevLett.115.231301} {\bibfield  {journal} {\bibinfo
  {journal} {Phys. Rev. Lett.}\ }\textbf {\bibinfo {volume} {115}},\ \bibinfo
  {pages} {231301} (\bibinfo {year} {2015})},\ \Eprint
  {http://arxiv.org/abs/1503.02641} {arXiv:1503.02641 [astro-ph.HE]}
  \BibitemShut {NoStop}%
\bibitem [{\citenamefont {Albert}\ \emph {et~al.}(2017)\citenamefont {Albert}
  \emph {et~al.}}]{Fermi-LAT:2016uux}%
  \BibitemOpen
  \bibfield  {author} {\bibinfo {author} {\bibfnamefont {A.}~\bibnamefont
  {Albert}} \emph {et~al.} (\bibinfo {collaboration} {Fermi-LAT, DES}),\ }\href
  {\doibase 10.3847/1538-4357/834/2/110} {\bibfield  {journal} {\bibinfo
  {journal} {Astrophys. J.}\ }\textbf {\bibinfo {volume} {834}},\ \bibinfo
  {pages} {110} (\bibinfo {year} {2017})},\ \Eprint
  {http://arxiv.org/abs/1611.03184} {arXiv:1611.03184 [astro-ph.HE]}
  \BibitemShut {NoStop}%
\bibitem [{\citenamefont {Linde}(1980)}]{Linde:1980ts}%
  \BibitemOpen
  \bibfield  {author} {\bibinfo {author} {\bibfnamefont {A.~D.}\ \bibnamefont
  {Linde}},\ }\href {\doibase 10.1016/0370-2693(80)90769-8} {\bibfield
  {journal} {\bibinfo  {journal} {Phys. Lett. B}\ }\textbf {\bibinfo {volume}
  {96}},\ \bibinfo {pages} {289} (\bibinfo {year} {1980})}\BibitemShut
  {NoStop}%
\bibitem [{\citenamefont {Arnold}(1992)}]{Arnold:1992fb}%
  \BibitemOpen
  \bibfield  {author} {\bibinfo {author} {\bibfnamefont {P.~B.}\ \bibnamefont
  {Arnold}},\ }\href {\doibase 10.1103/PhysRevD.46.2628} {\bibfield  {journal}
  {\bibinfo  {journal} {Phys. Rev. D}\ }\textbf {\bibinfo {volume} {46}},\
  \bibinfo {pages} {2628} (\bibinfo {year} {1992})},\ \Eprint
  {http://arxiv.org/abs/hep-ph/9204228} {arXiv:hep-ph/9204228} \BibitemShut
  {NoStop}%
\bibitem [{\citenamefont {Parwani}(1992)}]{Parwani:1991gq}%
  \BibitemOpen
  \bibfield  {author} {\bibinfo {author} {\bibfnamefont {R.~R.}\ \bibnamefont
  {Parwani}},\ }\href {\doibase 10.1103/PhysRevD.45.4695} {\bibfield  {journal}
  {\bibinfo  {journal} {Phys. Rev. D}\ }\textbf {\bibinfo {volume} {45}},\
  \bibinfo {pages} {4695} (\bibinfo {year} {1992})},\ \bibinfo {note}
  {[Erratum: Phys.Rev.D 48, 5965 (1993)]},\ \Eprint
  {http://arxiv.org/abs/hep-ph/9204216} {arXiv:hep-ph/9204216} \BibitemShut
  {NoStop}%
\bibitem [{\citenamefont {Curtin}\ \emph {et~al.}(2018)\citenamefont {Curtin},
  \citenamefont {Meade},\ and\ \citenamefont {Ramani}}]{Curtin:2016urg}%
  \BibitemOpen
  \bibfield  {author} {\bibinfo {author} {\bibfnamefont {D.}~\bibnamefont
  {Curtin}}, \bibinfo {author} {\bibfnamefont {P.}~\bibnamefont {Meade}}, \
  and\ \bibinfo {author} {\bibfnamefont {H.}~\bibnamefont {Ramani}},\ }\href
  {\doibase 10.1140/epjc/s10052-018-6268-0} {\bibfield  {journal} {\bibinfo
  {journal} {Eur. Phys. J. C}\ }\textbf {\bibinfo {volume} {78}},\ \bibinfo
  {pages} {787} (\bibinfo {year} {2018})},\ \Eprint
  {http://arxiv.org/abs/1612.00466} {arXiv:1612.00466 [hep-ph]} \BibitemShut
  {NoStop}%
\bibitem [{\citenamefont {Curtin}\ \emph {et~al.}(2024)\citenamefont {Curtin},
  \citenamefont {Roy},\ and\ \citenamefont {White}}]{Curtin:2022ovx}%
  \BibitemOpen
  \bibfield  {author} {\bibinfo {author} {\bibfnamefont {D.}~\bibnamefont
  {Curtin}}, \bibinfo {author} {\bibfnamefont {J.}~\bibnamefont {Roy}}, \ and\
  \bibinfo {author} {\bibfnamefont {G.}~\bibnamefont {White}},\ }\href
  {\doibase 10.1103/PhysRevD.109.116001} {\bibfield  {journal} {\bibinfo
  {journal} {Phys. Rev. D}\ }\textbf {\bibinfo {volume} {109}},\ \bibinfo
  {pages} {116001} (\bibinfo {year} {2024})},\ \Eprint
  {http://arxiv.org/abs/2211.08218} {arXiv:2211.08218 [hep-ph]} \BibitemShut
  {NoStop}%
\bibitem [{\citenamefont {Wainwright}\ \emph {et~al.}(2011)\citenamefont
  {Wainwright}, \citenamefont {Profumo},\ and\ \citenamefont
  {Ramsey-Musolf}}]{Wainwright:2011qy}%
  \BibitemOpen
  \bibfield  {author} {\bibinfo {author} {\bibfnamefont {C.}~\bibnamefont
  {Wainwright}}, \bibinfo {author} {\bibfnamefont {S.}~\bibnamefont {Profumo}},
  \ and\ \bibinfo {author} {\bibfnamefont {M.~J.}\ \bibnamefont
  {Ramsey-Musolf}},\ }\href {\doibase 10.1103/PhysRevD.84.023521} {\bibfield
  {journal} {\bibinfo  {journal} {Phys. Rev. D}\ }\textbf {\bibinfo {volume}
  {84}},\ \bibinfo {pages} {023521} (\bibinfo {year} {2011})},\ \Eprint
  {http://arxiv.org/abs/1104.5487} {arXiv:1104.5487 [hep-ph]} \BibitemShut
  {NoStop}%
\bibitem [{\citenamefont {Laine}(1995)}]{Laine:1994zq}%
  \BibitemOpen
  \bibfield  {author} {\bibinfo {author} {\bibfnamefont {M.}~\bibnamefont
  {Laine}},\ }\href {\doibase 10.1103/PhysRevD.51.4525} {\bibfield  {journal}
  {\bibinfo  {journal} {Phys. Rev. D}\ }\textbf {\bibinfo {volume} {51}},\
  \bibinfo {pages} {4525} (\bibinfo {year} {1995})},\ \Eprint
  {http://arxiv.org/abs/hep-ph/9411252} {arXiv:hep-ph/9411252} \BibitemShut
  {NoStop}%
\bibitem [{\citenamefont {Nielsen}(1975)}]{Nielsen:1975fs}%
  \BibitemOpen
  \bibfield  {author} {\bibinfo {author} {\bibfnamefont {N.~K.}\ \bibnamefont
  {Nielsen}},\ }\href {\doibase 10.1016/0550-3213(75)90301-6} {\bibfield
  {journal} {\bibinfo  {journal} {Nucl. Phys. B}\ }\textbf {\bibinfo {volume}
  {101}},\ \bibinfo {pages} {173} (\bibinfo {year} {1975})}\BibitemShut
  {NoStop}%
\bibitem [{\citenamefont {Fukuda}\ and\ \citenamefont
  {Kugo}(1976)}]{Fukuda:1975di}%
  \BibitemOpen
  \bibfield  {author} {\bibinfo {author} {\bibfnamefont {R.}~\bibnamefont
  {Fukuda}}\ and\ \bibinfo {author} {\bibfnamefont {T.}~\bibnamefont {Kugo}},\
  }\href {\doibase 10.1103/PhysRevD.13.3469} {\bibfield  {journal} {\bibinfo
  {journal} {Phys. Rev. D}\ }\textbf {\bibinfo {volume} {13}},\ \bibinfo
  {pages} {3469} (\bibinfo {year} {1976})}\BibitemShut {NoStop}%
\bibitem [{\citenamefont {Croon}\ \emph {et~al.}(2021)\citenamefont {Croon},
  \citenamefont {Gould}, \citenamefont {Schicho}, \citenamefont {Tenkanen},\
  and\ \citenamefont {White}}]{Croon:2020cgk}%
  \BibitemOpen
  \bibfield  {author} {\bibinfo {author} {\bibfnamefont {D.}~\bibnamefont
  {Croon}}, \bibinfo {author} {\bibfnamefont {O.}~\bibnamefont {Gould}},
  \bibinfo {author} {\bibfnamefont {P.}~\bibnamefont {Schicho}}, \bibinfo
  {author} {\bibfnamefont {T.~V.~I.}\ \bibnamefont {Tenkanen}}, \ and\ \bibinfo
  {author} {\bibfnamefont {G.}~\bibnamefont {White}},\ }\href {\doibase
  10.1007/JHEP04(2021)055} {\bibfield  {journal} {\bibinfo  {journal} {JHEP}\
  }\textbf {\bibinfo {volume} {04}},\ \bibinfo {pages} {055} (\bibinfo {year}
  {2021})},\ \Eprint {http://arxiv.org/abs/2009.10080} {arXiv:2009.10080
  [hep-ph]} \BibitemShut {NoStop}%
\bibitem [{\citenamefont {Gould}\ and\ \citenamefont
  {Tenkanen}(2021)}]{Gould:2021oba}%
  \BibitemOpen
  \bibfield  {author} {\bibinfo {author} {\bibfnamefont {O.}~\bibnamefont
  {Gould}}\ and\ \bibinfo {author} {\bibfnamefont {T.~V.~I.}\ \bibnamefont
  {Tenkanen}},\ }\href {\doibase 10.1007/JHEP06(2021)069} {\bibfield  {journal}
  {\bibinfo  {journal} {JHEP}\ }\textbf {\bibinfo {volume} {06}},\ \bibinfo
  {pages} {069} (\bibinfo {year} {2021})},\ \Eprint
  {http://arxiv.org/abs/2104.04399} {arXiv:2104.04399 [hep-ph]} \BibitemShut
  {NoStop}%
\bibitem [{\citenamefont {Athron}\ \emph {et~al.}(2023)\citenamefont {Athron},
  \citenamefont {Balazs}, \citenamefont {Fowlie}, \citenamefont {Morris},
  \citenamefont {White},\ and\ \citenamefont {Zhang}}]{Athron:2022jyi}%
  \BibitemOpen
  \bibfield  {author} {\bibinfo {author} {\bibfnamefont {P.}~\bibnamefont
  {Athron}}, \bibinfo {author} {\bibfnamefont {C.}~\bibnamefont {Balazs}},
  \bibinfo {author} {\bibfnamefont {A.}~\bibnamefont {Fowlie}}, \bibinfo
  {author} {\bibfnamefont {L.}~\bibnamefont {Morris}}, \bibinfo {author}
  {\bibfnamefont {G.}~\bibnamefont {White}}, \ and\ \bibinfo {author}
  {\bibfnamefont {Y.}~\bibnamefont {Zhang}},\ }\href {\doibase
  10.1007/JHEP01(2023)050} {\bibfield  {journal} {\bibinfo  {journal} {JHEP}\
  }\textbf {\bibinfo {volume} {01}},\ \bibinfo {pages} {050} (\bibinfo {year}
  {2023})},\ \Eprint {http://arxiv.org/abs/2208.01319} {arXiv:2208.01319
  [hep-ph]} \BibitemShut {NoStop}%
\bibitem [{\citenamefont {Kajantie}\ \emph
  {et~al.}(1996{\natexlab{b}})\citenamefont {Kajantie}, \citenamefont {Laine},
  \citenamefont {Rummukainen},\ and\ \citenamefont
  {Shaposhnikov}}]{Kajantie:1995dw}%
  \BibitemOpen
  \bibfield  {author} {\bibinfo {author} {\bibfnamefont {K.}~\bibnamefont
  {Kajantie}}, \bibinfo {author} {\bibfnamefont {M.}~\bibnamefont {Laine}},
  \bibinfo {author} {\bibfnamefont {K.}~\bibnamefont {Rummukainen}}, \ and\
  \bibinfo {author} {\bibfnamefont {M.~E.}\ \bibnamefont {Shaposhnikov}},\
  }\href {\doibase 10.1016/0550-3213(95)00549-8} {\bibfield  {journal}
  {\bibinfo  {journal} {Nucl. Phys. B}\ }\textbf {\bibinfo {volume} {458}},\
  \bibinfo {pages} {90} (\bibinfo {year} {1996}{\natexlab{b}})},\ \Eprint
  {http://arxiv.org/abs/hep-ph/9508379} {arXiv:hep-ph/9508379} \BibitemShut
  {NoStop}%
\bibitem [{\citenamefont {Farakos}\ \emph {et~al.}(1995)\citenamefont
  {Farakos}, \citenamefont {Kajantie}, \citenamefont {Rummukainen},\ and\
  \citenamefont {Shaposhnikov}}]{Farakos:1994xh}%
  \BibitemOpen
  \bibfield  {author} {\bibinfo {author} {\bibfnamefont {K.}~\bibnamefont
  {Farakos}}, \bibinfo {author} {\bibfnamefont {K.}~\bibnamefont {Kajantie}},
  \bibinfo {author} {\bibfnamefont {K.}~\bibnamefont {Rummukainen}}, \ and\
  \bibinfo {author} {\bibfnamefont {M.~E.}\ \bibnamefont {Shaposhnikov}},\
  }\href {\doibase 10.1016/0550-3213(95)80129-4} {\bibfield  {journal}
  {\bibinfo  {journal} {Nucl. Phys. B}\ }\textbf {\bibinfo {volume} {442}},\
  \bibinfo {pages} {317} (\bibinfo {year} {1995})},\ \Eprint
  {http://arxiv.org/abs/hep-lat/9412091} {arXiv:hep-lat/9412091} \BibitemShut
  {NoStop}%
\bibitem [{\citenamefont {Ekstedt}\ \emph {et~al.}(2023)\citenamefont
  {Ekstedt}, \citenamefont {Schicho},\ and\ \citenamefont
  {Tenkanen}}]{Ekstedt:2022bff}%
  \BibitemOpen
  \bibfield  {author} {\bibinfo {author} {\bibfnamefont {A.}~\bibnamefont
  {Ekstedt}}, \bibinfo {author} {\bibfnamefont {P.}~\bibnamefont {Schicho}}, \
  and\ \bibinfo {author} {\bibfnamefont {T.~V.~I.}\ \bibnamefont {Tenkanen}},\
  }\href {\doibase 10.1016/j.cpc.2023.108725} {\bibfield  {journal} {\bibinfo
  {journal} {Comput. Phys. Commun.}\ }\textbf {\bibinfo {volume} {288}},\
  \bibinfo {pages} {108725} (\bibinfo {year} {2023})},\ \Eprint
  {http://arxiv.org/abs/2205.08815} {arXiv:2205.08815 [hep-ph]} \BibitemShut
  {NoStop}%
\bibitem [{\citenamefont {Niemi}\ \emph
  {et~al.}(2021{\natexlab{a}})\citenamefont {Niemi}, \citenamefont {Schicho},\
  and\ \citenamefont {Tenkanen}}]{Niemi:2021qvp}%
  \BibitemOpen
  \bibfield  {author} {\bibinfo {author} {\bibfnamefont {L.}~\bibnamefont
  {Niemi}}, \bibinfo {author} {\bibfnamefont {P.}~\bibnamefont {Schicho}}, \
  and\ \bibinfo {author} {\bibfnamefont {T.~V.~I.}\ \bibnamefont {Tenkanen}},\
  }\href {\doibase 10.1103/PhysRevD.103.115035} {\bibfield  {journal} {\bibinfo
   {journal} {Phys. Rev. D}\ }\textbf {\bibinfo {volume} {103}},\ \bibinfo
  {pages} {115035} (\bibinfo {year} {2021}{\natexlab{a}})},\ \Eprint
  {http://arxiv.org/abs/2103.07467} {arXiv:2103.07467 [hep-ph]} \BibitemShut
  {NoStop}%
\bibitem [{\citenamefont {Schicho}\ \emph {et~al.}(2022)\citenamefont
  {Schicho}, \citenamefont {Tenkanen},\ and\ \citenamefont
  {White}}]{Schicho:2022wty}%
  \BibitemOpen
  \bibfield  {author} {\bibinfo {author} {\bibfnamefont {P.}~\bibnamefont
  {Schicho}}, \bibinfo {author} {\bibfnamefont {T.~V.~I.}\ \bibnamefont
  {Tenkanen}}, \ and\ \bibinfo {author} {\bibfnamefont {G.}~\bibnamefont
  {White}},\ }\href {\doibase 10.1007/JHEP11(2022)047} {\bibfield  {journal}
  {\bibinfo  {journal} {JHEP}\ }\textbf {\bibinfo {volume} {11}},\ \bibinfo
  {pages} {047} (\bibinfo {year} {2022})},\ \Eprint
  {http://arxiv.org/abs/2203.04284} {arXiv:2203.04284 [hep-ph]} \BibitemShut
  {NoStop}%
\bibitem [{\citenamefont {Ekstedt}\ and\ \citenamefont
  {L\"ofgren}(2020)}]{Ekstedt:2020abj}%
  \BibitemOpen
  \bibfield  {author} {\bibinfo {author} {\bibfnamefont {A.}~\bibnamefont
  {Ekstedt}}\ and\ \bibinfo {author} {\bibfnamefont {J.}~\bibnamefont
  {L\"ofgren}},\ }\href {\doibase 10.1007/JHEP12(2020)136} {\bibfield
  {journal} {\bibinfo  {journal} {JHEP}\ }\textbf {\bibinfo {volume} {12}},\
  \bibinfo {pages} {136} (\bibinfo {year} {2020})},\ \Eprint
  {http://arxiv.org/abs/2006.12614} {arXiv:2006.12614 [hep-ph]} \BibitemShut
  {NoStop}%
\bibitem [{\citenamefont {Niemi}\ \emph
  {et~al.}(2021{\natexlab{b}})\citenamefont {Niemi}, \citenamefont
  {Ramsey-Musolf}, \citenamefont {Tenkanen},\ and\ \citenamefont
  {Weir}}]{Niemi:2020hto}%
  \BibitemOpen
  \bibfield  {author} {\bibinfo {author} {\bibfnamefont {L.}~\bibnamefont
  {Niemi}}, \bibinfo {author} {\bibfnamefont {M.~J.}\ \bibnamefont
  {Ramsey-Musolf}}, \bibinfo {author} {\bibfnamefont {T.~V.~I.}\ \bibnamefont
  {Tenkanen}}, \ and\ \bibinfo {author} {\bibfnamefont {D.~J.}\ \bibnamefont
  {Weir}},\ }\href {\doibase 10.1103/PhysRevLett.126.171802} {\bibfield
  {journal} {\bibinfo  {journal} {Phys. Rev. Lett.}\ }\textbf {\bibinfo
  {volume} {126}},\ \bibinfo {pages} {171802} (\bibinfo {year}
  {2021}{\natexlab{b}})},\ \Eprint {http://arxiv.org/abs/2005.11332}
  {arXiv:2005.11332 [hep-ph]} \BibitemShut {NoStop}%
\bibitem [{\citenamefont {Gould}\ \emph {et~al.}(2022)\citenamefont {Gould},
  \citenamefont {G\"uyer},\ and\ \citenamefont {Rummukainen}}]{Gould:2022ran}%
  \BibitemOpen
  \bibfield  {author} {\bibinfo {author} {\bibfnamefont {O.}~\bibnamefont
  {Gould}}, \bibinfo {author} {\bibfnamefont {S.}~\bibnamefont {G\"uyer}}, \
  and\ \bibinfo {author} {\bibfnamefont {K.}~\bibnamefont {Rummukainen}},\
  }\href {\doibase 10.1103/PhysRevD.106.114507} {\bibfield  {journal} {\bibinfo
   {journal} {Phys. Rev. D}\ }\textbf {\bibinfo {volume} {106}},\ \bibinfo
  {pages} {114507} (\bibinfo {year} {2022})},\ \Eprint
  {http://arxiv.org/abs/2205.07238} {arXiv:2205.07238 [hep-lat]} \BibitemShut
  {NoStop}%
\bibitem [{\citenamefont {Niemi}\ \emph {et~al.}(2024)\citenamefont {Niemi},
  \citenamefont {Ramsey-Musolf},\ and\ \citenamefont {Xia}}]{Niemi:2024axp}%
  \BibitemOpen
  \bibfield  {author} {\bibinfo {author} {\bibfnamefont {L.}~\bibnamefont
  {Niemi}}, \bibinfo {author} {\bibfnamefont {M.~J.}\ \bibnamefont
  {Ramsey-Musolf}}, \ and\ \bibinfo {author} {\bibfnamefont {G.}~\bibnamefont
  {Xia}},\ }\href@noop {} {\  (\bibinfo {year} {2024})},\ \Eprint
  {http://arxiv.org/abs/2405.01191} {arXiv:2405.01191 [hep-ph]} \BibitemShut
  {NoStop}%
\bibitem [{\citenamefont {Coleman}\ and\ \citenamefont
  {Weinberg}(1973)}]{Coleman:1973jx}%
  \BibitemOpen
  \bibfield  {author} {\bibinfo {author} {\bibfnamefont {S.~R.}\ \bibnamefont
  {Coleman}}\ and\ \bibinfo {author} {\bibfnamefont {E.~J.}\ \bibnamefont
  {Weinberg}},\ }\href {\doibase 10.1103/PhysRevD.7.1888} {\bibfield  {journal}
  {\bibinfo  {journal} {Phys. Rev. D}\ }\textbf {\bibinfo {volume} {7}},\
  \bibinfo {pages} {1888} (\bibinfo {year} {1973})}\BibitemShut {NoStop}%
\bibitem [{\citenamefont {Weinberg}(1973)}]{Weinberg:1973am}%
  \BibitemOpen
  \bibfield  {author} {\bibinfo {author} {\bibfnamefont {E.~J.}\ \bibnamefont
  {Weinberg}},\ }\emph {\bibinfo {title} {{Radiative corrections as the origin
  of spontaneous symmetry breaking}}},\ \href@noop {} {Ph.D. thesis},\ \bibinfo
   {school} {Harvard U.} (\bibinfo {year} {1973}),\ \Eprint
  {http://arxiv.org/abs/hep-th/0507214} {arXiv:hep-th/0507214} \BibitemShut
  {NoStop}%
\bibitem [{\citenamefont {Dolan}\ and\ \citenamefont
  {Jackiw}(1974)}]{Dolan:1973qd}%
  \BibitemOpen
  \bibfield  {author} {\bibinfo {author} {\bibfnamefont {L.}~\bibnamefont
  {Dolan}}\ and\ \bibinfo {author} {\bibfnamefont {R.}~\bibnamefont {Jackiw}},\
  }\href {\doibase 10.1103/PhysRevD.9.3320} {\bibfield  {journal} {\bibinfo
  {journal} {Phys. Rev. D}\ }\textbf {\bibinfo {volume} {9}},\ \bibinfo {pages}
  {3320} (\bibinfo {year} {1974})}\BibitemShut {NoStop}%
\bibitem [{\citenamefont {Arnold}\ and\ \citenamefont
  {Espinosa}(1993)}]{Arnold:1992rz}%
  \BibitemOpen
  \bibfield  {author} {\bibinfo {author} {\bibfnamefont {P.~B.}\ \bibnamefont
  {Arnold}}\ and\ \bibinfo {author} {\bibfnamefont {O.}~\bibnamefont
  {Espinosa}},\ }\href {\doibase 10.1103/PhysRevD.47.3546} {\bibfield
  {journal} {\bibinfo  {journal} {Phys. Rev. D}\ }\textbf {\bibinfo {volume}
  {47}},\ \bibinfo {pages} {3546} (\bibinfo {year} {1993})},\ \bibinfo {note}
  {[Erratum: Phys.Rev.D 50, 6662 (1994)]},\ \Eprint
  {http://arxiv.org/abs/hep-ph/9212235} {arXiv:hep-ph/9212235} \BibitemShut
  {NoStop}%
\bibitem [{\citenamefont {Carrington}(1992)}]{Carrington:1991hz}%
  \BibitemOpen
  \bibfield  {author} {\bibinfo {author} {\bibfnamefont {M.~E.}\ \bibnamefont
  {Carrington}},\ }\href {\doibase 10.1103/PhysRevD.45.2933} {\bibfield
  {journal} {\bibinfo  {journal} {Phys. Rev. D}\ }\textbf {\bibinfo {volume}
  {45}},\ \bibinfo {pages} {2933} (\bibinfo {year} {1992})}\BibitemShut
  {NoStop}%
\bibitem [{\citenamefont {Athron}\ \emph {et~al.}(2020)\citenamefont {Athron},
  \citenamefont {Bal\'azs}, \citenamefont {Fowlie},\ and\ \citenamefont
  {Zhang}}]{Athron:2020sbe}%
  \BibitemOpen
  \bibfield  {author} {\bibinfo {author} {\bibfnamefont {P.}~\bibnamefont
  {Athron}}, \bibinfo {author} {\bibfnamefont {C.}~\bibnamefont {Bal\'azs}},
  \bibinfo {author} {\bibfnamefont {A.}~\bibnamefont {Fowlie}}, \ and\ \bibinfo
  {author} {\bibfnamefont {Y.}~\bibnamefont {Zhang}},\ }\href {\doibase
  10.1140/epjc/s10052-020-8035-2} {\bibfield  {journal} {\bibinfo  {journal}
  {Eur. Phys. J. C}\ }\textbf {\bibinfo {volume} {80}},\ \bibinfo {pages} {567}
  (\bibinfo {year} {2020})},\ \Eprint {http://arxiv.org/abs/2003.02859}
  {arXiv:2003.02859 [hep-ph]} \BibitemShut {NoStop}%
\bibitem [{\citenamefont {Mohapatra}\ and\ \citenamefont
  {Senjanovic}(1979)}]{Mohapatra:1979vr}%
  \BibitemOpen
  \bibfield  {author} {\bibinfo {author} {\bibfnamefont {R.~N.}\ \bibnamefont
  {Mohapatra}}\ and\ \bibinfo {author} {\bibfnamefont {G.}~\bibnamefont
  {Senjanovic}},\ }\href {\doibase 10.1103/PhysRevD.20.3390} {\bibfield
  {journal} {\bibinfo  {journal} {Phys. Rev. D}\ }\textbf {\bibinfo {volume}
  {20}},\ \bibinfo {pages} {3390} (\bibinfo {year} {1979})}\BibitemShut
  {NoStop}%
\bibitem [{\citenamefont {Masiero}\ \emph {et~al.}(1984)\citenamefont
  {Masiero}, \citenamefont {Nanopoulos},\ and\ \citenamefont
  {Yanagida}}]{Masiero:1983ux}%
  \BibitemOpen
  \bibfield  {author} {\bibinfo {author} {\bibfnamefont {A.}~\bibnamefont
  {Masiero}}, \bibinfo {author} {\bibfnamefont {D.~V.}\ \bibnamefont
  {Nanopoulos}}, \ and\ \bibinfo {author} {\bibfnamefont {T.}~\bibnamefont
  {Yanagida}},\ }\href {\doibase 10.1016/0370-2693(84)91879-3} {\bibfield
  {journal} {\bibinfo  {journal} {Phys. Lett. B}\ }\textbf {\bibinfo {volume}
  {138}},\ \bibinfo {pages} {91} (\bibinfo {year} {1984})}\BibitemShut
  {NoStop}%
\bibitem [{\citenamefont {Salomonson}\ and\ \citenamefont
  {Skagerstam}(1985)}]{Salomonson:1984px}%
  \BibitemOpen
  \bibfield  {author} {\bibinfo {author} {\bibfnamefont {P.}~\bibnamefont
  {Salomonson}}\ and\ \bibinfo {author} {\bibfnamefont {B.-S.~K.}\ \bibnamefont
  {Skagerstam}},\ }\href {\doibase 10.1016/0370-2693(85)91039-1} {\bibfield
  {journal} {\bibinfo  {journal} {Phys. Lett. B}\ }\textbf {\bibinfo {volume}
  {155}},\ \bibinfo {pages} {100} (\bibinfo {year} {1985})}\BibitemShut
  {NoStop}%
\bibitem [{\citenamefont {Kephart}\ \emph {et~al.}(1990)\citenamefont
  {Kephart}, \citenamefont {Weiler},\ and\ \citenamefont
  {Yuan}}]{Kephart:1988xh}%
  \BibitemOpen
  \bibfield  {author} {\bibinfo {author} {\bibfnamefont {T.~W.}\ \bibnamefont
  {Kephart}}, \bibinfo {author} {\bibfnamefont {T.~J.}\ \bibnamefont {Weiler}},
  \ and\ \bibinfo {author} {\bibfnamefont {T.~C.}\ \bibnamefont {Yuan}},\
  }\href {\doibase 10.1016/0550-3213(90)90128-Z} {\bibfield  {journal}
  {\bibinfo  {journal} {Nucl. Phys. B}\ }\textbf {\bibinfo {volume} {330}},\
  \bibinfo {pages} {705} (\bibinfo {year} {1990})}\BibitemShut {NoStop}%
\bibitem [{\citenamefont {Langacker}\ and\ \citenamefont
  {Pi}(1980)}]{Langacker:1980kd}%
  \BibitemOpen
  \bibfield  {author} {\bibinfo {author} {\bibfnamefont {P.}~\bibnamefont
  {Langacker}}\ and\ \bibinfo {author} {\bibfnamefont {S.-Y.}\ \bibnamefont
  {Pi}},\ }\href {\doibase 10.1103/PhysRevLett.45.1} {\bibfield  {journal}
  {\bibinfo  {journal} {Phys. Rev. Lett.}\ }\textbf {\bibinfo {volume} {45}},\
  \bibinfo {pages} {1} (\bibinfo {year} {1980})}\BibitemShut {NoStop}%
\bibitem [{\citenamefont {Farris}\ \emph {et~al.}(1992)\citenamefont {Farris},
  \citenamefont {Kephart}, \citenamefont {Weiler},\ and\ \citenamefont
  {Yuan}}]{Farris:1991rg}%
  \BibitemOpen
  \bibfield  {author} {\bibinfo {author} {\bibfnamefont {T.~H.}\ \bibnamefont
  {Farris}}, \bibinfo {author} {\bibfnamefont {T.~W.}\ \bibnamefont {Kephart}},
  \bibinfo {author} {\bibfnamefont {T.~J.}\ \bibnamefont {Weiler}}, \ and\
  \bibinfo {author} {\bibfnamefont {T.~C.}\ \bibnamefont {Yuan}},\ }\href
  {\doibase 10.1103/PhysRevLett.68.564} {\bibfield  {journal} {\bibinfo
  {journal} {Phys. Rev. Lett.}\ }\textbf {\bibinfo {volume} {68}},\ \bibinfo
  {pages} {564} (\bibinfo {year} {1992})}\BibitemShut {NoStop}%
\bibitem [{\citenamefont {Foreman-Mackey}\ \emph {et~al.}(2013)\citenamefont
  {Foreman-Mackey}, \citenamefont {Hogg}, \citenamefont {Lang},\ and\
  \citenamefont {Goodman}}]{ForemanMackey:2012ig}%
  \BibitemOpen
  \bibfield  {author} {\bibinfo {author} {\bibfnamefont {D.}~\bibnamefont
  {Foreman-Mackey}}, \bibinfo {author} {\bibfnamefont {D.~W.}\ \bibnamefont
  {Hogg}}, \bibinfo {author} {\bibfnamefont {D.}~\bibnamefont {Lang}}, \ and\
  \bibinfo {author} {\bibfnamefont {J.}~\bibnamefont {Goodman}},\ }\href
  {\doibase 10.1086/670067} {\bibfield  {journal} {\bibinfo  {journal} {Publ.
  Astron. Soc. Pac.}\ }\textbf {\bibinfo {volume} {125}},\ \bibinfo {pages}
  {306} (\bibinfo {year} {2013})},\ \Eprint {http://arxiv.org/abs/1202.3665}
  {arXiv:1202.3665 [astro-ph.IM]} \BibitemShut {NoStop}%
\bibitem [{\citenamefont {Zyla}\ \emph
  {et~al.}(2020{\natexlab{b}})\citenamefont {Zyla} \emph
  {et~al.}}]{ParticleDataGroup:2020ssz}%
  \BibitemOpen
  \bibfield  {author} {\bibinfo {author} {\bibfnamefont {P.~A.}\ \bibnamefont
  {Zyla}} \emph {et~al.} (\bibinfo {collaboration} {Particle Data Group}),\
  }\href {\doibase 10.1093/ptep/ptaa104} {\bibfield  {journal} {\bibinfo
  {journal} {PTEP}\ }\textbf {\bibinfo {volume} {2020}},\ \bibinfo {pages}
  {083C01} (\bibinfo {year} {2020}{\natexlab{b}})}\BibitemShut {NoStop}%
\bibitem [{\citenamefont {Caprini}\ \emph {et~al.}(2016)\citenamefont {Caprini}
  \emph {et~al.}}]{Caprini:2015zlo}%
  \BibitemOpen
  \bibfield  {author} {\bibinfo {author} {\bibfnamefont {C.}~\bibnamefont
  {Caprini}} \emph {et~al.},\ }\href {\doibase 10.1088/1475-7516/2016/04/001}
  {\bibfield  {journal} {\bibinfo  {journal} {JCAP}\ }\textbf {\bibinfo
  {volume} {04}},\ \bibinfo {pages} {001} (\bibinfo {year} {2016})},\ \Eprint
  {http://arxiv.org/abs/1512.06239} {arXiv:1512.06239 [astro-ph.CO]}
  \BibitemShut {NoStop}%
\bibitem [{\citenamefont {Jinno}\ and\ \citenamefont
  {Takimoto}(2017)}]{Jinno:2016vai}%
  \BibitemOpen
  \bibfield  {author} {\bibinfo {author} {\bibfnamefont {R.}~\bibnamefont
  {Jinno}}\ and\ \bibinfo {author} {\bibfnamefont {M.}~\bibnamefont
  {Takimoto}},\ }\href {\doibase 10.1103/PhysRevD.95.024009} {\bibfield
  {journal} {\bibinfo  {journal} {Phys. Rev. D}\ }\textbf {\bibinfo {volume}
  {95}},\ \bibinfo {pages} {024009} (\bibinfo {year} {2017})},\ \Eprint
  {http://arxiv.org/abs/1605.01403} {arXiv:1605.01403 [astro-ph.CO]}
  \BibitemShut {NoStop}%
\bibitem [{\citenamefont {Kosowsky}\ \emph
  {et~al.}(1992{\natexlab{a}})\citenamefont {Kosowsky}, \citenamefont
  {Turner},\ and\ \citenamefont {Watkins}}]{Kosowsky:1991ua}%
  \BibitemOpen
  \bibfield  {author} {\bibinfo {author} {\bibfnamefont {A.}~\bibnamefont
  {Kosowsky}}, \bibinfo {author} {\bibfnamefont {M.~S.}\ \bibnamefont
  {Turner}}, \ and\ \bibinfo {author} {\bibfnamefont {R.}~\bibnamefont
  {Watkins}},\ }\href {\doibase 10.1103/PhysRevD.45.4514} {\bibfield  {journal}
  {\bibinfo  {journal} {Phys. Rev. D}\ }\textbf {\bibinfo {volume} {45}},\
  \bibinfo {pages} {4514} (\bibinfo {year} {1992}{\natexlab{a}})}\BibitemShut
  {NoStop}%
\bibitem [{\citenamefont {Kosowsky}\ \emph
  {et~al.}(1992{\natexlab{b}})\citenamefont {Kosowsky}, \citenamefont
  {Turner},\ and\ \citenamefont {Watkins}}]{Kosowsky:1992rz}%
  \BibitemOpen
  \bibfield  {author} {\bibinfo {author} {\bibfnamefont {A.}~\bibnamefont
  {Kosowsky}}, \bibinfo {author} {\bibfnamefont {M.~S.}\ \bibnamefont
  {Turner}}, \ and\ \bibinfo {author} {\bibfnamefont {R.}~\bibnamefont
  {Watkins}},\ }\href {\doibase 10.1103/PhysRevLett.69.2026} {\bibfield
  {journal} {\bibinfo  {journal} {Phys. Rev. Lett.}\ }\textbf {\bibinfo
  {volume} {69}},\ \bibinfo {pages} {2026} (\bibinfo {year}
  {1992}{\natexlab{b}})}\BibitemShut {NoStop}%
\bibitem [{\citenamefont {Kosowsky}\ and\ \citenamefont
  {Turner}(1993)}]{Kosowsky:1992vn}%
  \BibitemOpen
  \bibfield  {author} {\bibinfo {author} {\bibfnamefont {A.}~\bibnamefont
  {Kosowsky}}\ and\ \bibinfo {author} {\bibfnamefont {M.~S.}\ \bibnamefont
  {Turner}},\ }\href {\doibase 10.1103/PhysRevD.47.4372} {\bibfield  {journal}
  {\bibinfo  {journal} {Phys. Rev. D}\ }\textbf {\bibinfo {volume} {47}},\
  \bibinfo {pages} {4372} (\bibinfo {year} {1993})},\ \Eprint
  {http://arxiv.org/abs/astro-ph/9211004} {arXiv:astro-ph/9211004} \BibitemShut
  {NoStop}%
\bibitem [{\citenamefont {Huber}\ and\ \citenamefont
  {Konstandin}(2008{\natexlab{b}})}]{Huber:2008hg}%
  \BibitemOpen
  \bibfield  {author} {\bibinfo {author} {\bibfnamefont {S.~J.}\ \bibnamefont
  {Huber}}\ and\ \bibinfo {author} {\bibfnamefont {T.}~\bibnamefont
  {Konstandin}},\ }\href {\doibase 10.1088/1475-7516/2008/09/022} {\bibfield
  {journal} {\bibinfo  {journal} {JCAP}\ }\textbf {\bibinfo {volume} {09}},\
  \bibinfo {pages} {022} (\bibinfo {year} {2008}{\natexlab{b}})},\ \Eprint
  {http://arxiv.org/abs/0806.1828} {arXiv:0806.1828 [hep-ph]} \BibitemShut
  {NoStop}%
\bibitem [{\citenamefont {Hindmarsh}\ \emph {et~al.}(2014)\citenamefont
  {Hindmarsh}, \citenamefont {Huber}, \citenamefont {Rummukainen},\ and\
  \citenamefont {Weir}}]{Hindmarsh:2013xza}%
  \BibitemOpen
  \bibfield  {author} {\bibinfo {author} {\bibfnamefont {M.}~\bibnamefont
  {Hindmarsh}}, \bibinfo {author} {\bibfnamefont {S.~J.}\ \bibnamefont
  {Huber}}, \bibinfo {author} {\bibfnamefont {K.}~\bibnamefont {Rummukainen}},
  \ and\ \bibinfo {author} {\bibfnamefont {D.~J.}\ \bibnamefont {Weir}},\
  }\href {\doibase 10.1103/PhysRevLett.112.041301} {\bibfield  {journal}
  {\bibinfo  {journal} {Phys. Rev. Lett.}\ }\textbf {\bibinfo {volume} {112}},\
  \bibinfo {pages} {041301} (\bibinfo {year} {2014})},\ \Eprint
  {http://arxiv.org/abs/1304.2433} {arXiv:1304.2433 [hep-ph]} \BibitemShut
  {NoStop}%
\bibitem [{\citenamefont {Giblin}\ and\ \citenamefont
  {Mertens}(2013)}]{Giblin:2013kea}%
  \BibitemOpen
  \bibfield  {author} {\bibinfo {author} {\bibfnamefont {J.~T.}\ \bibnamefont
  {Giblin}, \bibfnamefont {Jr.}}\ and\ \bibinfo {author} {\bibfnamefont
  {J.~B.}\ \bibnamefont {Mertens}},\ }\href {\doibase 10.1007/JHEP12(2013)042}
  {\bibfield  {journal} {\bibinfo  {journal} {JHEP}\ }\textbf {\bibinfo
  {volume} {12}},\ \bibinfo {pages} {042} (\bibinfo {year} {2013})},\ \Eprint
  {http://arxiv.org/abs/1310.2948} {arXiv:1310.2948 [hep-th]} \BibitemShut
  {NoStop}%
\bibitem [{\citenamefont {Giblin}\ and\ \citenamefont
  {Mertens}(2014)}]{Giblin:2014qia}%
  \BibitemOpen
  \bibfield  {author} {\bibinfo {author} {\bibfnamefont {J.~T.}\ \bibnamefont
  {Giblin}}\ and\ \bibinfo {author} {\bibfnamefont {J.~B.}\ \bibnamefont
  {Mertens}},\ }\href {\doibase 10.1103/PhysRevD.90.023532} {\bibfield
  {journal} {\bibinfo  {journal} {Phys. Rev. D}\ }\textbf {\bibinfo {volume}
  {90}},\ \bibinfo {pages} {023532} (\bibinfo {year} {2014})},\ \Eprint
  {http://arxiv.org/abs/1405.4005} {arXiv:1405.4005 [astro-ph.CO]} \BibitemShut
  {NoStop}%
\bibitem [{\citenamefont {Hindmarsh}\ \emph {et~al.}(2015)\citenamefont
  {Hindmarsh}, \citenamefont {Huber}, \citenamefont {Rummukainen},\ and\
  \citenamefont {Weir}}]{Hindmarsh:2015qta}%
  \BibitemOpen
  \bibfield  {author} {\bibinfo {author} {\bibfnamefont {M.}~\bibnamefont
  {Hindmarsh}}, \bibinfo {author} {\bibfnamefont {S.~J.}\ \bibnamefont
  {Huber}}, \bibinfo {author} {\bibfnamefont {K.}~\bibnamefont {Rummukainen}},
  \ and\ \bibinfo {author} {\bibfnamefont {D.~J.}\ \bibnamefont {Weir}},\
  }\href {\doibase 10.1103/PhysRevD.92.123009} {\bibfield  {journal} {\bibinfo
  {journal} {Phys. Rev. D}\ }\textbf {\bibinfo {volume} {92}},\ \bibinfo
  {pages} {123009} (\bibinfo {year} {2015})},\ \Eprint
  {http://arxiv.org/abs/1504.03291} {arXiv:1504.03291 [astro-ph.CO]}
  \BibitemShut {NoStop}%
\bibitem [{\citenamefont {Hindmarsh}\ \emph {et~al.}(2017)\citenamefont
  {Hindmarsh}, \citenamefont {Huber}, \citenamefont {Rummukainen},\ and\
  \citenamefont {Weir}}]{Hindmarsh:2017gnf}%
  \BibitemOpen
  \bibfield  {author} {\bibinfo {author} {\bibfnamefont {M.}~\bibnamefont
  {Hindmarsh}}, \bibinfo {author} {\bibfnamefont {S.~J.}\ \bibnamefont
  {Huber}}, \bibinfo {author} {\bibfnamefont {K.}~\bibnamefont {Rummukainen}},
  \ and\ \bibinfo {author} {\bibfnamefont {D.~J.}\ \bibnamefont {Weir}},\
  }\href {\doibase 10.1103/PhysRevD.96.103520} {\bibfield  {journal} {\bibinfo
  {journal} {Phys. Rev. D}\ }\textbf {\bibinfo {volume} {96}},\ \bibinfo
  {pages} {103520} (\bibinfo {year} {2017})},\ \bibinfo {note} {[Erratum:
  Phys.Rev.D 101, 089902 (2020)]},\ \Eprint {http://arxiv.org/abs/1704.05871}
  {arXiv:1704.05871 [astro-ph.CO]} \BibitemShut {NoStop}%
\bibitem [{\citenamefont {Cutting}\ \emph {et~al.}(2020)\citenamefont
  {Cutting}, \citenamefont {Hindmarsh},\ and\ \citenamefont
  {Weir}}]{Cutting:2019zws}%
  \BibitemOpen
  \bibfield  {author} {\bibinfo {author} {\bibfnamefont {D.}~\bibnamefont
  {Cutting}}, \bibinfo {author} {\bibfnamefont {M.}~\bibnamefont {Hindmarsh}},
  \ and\ \bibinfo {author} {\bibfnamefont {D.~J.}\ \bibnamefont {Weir}},\
  }\href {\doibase 10.1103/PhysRevLett.125.021302} {\bibfield  {journal}
  {\bibinfo  {journal} {Phys. Rev. Lett.}\ }\textbf {\bibinfo {volume} {125}},\
  \bibinfo {pages} {021302} (\bibinfo {year} {2020})},\ \Eprint
  {http://arxiv.org/abs/1906.00480} {arXiv:1906.00480 [hep-ph]} \BibitemShut
  {NoStop}%
\bibitem [{\citenamefont {Hindmarsh}(2018)}]{Hindmarsh:2016lnk}%
  \BibitemOpen
  \bibfield  {author} {\bibinfo {author} {\bibfnamefont {M.}~\bibnamefont
  {Hindmarsh}},\ }\href {\doibase 10.1103/PhysRevLett.120.071301} {\bibfield
  {journal} {\bibinfo  {journal} {Phys. Rev. Lett.}\ }\textbf {\bibinfo
  {volume} {120}},\ \bibinfo {pages} {071301} (\bibinfo {year} {2018})},\
  \Eprint {http://arxiv.org/abs/1608.04735} {arXiv:1608.04735 [astro-ph.CO]}
  \BibitemShut {NoStop}%
\bibitem [{\citenamefont {Hindmarsh}\ and\ \citenamefont
  {Hijazi}(2019)}]{Hindmarsh:2019phv}%
  \BibitemOpen
  \bibfield  {author} {\bibinfo {author} {\bibfnamefont {M.}~\bibnamefont
  {Hindmarsh}}\ and\ \bibinfo {author} {\bibfnamefont {M.}~\bibnamefont
  {Hijazi}},\ }\href {\doibase 10.1088/1475-7516/2019/12/062} {\bibfield
  {journal} {\bibinfo  {journal} {JCAP}\ }\textbf {\bibinfo {volume} {12}},\
  \bibinfo {pages} {062} (\bibinfo {year} {2019})},\ \Eprint
  {http://arxiv.org/abs/1909.10040} {arXiv:1909.10040 [astro-ph.CO]}
  \BibitemShut {NoStop}%
\bibitem [{\citenamefont {Caprini}\ and\ \citenamefont
  {Durrer}(2006)}]{Caprini:2006jb}%
  \BibitemOpen
  \bibfield  {author} {\bibinfo {author} {\bibfnamefont {C.}~\bibnamefont
  {Caprini}}\ and\ \bibinfo {author} {\bibfnamefont {R.}~\bibnamefont
  {Durrer}},\ }\href {\doibase 10.1103/PhysRevD.74.063521} {\bibfield
  {journal} {\bibinfo  {journal} {Phys. Rev. D}\ }\textbf {\bibinfo {volume}
  {74}},\ \bibinfo {pages} {063521} (\bibinfo {year} {2006})},\ \Eprint
  {http://arxiv.org/abs/astro-ph/0603476} {arXiv:astro-ph/0603476} \BibitemShut
  {NoStop}%
\bibitem [{\citenamefont {Kahniashvili}\ \emph
  {et~al.}(2008{\natexlab{a}})\citenamefont {Kahniashvili}, \citenamefont
  {Kosowsky}, \citenamefont {Gogoberidze},\ and\ \citenamefont
  {Maravin}}]{Kahniashvili:2008pf}%
  \BibitemOpen
  \bibfield  {author} {\bibinfo {author} {\bibfnamefont {T.}~\bibnamefont
  {Kahniashvili}}, \bibinfo {author} {\bibfnamefont {A.}~\bibnamefont
  {Kosowsky}}, \bibinfo {author} {\bibfnamefont {G.}~\bibnamefont
  {Gogoberidze}}, \ and\ \bibinfo {author} {\bibfnamefont {Y.}~\bibnamefont
  {Maravin}},\ }\href {\doibase 10.1103/PhysRevD.78.043003} {\bibfield
  {journal} {\bibinfo  {journal} {Phys. Rev. D}\ }\textbf {\bibinfo {volume}
  {78}},\ \bibinfo {pages} {043003} (\bibinfo {year} {2008}{\natexlab{a}})},\
  \Eprint {http://arxiv.org/abs/0806.0293} {arXiv:0806.0293 [astro-ph]}
  \BibitemShut {NoStop}%
\bibitem [{\citenamefont {Kahniashvili}\ \emph
  {et~al.}(2008{\natexlab{b}})\citenamefont {Kahniashvili}, \citenamefont
  {Campanelli}, \citenamefont {Gogoberidze}, \citenamefont {Maravin},\ and\
  \citenamefont {Ratra}}]{Kahniashvili:2008pe}%
  \BibitemOpen
  \bibfield  {author} {\bibinfo {author} {\bibfnamefont {T.}~\bibnamefont
  {Kahniashvili}}, \bibinfo {author} {\bibfnamefont {L.}~\bibnamefont
  {Campanelli}}, \bibinfo {author} {\bibfnamefont {G.}~\bibnamefont
  {Gogoberidze}}, \bibinfo {author} {\bibfnamefont {Y.}~\bibnamefont
  {Maravin}}, \ and\ \bibinfo {author} {\bibfnamefont {B.}~\bibnamefont
  {Ratra}},\ }\href {\doibase 10.1103/PhysRevD.78.123006} {\bibfield  {journal}
  {\bibinfo  {journal} {Phys. Rev. D}\ }\textbf {\bibinfo {volume} {78}},\
  \bibinfo {pages} {123006} (\bibinfo {year} {2008}{\natexlab{b}})},\ \bibinfo
  {note} {[Erratum: Phys.Rev.D 79, 109901 (2009)]},\ \Eprint
  {http://arxiv.org/abs/0809.1899} {arXiv:0809.1899 [astro-ph]} \BibitemShut
  {NoStop}%
\bibitem [{\citenamefont {Kahniashvili}\ \emph {et~al.}(2010)\citenamefont
  {Kahniashvili}, \citenamefont {Kisslinger},\ and\ \citenamefont
  {Stevens}}]{Kahniashvili:2009mf}%
  \BibitemOpen
  \bibfield  {author} {\bibinfo {author} {\bibfnamefont {T.}~\bibnamefont
  {Kahniashvili}}, \bibinfo {author} {\bibfnamefont {L.}~\bibnamefont
  {Kisslinger}}, \ and\ \bibinfo {author} {\bibfnamefont {T.}~\bibnamefont
  {Stevens}},\ }\href {\doibase 10.1103/PhysRevD.81.023004} {\bibfield
  {journal} {\bibinfo  {journal} {Phys. Rev. D}\ }\textbf {\bibinfo {volume}
  {81}},\ \bibinfo {pages} {023004} (\bibinfo {year} {2010})},\ \Eprint
  {http://arxiv.org/abs/0905.0643} {arXiv:0905.0643 [astro-ph.CO]} \BibitemShut
  {NoStop}%
\bibitem [{\citenamefont {Caprini}\ \emph {et~al.}(2009)\citenamefont
  {Caprini}, \citenamefont {Durrer},\ and\ \citenamefont
  {Servant}}]{Caprini:2009yp}%
  \BibitemOpen
  \bibfield  {author} {\bibinfo {author} {\bibfnamefont {C.}~\bibnamefont
  {Caprini}}, \bibinfo {author} {\bibfnamefont {R.}~\bibnamefont {Durrer}}, \
  and\ \bibinfo {author} {\bibfnamefont {G.}~\bibnamefont {Servant}},\ }\href
  {\doibase 10.1088/1475-7516/2009/12/024} {\bibfield  {journal} {\bibinfo
  {journal} {JCAP}\ }\textbf {\bibinfo {volume} {12}},\ \bibinfo {pages} {024}
  (\bibinfo {year} {2009})},\ \Eprint {http://arxiv.org/abs/0909.0622}
  {arXiv:0909.0622 [astro-ph.CO]} \BibitemShut {NoStop}%
\bibitem [{\citenamefont {Hindmarsh}\ \emph {et~al.}(2021)\citenamefont
  {Hindmarsh}, \citenamefont {L\"uben}, \citenamefont {Lumma},\ and\
  \citenamefont {Pauly}}]{Hindmarsh:2020hop}%
  \BibitemOpen
  \bibfield  {author} {\bibinfo {author} {\bibfnamefont {M.~B.}\ \bibnamefont
  {Hindmarsh}}, \bibinfo {author} {\bibfnamefont {M.}~\bibnamefont {L\"uben}},
  \bibinfo {author} {\bibfnamefont {J.}~\bibnamefont {Lumma}}, \ and\ \bibinfo
  {author} {\bibfnamefont {M.}~\bibnamefont {Pauly}},\ }\href {\doibase
  10.21468/SciPostPhysLectNotes.24} {\bibfield  {journal} {\bibinfo  {journal}
  {SciPost Phys. Lect. Notes}\ }\textbf {\bibinfo {volume} {24}},\ \bibinfo
  {pages} {1} (\bibinfo {year} {2021})},\ \Eprint
  {http://arxiv.org/abs/2008.09136} {arXiv:2008.09136 [astro-ph.CO]}
  \BibitemShut {NoStop}%
\bibitem [{\citenamefont {Athron}\ \emph {et~al.}(2024)\citenamefont {Athron},
  \citenamefont {Bal\'azs}, \citenamefont {Fowlie}, \citenamefont {Morris},\
  and\ \citenamefont {Wu}}]{Athron:2023xlk}%
  \BibitemOpen
  \bibfield  {author} {\bibinfo {author} {\bibfnamefont {P.}~\bibnamefont
  {Athron}}, \bibinfo {author} {\bibfnamefont {C.}~\bibnamefont {Bal\'azs}},
  \bibinfo {author} {\bibfnamefont {A.}~\bibnamefont {Fowlie}}, \bibinfo
  {author} {\bibfnamefont {L.}~\bibnamefont {Morris}}, \ and\ \bibinfo {author}
  {\bibfnamefont {L.}~\bibnamefont {Wu}},\ }\href {\doibase
  10.1016/j.ppnp.2023.104094} {\bibfield  {journal} {\bibinfo  {journal} {Prog.
  Part. Nucl. Phys.}\ }\textbf {\bibinfo {volume} {135}},\ \bibinfo {pages}
  {104094} (\bibinfo {year} {2024})},\ \Eprint
  {http://arxiv.org/abs/2305.02357} {arXiv:2305.02357 [hep-ph]} \BibitemShut
  {NoStop}%
\bibitem [{\citenamefont {Wainwright}(2012)}]{Wainwright:2011kj}%
  \BibitemOpen
  \bibfield  {author} {\bibinfo {author} {\bibfnamefont {C.~L.}\ \bibnamefont
  {Wainwright}},\ }\href {\doibase 10.1016/j.cpc.2012.04.004} {\bibfield
  {journal} {\bibinfo  {journal} {Comput. Phys. Commun.}\ }\textbf {\bibinfo
  {volume} {183}},\ \bibinfo {pages} {2006} (\bibinfo {year} {2012})},\ \Eprint
  {http://arxiv.org/abs/1109.4189} {arXiv:1109.4189 [hep-ph]} \BibitemShut
  {NoStop}%
\bibitem [{\citenamefont {Caprini}\ \emph {et~al.}(2020)\citenamefont {Caprini}
  \emph {et~al.}}]{Caprini:2019egz}%
  \BibitemOpen
  \bibfield  {author} {\bibinfo {author} {\bibfnamefont {C.}~\bibnamefont
  {Caprini}} \emph {et~al.},\ }\href {\doibase 10.1088/1475-7516/2020/03/024}
  {\bibfield  {journal} {\bibinfo  {journal} {JCAP}\ }\textbf {\bibinfo
  {volume} {03}},\ \bibinfo {pages} {024} (\bibinfo {year} {2020})},\ \Eprint
  {http://arxiv.org/abs/1910.13125} {arXiv:1910.13125 [astro-ph.CO]}
  \BibitemShut {NoStop}%
\bibitem [{\citenamefont {Durnford}\ and\ \citenamefont
  {Piro}(2021)}]{Durnford:2021mzg}%
  \BibitemOpen
  \bibfield  {author} {\bibinfo {author} {\bibfnamefont {D.}~\bibnamefont
  {Durnford}}\ and\ \bibinfo {author} {\bibfnamefont {M.-C.}\ \bibnamefont
  {Piro}} (\bibinfo {collaboration} {NEWS-G}),\ }\href {\doibase
  10.1088/1742-6596/2156/1/012059} {\bibfield  {journal} {\bibinfo  {journal}
  {J. Phys. Conf. Ser.}\ }\textbf {\bibinfo {volume} {2156}},\ \bibinfo {pages}
  {012059} (\bibinfo {year} {2021})},\ \Eprint
  {http://arxiv.org/abs/2111.02796} {arXiv:2111.02796 [physics.ins-det]}
  \BibitemShut {NoStop}%
\bibitem [{\citenamefont {Agnese}\ \emph {et~al.}(2017)\citenamefont {Agnese}
  \emph {et~al.}}]{Agnese:2016cpb}%
  \BibitemOpen
  \bibfield  {author} {\bibinfo {author} {\bibfnamefont {R.}~\bibnamefont
  {Agnese}} \emph {et~al.} (\bibinfo {collaboration} {SuperCDMS}),\ }\href
  {\doibase 10.1103/PhysRevD.95.082002} {\bibfield  {journal} {\bibinfo
  {journal} {Phys. Rev. D}\ }\textbf {\bibinfo {volume} {95}},\ \bibinfo
  {pages} {082002} (\bibinfo {year} {2017})},\ \Eprint
  {http://arxiv.org/abs/1610.00006} {arXiv:1610.00006 [physics.ins-det]}
  \BibitemShut {NoStop}%
\bibitem [{\citenamefont {Ma}\ \emph {et~al.}(2020)\citenamefont {Ma} \emph
  {et~al.}}]{Ma:2017nhc}%
  \BibitemOpen
  \bibfield  {author} {\bibinfo {author} {\bibfnamefont {H.}~\bibnamefont {Ma}}
  \emph {et~al.} (\bibinfo {collaboration} {CDEX}),\ }\href {\doibase
  10.1088/1742-6596/1342/1/012067} {\bibfield  {journal} {\bibinfo  {journal}
  {J. Phys. Conf. Ser.}\ }\textbf {\bibinfo {volume} {1342}},\ \bibinfo {pages}
  {012067} (\bibinfo {year} {2020})},\ \Eprint
  {http://arxiv.org/abs/1712.06046} {arXiv:1712.06046 [hep-ex]} \BibitemShut
  {NoStop}%
\bibitem [{\citenamefont {Hertel}\ \emph {et~al.}(2019)\citenamefont {Hertel},
  \citenamefont {Biekert}, \citenamefont {Lin}, \citenamefont {Velan},\ and\
  \citenamefont {McKinsey}}]{Hertel:2018aal}%
  \BibitemOpen
  \bibfield  {author} {\bibinfo {author} {\bibfnamefont {S.~A.}\ \bibnamefont
  {Hertel}}, \bibinfo {author} {\bibfnamefont {A.}~\bibnamefont {Biekert}},
  \bibinfo {author} {\bibfnamefont {J.}~\bibnamefont {Lin}}, \bibinfo {author}
  {\bibfnamefont {V.}~\bibnamefont {Velan}}, \ and\ \bibinfo {author}
  {\bibfnamefont {D.~N.}\ \bibnamefont {McKinsey}},\ }\href {\doibase
  10.1103/PhysRevD.100.092007} {\bibfield  {journal} {\bibinfo  {journal}
  {Phys. Rev. D}\ }\textbf {\bibinfo {volume} {100}},\ \bibinfo {pages}
  {092007} (\bibinfo {year} {2019})},\ \Eprint
  {http://arxiv.org/abs/1810.06283} {arXiv:1810.06283 [physics.ins-det]}
  \BibitemShut {NoStop}%
\bibitem [{\citenamefont {Wang}\ \emph
  {et~al.}(2023{\natexlab{b}})\citenamefont {Wang}, \citenamefont {Lei},
  \citenamefont {Ju}, \citenamefont {Liu}, \citenamefont {Zhou}, \citenamefont
  {Chen}, \citenamefont {Wang}, \citenamefont {Cui}, \citenamefont {Meng},\
  and\ \citenamefont {Zhao}}]{Wang:2023wrr}%
  \BibitemOpen
  \bibfield  {author} {\bibinfo {author} {\bibfnamefont {X.}~\bibnamefont
  {Wang}}, \bibinfo {author} {\bibfnamefont {Z.}~\bibnamefont {Lei}}, \bibinfo
  {author} {\bibfnamefont {Y.}~\bibnamefont {Ju}}, \bibinfo {author}
  {\bibfnamefont {J.}~\bibnamefont {Liu}}, \bibinfo {author} {\bibfnamefont
  {N.}~\bibnamefont {Zhou}}, \bibinfo {author} {\bibfnamefont {Y.}~\bibnamefont
  {Chen}}, \bibinfo {author} {\bibfnamefont {Z.}~\bibnamefont {Wang}}, \bibinfo
  {author} {\bibfnamefont {X.}~\bibnamefont {Cui}}, \bibinfo {author}
  {\bibfnamefont {Y.}~\bibnamefont {Meng}}, \ and\ \bibinfo {author}
  {\bibfnamefont {L.}~\bibnamefont {Zhao}},\ }\href {\doibase
  10.1088/1748-0221/18/05/P05028} {\bibfield  {journal} {\bibinfo  {journal}
  {JINST}\ }\textbf {\bibinfo {volume} {18}},\ \bibinfo {pages} {P05028}
  (\bibinfo {year} {2023}{\natexlab{b}})},\ \Eprint
  {http://arxiv.org/abs/2301.06044} {arXiv:2301.06044 [physics.ins-det]}
  \BibitemShut {NoStop}%
\bibitem [{\citenamefont {Aprile}\ \emph {et~al.}(2020)\citenamefont {Aprile}
  \emph {et~al.}}]{XENON:2020kmp}%
  \BibitemOpen
  \bibfield  {author} {\bibinfo {author} {\bibfnamefont {E.}~\bibnamefont
  {Aprile}} \emph {et~al.} (\bibinfo {collaboration} {XENON}),\ }\href
  {\doibase 10.1088/1475-7516/2020/11/031} {\bibfield  {journal} {\bibinfo
  {journal} {JCAP}\ }\textbf {\bibinfo {volume} {11}},\ \bibinfo {pages} {031}
  (\bibinfo {year} {2020})},\ \Eprint {http://arxiv.org/abs/2007.08796}
  {arXiv:2007.08796 [physics.ins-det]} \BibitemShut {NoStop}%
\bibitem [{\citenamefont {Aalseth}\ \emph {et~al.}(2018)\citenamefont {Aalseth}
  \emph {et~al.}}]{DarkSide-20k:2017zyg}%
  \BibitemOpen
  \bibfield  {author} {\bibinfo {author} {\bibfnamefont {C.~E.}\ \bibnamefont
  {Aalseth}} \emph {et~al.} (\bibinfo {collaboration} {DarkSide-20k}),\ }\href
  {\doibase 10.1140/epjp/i2018-11973-4} {\bibfield  {journal} {\bibinfo
  {journal} {Eur. Phys. J. Plus}\ }\textbf {\bibinfo {volume} {133}},\ \bibinfo
  {pages} {131} (\bibinfo {year} {2018})},\ \Eprint
  {http://arxiv.org/abs/1707.08145} {arXiv:1707.08145 [physics.ins-det]}
  \BibitemShut {NoStop}%
\bibitem [{\citenamefont {Aalbers}\ \emph {et~al.}(2016)\citenamefont {Aalbers}
  \emph {et~al.}}]{DARWIN:2016hyl}%
  \BibitemOpen
  \bibfield  {author} {\bibinfo {author} {\bibfnamefont {J.}~\bibnamefont
  {Aalbers}} \emph {et~al.} (\bibinfo {collaboration} {DARWIN}),\ }\href
  {\doibase 10.1088/1475-7516/2016/11/017} {\bibfield  {journal} {\bibinfo
  {journal} {JCAP}\ }\textbf {\bibinfo {volume} {11}},\ \bibinfo {pages} {017}
  (\bibinfo {year} {2016})},\ \Eprint {http://arxiv.org/abs/1606.07001}
  {arXiv:1606.07001 [astro-ph.IM]} \BibitemShut {NoStop}%
\bibitem [{\citenamefont {Essig}\ \emph {et~al.}(2022)\citenamefont {Essig},
  \citenamefont {Giovanetti}, \citenamefont {Kurinsky}, \citenamefont
  {McKinsey}, \citenamefont {Ramanathan}, \citenamefont {Stifter},\ and\
  \citenamefont {Yu}}]{Essig:2022dfa}%
  \BibitemOpen
  \bibfield  {author} {\bibinfo {author} {\bibfnamefont {R.}~\bibnamefont
  {Essig}}, \bibinfo {author} {\bibfnamefont {G.~K.}\ \bibnamefont
  {Giovanetti}}, \bibinfo {author} {\bibfnamefont {N.}~\bibnamefont
  {Kurinsky}}, \bibinfo {author} {\bibfnamefont {D.}~\bibnamefont {McKinsey}},
  \bibinfo {author} {\bibfnamefont {K.}~\bibnamefont {Ramanathan}}, \bibinfo
  {author} {\bibfnamefont {K.}~\bibnamefont {Stifter}}, \ and\ \bibinfo
  {author} {\bibfnamefont {T.-T.}\ \bibnamefont {Yu}},\ }in\ \href@noop {}
  {\emph {\bibinfo {booktitle} {{2022 Snowmass Summer Study}}}}\ (\bibinfo
  {year} {2022})\ \Eprint {http://arxiv.org/abs/2203.08297} {arXiv:2203.08297
  [hep-ph]} \BibitemShut {NoStop}%
\bibitem [{\citenamefont {Quiros}(1999)}]{Quiros:1999jp}%
  \BibitemOpen
  \bibfield  {author} {\bibinfo {author} {\bibfnamefont {M.}~\bibnamefont
  {Quiros}},\ }in\ \href@noop {} {\emph {\bibinfo {booktitle} {{ICTP Summer
  School in High-Energy Physics and Cosmology}}}}\ (\bibinfo {year} {1999})\
  pp.\ \bibinfo {pages} {187--259},\ \Eprint
  {http://arxiv.org/abs/hep-ph/9901312} {arXiv:hep-ph/9901312} \BibitemShut
  {NoStop}%
\bibitem [{\citenamefont {Anderson}\ and\ \citenamefont
  {Hall}(1992)}]{Anderson:1991zb}%
  \BibitemOpen
  \bibfield  {author} {\bibinfo {author} {\bibfnamefont {G.~W.}\ \bibnamefont
  {Anderson}}\ and\ \bibinfo {author} {\bibfnamefont {L.~J.}\ \bibnamefont
  {Hall}},\ }\href {\doibase 10.1103/PhysRevD.45.2685} {\bibfield  {journal}
  {\bibinfo  {journal} {Phys. Rev. D}\ }\textbf {\bibinfo {volume} {45}},\
  \bibinfo {pages} {2685} (\bibinfo {year} {1992})}\BibitemShut {NoStop}%
\bibitem [{\citenamefont {Espinosa}(1996)}]{Espinosa:1996qw}%
  \BibitemOpen
  \bibfield  {author} {\bibinfo {author} {\bibfnamefont {J.~R.}\ \bibnamefont
  {Espinosa}},\ }\href {\doibase 10.1016/0550-3213(96)00297-0} {\bibfield
  {journal} {\bibinfo  {journal} {Nucl. Phys. B}\ }\textbf {\bibinfo {volume}
  {475}},\ \bibinfo {pages} {273} (\bibinfo {year} {1996})},\ \Eprint
  {http://arxiv.org/abs/hep-ph/9604320} {arXiv:hep-ph/9604320} \BibitemShut
  {NoStop}%
\bibitem [{\citenamefont {Kastening}(1992)}]{Kastening:1991gv}%
  \BibitemOpen
  \bibfield  {author} {\bibinfo {author} {\bibfnamefont {B.~M.}\ \bibnamefont
  {Kastening}},\ }\href {\doibase 10.1016/0370-2693(92)90021-U} {\bibfield
  {journal} {\bibinfo  {journal} {Phys. Lett. B}\ }\textbf {\bibinfo {volume}
  {283}},\ \bibinfo {pages} {287} (\bibinfo {year} {1992})}\BibitemShut
  {NoStop}%
\bibitem [{\citenamefont {Bando}\ \emph {et~al.}(1993)\citenamefont {Bando},
  \citenamefont {Kugo}, \citenamefont {Maekawa},\ and\ \citenamefont
  {Nakano}}]{Bando:1992np}%
  \BibitemOpen
  \bibfield  {author} {\bibinfo {author} {\bibfnamefont {M.}~\bibnamefont
  {Bando}}, \bibinfo {author} {\bibfnamefont {T.}~\bibnamefont {Kugo}},
  \bibinfo {author} {\bibfnamefont {N.}~\bibnamefont {Maekawa}}, \ and\
  \bibinfo {author} {\bibfnamefont {H.}~\bibnamefont {Nakano}},\ }\href
  {\doibase 10.1016/0370-2693(93)90725-W} {\bibfield  {journal} {\bibinfo
  {journal} {Phys. Lett. B}\ }\textbf {\bibinfo {volume} {301}},\ \bibinfo
  {pages} {83} (\bibinfo {year} {1993})},\ \Eprint
  {http://arxiv.org/abs/hep-ph/9210228} {arXiv:hep-ph/9210228} \BibitemShut
  {NoStop}%
\bibitem [{\citenamefont {Ford}\ \emph {et~al.}(1993)\citenamefont {Ford},
  \citenamefont {Jones}, \citenamefont {Stephenson},\ and\ \citenamefont
  {Einhorn}}]{Ford:1992mv}%
  \BibitemOpen
  \bibfield  {author} {\bibinfo {author} {\bibfnamefont {C.}~\bibnamefont
  {Ford}}, \bibinfo {author} {\bibfnamefont {D.~R.~T.}\ \bibnamefont {Jones}},
  \bibinfo {author} {\bibfnamefont {P.~W.}\ \bibnamefont {Stephenson}}, \ and\
  \bibinfo {author} {\bibfnamefont {M.~B.}\ \bibnamefont {Einhorn}},\ }\href
  {\doibase 10.1016/0550-3213(93)90206-5} {\bibfield  {journal} {\bibinfo
  {journal} {Nucl. Phys. B}\ }\textbf {\bibinfo {volume} {395}},\ \bibinfo
  {pages} {17} (\bibinfo {year} {1993})},\ \Eprint
  {http://arxiv.org/abs/hep-lat/9210033} {arXiv:hep-lat/9210033} \BibitemShut
  {NoStop}%
\bibitem [{\citenamefont {Funakubo}\ and\ \citenamefont
  {Senaha}(2024{\natexlab{a}})}]{Funakubo:2023cyv}%
  \BibitemOpen
  \bibfield  {author} {\bibinfo {author} {\bibfnamefont {K.}~\bibnamefont
  {Funakubo}}\ and\ \bibinfo {author} {\bibfnamefont {E.}~\bibnamefont
  {Senaha}},\ }\href {\doibase 10.1103/PhysRevD.109.L071901} {\bibfield
  {journal} {\bibinfo  {journal} {Phys. Rev. D}\ }\textbf {\bibinfo {volume}
  {109}},\ \bibinfo {pages} {L071901} (\bibinfo {year} {2024}{\natexlab{a}})},\
  \Eprint {http://arxiv.org/abs/2307.02153} {arXiv:2307.02153 [hep-ph]}
  \BibitemShut {NoStop}%
\bibitem [{\citenamefont {Funakubo}\ and\ \citenamefont
  {Senaha}(2024{\natexlab{b}})}]{Funakubo:2023eic}%
  \BibitemOpen
  \bibfield  {author} {\bibinfo {author} {\bibfnamefont {K.}~\bibnamefont
  {Funakubo}}\ and\ \bibinfo {author} {\bibfnamefont {E.}~\bibnamefont
  {Senaha}},\ }\href {\doibase 10.1103/PhysRevD.109.056023} {\bibfield
  {journal} {\bibinfo  {journal} {Phys. Rev. D}\ }\textbf {\bibinfo {volume}
  {109}},\ \bibinfo {pages} {056023} (\bibinfo {year} {2024}{\natexlab{b}})},\
  \Eprint {http://arxiv.org/abs/2308.15876} {arXiv:2308.15876 [hep-ph]}
  \BibitemShut {NoStop}%
\bibitem [{\citenamefont {Staub}(2014)}]{Staub:2013tta}%
  \BibitemOpen
  \bibfield  {author} {\bibinfo {author} {\bibfnamefont {F.}~\bibnamefont
  {Staub}},\ }\href {\doibase 10.1016/j.cpc.2014.02.018} {\bibfield  {journal}
  {\bibinfo  {journal} {Comput. Phys. Commun.}\ }\textbf {\bibinfo {volume}
  {185}},\ \bibinfo {pages} {1773} (\bibinfo {year} {2014})},\ \Eprint
  {http://arxiv.org/abs/1309.7223} {arXiv:1309.7223 [hep-ph]} \BibitemShut
  {NoStop}%
\end{thebibliography}%


\end{document}